\documentclass{article}
\usepackage[totalwidth=480pt, totalheight=680pt]{geometry}

\usepackage[OT1]{RRA4}
\RRNo{6013}
\usepackage{hyperref}

\usepackage{stmaryrd}
\usepackage{bbm}
\usepackage{ifthen}

\usepackage{amssymb}
\usepackage{url}
\usepackage{amsmath}
\usepackage{latexsym}
\usepackage{paralist}
\usepackage{calrsfs}

\usepackage[curve,graph,matrix,poly]{xy}
\usepackage{derivtree}
\usepackage[three]{lutztheorem}

\usepackage{graphicx,pstricks,pst-node,pstricks-add}

\usepackage{lutzproofnetmacros}

\def\fbm#1{#1}

\newdimen\diagcol
\newdimen\diagrow

\def\diagsizenormal{
  \diagcol=4em
  \diagrow=5ex
}

\newcommand{\ndiagtriangleup}[9]{%
  \xymatrix@C=.5\diagcol@R=\diagrow{
    &#4 \ar@{#5}[rd]^{#6}\\
    #1 \ar@{#2}[ru]^{#3} \ar@{#7}[rr]_{#8} 
    & & #9
  }
}

\newcommand{\ndiagtriangledown}[9]{%
  \xymatrix@C=.5\diagcol@R=\diagrow{
    #1 \ar@{#2}[rd]_{#3} \ar@{#7}[rr]^{#8} 
    & & #9\\
    &#4 \ar@{#5}[ru]_{#6}
  }
}

\newcommand{\ndiagtriangleright}[9]{%
  \xymatrix@C=\diagcol@R=.5\diagrow{
    #1 \ar@{#2}[rd]^{#3} \ar@{#7}[dd]_{#8}\\ 
    & #4 \ar@{#5}[ld]^{#6}\\
    #9
  }
}

\newcommand{\ndiagtriangleleft}[9]{%
  \xymatrix@C=\diagcol@R=.5\diagrow{
    &#1 \ar@{#2}[ld]_{#3} \ar@{#7}[dd]^{#8}\\ 
    #4 \ar@{#5}[rd]_{#6}\\
    &#9
  }
}

\newcommand{\ndiagsquare}[8]{%
  \xymatrix@C=\diagcol@R=\diagrow{
    #1 \ar[r]^{#2} \ar[d]_{#5} 
    & #3 \ar[d]^{#4}\\
    #6 \ar[r]_{#7} 
    & #8
  }
}

\newcommand{\arbto}[3][r]{\fbm{#2}\ar[#1]^-{#3}}
\newcommand{\arpto}[3][r]{\fbm{#2}\ar[#1]_-{#3}}
\newcommand{\arbfrom}[3][r]{\fbm{#3}\ar@{<-}[#1]^-{#2}}
\newcommand{\arpfrom}[3][r]{\fbm{#3}\ar@{<-}[#1]_-{#2}}

\newcommand{\arbtofrom}[3][r]{#2\ar@{<->}[#1]^-{#3}}
\newcommand{\arptofrom}[3][r]{#2\ar@{<->}[#1]_-{#3}}
\newcommand{\arbfromto}[3][r]{#3\ar@{<->}[#1]^-{#2}}
\newcommand{\arpfromto}[3][r]{#3\ar@{<->}[#1]_-{#2}}

\newcommand{\arbtoid}[3][r]{\fbm{#2}\ar@{=}[#1]^-{#3}}
\newcommand{\arptoid}[3][r]{\fbm{#2}\ar@{=}[#1]_-{#3}}
\newcommand{\arbfromid}[3][r]{\fbm{#3}\ar@{=}[#1]^-{#2}}
\newcommand{\arpfromid}[3][r]{\fbm{#3}\ar@{=}[#1]_-{#2}}

\newcommand{\labupleft}[3]{\strut\smash{\raise#1pt\hbox{%
      $\scriptstyle#3\hskip#2ex$}}}
\newcommand{\labdownleft}[3]{\strut\smash{\lower#1pt\hbox{%
      $\scriptstyle#3\hskip#2ex$}}}
\newcommand{\labupright}[3]{\strut\smash{\raise#1pt\hbox{%
      $\scriptstyle\hskip#2ex#3$}}}
\newcommand{\labdownright}[3]{\strut\smash{\lower#1pt\hbox{%
      $\scriptstyle\hskip#2ex#3$}}}

\newcommand{\diagram}[1]{\xymatrix@C=\diagcol@R=\diagrow{#1}}
\newcommand{\diagramhc}[1]{\xymatrix@C=.5\diagcol@R=\diagrow{#1}}
\newcommand{\diagramhhc}[1]{\xymatrix@C=.25\diagcol@R=\diagrow{#1}}
\newcommand{\diagrammhc}[1]{\xymatrix@C=-.5\diagcol@R=\diagrow{#1}}
\newcommand{\diagrammtqc}[1]{\xymatrix@C=-.65\diagcol@R=\diagrow{#1}}
\newcommand{\diagramdc}[1]{\xymatrix@C=2\diagcol@R=\diagrow{#1}}
\newcommand{\diagramhr}[1]{\xymatrix@C=\diagcol@R=.5\diagrow{#1}}
\newcommand{\diagramdr}[1]{\xymatrix@C=\diagcol@R=2\diagrow{#1}}
\newcommand{\diagramhchr}[1]{\xymatrix@C=.5\diagcol@R=.5\diagrow{#1}}
\newcommand{\diagramzchr}[1]{\xymatrix@C=0.0\diagcol@R=.5\diagrow{#1}}
\newcommand{\diagrammqchr}[1]{\xymatrix@C=-.25\diagcol@R=.5\diagrow{#1}}
\newcommand{\diagramdchr}[1]{\xymatrix@C=2\diagcol@R=.5\diagrow{#1}}
\newcommand{\diagramdcdr}[1]{\xymatrix@C=2\diagcol@R=2\diagrow{#1}}
\newcommand{\diagramscsr}[1]{\xymatrix@C=1.5\diagcol@R=1.5\diagrow{#1}}

\newcommand{\diagtriangleup}[3]{%
  \xymatrix@C=.5\diagcol@R=\diagrow{
    &\arpto[dl]#1 \\
    \arpto[rr]#2 && \arpfrom[ul]#3 
    }
  }

\newcommand{\diagtriangledown}[3]{%
  \xymatrix@C=.5\diagcol@R=\diagrow{
    \arbto[rr]#1 &&\arbto[dl]#2\\
    &\arbfrom[ul]#3 
  }
}

\newcommand{\diagtriangleright}[3]{%
  \xymatrix@C=.5\diagcol@R=.5\diagrow{
    \arpto[dd]#1 \\
    &\arpfrom[ul]#3 \\
    \arpto[ur]#2
  }
}

\newcommand{\diagtriangleleft}[3]{%
  \xymatrix@C=.5\diagcol@R=.5\diagrow{
    &\arbto[dd]#2 \\
    \arbto[ur]#1 \\
    &\arbfrom[ul]#3
  }
}

\newcommand{\diagsquare}[4]{%
  \xymatrix@C=\diagcol@R=\diagrow{
    \arbto[r]#1   & \arbto[d]#2\\ 
    \arbfrom[u]#3 & \arbfrom[l]#4
  }
}

\newcommand{\diagsquaredown}[4]{%
  \xymatrix@C=\diagcol@R=\diagrow{
    \arpto[d]#1   & \arpfrom[l]#3\\ 
    \arpto[r]#2 & \arpfrom[u]#4
  }
}

\newcommand{\ddiagsquare}[5]{%
  \xymatrix@C=.7\diagcol@R=.7\diagrow{
    \arbto[rr]#1   && \arbto[dd]#2\\ 
    #3\\
    \arbfrom[uu]#4 && \arbfrom[ll]#5
  }
}

\newcommand{\diagpentagondown}[5]{%
  \xymatrix@C=.5\diagcol@R=\diagrow{
    \arbto[rr]#1   && \arbto[d]#2\\ 
    \arbfrom[u]#4 && \arbto[dl]#3\\
    &\arbfrom[ul]#5
  }
}

\newcommand{\diagpentagonright}[5]{%
  \xymatrix@C=.5\diagcol@R=.5\diagrow{
    \arpto[dd]#1   && \arpfrom[ll]#4 \\
    &&&\arpfrom[ul]#5 \\
    \arpto[rr]#2  && \arpto[ur]#3
  }
}

\newcommand{\diagpentagonleft}[5]{%
  \xymatrix@C=.5\diagcol@R=.5\diagrow{
    &\arbto[rr]#2   && \arbto[dd]#3 \\
    \arbto[ur]#1 \\
    &\arbfrom[ul]#4  && \arbfrom[ll]#5
  }
}

\newcommand{\diaghexagondown}[6]{%
  \xymatrix@C=\diagcol@R=\diagrow{
    \arpto[d]#1   & \arpfrom[l]#4\\
    \arpto[d]#2 & \arpfrom[u]#5\\
    \arpto[r]#3 & \arpfrom[u]#6
  }
}

\newcommand{\diaghexagonhexa}[6]{%
  \xymatrix@C=\diagcol@R=\diagrow{
    & \arpto[dl]#1 \\
    \arpto[d]#2 && \arpfrom[ul]#4\\
    \arpto[dr]#3 && \arpfrom[u]#5\\
     & \arpfrom[ur]#6
  }
}

\newcommand{\diaghexagondowndr}[6]{%
  \xymatrix@C=\diagcol@R=2\diagrow{
    \arpto[d]#1   & \arpfrom[l]#4\\
    \arpto[d]#2 & \arpfrom[u]#5\\
    \arpto[r]#3 & \arpfrom[u]#6
  }
}

\newcommand{\diaghexagonright}[6]{%
  \xymatrix@C=\diagcol@R=\diagrow{
    \arbto[r]#1   & \arbto[r]#2 & \arbto[d]#3\\
    \arbfrom[u]#4 & \arbfrom[l]#5 & \arbfrom[l]#6
  }
}

\newcommand{\diagoctagon}[8]{%
  \xymatrix@C=\diagcol@R=\diagrow{
    \arbto[r]#1   & \arbto[r]#2 & \arbto[d]#3\\
    \arbfrom[u]#5 && \arbto[d]#4\\
    \arbfrom[u]#6 & \arbfrom[l]#7 & \arbfrom[l]#8
  }
}

\newcommand{\diagoctagonslim}[8]{%
  \xymatrix@C=\diagcol@R=\diagrow{
    \arbto[r]#1   & \arbto[d]#2 \\
    \arbfrom[u]#5 & \arbto[d]#3 \\
    \arbfrom[u]#6 & \arbto[d]#4 \\
    \arbfrom[u]#7 & \arbfrom[l]#8
  }
}

\newcommand{\diagdecagonhexa}[9]{%
  \xymatrix@C=\diagcol@R=\diagrow{
    & \arbto[dr]#1 \\
    \arbfrom[ur]#6 && \arbto[d]#2 \\
    \arbfrom[u]#7 && \arbto[d]#3 \\
    \arbfrom[u]#8 && \arbto[d]#4 \\
    \arbfrom[u]#9 && \arbto[dl]#5 \\
    & \arbfrom[ul]{\lastarga}{\lastargb}
  }
}

\newcommand{\vcdiagtriangledown}[3]{\vcenter{\diagtriangledown{#1}{#2}{#3}}}%

\newcommand{\vcdiagsquaredown}[4]{%
  \vcenter{\diagsquaredown{#1}{#2}{#3}{#4}}}%

\newcommand{\vcdiagpentagondown}[5]{%
  \vcenter{\diagpentagondown{#1}{#2}{#3}{#4}{#5}}}%

\newcommand{\vcdiaghexagondown}[6]{%
  \vcenter{\diaghexagondown{#1}{#2}{#3}{#4}{#5}{#6}}}%

\diagsizenormal

\def\Pit{\tilde{\Pi}}

\def\cA{{\mathcal A}}

\def\cC{{\mathcal C}}

\def\cF{{\mathcal F}}

\def\SKS{\mathsf{SKS}}

\def\MLL{\mathsf{MLL}}
\def\MALL{\mathsf{MALL}}

\def\set#1{\{#1\}}

\def\cons#1{\{#1\}}
\def\conhole      {\cons{\enspace}}%

\def\fcomp{\mathbin{\circ}}

\def\grammareq {\mathrel{\raise.4pt\hbox{::}{=}}}%

\def\id{{\mathrm{1}}}

\newcommand{\dotto}[1][]{\mathrel{\!\xy\ar@{.>}^-{#1}(5,0)\endxy\!}}
\newcommand{\solto}[1][]{\mathrel{\!\xy\ar@{->}^-{#1}(5,0)\endxy\!}}
\newcommand{\longsolto}[1][]{\mathrel{\!\xy\ar@{->}^-{#1}(11,0)\endxy\!}}
\newcommand{\longdotto}[1][]{\mathrel{\!\xy\ar@{.>}^-{#1}(11,0)\endxy\!}}
\newcommand{\xldotto}[2][]{\mathrel{\!\xy\ar@{.>}^-{#1}(#2,0)\endxy\!}}

\def\Hom{}
\renewcommand{\Hom}[1][]{\mathrm{Hom}_{#1}}

\def\isom{\cong}

\def\sqn  #1{{\;\turnstile #1\;}}%

\def\cutr{\mathsf{cut}}
\def\idr{\mathsf{id}}
\def\mixr{\mathsf{mix}}
\def\mixzr{\mathsf{mix_0}}
\def\weakr{\mathsf{weak}}
\def\conr{\mathsf{cont}}
\def\exr{\mathsf{exch}}
\def\cyclr{\mathsf{cycl}}

\def\swir{\mathsf{s}}
\def\medr{\mathsf{m}}

\def\MLL{\mathsf{MLL}}

\def\ctrue{{\mathbf t}}
\def\cfalse{{\mathbf f}}
\def\cand{\wedge}
\def\cor{\vee}

\def\cneg#1{\bar{#1}}
\def\widecneg#1{\overline{#1}}
\def\cimp{\mathop{\Rightarrow}}
\def\fneg{\widecneg{(-)}}

\newbox\cutbox
\newdimen\cutwd
\newdimen\cutht
\newdimen\cutdp
\def\ccut{%
  \setbox\cutbox\hbox{$\lozenge$}
  \cutwd=\wd\cutbox
  \cutht=\ht\cutbox
  \cutdp=\dp\cutbox
  \setbox\cutbox\hbox to\cutwd{\hss\vrule width.3pt height\cutht depth\cutdp\hss}
  \mathbin{\lozenge\hskip-\cutwd\copy\cutbox}}
\def\scriptcut{%
  \setbox\cutbox\hbox{$\scriptstyle\lozenge$}
  \cutwd=\wd\cutbox
  \cutht=\ht\cutbox
  \cutdp=\dp\cutbox
  \setbox\cutbox\hbox to\cutwd{\hss\vrule width.3pt height\cutht depth\cutdp\hss}
  \mathord{\lozenge\hskip-\cutwd\copy\cutbox}}

\def\vccut{%
  \setbox\cutbox\hbox{$\lozenge$}
  \cutwd=\wd\cutbox
  \cutht=\ht\cutbox
  \cutdp=\dp\cutbox
  \setbox\cutbox\hbox to\cutwd{\hss\hskip.3pt\vrule width.3pt height\cutht depth\cutdp\hss}
  \mathbin{\lozenge\hskip-\cutwd\copy\cutbox}}

\let\turnstile=\vdash
\def\lone{1}
\def\lbot{\bot}
\def\lzero{0}
\def\ltop{\top}
\def\ltens{\mathop\varotimes}
\def\lcut{\mathop\varobar}
\def\lpar{\mathop\bindnasrepma}
\def\lplus{\mathop\varoplus}
\def\lwith{\mathop\binampersand}
\def\lwn{\mathord{?}}
\def\loc{\mathord{!}}
\def\lneg{^\bot}
\def\lnegneg{^{\bot\bot}}
\def\limp{\multimap}
\def\lseq{\mathop\vartriangleleft}
\def\lcoseq{\mathop\vartriangleright}

\def\lweakr{\mathord{\lwn}\mathsf{w}}
\def\lconr{\mathord{\lwn}\mathsf{c}}
\def\lderr{\mathord{\lwn}\mathsf{d}}

\def\quadfs {\rlap{\rm\quad.}}%
\def\quadcm {\rlap{\rm\quad,}}%
\def\quand {\quad\mbox{and}\quad}%
\def\qquand {\qquad\mbox{and}\qquad}%
\def\qqquand {\quad\qquad\mbox{and}\qquad\quad}%
\def\qqqquand {\qquad\qquad\mbox{and}\qquad\qquad}%
\def\qquor {\qquad\mbox{or}\qquad}%
\def\qqquor {\quad\qquad\mbox{or}\qquad\quad}%
\def\qqqquor {\qquad\qquad\mbox{or}\qquad\qquad}%

\def\clap#1{\hbox to 0pt{\hss#1\hss}}
\def\qlap#1{\hbox to 1em{\hss#1\hss}}
\def\qqlap#1{\hbox to 2em{\hss#1\hss}}
\def\qqqlap#1{\hbox to 3em{\hss#1\hss}}
\def\qqqqlap#1{\hbox to 4em{\hss#1\hss}}
\newcommand{\wlap}[2][10ex]{\hbox to#1{\hss#2\hss}}
\def\clapm#1{\clap{$#1$}}
\def\qlapm#1{\qlap{$#1$}}
\def\qqlapm#1{\qqlap{$#1$}}
\def\qqqlapm#1{\qqqlap{$#1$}}

\newcommand{\wlapm}[2][10ex]{\hbox to#1{\hss$#2$\hss}}
\def\rlapm#1{\hbox to 0pt{$#1$\hss}}
\def\llapm#1{\hbox to 0pt{\hss$#1$}}

\def\proofadjust{\vadjust{\nobreak\vskip-2.7ex\nobreak}}

\def\interdisplayskip{.5ex}
\newskip\mydisplaywidth
\newcommand{\twolinedisplay}[3][10pt]{%
  \mydisplaywidth=\displaywidth
  \advance\mydisplaywidth-#1
  \begin{array}{c}
    \clap{\hbox to\mydisplaywidth{$\displaystyle#2$\hss}}\\[\interdisplayskip]
    \clap{\hbox to\mydisplaywidth{\hss$\displaystyle#3$}}
  \end{array}
}

\def\ufMLL{\mathsf{MLL^-}}
\def\ufIMLL{\mathsf{IMLL^-}}
\def\ufMELL{\mathsf{MELL^-}}
\def\ufMALL{\mathsf{MALL^-}}
\def\MLL{\mathsf{MLL}}
\def\MELL{\mathsf{MELL}}
\def\MALL{\mathsf{MALL}}
\def\SBV{\mathsf{SBV}}
\def\BV{\mathsf{BV}}

\def\PN{\mathbf{PN}}

\newcommand{\idf}[1][]{{1_{#1}}}
\renewcommand{\id}[1][]{#1}

\def\atidr{\mathsf{atomic~id}}

\def\repr#1{h^{#1}}
\def\fakeone{\mathbbm{1}}
\newcommand{\ird}{\mathsf{i}\mathord{\downarrow}}
\newcommand{\iru}{\mathsf{i}\mathord{\uparrow}}
\newcommand{\atird}{\mathsf{ai}\mathord{\downarrow}}
\newcommand{\atiru}{\mathsf{ai}\mathord{\uparrow}}
\newcommand{\seqrd}{\mathsf{q}\mathord{\downarrow}}
\newcommand{\seqru}{\mathsf{q}\mathord{\uparrow}}
\newcommand{\conrd}{\mathsf{c}\mathord{\downarrow}}
\newcommand{\conru}{\mathsf{c}\mathord{\uparrow}}
\newcommand{\weakrd}{\mathsf{w}\mathord{\downarrow}}
\newcommand{\weakru}{\mathsf{w}\mathord{\uparrow}}
\renewcommand{\swir}[1][]{\mathsf{s}_{#1}}
\newcommand{\assrd}[1][]{\alpha\mathord{\downarrow}_{#1}}
\newcommand{\assru}[1][]{\alpha\mathord{\uparrow}_{#1}}
\newcommand{\sassrd}[1][]{\alpha^\triangleleft\mathord{\downarrow}_{#1}}
\newcommand{\sassru}[1][]{\alpha^\triangleleft\mathord{\uparrow}_{#1}}
\newcommand{\comrd}[1][]{\sigma\mathord{\downarrow}_{#1}}
\newcommand{\comru}[1][]{\sigma\mathord{\uparrow}_{#1}}
\newcommand{\assoc}[1][]{\alpha_{#1}}
\newcommand{\twist}[1][]{\sigma_{#1}}

\def\ltensl{\mathord{\ltens}\mathsf{L}}
\def\ltensr{\mathord{\ltens}\mathsf{R}}
\def\lparl{\mathord{\lpar}\mathsf{L}}
\def\lparr{\mathord{\lpar}\mathsf{R}}
\def\limpl{\mathord{\limp}\mathsf{L}}
\def\limpr{\mathord{\limp}\mathsf{R}}
\def\lnegl{\mathord{\cdot^\lbot}\mathsf{L}}
\def\lnegr{\mathord{\cdot^\lbot}\mathsf{R}}
\def\exrl{\mathsf{exchL}}
\def\exrr{\mathsf{exchR}}
\def\markl{\mathord{\mathsf{:}}\mathsf{L}}
\def\markr{\mathord{\mathsf{:}}\mathsf{R}}

\def\poll{\bullet}
\def\polr{\circ}
\def\markpl{^\poll}
\def\markpr{^\polr}
\def\lnegpl{^{\lbot\poll}}
\def\lnegpr{^{\lbot\polr}}

\def\impI{\mathord{\to}\mathsf{I}}
\def\impE{\mathord{\to}\mathsf{E}}

\def\absr{\mathsf{abs}}
\def\appr{\mathsf{app}}

\newcommand{\lunit}[1][]{\lambda_{#1}}
\newcommand{\runit}[1][]{\varrho_{#1}}
\newcommand{\hlunit}[1][]{\hat{\lambda}_{#1}}

\newcommand{\ruler}{\mathsf{r}}
\newcommand{\ruled}{\mathsf{r}\mathord{\downarrow}}
\newcommand{\ruleu}{\mathsf{r}\mathord{\uparrow}}

\def\un{\circ}

\def\wlpar{\;\lpar\;}

\newcommand{\rp}[1]{{\rm(}#1{\rm)}}

\def\lone{\mathbf{1}}
\def\arrayskip{4ex}

\def\scderione{%
  \dernote{\lpar}{}{\black
    \sqn{\na{11}\lneg\lpar(\na{21}\ltens \na{31}),
      \na{41}\lpar(\na{51}\lneg\ltens \na{61}\lneg)}}{
    \root{\exr}{
      \sqn{\na{12}\lneg\lpar(\na{22}\ltens \na{32}),
	\na{42},\na{52}\lneg\ltens \na{62}\lneg}}{
      \rroot{\ltens}{
	\sqn{\na{13}\lneg\lpar(\na{23}\ltens \na{33}),
	  \na{53}\lneg\ltens \na{63}\lneg,\na{43}}}{
	\root{\lpar}{\sqn{\na{14}\lneg\lpar(\na{24}\ltens \na{34}),
	    \na{54}\lneg}}{
	  \rroot{\ltens}{\sqn{\na{15}\lneg,\na{25}\ltens \na{35},
	      \na{55}\lneg}}{
	    \root{\idr}{\sqn{\na{16}\lneg,\na{26}}}{
	      \leaf{}}}{
	    \root{\idr}{\sqn{\na{36},\na{56}\lneg}}{
	      \leaf{}}}}}{
	\root{\idr}{\sqn{\na{64}\lneg,\na{44}}}{
	  \leaf{}}}}}%
}

\def\pathscderione{%
  \flowgraphmode
  \longuline{a}{11}{16}
  \longuline{a}{21}{26}
  \longuline{a}{31}{36}
  \longuline{a}{41}{44}
  \longuline{a}{51}{56}
  \longuline{a}{61}{64}
  \udline{a16}{a26}
  \udline{a64}{a44}
  \udline{a36}{a56}
}
  
\def\scderitwo{%
  \dernote{\lpar}{}{\black
    \sqn{\na{11}\lneg\lpar(\na{21}\ltens \na{31}),
      \na{41}\lpar(\na{51}\lneg\ltens \na{61}\lneg)}}{
    \rroot{\ltens}{
      \sqn{\na{12}\lneg,\na{22}\ltens \na{32},
	\na{42}\lpar(\na{52}\lneg\ltens \na{62}\lneg)}}{
      \root{\idr}{\sqn{\na{13}\lneg,\na{23}}}{
	\leaf{}}}{
      \root{\lpar}{\sqn{\na{33},
	  \na{43}\lpar(\na{53}\lneg\ltens \na{63}\lneg)}}{
	\root{\exr}{\sqn{\na{34},\na{44},\na{54}\lneg\ltens \na{64}\lneg}}{
	  \rroot{\ltens}{\sqn{\na{35},\na{55}\lneg\ltens \na{65}\lneg,
	      \na{45}}}{
	    \root{\idr}{\sqn{\na{36},\na{56}\lneg}}{
	      \leaf{}}}{
	    \root{\idr}{\sqn{\na{66}\lneg,\na{46}}}{
	      \leaf{}}}}}}}
}

\def\pathscderitwo{%
  \flowgraphmode
  \longuline{a}{11}{13}
  \longuline{a}{21}{23}
  \longuline{a}{31}{36}
  \longuline{a}{41}{46}
  \longuline{a}{51}{56}
  \longuline{a}{61}{66}
  \udline{a13}{a23}
  \udline{a66}{a46}
  \udline{a36}{a56}
}

\def\scderiplanar{%
  \dernote{\lpar}{}{\black
    \sqn{\na{11}\lneg\lpar(\na{21}\ltens \na{31}),
      \na{41}\lpar(\na{51}\lneg\ltens \na{61}\lneg)}}{
    \rroot{\ltens}{
      \sqn{\na{12}\lneg,\na{22}\ltens \na{32},
	\na{42}\lpar(\na{52}\lneg\ltens \na{62}\lneg)}}{
      \root{\idr}{\sqn{\na{13}\lneg,\na{23}}}{
	\leaf{}}}{
      \root{\lpar}{\sqn{\na{33},
	  \na{43}\lpar(\na{53}\lneg\ltens \na{63}\lneg)}}{
	\root{\cyclr}{\sqn{\na{34},\na{44},\na{54}\lneg\ltens \na{64}\lneg}}{
	  \rroot{\ltens}{\sqn{\na{45},\na{55}\lneg\ltens \na{65}\lneg,
	      \na{35}}}{
	    \root{\idr}{\sqn{\na{46},\na{56}\lneg}}{
	      \leaf{}}}{
	    \root{\idr}{\sqn{\na{66}\lneg,\na{36}}}{
	      \leaf{}}}}}}}
}

\def\pathscderiplanar{%
  \flowgraphmode
  \longuline{a}{11}{13}
  \longuline{a}{21}{23}
  \longuline{a}{31}{36}
  \longuline{a}{41}{46}
  \longuline{a}{51}{56}
  \longuline{a}{61}{66}
  \udline{a13}{a23}
  \udline{a66}{a36}
  \udline{a46}{a56}
}

\def\scderithree{%
  \dernote{\exr}{}{\black
    \sqn{\na{11}\lneg\lpar(\na{21}\ltens \na{31}),
      \na{41}\lpar(\na{51}\lneg\ltens \na{61}\lneg)}}{
    \root{\lpar}{
      \sqn{\na{42}\lpar(\na{52}\lneg\ltens \na{62}\lneg),
	\na{12}\lneg\lpar(\na{22}\ltens \na{32})}}{
      \rroot{\ltens}{
	\sqn{\na{43},\na{53}\lneg\ltens \na{63}\lneg,
	  \na{13}\lneg\lpar(\na{23}\ltens \na{33})}}{
	\root{\idr}{\sqn{\na{44},\na{54}\lneg}}{
	  \leaf{}}}{
	\root{\lpar}{\sqn{\na{64}\lneg,
	    \na{14}\lneg\lpar(\na{24}\ltens \na{34})}}{
	  \root{\exr}{\sqn{\na{65}\lneg,\na{15}\lneg,\na{25}\ltens \na{35}}}{
	    \rroot{\ltens}{\sqn{\na{66}\lneg,\na{26}\ltens \na{36},
		\na{16}\lneg}}{
	      \root{\idr}{\sqn{\na{67}\lneg,\na{27}}}{
		\leaf{}}}{
	      \root{\idr}{\sqn{\na{37},\na{17}\lneg}}{
		\leaf{}}}}}}}}
}

\def\pathscderithree{%
  \flowgraphmode
  \def\vertangleA{30}
  \def\vertangleB{150}
  \urline{a11}{a12}
  \urline{a21}{a22}
  \urline{a31}{a32}
  \ulline{a41}{a42}
  \ulline{a51}{a52}
  \ulline{a61}{a62}
  \longuline{a}{12}{17}
  \longuline{a}{22}{27}
  \longuline{a}{32}{37}
  \longuline{a}{42}{44}
  \longuline{a}{52}{54}
  \longuline{a}{62}{67}
  \udline{a37}{a17}
  \udline{a44}{a54}
  \udline{a67}{a27}
}

\def\cosderione{%
  \dernote{\swir}{}{{\black
    \na{11}\lneg\lpar(\na{21}\ltens\na{31})}
    \lpar
    {\black\na{41}\lpar(\na{51}\lneg\ltens \na{61}\lneg)}}{
    \root{\idr}{
      ((\na{12}\lneg\lpar\na{22})\ltens\na{32})\lpar
      \na{42}\lpar(\na{52}\lneg\ltens \na{62}\lneg)}{
      \root{\swir}{
	\na{33}\lpar\na{43}\lpar(\na{53}\lneg\ltens\na{63}\lneg)}{
	\root{\idr}{
	  \na{34}\lpar(\na{54}\lneg\ltens(\na{44}\lpar\na{64}\lneg))}{
	  \root{\idr}{
	    \na{35}\lpar\na{55}\lneg}{
	    \leaf{}}}}}}
}

\def\pathcosderione{%
  \flowgraphmode
  \longuline{a}{11}{12}
  \longuline{a}{21}{22}
  \longuline{a}{31}{35}
  \longuline{a}{41}{44}
  \longuline{a}{51}{55}
  \longuline{a}{61}{64}
  \udline{a12}{a22}
  \udline{a44}{a64}
  \udline{a35}{a55}
}
  
\def\cosderitwo{%
  \dernote{\swir}{}{{\black
    \na{11}\lneg\lpar(\na{21}\ltens\na{31})}
    \lpar
    {\black\na{41}\lpar(\na{51}\lneg\ltens \na{61}\lneg)}}{
    \root{\idr}{
      \na{12}\lneg\lpar(\na{22}\ltens\na{32})\lpar
      (\na{52}\lneg\ltens(\na{42}\lpar\na{62}\lneg))}{
      \root{\swir}{
	\na{13}\lneg\lpar(\na{23}\ltens\na{33})\lpar\na{53}\lneg}{
	\root{\idr}{
      	  \na{14}\lneg\lpar(\na{24}\ltens(\na{34}\lpar\na{54}\lneg))}{
	  \root{\idr}{
      	    \na{15}\lneg\lpar\na{25}}{
	    \leaf{}}}}}}
}

\def\pathcosderitwo{%
  \flowgraphmode
  \longuline{a}{11}{15}
  \longuline{a}{21}{25}
  \longuline{a}{31}{34}
  \longuline{a}{41}{42}
  \longuline{a}{51}{54}
  \longuline{a}{61}{62}
  \udline{a15}{a25}
  \udline{a42}{a62}
  \udline{a34}{a54}
}
  
\def\cosderiplanar{%
  \dernote{\swir}{}{{\black
    \na{11}\lneg\lpar(\na{21}\ltens\na{31})}
    \lpar
    {\black\na{41}\lpar(\na{51}\lneg\ltens \na{61}\lneg)}}{
    \root{\idr}{
      \na{12}\lneg\lpar(\na{22}\ltens\na{32})\lpar
      ((\na{42}\lpar\na{52}\lneg)\ltens\na{62}\lneg)}{
      \root{\swir}{
	\na{13}\lneg\lpar(\na{23}\ltens\na{33})\lpar\na{63}\lneg}{
	\root{\idr}{
      	  \na{14}\lneg\lpar(\na{24}\ltens(\na{34}\lpar\na{64}\lneg))}{
	  \root{\idr}{
      	    \na{15}\lneg\lpar\na{25}}{
	    \leaf{}}}}}}
}

\def\pathcosderiplanar{%
  \flowgraphmode
  \longuline{a}{11}{15}
  \longuline{a}{21}{25}
  \longuline{a}{31}{34}
  \longuline{a}{41}{42}
  \longuline{a}{51}{52}
  \longuline{a}{61}{64}
  \udline{a15}{a25}
  \udline{a42}{a52}
  \udline{a34}{a64}
}
  
\def\cosderithree{%
  \dernote{\swir}{}{{\black
    \na{11}\lneg\lpar(\na{21}\ltens\na{31})}
    \lpar
    {\black\na{41}\lpar(\na{51}\lneg\ltens \na{61}\lneg)}}{
    \root{\idr}{
      \na{12}\lneg\lpar(\na{22}\ltens\na{32})\lpar
      ((\na{42}\lpar\na{52}\lneg)\ltens \na{62}\lneg)}{
      \root{\swir}{
	\na{13}\lneg\lpar(\na{23}\ltens\na{33})\lpar\na{63}\lneg}{
	\root{\idr}{
	  \na{14}\lneg\lpar((\na{24}\lpar\na{64}\lneg)\ltens\na{34})}{
	  \root{\idr}{
	    \na{15}\lneg\lpar\na{35}}{
	    \leaf{}}}}}}
}

\def\pathcosderithree{%
  \flowgraphmode
  \longuline{a}{11}{15}
  \longuline{a}{21}{24}
  \longuline{a}{31}{35}
  \longuline{a}{41}{42}
  \longuline{a}{51}{52}
  \longuline{a}{61}{64}
  \udline{a15}{a35}
  \udline{a24}{a64}
  \udline{a42}{a52}
}
  
\def\thefirstcomma{,}
\def\thesecondcomma{,}
\def\theblack{}
\def\scderifour{%
  \dernote{\lpar}{}{\black
    \sqn{\na{11}\lneg\lpar(\na{21}\ltens \na{31}),
      \na{41}\lpar(\na{51}\lneg\ltens \na{61}\lneg)}}{
    \rroot{\cutr}{
      \sqn{\na{12}\lneg\lpar(\na{22}\ltens \na{32}),
	\na{42}\thefirstcomma\na{52}\lneg\ltens \na{62}\lneg}}{
      \root{\exr}{
	\sqn{\na{13}\lneg\lpar(\na{23}\ltens \na{33}),\na{43},
	\theblack
	((\na{73}\lneg\lpar\na{83}\lneg)\ltens\na{93})\ltens\na{103}\lneg}}{
	\rroot{\ltens}{
	  \sqn{\na{14}\lneg\lpar(\na{24}\ltens \na{34}),
	    ((\na{74}\lneg\lpar\na{84}\lneg)\ltens\na{94})
	    \ltens\na{104}\lneg,\na{44}}}{
	  \root{\idr}{
	    \sqn{\na{15}\lneg\lpar(\na{25}\ltens \na{35}),
	      (\na{75}\lneg\lpar\na{85}\lneg)\ltens\na{95}}}{
	    \leaf{}}}{
	  \root{\idr}{
	    \sqn{\na{105}\lneg,\na{45}}}{
	    \leaf{}}}}}{
      \root{\lpar}{
	\sqn{{\theblack
	    \na{113}\lpar(\na{123}\lneg\lpar(\na{133}\ltens\na{143}))},
	  \na{53}\lneg\ltens \na{63}\lneg}}{
	\root{\exr}{
	  \sqn{\na{114},\na{124}\lneg\lpar(\na{134}\ltens\na{144}),
	    \na{54}\lneg\ltens \na{64}\lneg}}{
	  \root{\lpar}{
	    \sqn{\na{125}\lneg\lpar(\na{135}\ltens\na{145}),\na{115},
	      \na{55}\lneg\ltens \na{65}\lneg}}{
	    \rroot{\cutr}{
	      \sqn{\na{126}\lneg\thesecondcomma
		\na{136}\ltens\na{146},\na{116},
		\na{56}\lneg\ltens \na{66}\lneg}}{
	      \root{\idr}{
		\sqn{\na{127}\lneg,\theblack\na{157}}}{
		\leaf{}}}{
	      \rroot{\ltens}{
		\sqn{{\theblack\na{167}\lneg},\na{137}\ltens\na{147},\na{117},
		  \na{57}\lneg\ltens \na{67}\lneg}}{
		\root{\idr}{
		  \sqn{\na{168}\lneg,\na{138}}}{
		  \leaf{}}}{
		\root{\exr}{
		  \sqn{\na{148},\na{118},
		    \na{58}\lneg\ltens \na{68}\lneg}}{
		  \rroot{\ltens}{
		    \sqn{\na{149},
		      \na{59}\lneg\ltens \na{69}\lneg,\na{119}}}{
		    \root{\idr}{
		      \sqn{\na{150},\na{60}\lneg}}{
		      \leaf{}}}{
		    \root{\idr}{
		      \sqn{\na{70}\lneg,\na{120}}}{
		      \leaf{}}}}}}}}}}}
}

\def\thecutsinpathscderifour{%
  \def\loopangleA{50}
  \def\loopangleB{130}
  \duline{a157}{a167}
  \def\loopangleA{23}
  \def\loopangleB{158}
  \def\loopvecheight{.6}
  \duline{a73}{a143}
  \duline{a83}{a133}
  \duline{a93}{a123}
  \duline{a103}{a113}
}

\def\pathscderifour{%
  \flowgraphmode
  \def\vertangleA{23}
  \def\vertangleB{158}
  \longulline{a}{11}{13}\longuline{a}{13}{15}
  \longulline{a}{21}{23}\longuline{a}{23}{25}
  \longulline{a}{31}{33}\longuline{a}{33}{35}
  \longulline{a}{41}{43}\longurline{a}{43}{45}
  \longulline{a}{73}{75}
  \longulline{a}{83}{85}
  \longuline{a}{93}{95}
  \longuline{a}{103}{105}
  \longurline{a}{113}{120}
  \longuline{a}{123}{127}
  \longuline{a}{133}{138}
  \longuline{a}{143}{145}\longurline{a}{145}{148}\longuline{a}{148}{150}
  \longuline{a}{167}{168}
  \longurline{a}{51}{53}\longuline{a}{53}{55}
  \longurline{a}{55}{58}\longuline{a}{58}{60}
  \longurline{a}{61}{63}\longuline{a}{63}{65}
  \longurline{a}{65}{68}\longuline{a}{68}{70}
  \udline{a105}{a45}
  \udline{a127}{a157}
  \udline{a168}{a138}
  \udline{a150}{a60}
  \udline{a70}{a120}
  \def\loopvecheight{.7}
  \udline{a15}{a95}
  \udline{a25}{a85}
  \udline{a35}{a75}
  \thecutsinpathscderifour
}

\def\cosderifour{%
  \dernote{\iru}{}{{\black
      \na{10}\lneg\lpar(\na{20}\ltens\na{30})}
    \lpar
	{\black\na{40}
	  \qqqquad\lpar\qqqquad
	  (\na{50}\lneg\ltens \na{60}\lneg)}}{
    \root{\swir}{
      \na{11}\lneg\lpar(\na{21}\ltens\na{31})\lpar\na{41}\lpar
      ({\theblack
	(((\na{71}\lneg\lpar\na{81}\lneg)\ltens\na{91})\ltens\na{101}\lneg)}
      \ltens
      {\theblack
      (\na{111}\lpar(\na{121}\lneg\lpar(\na{131}\ltens\na{141})))})
      \lpar
      (\na{51}\lneg\ltens\na{61}\lneg)}{
      \root{\ird}{
	\na{12}\lneg\lpar(\na{22}\ltens\na{32})\lpar
	((((\na{72}\lneg\lpar\na{82}\lneg)\ltens\na{92})\ltens
	(\na{42}\lpar\na{102}\lneg))
	\ltens
	(\na{112}\lpar\na{122}\lneg\lpar(\na{132}\ltens\na{142})))  
	\lpar
	(\na{52}\lneg\ltens \na{62}\lneg)}{
	\root{\iru}{
	  \na{13}\lneg\lpar(\na{23}\ltens\na{33})\lpar
	  (((\na{73}\lneg\lpar\na{83}\lneg)\ltens\na{93})\ltens
	  (\na{113}\lpar\na{123}\lneg
	  \wlpar
	  (\na{133}\ltens\na{143})))    
	  \lpar
	  (\na{53}\lneg\ltens \na{63}\lneg)}{
	  \root{\swir}{
	    \na{14}\lneg\lpar(\na{24}\ltens\na{34})\lpar
	    (((\na{74}\lneg\lpar\na{84}\lneg)\ltens\na{94})\ltens
	    (\na{114}\lpar\na{124}\lneg\lpar
	    ({\theblack\na{154}}\ltens{\theblack\na{164}\lneg})\lpar
	    (\na{134}\ltens\na{144})))    
	    \lpar
	    (\na{54}\lneg\ltens \na{64}\lneg)}{
	    \root{\ird}{
	      \na{15}\lneg\lpar(\na{25}\ltens\na{35})\lpar
	      (((\na{75}\lneg\lpar\na{85}\lneg)\ltens\na{95})\ltens
	      (\na{115}\lpar
	      ((\na{125}\lneg\lpar\na{155})\ltens\na{165}\lneg)\lpar
	      (\na{135}\ltens\na{145})))    
	      \lpar
	      (\na{55}\lneg\ltens \na{65}\lneg)}{
	      \root{\swir}{
		\na{16}\lneg\lpar(\na{26}\ltens\na{36})\lpar
		(((\na{76}\lneg\lpar\na{86}\lneg)\ltens\na{96})\ltens
		(\na{116}\wlpar\na{166}\lneg\lpar
		(\na{136}\ltens\na{146})))    
		\lpar
		(\na{56}\lneg\ltens \na{66}\lneg)}{
		\root{\ird}{
		  \na{17}\lneg\lpar(\na{27}\ltens\na{37})\lpar
		  (((\na{77}\lneg\lpar\na{87}\lneg)\ltens\na{97})\ltens
		  (\na{117}\lpar
		  ((\na{167}\lneg\lpar\na{137})\ltens\na{147})))    
		  \lpar
		  (\na{57}\lneg\ltens \na{67}\lneg)}{
		  \root{\swir}{
		    \na{18}\lneg\lpar(\na{28}\ltens\na{38})\lpar
		    (((\na{78}\lneg\lpar\na{88}\lneg)\ltens\na{98})\ltens
		    (\na{118}\wlpar\na{148}))    
		    \lpar
		    (\na{58}\lneg\ltens \na{68}\lneg)}{
		    \root{\ird}{
		      \na{19}\lneg\lpar(\na{29}\ltens\na{39})\lpar
		      (((\na{79}\lneg\lpar\na{89}\lneg)\ltens\na{99})\ltens
		      (\na{119}\lpar\na{149}\lpar
		      (\na{59}\lneg\ltens \na{69}\lneg)))}{
		      \root{\ird}{
			\na{220}\lneg\lpar(\na{230}\ltens\na{240})\lpar
			((\na{80}\lneg\lpar\na{90}\lneg)\ltens\na{100})}{
			\leaf{}}}}}}}}}}}}%
}

\def\thecutsinpathcosderifour{%
  \def\loopangleA{50}
  \def\loopangleB{130}
  \duline{a154}{a164}
  \def\loopangleA{23}
  \def\loopangleB{158}
  \def\loopvecheight{.6}
  \duline{a71}{a141}
  \duline{a81}{a131}
  \duline{a91}{a121}
  \duline{a101}{a111}%
}

\def\pathcosderifour{%
  \flowgraphmode
  \def\vertangleA{23}
  \def\vertangleB{158}
  \ulline{a10}{a11}\longuline{a}{11}{19}\urline{a19}{a220}
  \ulline{a20}{a21}\longuline{a}{21}{29}\urline{a29}{a230}
  \ulline{a30}{a31}\longuline{a}{31}{39}\urline{a39}{a240}
  \ulline{a40}{a41}\urline{a41}{a42}
  \longuline{a}{71}{79}\urline{a79}{a80}
  \longuline{a}{81}{89}\urline{a89}{a90}
  \longuline{a}{91}{99}\urline{a99}{a100}
  \longuline{a}{101}{102}
  \longuline{a}{111}{112}\longulline{a}{112}{114}\longuline{a}{114}{119}
  \longuline{a}{121}{122}\longulline{a}{122}{124}\longuline{a}{124}{125}
  \longuline{a}{131}{137}
  \longuline{a}{141}{149}
  \longuline{a}{154}{155}
  \longuline{a}{164}{167}
  \urline{a50}{a51}\longuline{a}{51}{59}
  \urline{a60}{a61}\longuline{a}{61}{69}
  \def\loopangleA{50}
  \def\loopangleB{130}
  \udline{a42}{a102}
  \udline{a125}{a155}
  \udline{a167}{a137}
  \udline{a149}{a59}
  \def\loopvecheight{.6}
  \udline{a119}{a69}
  \udline{a240}{a80}
  \def\loopangleA{40}
  \def\loopangleB{140}
  \udline{a220}{a100}
  \udline{a230}{a90}
  \thecutsinpathcosderifour%
}

\def\pathpnone{%
  \def\pathderimode{b}
  \begin{psmatrix}[rowsep=3ex,colsep=1ex]
    \na1\lneg & & \na2 & & \na3 &\qqquad & \na4 & & \na5\lneg & & \na6\lneg\\
    & & & \nltens1 &&&&&& \nltens2\\
    &\nlpar1 &&&&&&\nlpar2
    \pntreemode
    \ncline{a1}{lpar1}
    \ncline{a2}{ltens1}
    \ncline{a3}{ltens1}
    \ncline{lpar1}{ltens1}
    \ncline{a4}{lpar2}
    \ncline{a5}{ltens2}
    \ncline{a6}{ltens2}
    \ncline{lpar2}{ltens2}
    \pnlinkmode
    \udline{a1}{a2}
    \udline{a3}{a5}
    \udline{a4}{a6}
  \end{psmatrix}
}

\def\pathpnplanar{%
  \def\pathderimode{b}
  \begin{psmatrix}[rowsep=3ex,colsep=1ex]
  \\
    \na1\lneg & & \na2 & & \na3 &\qqquad & \na4 & & \na5\lneg & & \na6\lneg\\
    & & & \nltens1 &&&&&& \nltens2\\
    &\nlpar1 &&&&&&\nlpar2
    \pntreemode
    \ncline{a1}{lpar1}
    \ncline{a2}{ltens1}
    \ncline{a3}{ltens1}
    \ncline{lpar1}{ltens1}
    \ncline{a4}{lpar2}
    \ncline{a5}{ltens2}
    \ncline{a6}{ltens2}
    \ncline{lpar2}{ltens2}
    \pnlinkmode
    \udline{a1}{a2}
    \udline{a3}{a6}
    \udline{a4}{a5}
  \end{psmatrix}
}

\def\pathpnthree{%
  \def\pathderimode{b}
  \begin{psmatrix}[rowsep=3ex,colsep=1ex]
    \na1\lneg & & \na2 & & \na3 &\qqquad & \na4 & & \na5\lneg & & \na6\lneg\\
    & & & \nltens1 &&&&&& \nltens2\\
    &\nlpar1 &&&&&&\nlpar2
    \pntreemode
    \ncline{a1}{lpar1}
    \ncline{a2}{ltens1}
    \ncline{a3}{ltens1}
    \ncline{lpar1}{ltens1}
    \ncline{a4}{lpar2}
    \ncline{a5}{ltens2}
    \ncline{a6}{ltens2}
    \ncline{lpar2}{ltens2}
    \pnlinkmode
    \udline{a1}{a3}
    \udline{a4}{a5}
    \def\loopvecheight{.7}
    \udline{a2}{a6}
  \end{psmatrix}
}

\def\pathpnfour{%
  \def\pathderimode{b}
  \begin{psmatrix}[rowsep=3ex,colsep=1ex]
    \na1\lneg & & \na2 & & \na3 & \phantom{a}  &\na4 & \phantom{a} & 
    \na7\lneg & & \na8\lneg && \na9 && \na{10}\lneg & \hskip2em &
    \na{11} && \na{12}\lneg && \na{15} &\qquad& \na{16}\lneg && 
    \na{13} && \na{14} & \qqquad & 
    \na5\lneg & & \na6\lneg\\
    &  && \nltens1 && && && \nlpar5 &&&&&&&&&&&&&&&&\nltens3 &&&& \nltens2\\
    &\nlpar1 && && && && & \nltens5 &&&&&&&&&&&\nlpar3\\
    &        && && && && && & \nltens6&&&&&&\nlpar4\\ \\
    &&&&&&&&&&&&&&\nlpar2 
    \pntreemode
    \ncline{a1}{lpar1}
    \ncline{a2}{ltens1}
    \ncline{a3}{ltens1}
    \ncline{lpar1}{ltens1}
    \ncline{a4}{lpar2}
    \ncline{a5}{ltens2}
    \ncline{a6}{ltens2}
    \ncline{lpar2}{ltens2}
    \ncline{a7}{lpar5}
    \ncline{a8}{lpar5}
    \ncline{lpar5}{ltens5}
    \ncline{a9}{ltens5}
    \ncline{ltens5}{ltens6}
    \ncline{a10}{ltens6}
    \ncline{a11}{lpar4}
    \ncline{a12}{lpar3}
    \ncline{lpar3}{lpar4}
    \ncline{ltens3}{lpar3}
    \ncline{a13}{ltens3}
    \ncline{a14}{ltens3}
    \pnlinkmode
    \udline{a3}{a7}
    \udline{a12}{a15}
    \duline{a15}{a16}
    \udline{a16}{a13}
    \def\loopangleA{60}
    \def\loopangleB{120}
    \udline{a14}{a5}
    \def\loopvecheight{.7}    
    \udline{a2}{a8}
    \duline{ltens6}{lpar4}
    \def\loopvecheight{.6}    
    \udline{a1}{a9}
    \udline{a4}{a10}
    \def\loopvecheight{.5}    
    \udline{a11}{a6}
  \end{psmatrix}
}

\def\switchingexamplecorrect{
  \ncline{lpar1}{a3}
  \ncline{lpar1}{a4}
  \ncline{lpar2}{c1}
  \ncline{lpar2}{c2}
}

\def\examplecorrect{%
  \vcnpn{
  \quad\\
    \na1 && \na2\lneg && \na3 && \na4\lneg && \nc1 && \nc2\lneg\\
    &\pa&& \pa && \nlpar1 && \pa && \nlpar2\\
    &   &&     &&         &&\nltens1
  }{
    \switchingexamplecorrect
    \ncline{lpar1}{ltens1}
    \ncline{lpar2}{ltens1}
  }{
    \udline{a1}{a4}
    \udline{a2}{a3}
    \udline{c1}{c2}
    \duline{a1}{a2}
  }
}

\def\examplecorrectcut{%
  \vcnpn{
  \quad\\
    \na1 && \na2\lneg && \na3 && \na4\lneg && \nc1 && \nc2\lneg\\
    &\nlcut1&& \pa && \nlpar1 && \pa && \nlpar2\\
    &   &&     &&         &&\nltens1
  }{
    \switchingexamplecorrect
    \ncline{lcut1}{a1}
    \ncline{lcut1}{a2}
    \ncline{lpar1}{ltens1}
    \ncline{lpar2}{ltens1}
  }{
    \udline{a1}{a4}
    \udline{a2}{a3}
    \udline{c1}{c2}
  }
}

\def\examplecorrectlong{%
  \vcnpn{
  \quad\\
    \na1 && \na2\lneg && \na3 && \na4\lneg && \nc1 && \nc2\lneg\\
    &\qlapnode{lcut1}{a\lcut a\lneg}&& \pa && 
    \qlapnode{lpar1}{a\lpar a\lneg} && \pa && 
    \qlapnode{lpar2}{c\lpar c\lneg}\\
    &   &&     &&         &&
    \qlapnode{ltens1}{(a\lpar a\lneg)\ltens(c\lpar c\lneg)}
  }{
    \switchingexamplecorrect
    \ncline{lcut1}{a1}
    \ncline{lcut1}{a2}
    \ncline{lpar1}{ltens1}
    \ncline{lpar2}{ltens1}
  }{
    \udline{a1}{a4}
    \udline{a2}{a3}
    \udline{c1}{c2}
  }
}

\def\scderitwosided{
  \dernote{\lparr}{}{\black\ssqn{\na{11}\ltens(\na{21}\limp\na{31}\lneg)}
    {\na{41}\lpar(\na{51}\lpar\na{61})\lneg}}{
    \root{\ltensl}{\ssqn{\na{12}\ltens(\na{22}\limp\na{32}\lneg)}
      {\na{42},(\na{52}\lpar\na{62})\lneg}}{
      \root{\lnegr}{\ssqn{\na{13},\na{23}\limp\na{33}\lneg}
	{\na{43},(\na{53}\lpar\na{63})\lneg}}{
	\rroot{\lparl}{\ssqn{\na{14},\na{24}\limp\na{34}\lneg,
	    \na{54}\lpar\na{64}}
	  {\na{44}}}{
	  \rroot{\limpl}{\ssqn{\na{15},\na{25}\limp\na{35}\lneg,\na{55}}{}}{
	    \root{\idr}{\ssqn{\na{16}}{\na{26}}}{
	      \leaf{}}}{
	    \root{\lnegl}{\ssqn{\na{36}\lneg,\na{56}}{}}{
	      \root{\idr}{\ssqn{\na{57}}{\na{37}}}{
		\leaf{}}}}}{
	  \root{\idr}{\ssqn{\na{65}}{\na{45}}}{
	    \leaf{}}}}}}
}

\def\pathscderitwosided{%
  \flowgraphmode
  \longuline{a}{11}{16}
  \longuline{a}{21}{26}
  \longuline{a}{31}{37}
  \longuline{a}{41}{45}
  \longuline{a}{51}{57}
  \longuline{a}{61}{65}
  \udline{a16}{a26}
  \udline{a65}{a45}
    \udline{a57}{a37}
}
  
\def\pathpntwosided{%
  \def\pathderimode{b}
  \begin{psmatrix}[rowsep=3ex,colsep=1ex]
  \\
    \na1\markpl & & \na2\markpr & & \na3\markpr &\qqquad & 
    \na4\markpr & & \na5\markpl & & \na6\markpl\\
    & & & &\nlneg1\markpl &&&&& \nlpar3\markpl\\
    & & & \nlimp1\markpl &&&&&& \nlneg2\markpr\\
    &\nltens1\markpl &&&&&&\nlpar2\markpr
    \pntreemode
    \ncline{a1}{ltens1}
    \ncline{a2}{limp1}
    \ncline{a3}{lneg1}
    \ncline{lneg1}{limp1}
    \ncline{limp1}{ltens1}
    \ncline{a4}{lpar2}
    \ncline{a5}{lpar3}
    \ncline{a6}{lpar3}
    \ncline{lpar3}{lneg2}
    \ncline{lneg2}{lpar2}
    \pnlinkmode
    \udline{a1}{a2}
    \udline{a3}{a5}
    \udline{a4}{a6}
  \end{psmatrix}
}

\def\exacostwosideda{%
  \dernote{\swir}{}{\na{11}\lpar(\nb{21}\lneg\ltens\na{31}\lneg)}{
    \root{\ird}{\nb{22}\lneg\ltens(\na{12}\lpar\na{32}\lneg)}{
      \root{\iru}{\nb{23}\lneg}{
        \root{\swir}{\nb{24}\lneg\lpar(\nc{44}\lneg\ltens\nc{54})}{
	  \leaf{(\nb{25}\lneg\lpar\nc{45}\lneg)\ltens\nc{55}}}}}}
}

\def\pathexacostwosideda{%
  \flowgraphmode
  \longuline{a}{11}{12}
  \longuline{b}{21}{25}
  \longuline{a}{31}{32}
  \longuline{c}{44}{45}
  \longuline{c}{54}{55}
  \udline{a12}{a32}
  \duline{c44}{c54}
}

\def\exacostwosidedb{%
  \dernote{\swir}{}{\nc{11}\lneg\lpar(\nc{21}\ltens\nb{31})}{
    \root{\ird}{(\nc{12}\lneg\lpar\nc{22})\ltens\nb{32}}{
      \root{\iru}{\nb{33}}{
	\root{\swir}{(\na{44}\ltens\na{54}\lneg)\lpar\nb{34}}{
	  \leaf{(\na{45}\lpar\nb{35})\ltens\na{55}\lneg}}}}}
}

\def\pathexacostwosidedb{%
  \flowgraphmode
  \longuline{c}{11}{12}
  \longuline{c}{21}{22}
  \longuline{b}{31}{35}
  \longuline{a}{44}{45}
  \longuline{a}{54}{55}
  \udline{c12}{c22}
  \duline{a44}{a54}
}

\def\exapntwosideda{%
  \vcnpn{
    &  & & \nltens1\markpl\\
    & \nlpar1\markpl \\
    \mb3\lnegpl & & \nc2\lnegpl & & \nc1\markpl\\ \\
    \na4\markpr & & \mb5\lnegpr & & \na6\lnegpr\\
    &  & & \nltens2\markpr\\
    & \nlpar2\markpr
  }{
    \ncline{ltens1}{lpar1}
    \ncline{ltens1}{c1}
    \ncline{lpar1}{b3}
    \ncline{lpar1}{c2}
    \ncline{ltens2}{lpar2}
    \ncline{lpar2}{a4}
    \ncline{ltens2}{b5}
    \ncline{ltens2}{a6}
  }{
    \uline{b5}{b3}
    \udline{a4}{a6}
    \duline{c2}{c1}
  }
}

\def\exapntwosidedavar{%
  \vcnpn{
    &  & & \nlpar1\\
    & \nltens1 \\
    \mb3 & & \nc2 & & \nc1\lneg\\ \\
    \na4 & & \mb5\lneg & & \na6\lneg\\
    &  & & \nltens2\\
    & \nlpar2
  }{
    \ncline{ltens1}{lpar1}
    \ncline{lpar1}{c1}
    \ncline{ltens1}{b3}
    \ncline{ltens1}{c2}
    \ncline{ltens2}{lpar2}
    \ncline{lpar2}{a4}
    \ncline{ltens2}{b5}
    \ncline{ltens2}{a6}
  }{
    \uline{b5}{b3}
    \udline{a4}{a6}
    \duline{c2}{c1}
  }
}

\def\exapntwosidedb{%
  \vcnpn{
    &  & & \nltens2\markpl\\
    & \nlpar2\markpl \\
    \na6\markpl & & \mb5\markpl & & \na4\lnegpl\\ \\
    \nc1\lnegpr & & \nc2\markpr & & \mb3\markpr\\
    &  & & \nltens1\markpr\\
    & \nlpar1\markpr
  }{
    \ncline{ltens1}{lpar1}
    \ncline{lpar1}{c1}
    \ncline{ltens1}{b3}
    \ncline{ltens1}{c2}
    \ncline{ltens2}{lpar2}
    \ncline{ltens2}{a4}
    \ncline{lpar2}{b5}
    \ncline{lpar2}{a6}
  }{
    \uline{b3}{b5}
    \duline{a6}{a4}
    \udline{c1}{c2}
  }
}
	
\def\exapntwosidedbvar{%
  \vcnpn{
    &  & & \nlpar2\\
    & \nltens2 \\
    \na6\lneg & & \mb5\lneg & & \na4\\ \\
    \nc1\lneg & & \nc2 & & \mb3\\
    &  & & \nltens1\\
    & \nlpar1
  }{
    \ncline{ltens1}{lpar1}
    \ncline{lpar1}{c1}
    \ncline{ltens1}{b3}
    \ncline{ltens1}{c2}
    \ncline{ltens2}{lpar2}
    \ncline{lpar2}{a4}
    \ncline{ltens2}{b5}
    \ncline{ltens2}{a6}
  }{
    \uline{b3}{b5}
    \duline{a6}{a4}
    \udline{c1}{c2}
  }
}

\def\mback{\hskip-.4em\strut}
\def\exapnonesideda{%
  \vcnpn{
  \\
    \nc1\markpl & & \nc2\lnegpl\mback & & \mb3\lnegpl\;\;\strut & & 
    \na4\markpr\mback & & \mb5\lnegpr\mback & & \na6\lnegpr\mback\\
    & & & \nlpar1\markpl\mback &&&&&& \nltens2\markpr\mback\\
    &\nltens1\markpl\mback &&&&&&\nlpar2\markpr\mback
  }{
    \ncline{c1}{ltens1}
    \ncline{c2}{lpar1}
    \ncline{b3}{lpar1}
    \ncline{lpar1}{ltens1}
    \ncline{a4}{lpar2}
    \ncline{b5}{ltens2}
    \ncline{a6}{ltens2}
    \ncline{lpar2}{ltens2}
  }{
    \udline{c1}{c2}
    \udline{b3}{b5}
    \udline{a4}{a6}
  }
}

\def\exapnonesidedavar{%
  \vcnpn{
  \\
    \nc1\lneg & & \nc2 & & \mb3 &\qqquad & \na4 & & \mb5\lneg & & \na6\lneg\\
    & & & \nltens1 &&&&&& \nltens2\\
    &\nlpar1 &&&&&&\nlpar2
  }{
    \ncline{c1}{lpar1}
    \ncline{c2}{ltens1}
    \ncline{b3}{ltens1}
    \ncline{lpar1}{ltens1}
    \ncline{a4}{lpar2}
    \ncline{b5}{ltens2}
    \ncline{a6}{ltens2}
    \ncline{lpar2}{ltens2}
  }{
    \udline{c1}{c2}
    \udline{b3}{b5}
    \udline{a4}{a6}
  }
}

\def\exascMALL{%
  \dernote{\lpar}{}{\black
    \sqn{(\na{11}\ltens\na{21})\lpar\na{31}\lneg,
      \na{41}\lneg\lwith(\na{51}\lneg\lplus\nb{61})}}{
    \rroot{\lwith}{\sqn{\na{12}\;\ltens\;\na{22}\;,\;\na{32}\lneg\;,\;
	\na{42}\lneg\lwith(\na{52}\lneg\lplus\nb{62})}}{
      \rroot{\ltens}{\sqn{\na{13}\;\ltens\;\na{23}\;,\;\na{33}\lneg\;,\;
	  \na{43}\lneg}}{
	\root{\idr}{\sqn{\na{14},\na{34}\lneg}}{
	  \leaf{}}}{
	\root{\idr}{\sqn{\na{24},\na{44}\lneg}}{
	  \leaf{}}}}{
      \rroot{\ltens}{\sqn{\na{73}\;\ltens\;\na{83}\;,\;\na{93}\lneg\;,\;
	  \na{53}\lneg\lplus\nb{63}}}{
	\root{\lplus_1}{\sqn{\na{74},\na{54}\lneg\lplus\nb{64}}}{
	  \root{\idr}{\sqn{\na{75},\na{55}\lneg}}{
	    \leaf{}}}}{
	\root{\idr}{\sqn{\na{84},\na{94}\lneg}}{
	  \leaf{}}}}}
}

\def\pathexascMALL{%
  \flowgraphmode
  \def\vertangleA{23}
  \def\vertangleB{158}
  \longuline{a}{11}{12}\urline{a12}{a73}\longuline{a}{73}{75}
  \longuline{a}{21}{22}\urline{a22}{a83}\longurline{a}{83}{84}
  \longuline{a}{31}{32}\urline{a32}{a93}\longurline{a}{93}{94}
  \longuline{a}{51}{52}\longurline{a}{52}{53}
  \longulline{a}{53}{54}\longuline{a}{54}{55}
  \longuline{b}{61}{62}\longurline{b}{62}{63}\longulline{b}{63}{64}
  \udline{a75}{a55}
  \udline{a84}{a94}
  \varflowgraphmode
  \longuline{a}{11}{12}\longulline{a}{12}{14}
  \longuline{a}{21}{22}\longulline{a}{22}{23}\longurline{a}{23}{24}
  \longuline{a}{31}{32}\longulline{a}{32}{34}
  \longuline{a}{41}{42}\longulline{a}{42}{43}\longuline{a}{43}{44}
  \udline{a14}{a34}
  \udline{a24}{a44}
}

\def\pathpnexaMALL{%
  \vcnpn{
    \\
    \na1&&\na2&&\na3\lneg&&\na4\lneg\mback&&\na5\lneg\mback&&\mb1\\
    &\nltens1&&\qlapm{}&&\qlapm{}&&\qlapm{}&&\nlplus1\\
    &&\nlpar1&&&&&&\nlwith1
  }{
    \ncline{a1}{ltens1}
    \ncline{a2}{ltens1}
    \ncline{a3}{lpar1}
    \ncline{ltens1}{lpar1}
    \ncline{a5}{lplus1}
    \ncline{b1}{lplus1}
    \ncline{a4}{lwith1}
    \ncline{lwith1}{lplus1}
  }{
    \lineanglesheight{a2}{a3}{55}{125}{.6}
    \lineanglesheight{a1}{a5}{85}{125}{.6}
    \varpnlinkmode
    \lineanglesheight{a1}{a3}{50}{95}{.8}
    \lineanglesheight{a2}{a4}{85}{125}{.8}
  }
}

\def\bvderione{%
  \dernote{=}{}{\black
  \nb{11}\lpar\na{21}\lpar(\na{31}\lneg\lseq\nb{41}\lneg)}{
    \root{\seqrd}{\nb{12}\lpar(\na{22}\lseq\un)\lpar
      (\na{32}\lneg\lseq\nb{42}\lneg)}{
      \root{\ird}{\nb{13}\lpar((\na{23}\lpar\na{33}\lneg)\lseq
	(\un\lpar\nb{43}\lneg))}{
	\root{=}{\nb{14}\lpar(\un\lseq(\un\lpar\nb{44}\lneg))}{
	  \root{\ird}{\nb{15}\lpar\nb{45}}{
	    \leaf{\un}}}}}}
}

\def\pathbvderione{%
  \flowgraphmode
  \longuline{b}{11}{15}
  \longuline{a}{21}{23}
  \longuline{a}{31}{33}
  \longuline{b}{41}{45}
  \udline{b15}{b45}
  \udline{a23}{a33}
}

\def\pathbvpnone{%
  \vcnpn{
    \\
    \mb1&&\na2&&\na3\lneg\mback&&\nb4\lneg\\
    &\nlpar1&&\qlapm{}&&\nlseq1
  }{
    \ncline{b1}{lpar1}
    \ncline{a2}{lpar1}
    \ncline{a3}{lseq1}
    \ncline{b4}{lseq1}
  }{
    \udline{b1}{b4}
    \udline{a2}{a3}
  }
}

\def\bvderitwo{%
  \dernote{=}{}{\black
  (\nb{11}\lseq\na{21})\lpar(\na{31}\lneg\lcoseq\nb{41}\lneg)}{
    \root{\seqrd}{(\nb{12}\lseq\na{22})\lpar(\nb{42}\lneg\lseq\na{32}\lneg)}{
      \root{\ird}{(\nb{13}\lpar\nb{43}\lneg)\lseq(\na{23}\lpar\na{33}\lneg)}{
	\root{=}{(\nb{14}\lpar\nb{44}\lneg)\lseq\un}{
	  \root{\ird}{\nb{15}\lpar\nb{45}}{
	    \leaf{\un}}}}}}
}

\def\pathbvderitwo{%
  \flowgraphmode
  \longuline{b}{11}{15}
  \longuline{a}{21}{23}
  \longuline{a}{31}{33}
  \longuline{b}{41}{45}
  \udline{b15}{b45}
  \udline{a23}{a33}
}

\def\pathbvpntwo{%
  \vcnpn{
    \\
    \mb1&&\na2&&\na3\lneg\mback&&\nb4\lneg\\
    &\nlseq1&&\qlapm{}&&\nlcoseq1
  }{
    \ncline{b1}{lseq1}
    \ncline{a2}{lseq1}
    \ncline{a3}{lcoseq1}
    \ncline{b4}{lcoseq1}
  }{
    \udline{b1}{b4}
    \udline{a2}{a3}
  }
}

\def\exaclflow{%
  \dernote{\conr^3}{}{\black
    \sqn{\cneg{\na{11}},\na{21}\cand\cneg{\na{31}},\na{41}}}{
    \root{\exr^5}{\sqn{\cneg{\na{12}},\cneg{\na{52}},
	\na{22}\cand\cneg{\na{32}},\na{62}\cand\cneg{\na{72}},
	\na{42},\na{82}}}{
      \rroot{\cand}{\sqn{\cneg{\na{13}},\na{43},
	  \na{23}\cand\cneg{\na{33}},\na{63}\cand\cneg{\na{73}},
	  \cneg{\na{53}},\na{83}}}{
	\root{\weakr}{\sqn{\cneg{\na{14}},\na{44},\na{24}}}{
	  \root{\idr}{\sqn{\cneg{\na{15}},\na{45}}}{
	    \leaf{}}}}{
	\rroot{\cand}{\sqn{\cneg{\na{34}},\na{64}\cand\cneg{\na{74}},
	    \cneg{\na{54}},\na{84}}}{
	  \root{\idr}{\sqn{\cneg{\na{35}},\na{65}}}{
	    \leaf{}}}{
	  \root{\weakr}{\sqn{\cneg{\na{75}},\cneg{\na{55}},\na{85}}}{
	    \root{\idr}{\sqn{\cneg{\na{56}},\na{86}}}{
	      \leaf{}}}}}}}
}

\def\pathexaclflow{%
  \flowgraphmode
  \def\vertangleA{23}
  \def\vertangleB{158}
  \longuline{a}{11}{13}\longulline{a}{13}{14}\longuline{a}{14}{15}
  \longuline{a}{21}{23}\longulline{a}{23}{24}
  \longuline{a}{31}{35}
  \longuline{a}{41}{42}\longulline{a}{42}{44}\longuline{a}{44}{45}
  \uline{a11}{a52}\longurline{a}{52}{55}\longuline{a}{55}{56}
  \uline{a21}{a62}\longuline{a}{62}{65}
  \uline{a31}{a72}\longuline{a}{72}{73}\longurline{a}{73}{75}
  \uline{a41}{a82}\longuline{a}{82}{83}\longurline{a}{83}{85}
  \longuline{a}{85}{86}
  \udline{a15}{a45}
  \udline{a35}{a65}
  \udline{a56}{a86}
}

\def\pathpnclflow{%
  \vcnpn{
  \quad\\
    \cneg{\na1} && \na2 && \cneg{\na3} && \na4\\
    &\qlapm{}  && \ncand1 && \qlapm{}
  }{
    \ncline{cand1}{a2}
    \ncline{cand1}{a3}
  }{
    \udline{a2}{a3}
    \def\loopvecheight{.6}    
    \udline{a1}{a4}
    \def\loopvecheight{.9}    
    \def\loopangleA{90}
    \def\loopangleB{90}
    \udline{a1}{a4}
  }
}

\def\pathpnclflowsimple{%
  \vcnpn{
  \quad\\
    \cneg{\na1} && \na2 && \cneg{\na3} && \na4\\
    &\qlapm{}  && \ncand1 && \qlapm{}
  }{
    \ncline{cand1}{a2}
    \ncline{cand1}{a3}
  }{
    \udline{a2}{a3}
    \def\loopvecheight{.75}    
    \udline{a1}{a4}
  }
}

\def\derpfeilimmedial{}

\def\medialderi{%
  \dernote{\swir^4}{}{\black
    (\nnb{11}\cand\na{11})\cor(\nna{21}\cand\nnb{21})\cor
    (\nb{31}\cand\na{31})\cor(\nna{41}\cand\nb{41})}{
    \root{\derpfeilimmedial\ird^2}{
      (\nnb{12}\cand(\na{12}\cor\nna{22})\cand\nnb{22})\cor
      (\nb{32}\cand(\na{32}\cor\nna{42})\cand\nb{42})}{
      \root{\conrd^4}{
	(\nnb{13}\;\cand\;\nnb{23})\;\cor\;(\nb{33}\;\cand\;\nb{43})}{
        \root{\medr}{((\nnb{14}\cor\nnb{54})\cand(\nnb{24}\cor\nnb{64}))
	  \cor
	  ((\nb{34}\cor\nb{74})\cand(\nb{44}\cor\nb{84}))}{
	  \root{\ird}{(\nnb{15}\cand\nnb{25})\cor(\nnb{55}\cand\nnb{65})
	  \cor
	  ((\nb{35}\cor\nb{75})\cand(\nb{45}\cor\nb{85}))}{
	    \leaf{\vbox to 3ex{}}}}}}}
}

\def\pathmedialderi{%
  \flowgraphmode
  \longuline{a}{11}{12}
  \longuline{a}{21}{22}
  \longuline{a}{31}{32}
  \longuline{a}{41}{42}
  \udline{a12}{a22}
  \udline{a32}{a42}
  \longuline{b}{11}{15}
  \longuline{b}{21}{25}
  \longuline{b}{31}{35}
  \longuline{b}{41}{45}
  \uline{b13}{b54}\longuline{b}{54}{55}
  \uline{b23}{b64}\longuline{b}{64}{65}
  \uline{b33}{b74}\longuline{b}{74}{75}
  \uline{b43}{b84}\longuline{b}{84}{85}
  \def\loopvecheight{.6}    
  \def\loopangleA{80}
  \def\loopangleB{100}
  \udline{b15}{b85}
  \udline{b25}{b45}
  \udline{b55}{b75}
  \udline{b65}{b35}
}

\RRdate{October 2006}

\RRauthor{Lutz Stra\ss burger}

\authorhead{Lutz Stra\ss burger}

\RRtitle{Reseaux de d\'emonstration et l'identit\'e des d\'emonstrations}

\RRetitle{Proof Nets and the Identity of Proofs}

\titlehead{Proof Nets and the Identity of Proofs}

\RRabstract{
These are the notes for a 5-lecture-course given at ESSLLI 2006 in
Malaga, Spain. The URL of the school is
\url{http://esslli2006.lcc.uma.es/}. This version slightly differs
from the one which has been distributed at the school because typos
have been removed and comments and suggestions by students have been
worked in.

The course is intended to be introductory. That means no prior
knowledge of proof nets is required. However, the student should be
familiar with the basics of propositional logic, and should have seen
formal proofs in some formal deductive system (e.g., sequent calculus,
natural deduction, resolution, tableaux, calculus of structures,
Frege-Hilbert-systems, etc.). It is probably helpful if the student
knows already what cut elimination is, but this is not strictly
necessary.

In these notes, I will introduce the concept of ``proof nets'' from
the viewpoint of the problem of the identity of proofs.  I will
proceed in a rather informal way. The focus will be more on presenting
ideas than on presenting technical details. The goal of the course is
to give the student an overview of the theory of proof nets and make
the vast amount of literature on the topic easier accessible to the
beginner.
 
For introducing the basic concepts of the theory, I will in the first
part of the course stick to the unit-free multiplicative fragment of
linear logic because of its rather simple notion of proof nets. In the
second part of the course we will see proof nets for more
sophisticated logics.

This is a basic introduction into proof nets from the
perspective of the identity of proofs. We discuss how deductive proofs
can be translated into proof nets and what a correctness criterion
is.}

\RRmotcle{Reseaux de d\'emonstration, \'elimination des coupures,
identit\'e des d\'emonstrations, logique lineaire, logique classique}

\RRkeyword{Proof nets, cut elimination, identity of proofs,
correctness criterion, multiplicative linear logic, pomset logic, 
classical logic}

\RRprojet{Parsifal}

\RRtheme{\THSym} 

\URFuturs 

\begin{document}

\makeRR   

\sloppy

\tableofcontents


\section{Introduction}\label{sec:intro}

\subsection{The problem of the identity of proofs}

Whenever we study mathematical objects within a certain mathematical
theory, we normally know when two of these objects are considered to
be the same, i.e., are indistinguishable within the theory. For
example in group theory two groups are indistinguishable if they are
isomorphic, in topology two spaces are considered the same if they are
homeomorphic, and in graph theory we have the notion of graph
isomorphism. However, in proof theory the situation is
different. Although we are able to manipulate and transform proofs in
various ways, we have no satisfactory notion telling us when two
proofs are the same, in the sense that they use the same argument.
The main reason is the lack of understanding of the essence of a
proof, which in turn is caused by the bureaucracy involved in the
syntactic presentation of proofs.  It is therefore essential to find
new presentations of proofs that are ``syntax-free'', in the sense
that they avoid ``unnecessary bureaucracy''.

Finding such presentations is, of course, of utter importance for
logic and proof theory in its own sake. We can only speak of a real
\emph{theory of proofs}, if we are able to identify its objects. Apart
from that, the problem is of relevance not only for philosophy and
mathematics, but also for computer science, because many logical
systems permit a close correspondence between proofs and programs.  In
the view of this so-called \emph{Curry-Howard correspondence}, the
question of the identity of proofs becomes the question of the
identity of programs, i.e., when are two programs just different
presentations of the same algorithm. In other words, the fundamental
proof theoretical question on the nature of proofs is closely related
to the fundamental question on the nature of algorithms. In both cases
the problem is finding the right presentation that is able to avoid
unnecessary syntactic bureaucracy.

\subsection{Historical overview}

Interestingly, the problem of the identity of proofs can be considered
older than proof theory itself.  Already in 1900, when Hilbert was
preparing his celebrated lecture, he considered to add a 24th problem,
asking to develop a theory of proofs that allows to compare proofs and
provide criteria for judging which is the simplest proof of a
theorem.\footnote{This has been discovered by the historian R\"udiger
Thiele while studying the original notebooks of Hilbert
\cite{thiele:03}. The history of proof theory might have taken a
different development if Hilbert had included his 24th problem into
the lecture.}  But only in the early 1920s, when Hilbert launched his
famous program to give a formalization of mathematics in an axiomatic
form, together with a proof that this axiomatization is consistent,
formal proof theory as we know it today came into existence.

It was G\"odel \cite{godel:31}, who first considered formal proofs as
first-class citizens of the logical world, by assigning a unique
number to each of them. Even though this work destroyed Hilbert's
program, the idea of treating proofs as mathematical objects---in the
very same way as it is done with formulas---led eventually to our
modern understanding of formal proofs. Only a few years later, Gentzen
\cite{gentzen:34} provided the first structural analysis of formal
proofs and introduced methods of transforming them. His concept of cut
elimination is still the most central target of investigation in
structural proof theory. But even after Gentzen's work, the natural
question of asking for a notion of identity between proofs seemed
silly because there are only two trivial answers: two proofs are the
same if they prove the same formula, or, two proofs are the same if
they are syntactically equal.

In \cite{prawitz:71}, Prawitz proposed the notion of normalization in
natural deduction for determining the identity of proofs: two proofs
are the same (in the sense that they stand for the same argument) if
and only if they have the same normal form. The normalization process
in natural deduction corresponds to Gentzen's cut elimination in the
sequent calculus: All auxiliary lemmas are removed from the proof,
which then uses only material that does already appear in the formula
to be proved. However, normalization does not respect any complexity
issues because it is hyper-exponential. This means in particular that
all so-called speed-up theorems are ignored. In fact, it can happen
that a proof with cuts that fits a page is identified with a cut-free
proof that exhausts the size of the universe \cite{boolos:dont}.
Furthermore, it is probably quite difficult to convince a working
mathematician of the idea that a cunningly short proof using three
clever lemmas should be the same as an extraordinarily long proof that
does not use these lemmas, even if it can be obtained from the first
one via cut elimination. After all, the main part of the
mathematician's work consists of finding the right auxiliary lemmas in
the first place.

From the viewpoint of computer science the situation looks similar.
Through the Curry-Howard correspondence, formulas become types and
proofs become programs. The normalization of the proof corresponds to
the computation of the program. Translating Prawitz' idea into this
setting means that two programs are the same if and only if they have
the same input-output-behavior, which completely disregards any
reasonable complexity property.

Independently, Lambek \cite{lambek:68,lambek:69} proposed an idea for
identifying proofs that is based on commuting diagrams in categories
seen as deductive systems: two proofs are the same, if they constitute
the same morphism in the category. For propositional intuitionistic
logic on the one side, and Cartesian closed categories on the other
side, the two notions coincide. Similarly, by using *-autonomous
categories, one can make the two notions coincide for linear logic
(see Section~\ref{sec:star}).  But for classical logic, the logic of
our every day reasoning, neither notion has a commonly agreed
definition (see Section~\ref{sec:cl}).\footnote{See also
\cite{dosen:03} for a comparison of the two notions.}

Unfortunately, the problem of identifying proofs has not received much
attention since the work by Prawitz and Lambek.  Probably one of the
reasons is that the fundamental problem of the bureaucracy involved in
deductive systems (in which formal proofs are carried out) seemed to
be an insurmountable obstacle. In fact, the problem seems so
difficult, that it is widely considered to be ``only
philosophical''. However, behind the undeniable philosophical aspects,
the problem clearly is a mathematical one and deserves a rigorous
mathematical treatment. The developments in logic, proof theory, and
related fields within the last two decades suggest that it is
worthwhile to give it a new attack. In these notes we will see some
ideas in that direction.

\subsection{Proof nets}

The term ``proof net'' has been coined by Girard \cite{girard:87} for
his ``bureaucracy-free'' presentation of proofs in linear logic. He
used the term ``bureaucracy'' for the phenomenon of ``trivial rule
permutations'' in the sequent calculus that do not change the essence
of a proof.

In these notes, we will use the term ``proof net'' is a broader sense:
A proof net is a graph theoretical or geometric presentation that
captures the essence of a proof and is free of any syntactic
bureaucracy. Of course, for making this precise, it is necessary to
say what is meant by ``the essence of a proof'' and by ``syntactic
bureaucracy''. This is far from clear and is the target of most of
today's research efforts on the problem of the identity of proofs.

Although we use Girard's terminology, some of the ideas and technical
breakthroughs that we are going to present are much older. The proof
nets for unit-free multiplicative logic (that we use as playground for
introducing the theory) are essentially the \emph{coherence graphs} of
Eilenberg, Kelly, and Mac~Lane
\cite{eilenberg:kelly:fun-cal,kelly:maclane:71}. For the case of
classical logic, the same idea has been rediscovered under the name of
\emph{logical flow-graph} by Buss \cite{buss:91}.  A very simplified
version of proof nets for classical logic is based on Andrews'
\emph{matings} \cite{andrews:76} and Bibel's \emph{connections}
\cite{bibel:81}. If the proof presentations by Andrews and Bibel
(which are identical) are restricted to ``a linear version'' we can
(by using the right notion of correctness) get back Girard's
\emph{proof nets}. But these linear proof nets have a much better
proof theoretical behavior than the nonlinear (i.e., classical)
version. This is the reason why we start our survey with proof nets
for linear logic.

\section{Unit-free multiplicative linear logic}\label{sec:mll}

Unit-free multiplicative linear logic ($\ufMLL$) is a very simple
logic, that has nonetheless a well-developed theory of proof
nets.\footnote{One could even say the best developed theory of proof
nets among all logics that are out there...}  For this reason I will
use $\ufMLL$ to introduce the concept of proof nets.

\subsection{Sequent calculus for $\ufMLL$}\label{sec:sc-mll}

When we define a logic in terms of a deductive system, we have to do
two things. First, we have to define the set of well-formed formulas,
and second, we have to define the subset of derivable (or provable)
formulas, which is done via a set of inference rules\footnote{To be
precise, one should say \emph{axioms and inference rules}. But we
consider here axioms as special kinds of inference rules, namely,
those without premises.}. Here is the necessary data for $\ufMLL$: The
set of formulas is defined via
$$
\cF\grammareq\cA\mid\cA\lneg\mid\cF\lpar\cF\mid\cF\ltens\cF
$$ where $\cA=\set{a,b,c,\ldots}$ is a countable set of propositional
variables, and $\cA\lneg=\set{a\lneg,b\lneg,c\lneg,\ldots}$ are their
duals. In the following, we will call the elements of the set
$\cA\cup\cA\lneg$ \emph{atoms}.

The \emph{\rp{linear} negation} of a formula is defined
inductively via
$$
a\lnegneg=a
\qqquad
(A\ltens B)\lneg=B\lneg\lpar A\lneg
\qqquad
(A\lpar B)\lneg=B\lneg\ltens A\lneg
$$ Note that we invert the order of the arguments when we take the
negation of a binary connective. This is not strictly necessary (since
for the time being we stay in the commutative world) but will simplify
our life when it comes to drawing pictures of proof nets in later
sections.

Here is a set of inference rules for $\ufMLL$ given in the formalism of the
\emph{sequent calculus}:
\begin{equation}\label{eq:sc-mll}
  \begin{array}{c}
    \vcinf{\idr}{\sqn{A\lneg,A}}{}
    \qquad
    \vcinf{\exr}{\sqn{\Gamma,B,A,\Delta}}{\sqn{\Gamma,A,B,\Delta}}
    \qquad
    \vcinf{\lpar}{\sqn{\Gamma,A\lpar B,\Delta}}{\sqn{\Gamma,A,B,\Delta}}
    \\[\arrayskip]
    \vciinf{\ltens}{\sqn{\Gamma,A\ltens B,\Delta}}{
      \sqn{\Gamma,A}} {\sqn{B,\Delta}}
    \qquad
    \vciinf{\cutr}{\sqn{\Gamma,\Delta}}{
      \sqn{\Gamma,A}}{\sqn{A\lneg,\Delta}}
  \end{array}
\end{equation}
Note that the sequent calculus needs (apart from the concept of
formula) another kind of syntactic entity, called \emph{sequent}. Very
often these are just sets or multisets of formulas. But depending on
the logic in question, sequents can be more sophisticated structures
like lists or partial orders (or whatever) of formulas. For us,
throughout these lecture notes, sequents will be finite lists of
formulas, separated by a comma, and written with a $\sqn{}$ at the
beginning. Usually they are denoted by $\Gamma$ or $\Delta$.

\begin{example}
  If $A$ and $B$ are two different formulas, then
  $$
  \sqn{A,B}\qqquad\sqn{A,B,A}\qqquad\sqn{A,A,B}
  $$
  are three different sequents.
\end{example}

We say a sequent $\sqn{\Gamma}$ is \emph{derivable} (or
\emph{provable}) if there is a \emph{derivation} (or \emph{proof
tree}) with $\sqn{\Gamma}$ as conclusion. Defining this formally
precise tends to be messy. Since the basic concept should be familiar
for the reader, we content ourselves here by giving some examples.

\begin{example}
  The two sequents 
  $
  \sqn{a\lneg,a\ltens b\lneg, b\ltens c\lneg,c}
  $ and $
  \sqn{((a\lpar a\lneg)\ltens b)\lpar b\lneg}
  $ are provable:
  \begin{equation}\label{eq:exa1}
    \vcenter{\hbox{\hskip-2em$
      \ddernote{\ltens}{}{
      \sqn{a\lneg,a\ltens b\lneg, b\ltens c\lneg,c}} {
      \root{\idr}{\sqn{a\lneg,a}}{\leaf{}}}{
      \rroot{\ltens}{\sqn{b\lneg, b\ltens c\lneg,c}} {
	\root{\idr}{\sqn{b\lneg, b}}{\leaf{}}}{
      	\root{\idr}{\sqn{c\lneg, c}}{\leaf{}}}}
    \qqqquad
    \dernote{\lpar}{}{\sqn{((a\lpar a\lneg)\ltens b)\lpar b\lneg}}{
      \rroot{\ltens}{\sqn{(a\lpar a\lneg)\ltens b,b\lneg}}{
	\root{\lpar}{\sqn{a\lpar a\lneg}}{
	  \root{\idr}{\sqn{a\lneg,a}}{\leaf{}}}}{
	\root{\idr}{\sqn{b\lneg, b}}{\leaf{}}}}
    $}}
  \end{equation}
\end{example}
Of course it can happen that a sequent or a formula has more than one
proof. This is where things get interesting. At least for these course
notes. Here are four different proofs of the sequent $\sqn{a\lneg\lpar(a\ltens
a),a\lpar(a\lneg\ltens a\lneg)}$, three of them do not contain the
$\cutr$-rule: 
\begin{equation}\label{eq:3sq1}
  \vcdernote{\lpar}{}{
      \sqn{a\lneg\lpar(a\ltens a),a\lpar(a\lneg\ltens a\lneg)}}{
      \root{\exr}{
	\sqn{a\lneg\lpar(a\ltens a),a,a\lneg\ltens a\lneg}}{
	\rroot{\ltens}{
	  \sqn{a\lneg\lpar(a\ltens a),a\lneg\ltens a\lneg,a}}{
	  \root{\lpar}{\sqn{a\lneg\lpar(a\ltens a),a\lneg}}{
	    \rroot{\ltens}{\sqn{a\lneg,a\ltens a,a\lneg}}{
	      \root{\idr}{\sqn{a\lneg,a}}{
		\leaf{}}}{
	      \root{\idr}{\sqn{a,a\lneg}}{
		\leaf{}}}}}{
	  \root{\idr}{\sqn{a\lneg,a}}{
	    \leaf{}}}}}
\end{equation}
\begin{equation}\label{eq:3sq2}
  \vcdernote{\lpar}{}{
    \sqn{a\lneg\lpar(a\ltens a),a\lpar(a\lneg\ltens a\lneg)}}{
    \rroot{\ltens}{
      \sqn{a\lneg,a\ltens a,a\lpar(a\lneg\ltens a\lneg)}}{
      \root{\idr}{\sqn{a\lneg,a}}{
	\leaf{}}}{
      \root{\lpar}{\sqn{a,a\lpar(a\lneg\ltens a\lneg)}}{
	\root{\exr}{\sqn{a,a,a\lneg\ltens a\lneg}}{
	  \rroot{\ltens}{\sqn{a,a\lneg\ltens a\lneg,a}}{
	    \root{\idr}{\sqn{a,a\lneg}}{
	      \leaf{}}}{
	    \root{\idr}{\sqn{a\lneg,a}}{
	      \leaf{}}}}}}}
\end{equation}
\begin{equation}\label{eq:3sq3}
  \vcdernote{\exr}{}{
    \sqn{a\lneg\lpar(a\ltens a),a\lpar(a\lneg\ltens a\lneg)}}{
    \root{\lpar}{
      \sqn{a\lpar(a\lneg\ltens a\lneg),a\lneg\lpar(a\ltens a)}}{
      \rroot{\ltens}{
	\sqn{a,a\lneg\ltens a\lneg,a\lneg\lpar(a\ltens a)}}{
	\root{\idr}{\sqn{a,a\lneg}}{
	  \leaf{}}}{
	\root{\lpar}{\sqn{a\lneg,a\lneg\lpar(a\ltens a)}}{
	  \root{\exr}{\sqn{a\lneg,a\lneg,a\ltens a}}{
	    \rroot{\ltens}{\sqn{a\lneg,a\ltens a,a\lneg}}{
	      \root{\idr}{\sqn{a\lneg,a}}{
		\leaf{}}}{
	      \root{\idr}{\sqn{a,a\lneg}}{
		\leaf{}}}}}}}}
\end{equation}
and one does contain the
$\cutr$-rule: 
\begin{equation}\label{eq:3sq4}
  \vcenter{\scderifour}
\end{equation}
Are these proofs really different? Or are they just different ways of
writing down the same proof, i.e., they only seem different because of
the syntactic bureaucracy that the sequent calculus forces upon us?
In the following, we will try to give a sensible answer to this
question, and proof nets are a way to do so.

\begin{exercise}\label{ex:sc}
  Give at least two more sequent calculus proofs for the sequent
  $$
  \sqn{a\lneg\lpar(a\ltens a),a\lpar(a\lneg\ltens a\lneg)}
  \quadfs
  $$
\end{exercise}

\subsection{From sequent calculus to proof nets, 1st way 
(sequent calculus rule based)}\label{sec:sc-to-pn-1}

Although, morally, the concept of proof net should stand
independently from any deductive formalism, the proof nets introduced by
Girard very much depend on the sequent calculus. The ideology is the
following: 

\begin{ideology}\label{ideo:girard}
  A proof net is a graph in which every vertex represents
  an inference rule application in the corresponding sequent calculus
  proof, and every edge of the graph stands for a formula appearing in
  the proof. A sequent calculus proof with conclusion
  $\sqn{A_1,A_2,\ldots,A_n}$, written as
  \semiproofadjust
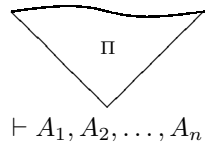
  $$
  \sqnbigderi{A_1,A_2,\ldots,A_n}{\Pi}
  $$ is translated into a proof net with conclusions
  $A_1,A_2,\ldots,A_n$, written as
  $$
  \vcgpnonlybox{\pi}{
    \gpnboxout1\qquad\gpnboxout2\qquad\gpnboxout3\qquad\gpnboxout4}{
    \gpnout1\qquad\gpnout2\qquad\gpnout3\qquad\gpnout4}{	
    \boxoutline{1}{A_1}
    \boxoutline{2}{A_3}
    \boxoutnoline{3}{\ldots}
    \boxoutline{4}{A_n}
  }
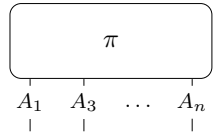
  $$ 
\end{ideology}

\begin{figure}[!t]
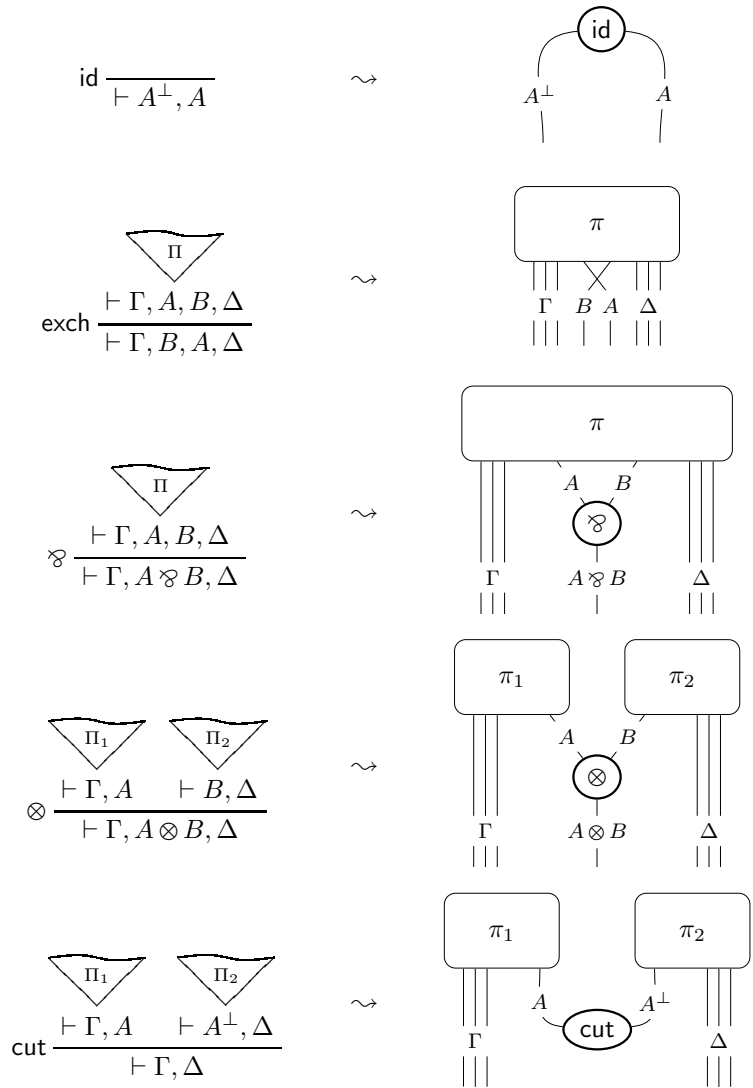

  \def\arrayskip{8ex}
  \vskip-2ex
  \begin{center}
    $$
    \begin{array}{c@{\qquad\leadsto\qquad}c}
      \vcinf{\idr}{\sqn{A\lneg,A}}{}
      &
      \idrulenet{A\lneg}{A}
      \\[\arrayskip]
      \vcinf{\exr}{\sqn{\Gamma,B,A,\Delta}}{
	\sqnsmallderi{\Gamma,A,B,\Delta}{\Pi}}
      &
       \vcgpn{
	 \begin{array}{c}
	  \gpnbox{\pi}{
	    \gpbbunchboxout{1}\quad\gpnboxout2\quad\gpnboxout3
	    \quad\gpbbunchboxout{4}}\\[\halfpnrowsep]
	  \fakebox{\gpnout{8}\quad\gpnout{9}}\\[\halfpnrowsep]
	    \gpbbunchout{1}\quad\gpnout2\quad\gpnout3
	    \quad\gpbbunchout{4}
 	 \end{array}
       }{
 	\mybunchboxoutline{.45}{1}{\Gamma}
 	\mybunchboxoutline{.45}{4}{\Delta}
 	\ncline{box2}{out9}
 	\ncline{box3}{out8}
	\myoutline{0.15}{out9}{out3}{A}
 	\myoutline{0.15}{out8}{out2}{B}
       }
      \\[\arrayskip]
      \vcinf{\lpar}{\sqn{\Gamma,A\lpar B,\Delta}}{
	\sqnsmallderi{\Gamma,A,B,\Delta}{\Pi}}
      &\parrulenet{\pi}{\Gamma}{A}{B}{\Delta}
      \\[\arrayskip]
      \vciinf{\ltens}{\sqn{\Gamma,A\ltens B,\Delta}}{
	\sqnsmallderi{\Gamma,A}{\Pi_1}} {\sqnsmallderi{B,\Delta}{\Pi_2}}
      &\tensorrulenet{\pi_1}{\Gamma}{A}{\pi_2}{B}{\Delta}
      \\[\arrayskip]
      \vciinf{\cutr}{\sqn{\Gamma,\Delta}}{
	\sqnsmallderi{\Gamma,A}{\Pi_1}}{\sqnsmallderi{A\lneg,\Delta}{\Pi_2}}
      &
      \cutrulenet{\pi_1}{\Gamma}{A}{\pi_2}{A\lneg}{\Delta}
    \end{array}  
    $$
    \caption{From sequent calculus to proof nets (sequent calculus
    rule driven)}
    \label{fig:sc-to-pn-1}
  \end{center}
\end{figure}

This is done inductively, rule instance by rule instance, as shown in
Figure~\ref{fig:sc-to-pn-1}. Note that the $\exr$-rule does not
exactly follow the ideology.

Let us see, what happens if we apply this translation to our four
different proofs \eqref{eq:3sq1}--\eqref{eq:3sq4}: The first two, i.e., 
\eqref{eq:3sq1} and \eqref{eq:3sq2}, both
yield the same proof net, namely
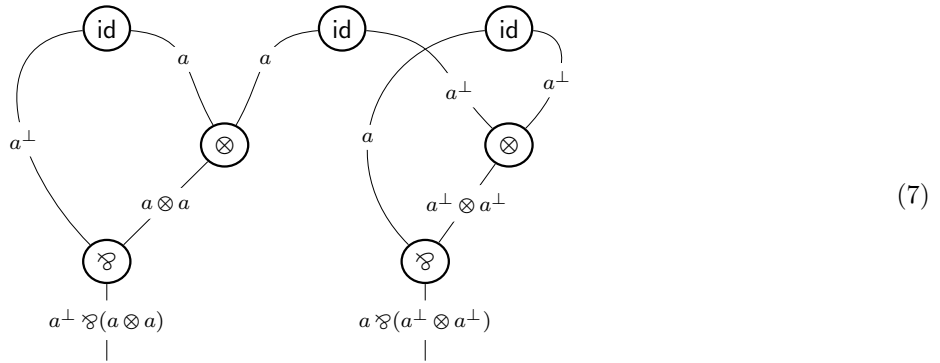
\begin{equation}\label{eq:3sq-pn1}
  \vcgpn{
    & \gpnid{1}&\quad&         &\quad&\gpnid{2}&      &\gpnid3\\
    &          &     &\gpntens1&     &         &      &\gpntens2\\
    &\gpnpar1  &     &         &     &         &\gpnpar2\\
    &\gpnout1  &     &         &     &         &\gpnout2
  }{
    \idleftline{lpar1}{id1}{a\lneg}
    \treeline{lpar1}{ltens1}{a\ltens a}
    \treeoutline{lpar1}{out1}{a\lneg\lpar(a\ltens a)}
    \gpncurve{ltens1}{id1}{120}{-5}{a}
    \gpncurve{ltens1}{id2}{60}{185}{a}
    \gpnmycurve{ltens2}{id2}{135}{-5}{.25}{a\lneg}
    \idrightline{ltens2}{id3}{a\lneg}
    \idleftline{lpar2}{id3}{a}
    \treeline{lpar2}{ltens2}{a\lneg\ltens a\lneg}
    \treeoutline{lpar2}{out2}{a\lpar(a\lneg\ltens a\lneg)}
  }
\end{equation}
The proof in \eqref{eq:3sq3} yields a different proof net:
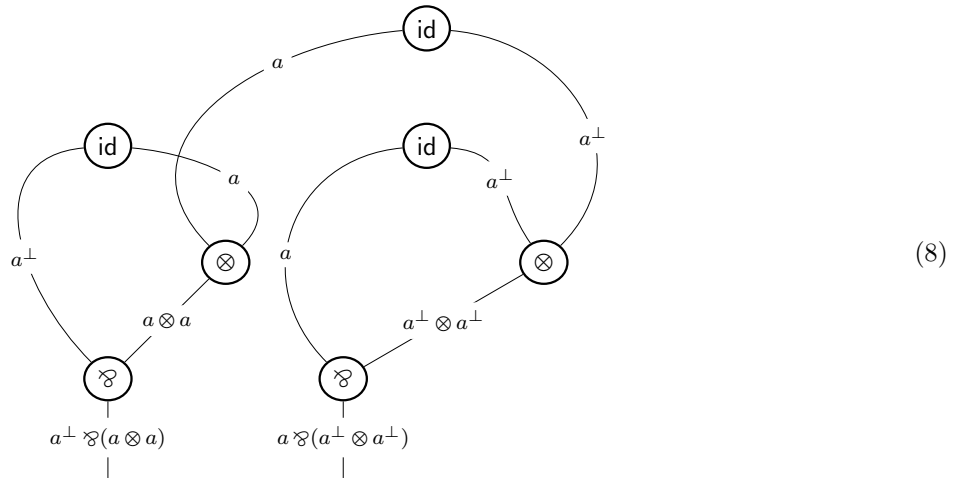
\begin{equation}\label{eq:3sq-pn3}
  \vcgpn{
    &        &     &         &     &     &\gpnid3\\
    &\gpnid1 &\quad&         &     &\quad&\gpnid2\\
    &        &     &\gpntens1&     &     &      &\quad&\gpntens2\\
    &\gpnpar1&     &         &     &\gpnpar2\\
    &\gpnout1&     &         &     &\gpnout2
  }{
    \idleftline{lpar1}{id1}{a\lneg}
    \treeline{lpar1}{ltens1}{a\ltens a}
    \treeoutline{lpar1}{out1}{a\lneg\lpar(a\ltens a)}
    \gpncurve{ltens1}{id1}{45}{-5}{a}
    \gpnmycurve{ltens1}{id3}{135}{185}{.7}{a}
    \gpncurve{ltens2}{id2}{125}{-5}{a\lneg}
    \idrightline{ltens2}{id3}{a\lneg}
    \idleftline{lpar2}{id2}{a}
    \treeline{lpar2}{ltens2}{a\lneg\ltens a\lneg}
    \treeoutline{lpar2}{out2}{a\lpar(a\lneg\ltens a\lneg)}
  }
\end{equation}
And for the proof in \eqref{eq:3sq4}, we get
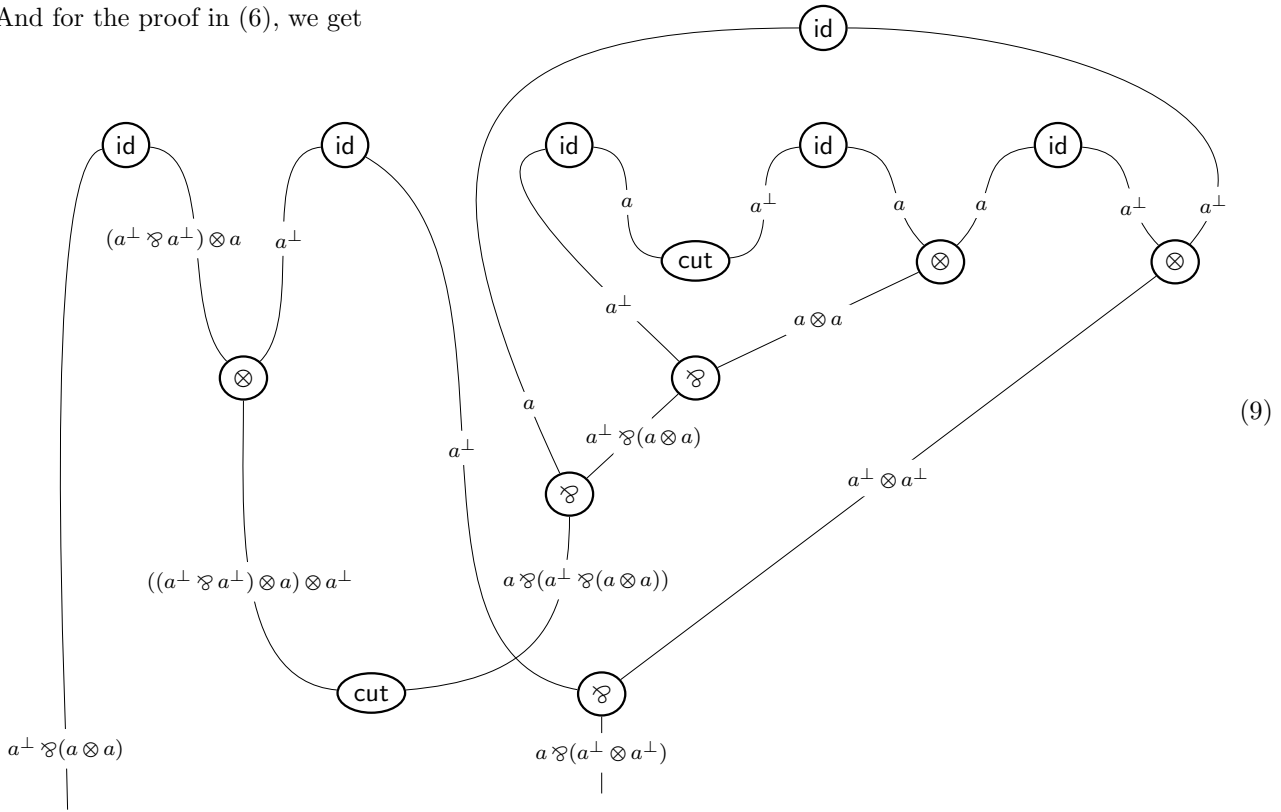
\proofadjust
\begin{equation}\label{eq:3sq-pn4}
  \qlapm{
  \vcgpn{
    &&&&&&&&&&&&&\gpnid6\\
    \quad&\gpnid1&&&&\qlapm{\gpnid2\hskip2em}&&&&
    \clapm{\hskip2em\gpnid3}&&&&\gpnid4&&&&\gpnid5\\
    &&&         &&&&&&&&\gpncut1&&&&\gpntens3&&&&\gpntens2\\
    &&&\gpntens1&&&&&&&&\gpnpar3\\
    &&&         &&&&&&\clapm{\hskip2em\gpnpar4}\\ \\
    &&&         &&\gpncut2&&&&&        \gpnpar2\\
    \gpnout1&&& &&&&&        &&        \gpnout2
  }{
    \gpnmycurveh{out1}{id1}{90}{190}{.2}{.15}{a\lneg\lpar(a\ltens a)}
    \gpncurveh{ltens1}{id1}{135}{-5}{.6}{
      \qlapm{(a\lneg\lpar a\lneg)\ltens a\qquad}}
    \gpncurveh{ltens1}{id2}{45}{185}{.6}{\qlapm{\;\;a\lneg}}
    \gpncurveh{lpar2}{id2}{170}{-30}{.7}{a\lneg}
    \gpncurveh{ltens1}{cut2}{-90}{175}{.7}{
      ((a\lneg\lpar a\lneg)\ltens a)\ltens a\lneg}
    \gpnmycurve{lpar4}{cut2}{-90}{5}{.2}{
      \qlapm{\qquad a\lpar(a\lneg\lpar(a\ltens a))}}
    \gpnmycurveh{lpar4}{id6}{115}{180}{1.4}{.075}{a}
    \gpnmycurveh{ltens2}{id6}{50}{0}{.9}{.1}{a\lneg}
    \gpnmycurveh{lpar3}{id3}{135}{190}{.7}{.25}{a\lneg}
    \gpncurve{cut1}{id3}{175}{-5}{a}
    \gpncurve{cut1}{id4}{5}{185}{a\lneg}
    \gpnmycurve{ltens3}{id4}{135}{-5}{.3}{a}
    \gpnmycurve{ltens3}{id5}{45}{185}{.3}{a}
    \gpnmycurve{ltens2}{id5}{135}{-5}{.3}{a\lneg}
    \treeline{lpar3}{ltens3}{a\ltens a}
    \treeline{lpar4}{lpar3}{\qlapm{\quad a\lneg\lpar(a\ltens a)}}
    \treeline{lpar2}{ltens2}{a\lneg\ltens a\lneg}
    \treeoutline{lpar2}{out2}{a\lpar(a\lneg\ltens a\lneg)}
  }}
\end{equation}

The proof nets for \eqref{eq:3sq1} and \eqref{eq:3sq2} are the same,
and the ones for \eqref{eq:3sq3} and \eqref{eq:3sq4} are
different. The big question is now:

\begin{para}{Big Question:}
  Is it reasonable to conclude that the two proofs in \eqref{eq:3sq1}
  and \eqref{eq:3sq2} are the same, while the one in \eqref{eq:3sq3}
  is different, and the one in \eqref{eq:3sq4} is even more different?
\end{para}

For finding an answer, consider the following two (partial) sequent
calculus proofs.
\proofadjust
\begin{equation}\label{eq:rule-permut}
  \vcdernote{\lpar}{}{\sqn{\Gamma,A\ltens B,\Delta,C\lpar D,\Sigma}}{
    \rroot{\ltens}{\sqn{\Gamma,A\ltens B,\Delta,C,D,\Sigma}}{
      \leaf{\sqnsmallderi{\Gamma,A}{\Pi_1}}}{
      \leaf{\sqnsmallderi{B,\Delta,C,D,\Sigma}{\Pi_2}}}}
  \qquad
  \mbox{vs.}
  \qquad
  \vcddernote{\ltens}{}{\sqn{\Gamma,A\ltens B,\Delta,C\lpar D,\Sigma}}{
    \leaf{\sqnsmallderi{\Gamma,A}{\Pi_1}}}{
    \root{\lpar}{\sqn{B,\Delta,C\lpar D,\Sigma}}{
      \leaf{\sqnsmallderi{B,\Delta,C,D,\Sigma}{\Pi_2}}}}
\end{equation}
It is certainly reasonable to say that the two proofs in
\eqref{eq:rule-permut} are essentially the same---whether we apply
first the $\ltens$-rule or first the $\lpar$-rule does not matter in
this situation.

The phenomenon shown in \eqref{eq:rule-permut} is called a
\emph{trivial rule permutation}. There are clearly more of them, and
we will not give a complete list here. The point to note here is that
these trivial rule permutation are a type of bureaucracy, which is
typical for the sequent calculus, and which was one of Girard's
original motivations for the introduction of proof nets. In fact,
there is the following theorem:

\begin{theorem}\label{thm:trivial}
  Two sequent calculus proofs using the rules in \eqref{eq:sc-mll}
  translate into the same proof net if and only if they can be
  transformed into each other via trivial rule permutations.
\end{theorem}

We will not give a proof here because it won't give any new
insights. But to illustrate the main point, consider the following
three derivations\doubleproofadjust
$$
\vcdernote{\exr}{}{\sqn{B,B\lpar C}}{
  \root{\lpar}{\sqn{B\lpar C,B}}{
    \leaf{\sqnsmallderi{B,C,B}{}}}}
\qqqqqquad
\vcdernote{\lpar}{}{\sqn{B,B\lpar C}}{
  \root{\exr}{\sqn{B,B,C}}{
    \leaf{\sqnsmallderi{B,C,B}{}}}}
\qqqqqquad
\vcdernote{\lpar}{}{\sqn{B,B\lpar C}}{
  \root{\exr}{\sqn{B,B,C}}{
    \root{\exr}{\sqn{B,B,C}}{
      \leaf{\sqnsmallderi{B,C,B}{}}}}}
$$ where $B$ and $C$ are arbitrary formulae, and the second instance
of $\exr$ in the rightmost derivation exchanges the two occurrences of
$B$. Note that the leftmost and the middle derivation both consist of
one instance of the $\lpar$-rule and one instance of $\exr$, but are
{\bf not} a trivial rule permutation.  When we ``trivialy'' permute
the $\lpar$-rule and the $\exr$-rule in the leftmost example, then we
obtain the rightmost derivation, which contains one $\lpar$-rule and
two exchanges.

Now the reader is invited to do the
following exercise:

\begin{exercise}
  Find a sequence of trivial rule permutations that transforms the
  proof in \eqref{eq:3sq1} into the one in \eqref{eq:3sq2}. Convince
  yourselves that it is impossible to find a series of rule
  permutations that converts the proof in \eqref{eq:3sq2} into the
  one in \eqref{eq:3sq3}.
\end{exercise}

\begin{exercise}
  Give the proof nets for the two sequent proofs in \eqref{eq:exa1}.
\end{exercise}

\begin{exercise}
  Give the proof nets for the sequent proofs that you found in
  Exercise~\ref{ex:sc}, and compare them with \eqref{eq:3sq-pn1},
  \eqref{eq:3sq-pn3}, and \eqref{eq:3sq-pn4}.
\end{exercise}

\subsection{From sequent calculus to proof nets, 2nd way 
(coherence graph based)}\label{sec:sc-to-pn-2}

Let us now discuss a second method for obtaining a proof net from a sequent
calculus proof. Here the ideology is: 

\begin{ideology}\label{ideo:coherence}
  A proof net consists of the formula tree/sequent forest of the
  conclusion of the proof, together with some additional graph
  structure capturing the ``essence'' of the proof.
\end{ideology}

\begin{figure}[!t]
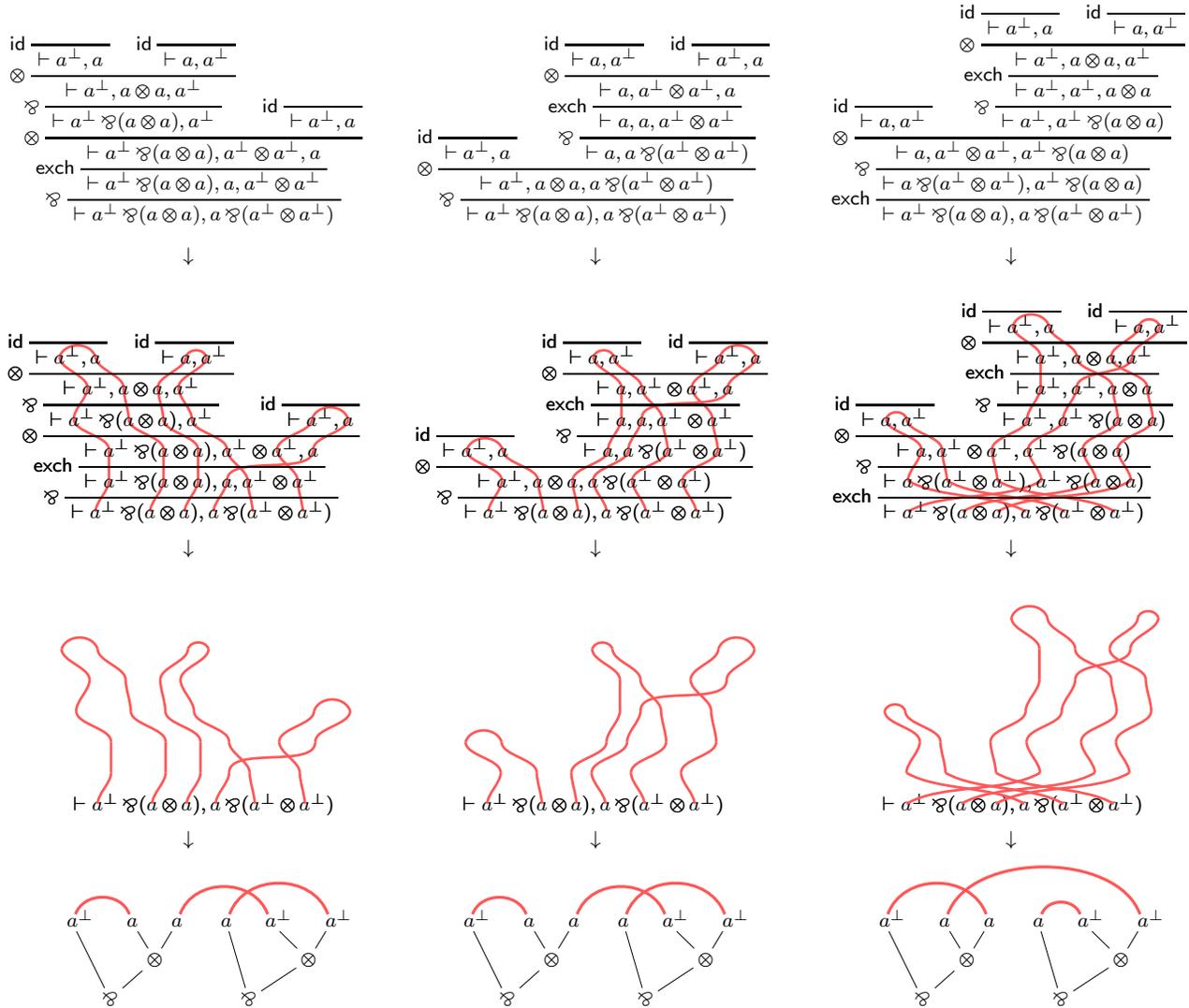

  \footnotesize
  \def\arrayskip{2ex}
  \begin{center}
    $$
    \qlapm{\begin{array}{c@{\hskip-1.5em}c@{\hskip-1.5em}c}
      \scderione
      &
      \scderitwo
      &
      \scderithree
      \\[\arrayskip]
      \downarrow&\downarrow&\downarrow
      \\[\arrayskip]
      \tpathderivation{\scderione}{\pathscderione}
      &
      \tpathderivation{\scderitwo}{\pathscderitwo}
      &
      \tpathderivation{\scderithree}{\pathscderithree}
      \\[\arrayskip]
      \downarrow&\downarrow&\downarrow
      \\[\arrayskip]
      \tpathonlyderivation{\scderione}{\pathscderione}
      &
      \tpathonlyderivation{\scderitwo}{\pathscderitwo}
      &
      \tpathonlyderivation{\scderithree}{\pathscderithree}
      \\[\arrayskip]
      \downarrow&\downarrow&\downarrow
      \\[\arrayskip]
      \\[\arrayskip]
      \qquad\pathpnone
      &
      \quad\pathpnone
      &
      \quad\pathpnthree
    \end{array}}
    $$
    \caption{From sequent calculus to proof nets via coherence graphs}
    \label{fig:sc-to-pn-2}
  \end{center}
\end{figure}

It turns out that for $\ufMLL$ the ``essence'' of a proof is captured by the
axiom links. More precisely, the proof net is obtained by drawing the
``flow-graph'' (or ``coherence-graph'') through the sequent calculus proof.
This means that we trace all atom occurrences through the proof.  The idea is
quite simple, but again, the formal definitions tend to be messy.  For these
lecture notes, I decided not to give the detailed definitions but to show the
idea via examples. Figure~\ref{fig:sc-to-pn-2} shows how it is done for the
examples \eqref{eq:3sq1},\eqref{eq:3sq2}, and \eqref{eq:3sq3}.

\begin{para}{Trivial Observation}\label{obs:exchange}
  The flow-graphs drawn inside the proofs have crossings exactly at
  those places where the exchange rule is applied.
\end{para}

\begin{figure}[!t]
  \footnotesize
  \def\arrayskip{2ex}
  \def\thefirstcomma{\qqqqquad\qquad,\qqqqquad\qquad}
  \def\thesecondcomma{\hskip1.5em,\hskip1.5em}
  \begin{center}
    \vskip-1.9pt 
    $$
    \qlapm{\begin{array}{c}
      \scderifour
      \\[\arrayskip]
      \downarrow
      \\[-4ex]
      \tpathderivation{\scderifour}{\pathscderifour}
      \\[\arrayskip]
      \downarrow
      \\[-4ex]
      \tpathonlyderivation{\scderifour}{\pathscderifour}
      \\[\arrayskip]
      \downarrow
      \\[\arrayskip]
      \\[\arrayskip]
      \pathpnone
    \end{array}}
    $$
    \caption{From sequent calculus to proof nets via coherence graphs}
    \label{fig:sc-to-pn-2-cut}
  \end{center}
\end{figure}

\begin{figure}[!t]
  \footnotesize
  \def\arrayskip{2.2ex}
  \def\thefirstcomma{\qqqqquad\qquad,\qqqqquad\qquad}
  \def\thesecondcomma{\hskip1.5em,\hskip1.5em}
  \def\theblack{\black}
  \def\thecutsinpathscderifour{}
  \begin{center}
    $$
    \qlapm{\begin{array}{c}
      \tpathderivation{\scderifour}{\pathscderifour}
      \\[\arrayskip]
      \downarrow
      \\[-4ex]
      \tpathonlyderivation{\scderifour}{\pathscderifour}
      \\[\arrayskip]
      \downarrow
      \\[\arrayskip]
      \\[\arrayskip]
      \pathpnfour
    \end{array}}
    $$
    \caption{From sequent calculus to proof nets with cuts via
    coherence graphs}
    \label{fig:sc-to-pn-2-cutcut}
  \end{center}
\end{figure}

Note that if the $\idr$-rule is applied only to atoms, and there is no
$\cutr$-rule present, then we get (up to some trivial change of
notation) \emph{exactly the same proof nets} as with the first
method. The reason is that in $\ufMLL$ there is a one-to-one
correspondence between the binary connectives $\lpar$ and $\ltens$
appearing in the sequent and the instances of the inference rules for
$\lpar$ and $\ltens$ appearing in the proof. Further, every atom
occurrence in the sequent is killed by an instance of the $\idr$-rule.

\begin{exercise}
  Convince yourselves that indeed the two methods always yield the
  same result for cut-free proofs with only atomic instances of
  identity. Draw the flow graphs for the examples in \eqref{eq:exa1}
  and the sequent calculus proofs that you found in
  Exercise~\ref{ex:sc}.
\end{exercise}

In Figure~\ref{fig:sc-to-pn-2-cut} we convert the example in
\eqref{eq:3sq4} into a proof net via the flow-graph method. Note that
this time the result is different from the proof net in
\eqref{eq:3sq-pn4}. There are two reasons. First, the non-atomic
instance of the $\idr$-rule, and second, the presence of the
$\cutr$-rule.

The non-atomic identity rule is not a problem because
an important fact about the sequent calculus system in
\eqref{eq:sc-mll} is that the $\idr$-rule can be replaced by its
atomic version without changing derivability:
  $$
  \vcinf{\idr}{\sqn{A\lneg,A}}{}
  \qquad\leadsto\qquad
  \vcinf{\atidr}{\sqn{a\lneg,a}}{}
  $$
This is done inductively by systematically replacing 
\begin{equation}\label{eq:sc-id-to-atom}
  \vcinf{\idr}{\sqn{A\lneg\lpar B\lneg,B\ltens A}}{}
  \qquad\mbox{by}\qquad
  \vcdernote{\lpar}{}{\sqn{A\lneg\lpar B\lneg,B\ltens A}}{
    \root{\exr}{\sqn{A\lneg, B\lneg,B\ltens A}}{
      \root{\exr}{\sqn{B\lneg,A\lneg, B\ltens A}}{
	\rroot{\ltens}{\sqn{B\lneg,B\ltens A,A\lneg}}{
	  \root{\idr}{\sqn{B\lneg,B}}{\leaf{}}}{
	  \root{\idr}{\sqn{A,A\lneg}}{\leaf{}}}}}}
\end{equation}

\begin{exercise}\label{ex:sc-to-pn-atid}
  Reduce in \eqref{eq:3sq4} the leftmost instance of $\idr$ to atomic
  version. And draw the proof net according to the method in
  Figure~\ref{fig:sc-to-pn-1}. What does change compared to the net in
  \eqref{eq:3sq-pn4}?
\end{exercise}

For dealing with cuts (without forgetting them!), we can prevent the
flow-graph from flowing through the cut, i.e., by keeping the
information that there is a cut. What is meant by this is shown in
Figure~\ref{fig:sc-to-pn-2-cutcut}.

\begin{exercise}
  Compare the net obtained in Figure~\ref{fig:sc-to-pn-2-cutcut} with
  your result of Exercise~\ref{ex:sc-to-pn-atid}.
\end{exercise}

Now, we indeed get the same result with both methods, and it might
seem foolish to emphasize the different nature of the two methods if
they yield the same notion of proof net. The point to make here is
that this is the case only for $\ufMLL$, which is a very fortunate
coincidence. For any other logic, which is more sophisticated, like
classical logic or larger fragments of linear logic, the two methods
yield different notions of proof nets. We will come back to this in
later sections when we discuss these logics.

\subsection{From deep inference to proof nets}\label{sec:cos-to-pn}

\begin{figure}[t]
  \begin{center}
    $$
    \begin{array}{c@{\qqquad}c}
      \vcinf{\ird}{A\lneg\lpar A}{} 
      &
      \vcinf{\iru}{}{A\ltens A\lneg} 
      \\[\arrayskip]
      \vcinf{\ird}{S\cons{(A\lneg\lpar A)\ltens B}}{S\cons{B}} 
      &
      \vcinf{\iru}{S\cons{B}}{S\cons{B\lpar(A\ltens A\lneg)}} 
      \\[\arrayskip]
      \vcinf{\comrd}{S\cons{B\lpar A}}{S\cons{A\lpar B}}
      &
      \vcinf{\comru}{S\cons{B\ltens A}}{S\cons{A\ltens B}}
      \\[\arrayskip]
      \vcinf{\assrd}{S\cons{(A\lpar B)\lpar C}}{S\cons{A\lpar(B\lpar C)}}
      &
      \vcinf{\assru}{S\cons{(A\ltens B)\ltens C}}{S\cons{A\ltens (B\ltens C)}}
      \\[\arrayskip]
      \multicolumn{2}{c}{
	\vcinf{\swir}{S\cons{(A\ltens B)\lpar C}}{S\cons{A\ltens(B\lpar C)}}
      }
    \end{array}
    $$
    \caption{A system for $\ufMLL$ in the calculus of structures}
    \label{fig:cos-mll}
  \end{center}
\end{figure}

The flow graph method has the advantage of being independent from the
formalism that is used for describing the deductive system for the logic. We
will now repeat exactly the same exercise we did for the sequent calculus
system for $\ufMLL$ in the previous section. But this time we start from a
different deductive system for $\ufMLL$, which is given in the formalism of
calculus of structures. It is shown in Figure~\ref{fig:cos-mll}, where
$S\conhole$ stands for an arbitrary (positive) formula context. Because of the
possibility of applying inference rules deep inside any context, the name
``deep inference'' is used.

Because of this possibility and because we do not have units in
the system, we need two variants of the $\ird$-rule (doing the same job
as the $\idr$-rule in the sequent calculus). Similarly, the cut, i.e., the
rule $\iru$ comes in two versions. This also means, that in principle a
derivation could have nothing as conclusion. Then we do not have a proof, but
a refutation. This leads to an important property of the system in
Figure~\ref{fig:cos-mll}, namely its up-down symmetry. For every rule there is
a dual co-rule, which is obtained by negating and exchanging premise and
conclusion. This flipping around can then be done also for whole
derivations. A derivation from $A$ to $B$ becomes a derivation from $B\lneg$
to $A\lneg$, and a proof of a formula $A$ becomes a refutation of $A\lneg$.

The $\swir$-rule (called \emph{switch}) is dual to itself. With the
help of commutativity and associativity, i.e., the rule $\comrd$,
$\comru$, $\assrd$, and $\assru$ we can obtain the following variants
that we also call \emph{switch}:
\begin{equation}\label{eq:switch}
  \vcinf{\swir}{S\cons{A\lpar(B\ltens C)}}{S\cons{(A\lpar B)\ltens C}}
  \qqquad
  \vcinf{\swir}{S\cons{A\lpar(B\ltens C)}}{S\cons{B\ltens(A\lpar C)}}
  \qqquad
  \vcinf{\swir}{S\cons{(A\ltens B)\lpar C}}{S\cons{(A\lpar C)\ltens B}}
\end{equation}
Similarly, with the help of $\comrd$ and $\comru$, we can get two other
versions if $\ird$ and $\iru$:
\begin{equation}\label{eq:ir}
  \vcinf{\ird}{S\cons{B \ltens(A\lneg\lpar A)}}{S\cons{B}} 
  \qquand
  \vcinf{\iru}{S\cons{B}}{S\cons{(A\ltens A\lneg)\lpar B}} 
\end{equation}

Here are three examples of derivations in the calculus of structures: 
\begin{equation}\label{eq:exa-cos}
  \vcdernote{\ird}{}{((a\lpar a\lneg)\ltens b)\lpar b\lneg}{
    \root{\ird}{b\lpar b\lneg}{
      \leaf{}}}
  \qqqquad
  \vcenter{\exacostwosideda}
  \qqqquad
  \vcenter{\exacostwosidedb}
\end{equation}
The last two are dual to each other.

In order to go on, we have to make sure that the deductive system
shown in Figure~\ref{fig:cos-mll} speaks about the same logic, as the
one shown in \eqref{eq:sc-mll}. This does the following theorem:

\begin{theorem}\label{thm:sq-cos}
  Let $A$ be an $\ufMLL$ formula. There is a
  sequent calculus proof of $A$, 
  denoted by
  $$
  \sqnbigderi{A}{\Pi}
  $$ in the system shown in \eqref{eq:sc-mll},
  if and only if there is a proof of $A$ in the calculus of
  structures 
  denoted by
  $$
  \strpr{}{\Pit}{A}
  $$ in the system shown in Figure~\ref{fig:cos-mll}.
\end{theorem}

\begin{proof}
  For going from sequent calculus to calculus of structures, we show
  that every rule in \eqref{eq:sc-mll} can be simulated by the rules
  in Figure~\ref{fig:cos-mll}. The only interesting cases are $\ltens$
  and $\cutr$, which are simulated as follows:
  $$
  \vcdernote{\swir}{}{\Gamma\lpar(A\ltens B)\lpar\Delta}{
    \root{\swir}{\Gamma\lpar(A\ltens(B\lpar\Delta))}{
      \leaf{(\Gamma\lpar A)\ltens(B\lpar\Delta)}}}
  \qquand
  \vcdernote{\iru}{}{\Gamma\lpar\Delta}{
    \root{\swir}{\Gamma\lpar(A\ltens A\lneg)\lpar\Delta}{
    \root{\swir}{\Gamma\lpar(A\ltens(A\lneg\lpar\Delta))}{
      \leaf{(\Gamma\lpar A)\ltens(A\lneg\lpar\Delta)}}}}
  $$ For the rules $\lpar$ and $\exr$ it is trivial. Now we can mimic
  the sequent calculus proof such that different branches in the proof
  tree, say
  $$
  \sqn{\Gamma_1}\qquad \sqn{\Gamma_2}\qquad\ldots\qquad\sqn{\Gamma_n}
  $$
  are kept together in a single formula 
  $$
  \Gamma_1\ltens\Gamma_2\ltens\cdots\ltens\Gamma_n
  $$ In the end, we have for every instance of $\idr$ in the sequent
  proof tree and instance of $\ird$ in the calculus of structures
  proof.

  For the other direction, we first have to show for every rule 
  $$
  \vcinf{\ruler}{S\cons{B}}{S\cons{A}}
  $$ in Figure~\ref{fig:cos-mll}, the sequent $\sqn{A\lneg,B}$ is
  provable with the rules in~\eqref{eq:sc-mll}. Here we do it only for
  the switch rule:\proofadjust
  $$
  \vcdernote{\lpar}{}{
    \sqn{(C\lneg\ltens B\lneg)\lpar A\lneg,(A\ltens B)\lpar C}}{
    \root{\exr}{\sqn{(C\lneg\ltens B\lneg)\lpar A\lneg,A\ltens B,C}}{
      \root{\exr}{\sqn{(C\lneg\ltens B\lneg)\lpar A\lneg,C,A\ltens B}}{
	\root{\lpar}{\sqn{C,(C\lneg\ltens B\lneg)\lpar A\lneg,A\ltens B}}{
	  \rroot{\ltens}{\sqn{C,C\lneg\ltens B\lneg,A\lneg,A\ltens B}}{
	    \root{\idr}{\sqn{C,C\lneg}}{
	      \leaf{}}}{
	    \root{\exr}{\sqn{B\lneg,A\lneg,A\ltens B}}{
	      \root{\exr}{\sqn{A\lneg,B\lneg,A\ltens B}}{
		\rroot{\ltens}{\sqn{A\lneg,A\ltens B,B\lneg}}{
		  \root{\idr}{\sqn{A\lneg,A}}{
		    \leaf{}}}{
		  \root{\idr}{\sqn{B,B\lneg}}{
		    \leaf{}}}}}}}}}}
  $$ for the other rules it is similar. Now we show that in fact the
  sequent $\sqn{S\cons{A}\lneg,S\cons{B}}$ (for every positive context
  $S\conhole$) can be proved with the rules in~\eqref{eq:sc-mll}. This
  is done by structural induction on $S\conhole$. If
  $S\conhole=\conhole$, we are done. If $S\conhole=C\ltens S'\conhole$
  for some formula $C$ and some (smaller) context $S'\conhole$, then we
  have\proofadjust
  $$
  \vcdernote{\lpar}{}{
    \sqn{S'\cons{A}\lneg\lpar C\lneg,C\ltens S'\cons{B}}}{
    \root{\exr}{\sqn{S'\cons{A}\lneg,C\lneg,C\ltens S'\cons{B}}}{
      \root{\exr}{\sqn{C\lneg,S'\cons{A}\lneg,C\ltens S'\cons{B}}}{
	\rroot{\ltens}{\sqn{C\lneg,C\ltens S'\cons{B},S'\cons{A}\lneg}}{
	  \root{\idr}{\sqn{C\lneg,C}}{
	    \leaf{}}}{
	  \leaf{\sqnsmallderi{S'\cons{B},S'\cons{A}\lneg}{\Pi'}}}}}}
  $$ where $\Pi'$ exists by induction hypothesis. The other cases,
  i.e., $S\conhole=C\lpar S'\conhole$, $S\conhole=S'\conhole\ltens C$,
  and $S\conhole=S'\conhole\lpar C$, are similar.  Now we are ready to
  simulate the whole derivation. We proceed by induction om the length
  of $\Pit$ (i.e., the number of rule instances). If the length is
  $1$, then it must be an instance of $\ird$, and we get a sequent
  calculus proof with $\idr$. Now assume the length of $\Pit$ is $>1$,
  i.e., $\Pit$ is of shape
  $$
  \dernote{\ruler}{}{S\cons{B}}{
    \stempr{}{\Pit'}{S\cons{A}}}
  $$
  Then we can build\proofadjust
  $$
  \vcddernote{\cutr}{}{\sqn{S\cons{B}}}{
    \leaf{\sqnsmallderi{S\cons{A}}{\Pi'}}}{
    \leaf{\sqnsmallderi{S\cons{A}\lneg,S\cons{B}}{\Pi''}}}
  $$ where $\Pi'$ exists by induction hypothesis and $\Pi''$ by what
  has been said above.
  \qed
\end{proof}

\begin{figure}[t]
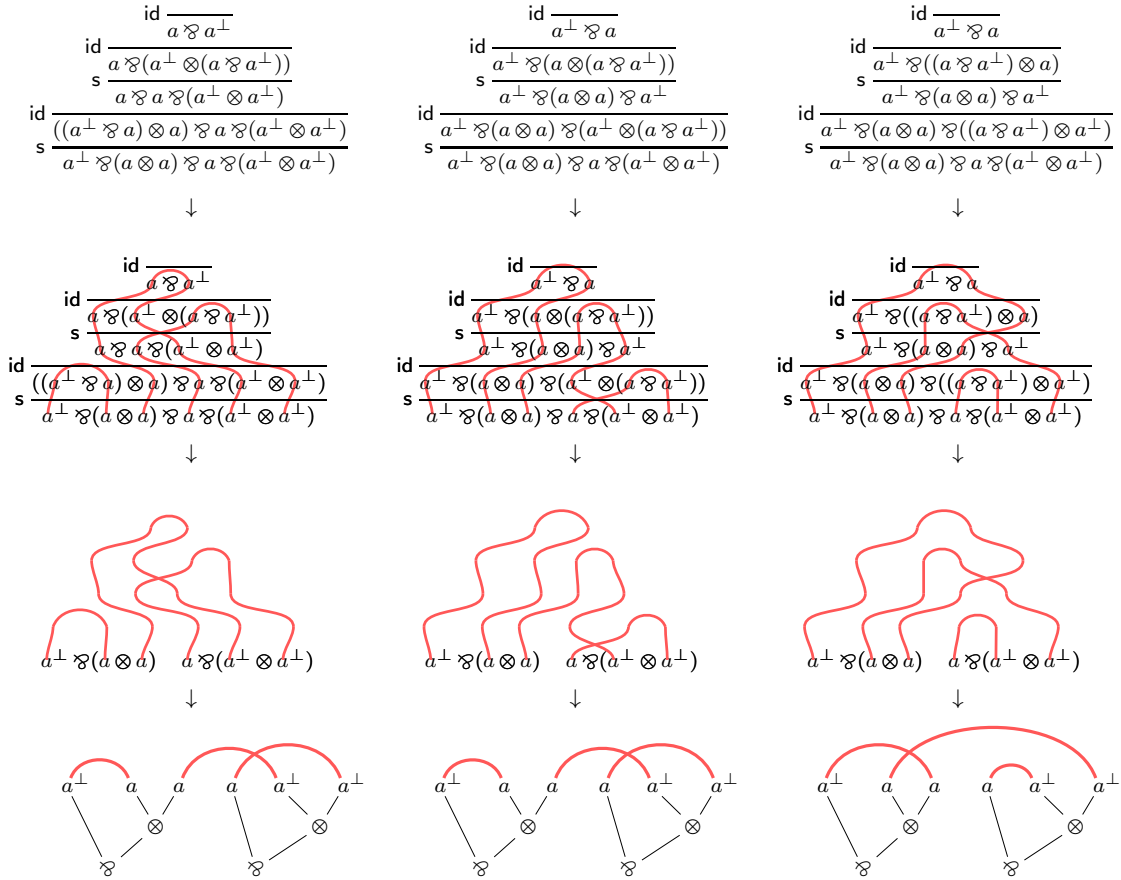

  \footnotesize
  \def\arrayskip{2ex}
  \begin{center}
    $$
    \qlapm{\begin{array}{c@{\hskip.5em}c@{\hskip.5em}c}
      \cosderione
      &
      \cosderitwo
      &
      \cosderithree
      \\[\arrayskip]
      \downarrow&\downarrow&\downarrow
      \\[\arrayskip]
      \tpathderivation{\cosderione}{\pathcosderione}
      &
      \tpathderivation{\cosderitwo}{\pathcosderitwo}
      &
      \tpathderivation{\cosderithree}{\pathcosderithree}
      \\[\arrayskip]
      \downarrow&\downarrow&\downarrow
      \\[\arrayskip]
      \tpathonlyderivation{\cosderione}{\pathcosderione}
      &
      \tpathonlyderivation{\cosderitwo}{\pathcosderitwo}
      &
      \tpathonlyderivation{\cosderithree}{\pathcosderithree}
      \\[\arrayskip]
      \downarrow&\downarrow&\downarrow
      \\[\arrayskip]
      \\[\arrayskip]
      \qquad\pathpnone
      &
      \quad\pathpnone
      &
      \quad\pathpnthree
    \end{array}}
    $$
    \caption{From calculus of structures to proof nets}
    \label{fig:cos-to-pn}
  \end{center}
\end{figure}

\begin{figure}[!t]
  \footnotesize
  \def\arrayskip{1ex}
  \begin{center}
    \vskip-4ex
    $$
    \qlapm{\begin{array}{c}
      \cosderifour
      \\[\arrayskip]
      \downarrow
      \\[\arrayskip]
      \tpathderivation{\cosderifour}{\pathcosderifour}
      \\[\arrayskip]
      \downarrow
      \\[\arrayskip]
      \tpathonlyderivation{\cosderifour}{\pathcosderifour}
      \\[\arrayskip]
      \downarrow
      \\[\arrayskip]
      \\[\arrayskip]
      \pathpnone
    \end{array}}
    $$
    \caption{From calculus of structures to proof nets}
    \label{fig:cos-to-pn-cut}
  \vskip-1.8cm 
  \end{center}
\end{figure}

\begin{figure}[!t]
  \footnotesize
  \def\arrayskip{1ex}
  \def\theblack{\black}
  \def\thecutsinpathcosderifour{}
  \begin{center}
    \vskip-4ex
    $$
    \qlapm{\begin{array}{c}
      \tpathderivation{\cosderifour}{\pathcosderifour}
      \\[\arrayskip]
      \downarrow
      \\[\arrayskip]
      \tpathonlyderivation{\cosderifour}{\pathcosderifour}
      \\[\arrayskip]
      \downarrow
      \\[\arrayskip]
      \\[\arrayskip]
      \\[\arrayskip]
      \pathpnfour
    \end{array}}
    $$
    \caption{From calculus of structures to proof nets}
    \label{fig:cos-to-pn-cutcut}
  \end{center}
\end{figure}

\begin{corollary}\label{cor:sq-cos}
  Let $A_1,A_2,\ldots,A_n$ be a list of $\ufMLL$ formulas. There is a
  sequent calculus derivation 
  \proofadjust
  $$
  \sqnbigderi{A_1,A_2,\ldots,A_n}{\Pi}
  $$ 
  if and only if there is a derivation
  $$
  \strpr{}{\Pit}{A_1\lpar A_2\lpar\cdots\lpar A_n}
  $$ 
  in the calculus of
  structures.
\end{corollary}

Let us now see how derivations in the calculus of structures are translated
into proof nets. The method is exactly the same as in the previous section: We
simply trace the atoms through the
derivation. Figures~\ref{fig:cos-to-pn}--\ref{fig:cos-to-pn-cutcut} show the
calculus of structures version of
Figures~\ref{fig:sc-to-pn-2}--\ref{fig:sc-to-pn-2-cutcut}.

\begin{para}{Trivial Observation}
  We get the same proof nets as before.
\end{para}

\begin{para}{Subtle Observation}\label{obs:subtle}
  In Figures \ref{fig:cos-to-pn} and~\ref{fig:cos-to-pn-cutcut} the
  flow-graphs drawn inside the proofs have the same crossings as the
  resulting proof nets, while in
  Figures~\ref{fig:sc-to-pn-2} and~\ref{fig:sc-to-pn-2-cutcut}, the
  flow-graphs drawn inside the proofs have more (seemingly
  unnecessary) crossings.
\end{para}

\newpage 

\subsection{Correctness criteria}\label{sec:correctness}

We have seen how we can obtain a proof net out of a formal proof in
some deductive system. But what about the other way around? Suppose we
have such a graph that looks like a proof net. Can we decide whether
it really comes from a proof, and if so, can we recover this proof?
Of course the answer is trivially yes because the graph is finite and
we just need to check all proofs of that size. The interesting question
is therefore, whether we can do it efficiently.

The answer is still yes, and it is done via so-called
\emph{correctness criteria}.  For introducing the idea, we take the
following graphs as running examples 
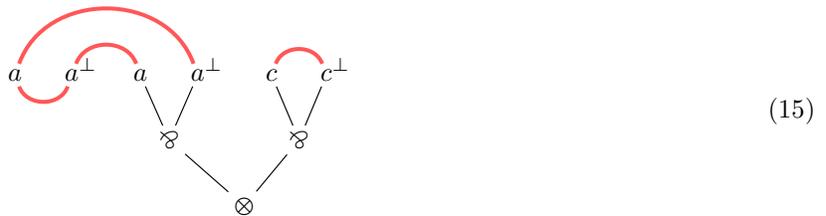
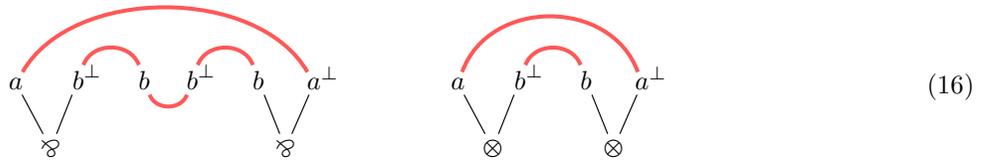
\begin{equation}\label{eq:exa-pre-pn}
  \examplecorrect
\end{equation}
\begin{equation}\label{eq:exa-pre-pn-2}
  \vcnpn{
    \quad\\
    \na1 && \mb1\lneg && \mb2 && \mb3\lneg && \mb4 && \na2\lneg\\
    &\nlpar1&&\pa && \pa && \pa && \nlpar2
  }{
    \ncline{lpar1}{a1}
    \ncline{lpar1}{b1}
    \ncline{lpar2}{a2}
    \ncline{lpar2}{b4}
  }{
    \udline{b1}{b2}
    \duline{b2}{b3}
    \udline{b3}{b4}
    \def\loopangleA{60}
    \def\loopangleB{120}
    \def\loopvecheight{.7}    
    \udline{a1}{a2}
  }
  \qqqquad
  \vcnpn{
    \quad\\
    \na1 && \mb1\lneg && \mb2 && \na2\lneg\\
    &\nltens1&&\pa&&\nltens2
    }{
    \ncline{ltens1}{a1}
    \ncline{ltens1}{b1}
    \ncline{ltens2}{a2}
    \ncline{ltens2}{b2}
    }{
    \udline{b1}{b2}
    \udline{a1}{a2}
  }
\end{equation}
By playing around, you will notice that it is quite easy to find a proof (in
sequent calculus or calculus of structures) that translates into the net
in~\eqref{eq:exa-pre-pn}, but it seems impossible to find such proofs for the
two examples in~\eqref{eq:exa-pre-pn-2}. We are now going to show that this
is indeed impossible. For doing so, we need some formal definitions.

\begin{definition}\label{def:pre-pn}
  A \emph{pre-proof net} is a sequent forest $\Gamma$, possibly with
  cuts, together with a perfect matching of the set of leaves (i.e.,
  the set of occurrences of propositional variables and their duals),
  such that only dual pairs are matched.
\end{definition}

In this context, a cut must be seen as a special kind of formula
$A\lcut A\lneg$, where $\lcut$ is a special connective which may occur
only at the root of a formula tree in which the two direct subformulas
are dual to each other. For example, \eqref{eq:exa-pre-pn} 
should be read as 
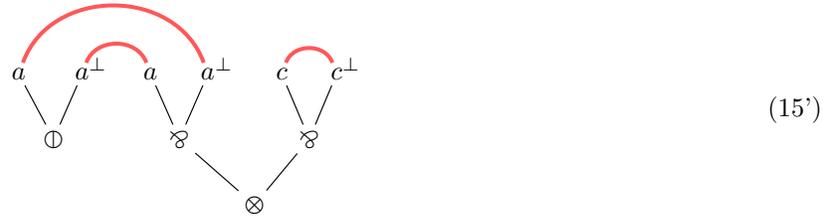
\begin{equation}\label{eq:exa-pre-pn-cut}
  \examplecorrectcut
  \tag{\ref{eq:exa-pre-pn}'}
\end{equation}
Clearly, the examples in \eqref{eq:exa-pre-pn} and
\eqref{eq:exa-pre-pn-2} are all pre-proof nets.
In the following, we will think of an inner node (i.e., a non-leaf node) of
the sequent forest labeled not only by the connective but by the whole
subformula rooted by that connective. Our favorite
example~\eqref{eq:exa-pre-pn} should then be read as  
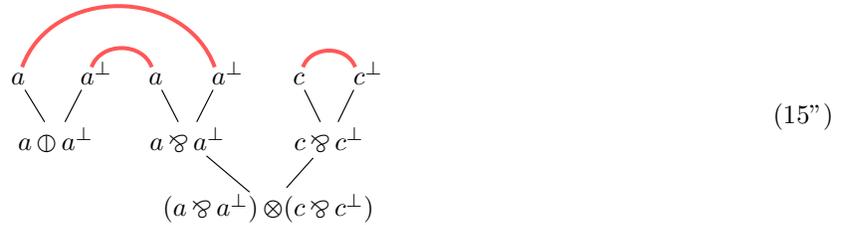
\begin{equation}\label{eq:exa-pre-pn-long}
  \examplecorrectlong
  \tag{\ref{eq:exa-pre-pn}''}
\end{equation}
Although sometimes we think of pre-proof nets to be written as
in~\eqref{eq:exa-pre-pn-long}, we will keep writing them as
in~\eqref{eq:exa-pre-pn} for better readability.

\begin{definition}
  A pre-proof net $\pi$ is called \emph{sequentializable} iff there is a proof
  in the sequent calculus or in the calculus of structures that translates
  into~$\pi$.
\end{definition}

Originally, the term ``sequentializable'' was motivated by the name ``sequent
calculus''. But we use it here also if the ``sequentialization'' is done in
the calculus of structures.

\begin{definition}\label{def:switching}
  Let $\pi$ be a pre-proof net. A \emph{switching} for $\pi$ is a
  graph obtained from $\pi$ by removing for every $\lpar$-node one of
  the two edges connecting it to its children.
\end{definition}

Clearly, if a pre-proof net contains $n$ $\lpar$-nodes, then there are
$2^n$ switchings. Here are all 4 switchings for the example
in~\eqref{eq:exa-pre-pn}:
\begin{equation}\label{exa:switching}
  \begin{array}{c@{\qqqquad}c}
    \def\switchingexamplecorrect{\ncline{lpar1}{a3}\ncline{lpar2}{c1}}
    \examplecorrect &
    \def\switchingexamplecorrect{\ncline{lpar1}{a3}\ncline{lpar2}{c2}}
    \examplecorrect \\[8ex]
    \def\switchingexamplecorrect{\ncline{lpar1}{a4}\ncline{lpar2}{c1}}
    \examplecorrect &
    \def\switchingexamplecorrect{\ncline{lpar1}{a4}\ncline{lpar2}{c2}}
    \examplecorrect
  \end{array}
\end{equation}

\begin{definition}\label{def:S-correct}
  A pre-proof net \emph{obeys the switching criterion} (or, shortly, is
  \emph{correct}) iff all its switchings are connected and acyclic.
\end{definition}

As \eqref{exa:switching} shows, the pre-proof net
in~\eqref{eq:exa-pre-pn} is correct. The two pre-proof nets
in~\eqref{eq:exa-pre-pn-2} are not, as the following switchings show:
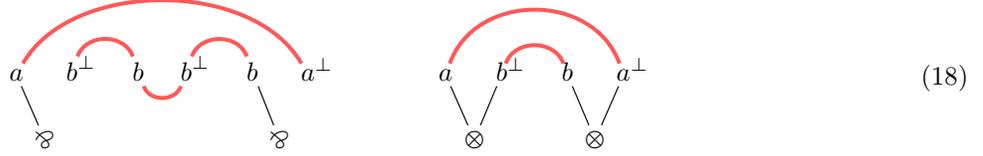
\begin{equation}\label{exa:switching-incorrect}
  \vcnpn{
    \\
    \na1 && \mb1\lneg && \mb2 && \mb3\lneg && \mb4 && \na2\lneg\\
    &\nlpar1&&\pa && \pa && \pa && \nlpar2
  }{
    \ncline{lpar1}{a1}
    \ncline{lpar2}{b4}
  }{
    \udline{b1}{b2}
    \duline{b2}{b3}
    \udline{b3}{b4}
    \def\loopangleA{60}
    \def\loopangleB{120}
    \def\loopvecheight{.7}    
    \udline{a1}{a2}
  }
  \qqqquad
  \vcnpn{
    \\
    \na1 && \nb1\lneg && \nb2 && \na2\lneg\\
    &\nltens1&&\pa&&\nltens2
    }{
    \ncline{ltens1}{a1}
    \ncline{ltens1}{b1}
    \ncline{ltens2}{a2}
    \ncline{ltens2}{b2}
    }{
    \udline{b1}{b2}
    \udline{a1}{a2}
  }
\end{equation}
The first is not connected, and the second is cyclic.

In the following, we use the term \emph{proof net} for those pre-proof nets
which are correct, i.e., obey the switching criterion. The following theorem
says that the proof nets are exactly those pre-proof nets that represent an
actual proof.

\begin{theorem}\label{thm:S-correct}
  A pre-proof net is correct if and only if it is sequentializable.
\end{theorem}

We will give two proofs of this theorem. The first uses the sequent
calculus, and the second the calculus of structures.
For the first proof, we need the following lemma:

\begin{lemma}\label{lem:splittingtensor}
  Let $\pi$ be a proof net with conclusions $A_1,\ldots,A_n$. If all
  $A_i$ have a $\ltens$ or a cut as root, then one of them is
  splitting, \ie by removing that $\ltens$ \rp{or $\lcut$}, the net
  becomes disconnected.
\end{lemma}

For proving this lemma, we need some more concepts. 

\begin{definition}\label{def:subnet}
  Let $\sigma$ and $\pi$ be pre-proof nets. We say $\sigma$ is a
  \emph{subprenet} of $\pi$, written as $\sigma\subseteq\pi$ if all
  formulas/cuts appearing in $\sigma$ are subformulas of the formulas/cuts
  appearing in $\pi$, and the linking of $\sigma$ is the restriction of the
  linking of $\pi$ to the formulas/cuts in $\sigma$. We say $\sigma$ is a
  \emph{subnet} of $\pi$ if $\sigma\subseteq\pi$, and $\sigma$ and $\pi$ are
  both correct. A \emph{door} of $\sigma$ is any formula that
  appears as conclusion of $\sigma$.
\end{definition}

\begin{example}\label{exa:subnet}
  Consider the following three graphs:
  $$
  \vcnpn{
    \\
    \na1 && \na2\lneg && \na3 && \na4\lneg \\
    &\pa&& \pa && \nlpar1
  }{
    \ncline{lpar1}{a3}
    \ncline{lpar1}{a4}
  }{
    \udline{a1}{a4}
    \udline{a2}{a3}
  }
  \qqqquad
  \vcnpn{
    \\
    \na1 && \na2\lneg && \na3 && \na4\lneg \\
    &\pa&& \pa && \nlpar1
  }{
    \ncline{lpar1}{a3}
    \ncline{lpar1}{a4}
  }{
    \udline{a1}{a4}
    \udline{a2}{a3}
    \duline{a1}{a2}
  }
  \qqqquad
  \vcnpn{
    \\
    \na1 && \na2\lneg && \na3 && \na4\lneg \\
    &\pa&& \pa && \nlpar1
  }{
    \ncline{lpar1}{a3}
    \ncline{lpar1}{a4}
  }{
    \udline{a1}{a4}
  }
  $$ The first two are subprenets of \eqref{eq:exa-pre-pn}, the third one is
  not (because a link is missing). The second one is in fact a subnet of
  \eqref{eq:exa-pre-pn}, but the first one is not (because it is not
  correct). The doors of the first example are $a$, $a\lneg$, and $a\lpar
  a\lneg$. The doors of the second example are $a\lcut a\lneg$ and $a\lpar
  a\lneg$.
\end{example}

\begin{lemma}\label{lem:subnet}
  Let $\sigma$ and $\rho$ be subnets of some proof net $\pi$.
  \begin{enumerate}[\rm(i)]
  \item The subprenet $\sigma\cup\rho$ is a subnet of $\pi$ if and only
    if $\sigma\cap\rho\neq\emptyset$.
  \item If $\sigma\cap\rho\neq\emptyset$ then $\sigma\cap\rho$ is a subnet of
    $\pi$.
  \end{enumerate}
\end{lemma}

\begin{proof}
  Intersection and union in the statement of that lemma have to be understood
  in the canonical sense: An edge/node/link appears in in $\sigma\cap\rho$
  (resp.~$\sigma\cup\rho$) if it appears in both, $\sigma$ and $\rho$ (resp.\
  in at least one of $\sigma$ or $\rho$). For giving the proof, let us first
  note that because in $\pi$ every switching is acyclic, also in every
  subprenet of $\pi$ every switching is acyclic, in particular also in
  $\sigma\cup\rho$ and $\sigma\cap\rho$. Therefore, we need only to consider
  the connectedness condition. 
  \begin{enumerate}[\rm(i)]
  \item If $\sigma\cap\rho=\emptyset$ then every switching of $\sigma\cup\rho$
    must be disconnected. Conversely, if $\sigma\cap\rho\neq\emptyset$, then
    every switching of $\sigma\cup\rho$ must be connected (in every switching
    of $\sigma\cup\rho$ every node in $\sigma\cap\rho$ must be connected to
    every node in~$\sigma$ and to every node in $\rho$, because $\sigma$ and
    $\rho$ are both correct).
  \item Let $\sigma\cap\rho\neq\emptyset$ and let $s$ be a switching for
    $\sigma\cup\rho$. Then $s$ is connected and acyclic by (i). Let
    $s_\sigma$, $s_\rho$, and $s_{\sigma\cap\rho}$, be the restrictions of $s$
    to $\sigma$, $\rho$, and $\sigma\cap\rho$, respectively. Now let $A$ and
    $B$ be two vertices in $\sigma\cap\rho$. Then $A$ and $B$ are connected by
    a path in $s_\sigma$ because $\sigma$ is correct, and by a path in
    $s_\rho$ because $\rho$ is correct. Since $s$ is acyclic, the two paths
    must be the same and therefore be contained in $s_{\sigma\cap\rho}$. \qed
  \end{enumerate}
\end{proof}

\begin{lemma}\label{lem:exists}
  Let $\pi$ be a proof net, and let $A$ be
  a subformula of some formula/cut appearing in $\pi$. Then there is a subnet
  $\sigma$ of $\pi$, that has $A$ as a door.
\end{lemma}

\begin{proof}
  \def\swing#1#2#3{#1(#2,#3)}
  For proving this lemma, we need the following notation. Let $\pi$ be a proof
  net, let $A$ be some formula occurrence in $\pi$, and let $s$ be a switching
  for $\pi$. Then we write $\swing{s}{\pi}{A}$ for the graph obtained as
  follows: 
  \begin{itemize}
  \item If $A$ is an immediate subformula of a formula occurrence $B$ in
    $\pi$, and there is an edge from $B$ to $A$ in $s$, then remove that edge
    and let $\swing{s}{\pi}{A}$ be the connected component of (the remainder
    of) $s$ that contains $A$.
  \item Otherwise let $\swing{s}{\pi}{A}$ be just $s$.
  \end{itemize}
  Now let $$\sigma=\bigcap_s \swing{s}{\pi}{A}$$ where $s$ ranges over all
  possible switchings of $\pi$. (Note that it could happen that formally
  $\sigma$ is not a subprenet because some edges in the formula trees might be
  missing. We graciously add these missing edges to $\sigma$ such that it
  becomes a subprenet.) Clearly, $A$ is in $\sigma$. We are now going to show
  that $A$ is a door of $\sigma$.  \Bwoc, assume it is not. This means there
  is ancestor $B$ of $A$ that is in $\bigcap_s \swing{s}{\pi}{A}$. Now choose
  a switching $\hat{s}$ such that whenever there is a $\lpar$ node between $A$
  and $B$, i.e.,
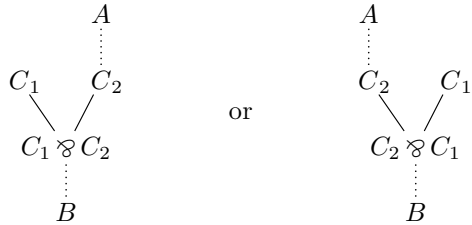
  $$
  \vcnpn{
    \pa&&\nA1\phantom{_0}\\
    \nC1_1&&\nC2_2\\
    &\qlapnode{lpar1}{C_1\lpar C_2}\\
    &\nB1
  }{
    \ncline{lpar1}{C1}
    \ncline{lpar1}{C2}
    \pntreemodedotted
    \ncline{B1}{lpar1}
    \ncline{C2}{A1}
  }{}
  \qqqquor
  \vcnpn{
    \nA1\phantom{_0}&&\pa\\
    \nC2_2&&\nC1_1\\
    &\qlapnode{lpar1}{C_2\lpar C_1}\\
    &\nB1
  }{
    \ncline{lpar1}{C1}
    \ncline{lpar1}{C2}
    \pntreemodedotted
    \ncline{B1}{lpar1}
    \ncline{C2}{A1}
  }{}
  $$ then $\hat{s}$ chooses $C_1$ (i.e., removes the edge between $C_2$ and
  its parent). Then there must be a $\ltens$ between $A$ and $B$:
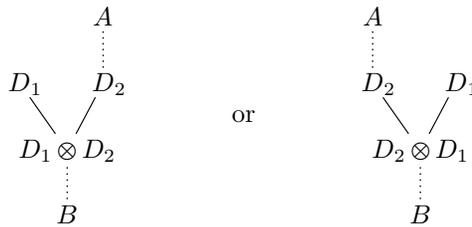
  $$
  \vcnpn{
    \pa&&\nA1\phantom{_0}\\
    \nD1_1&&\nD2_2\\
    &\qlapnode{ltens1}{D_1\ltens D_2}\\
    &\nB1
  }{
    \ncline{ltens1}{D1}
    \ncline{ltens1}{D2}
    \pntreemodedotted
    \ncline{B1}{ltens1}
    \ncline{D2}{A1}
  }{}
  \qqqquor
  \vcnpn{
    \nA1\phantom{_0}&&\pa\\
    \nD2_2&&\nD1_1\\
    &\qlapnode{ltens1}{D_2\ltens D_1}\\
    &\nB1
  }{
    \ncline{ltens1}{D1}
    \ncline{ltens1}{D2}
    \pntreemodedotted
    \ncline{B1}{ltens1}
    \ncline{D2}{A1}
  }{}
  $$ Otherwise $B$ would not be in $\sigma$. Now suppose we have chosen the
  uppermost such $\ltens$. Then the path connecting $A$ and $D_1$ in
  $\swing{\hat{s}}{\pi}{A}$ cannot pass through $D_2$ (by the definition of
  $\swing{\hat{s}}{\pi}{A}$). But this means that in $\hat{s}$ there are two
  distinct paths connecting $A$ and $D_1$, which contradicts the acyclicity of
  $\hat{s}$.

  Now we have to show that $\sigma$ is a subnet. Let $s$ be a switching for
  $\sigma$. Since $\sigma$ is a subprenet of $\pi$, we have that $s$ is
  acyclic. Now let $\tilde{s}$ be an extension of $s$ to $\pi$. Then $s$ is
  the restriction of $\swing{\tilde{s}}{\pi}{A}$ to $\sigma$, and hence
  connected.
  \qed
\end{proof}

\begin{definition}\label{def:kingdom}
  Let $\pi$ be a proof net, and let $A$ be
  a subformula of some formula/cut appearing in $\pi$. The \emph{kingdom of
  $A$ in $\pi$}, denoted by $kA$, is the smallest subnet of $\pi$, that has
  $A$ as a door. Similarly, the \emph{empire of
  $A$ in $\pi$}, denoted by $eA$, is the largest subnet of $\pi$, that has
  $A$ as a door. We define $A\ll B$ iff $A\in kB$, where $A$ and $B$ can be
  any (sub)formula/cut occurrences in~$\pi$. 
\end{definition}

An immediate consequence of Lemmas \ref{lem:subnet} and~\ref{lem:exists} is
that kingdom and empire always exist.

\begin{exercise}
  Why?
\end{exercise}

\begin{remark}
  The subnet $\sigma$ constructed in the proof of Lemma~\ref{lem:exists} is in
  fact the empire of $A$. But we will not need this fact later and will not
  prove it here.
\end{remark}

\begin{lemma}\label{lem:nesting}
  Let $\pi$ be a proof net, and let $A$, $A'$, $B$, and $B'$ be subformula
  occurrences appearing in $\pi$, such that $A$ and $B$ are distinct, $A'$
  is immediate subformula of $A$, and $B'$ is immediate subformula of $B$. Now
  suppose that $B'\in eA'$. Then we have that $B\notin eA'$ if and only if
  $A\in kB$.
\end{lemma}

\begin{proof}
  We have $B'\in eA'\cap kB$. Hence, $\sigma=eA'\cap kB$ and $\rho=eA'\cup kB$
  are subnets of $\pi$. \Bwoc, let $B\notin eA'$ and $A\notin kB$. Then $\rho$
  has $A'$ as door and is larger than $eA'$ because it contains $B$. This
  contradicts the definition of $eA'$. On the other hand, if $B\in eA'$ and
  $A\in kB$ then $\sigma$ has $B$ as door and is smaller than $kB$ because it
  does not contain $A$. This
  contradicts the definition of $kB$. \qed
\end{proof}

\begin{lemma}\label{lem:order}
  Let $\pi$ be a proof net, and let $A$ and $B$ be subformulas appearing in
  $\pi$. If $A\ll B$ and $B\ll A$, then either $A$ and $B$ are the same
  occurrence or they are dual atoms connected via an identity link. 
\end{lemma}

\begin{proof}
  If $a$ and $a\lneg$ are two dual atom occurrences connected by a link, then
  clearly $ka=ka\lneg$. Now let $A$ and $B$ be two distinct non-atomic formula
  occurrences in $\pi$ with $A\in kB$ and $B\in kA$. Then $kA\cap kB$ is a
  subnet and hence $kA=kA\cap kB=kB$. We have two cases:
  \begin{itemize}
  \item If $A=A'\lpar A''$ then the result of removing $A$ from $kB$ is still
    a subnet, contradicting the minimality of $kB$.
  \item If $A=A'\ltens A''$ then $kA=kA'\cup kA''\cup \set{A'\ltens
    A''}$. Hence $B\in kA'$ or $B\in kA''$. This contradicts
    Lemma~\ref{lem:nesting}, which says that $B\notin eA'$ and $B\notin eA''$.
    \qed
  \end{itemize}
\end{proof}

From Lemma~\ref{lem:order} it immediately follows that $\ll$ is a partial
order on the non-atomic subformula occurrences in $\pi$. We make crucial use of
this fact in the following:

\begin{proof}[ of Lemma~\ref{lem:splittingtensor}]
  Choose among the conclusions $A_1,\ldots,A_n$ (including the cuts) of $\pi$
  one which is maximal w.r.t.~$\ll$. \Wolg, assume it is $A_i=A'_i\ltens
  A''_i$. We will now show that it is splitting, i.e., $\pi=\set{A'_i\ltens
  A''_i}\cup eA'_i\cup eA''_i$. \Bwoc, assume $A'_i\ltens A''_i$ is not
  splitting. This means we have somewhere in $\pi$ a formula occurrence $B$
  with immediate subformula $B'$ such that (\wolg) $B'\in eA'_i$ and $B\notin
  eA'_i$. We also know that $B$ must occur at or above some other conclusion,
  say $A_j=A'_j\ltens A''_j$. Hence $B\in kA_j$ and therefore $kB\subseteq
  kA_j$. But by Lemma~\ref{lem:nesting} we have $A_i\in kB$ and therefore
  $A_i\in kA_j$, which contradicts the maximality of $A_i$ w.r.t.~$\ll$.
  \qed
\end{proof}

Finally, we can prove Theorem~\ref{thm:S-correct}.

\begin{proofname}[First Proof of Theorem~\ref{thm:S-correct}]
  Let us first show that the (in the sequent calculus) sequentializable
  pre-proof nets are indeed correct. This is done by verifying that the
  $\idr$-rule yields correct nets (which is obvious) and that all other
  inference rules preserve correctness. For the $\exr$-rule this is
  obvious. Let us now consider the $\ltens$-rule. \Bwoc, assume that
  $$
  \vcgpnonlybox{\pi_1}{
    \gpbbunchboxout{1}\quad\gpnboxout{2}}{
    \gpbbunchout{1}\quad\gpnout{2}}{
    \bunchboxoutline{1}{\Gamma}
    \boxoutline{2}{A}}
  \qquand
  \vcgpnonlybox{\pi_2}{
    \gpnboxout{1}\quad\gpbbunchboxout{2}}{
    \gpnout{1}\quad\gpbbunchout{2}}{
    \bunchboxoutline{2}{\Delta}
    \boxoutline{1}{B}}
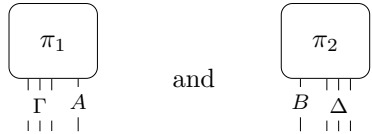
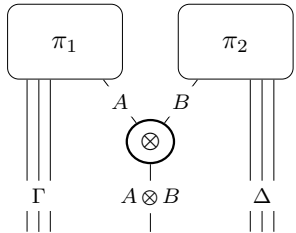
  $$
  are correct, but
  $$
  \tensorrulenet{\pi_1}{\Gamma}{A}{\pi_2}{B}{\Delta}
  $$ is not correct. This means there is a switching that is either
  disconnected or contains a cycle. Since a $\ltens$-node does not
  affect switchings, we conclude that the property of being
  disconnected or cyclic must hold for the same switching in one of
  $\pi_1$ or $\pi_2$. But this is a contradiction to the correctness
  of $\pi_1$ and $\pi_2$.  For the the $\lpar$-rule and the
  $\cutr$-rule we proceed similarly.

  Conversely, let $\pi$ be a correct pre-proof net. We proceed by
  induction on the size of $\pi$, i.e., the number $n$ of $\lpar$-,
  $\ltens$-, and $\cutr$-nodes in $\pi$, to construct a sequent
  calculus proof $\Pi$, that translates into $\pi$. If $n$ is $0$,
  then $\pi$ must be of the shape
  $$
  \idrulenet{A\lneg}{A}
  $$ and we can apply the $\idr$-rule. Now let $n>0$. If one of the
  conclusion formulas of $\pi$ has a $\lpar$-root, we can apply the
  $\lpar$-rule and proceed by induction hypothesis. Now suppose all
  roots are $\ltens$ or cuts. Then we apply
  Lemma~\ref{lem:splittingtensor}, which tells us, that there is one
  of them which splits the net. Assume, \wolg, that it is a
  $\ltens$-root, say $A_i=A_i'\ltens A_i''$. This means, the net is of the
  shape
  $$
  \tensorrulenet{\pi_1}{\Gamma}{A_i'}{\pi_2}{A_i''}{\Delta}
  $$ and we can apply the $\ltens$-rule and proceed by induction
  hypothesis for $\pi_1$ and $\pi_2$. If the splitting root is a cut,
  we apply the $\cutr$-rule instead. \qed
\end{proofname}

Let us now see the second proof. For this, we need the following lemma:

\begin{lemma}\label{lem:pair}
  Let $\pi$ be a proof net with conclusion 
  $$S\cons{(A\ltens B\cons{a})\lpar(C\cons{a\lneg}\ltens D)}\quadcm$$
  such that the $a$ and the $a\lneg$ are paired up in the linking.
  Let $\pi'$ and $\pi''$ be pre-proof nets with conclusions
  $$
  S\cons{A\ltens(B\cons{a}\lpar(C\cons{a\lneg}\ltens D))}
  \qquand
  S\cons{((A\ltens B\cons{a})\lpar C\cons{a\lneg})\ltens D}
  $$
  respectively, such that the linkings of $\pi'$ and $\pi''$ 
  \rp{i.e., the pairing of dual atoms} are the same as the linking of $\pi$.
  Then at least one of $\pi'$ and $\pi''$ is also correct.
\end{lemma}

\begin{proof}
  Let us visualize the information we have about $\pi$, $\pi'$, and $\pi''$ as
  follows:
  $$
  \begin{array}{ccccc}
    \pi':&&\pi:&&\pi'':\\[2.5ex]
    \vcnpnmark{
      &&&&\na2\rlap{$\lneg$}\\
      &&\na1&&\nC1&&\nD1\\
      &&\nB1&&&\nltens2\\
      \nA1&&&&\nlpar1\\
      &&\nltens1
    }{
      \ncline{ltens1}{A1}
      \ncline{ltens1}{lpar1}
      \ncline{ltens2}{lpar1}
      \ncline{lpar1}{B1}
      \ncline{ltens2}{C1}
      \ncline{ltens2}{D1}
      \pntreemodedotted
      \ncline{B1}{a1}
      \ncline{C1}{a2}
    }{
      \lineanglesheight{a1}{a2}{100}{120}{1}
    }{}
    &
    \qquad\leftarrow\qquad
    &
    \vcnpn{
      &&\na1&&\na2\rlap{$\lneg$}\\
      \nA1&&\nB1&&\nC1&&\nD1\\
      &\nltens1&&&&\nltens2\\
      &&&\nlpar1      
    }{
      \ncline{ltens1}{A1}
      \ncline{ltens1}{lpar1}
      \ncline{ltens2}{lpar1}
      \ncline{ltens1}{B1}
      \ncline{ltens2}{C1}
      \ncline{ltens2}{D1}
      \pntreemodedotted
      \ncline{B1}{a1}
      \ncline{C1}{a2}
    }{
      \udline{a1}{a2}
    }
    &
    \qquad\rightarrow\qquad
    &
    \vcnpn{
      &&\na1\\
      \nA1&&\nB1&&\na2\rlap{$\lneg$}\\
      &\nltens1&&&\nC1\\
      &&\nlpar1&&&\nD1\\
      &&&&\nltens2
    }{
      \ncline{ltens1}{A1}
      \ncline{ltens1}{lpar1}
      \ncline{ltens2}{lpar1}
      \ncline{ltens1}{B1}
      \ncline{lpar1}{C1}
      \ncline{ltens2}{D1}
      \pntreemodedotted
      \ncline{B1}{a1}
      \ncline{C1}{a2}
    }{
      \lineanglesheight{a1}{a2}{60}{80}{1}
    }
  \end{array}
  $$
  We proceed \bwoc, and assume that $\pi$ is correct and that $\pi'$ and
  $\pi''$ are both incorrect. If there is a switching $s$ for $\pi'$ (or
  $\pi''$) that is disconnected, then the same switching is also disconnected
  in $\pi$. Hence, we need to consider only the acyclicity condition. Suppose
  that there is a switching $s'$ for $\pi'$ that is cyclic. Then, in $s'$ the
  $\lpar$ below $B$ must be switched to the right, and the cycle must pass
  through $A$, the root $\ltens$ and the $\lpar$ as follows:
  $$
  \vcnpnmark{
    &&&&\na2\rlap{$\lneg$}\\
    &&\na1&&\nC1&&\nD1\\
    &&\nB1&&&\nltens2\\
    \nA1&&&&\nlpar1\\
    &&\nltens1
  }{
    \ncline{ltens1}{A1}
    \ncline{ltens1}{lpar1}
    \ncline{ltens2}{lpar1}
    \ncline{ltens2}{C1}
    \ncline{ltens2}{D1}
    \pntreemodedotted
    \ncline{B1}{a1}
    \ncline{C1}{a2}
  }{
    \lineanglesheight{a1}{a2}{100}{120}{1}
  }{
    \ncline{ltens1}{A1}
    \ncline{ltens1}{lpar1}
    \ncline{ltens2}{lpar1}
  }
  $$
  Otherwise we could construct a switching with the same cycle in $\pi$. If
  our cycle continues through $D$, i.e.,
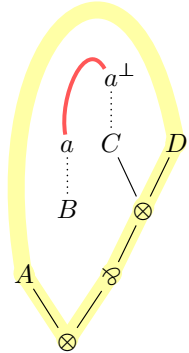
  \begin{equation}\label{eq:cyclAD}
    \vcnpnmark{
      \\
      &&&&\na2\rlap{$\lneg$}\\
      &&\na1&&\nC1&&\nD1\\
      &&\nB1&&&\nltens2\\
      \nA1&&&&\nlpar1\\
      &&\nltens1
    }{
      \ncline{ltens1}{A1}
      \ncline{ltens1}{lpar1}
      \ncline{ltens2}{lpar1}
      \ncline{ltens2}{C1}
      \ncline{ltens2}{D1}
      \pntreemodedotted
      \ncline{B1}{a1}
      \ncline{C1}{a2}
    }{
      \lineanglesheight{a1}{a2}{100}{120}{1}
    }{
      \ncline{ltens1}{A1}
      \ncline{ltens1}{lpar1}
      \ncline{ltens2}{lpar1}
      \ncline{ltens2}{D1}
      \lineanglesheight{A1}{D1}{100}{100}{2.5}
    }
  \end{equation}
  then we can use the path from $A$ to $D$ (that does not go through $B$ or
  $C$, see Exercise~\ref{ex:cyclAD}) to construct a cyclic switching $s$ in
  $\pi$ as follows:
  $$
  \vcnpnmark{
    \\
    &&\na1&&\na2\rlap{$\lneg$}\\
    \nA1&&\nB1&&\nC1&&\nD1\\
    &\nltens1&&&&\nltens2\\
    &&&\nlpar1
  }{
    \ncline{ltens1}{A1}
    \ncline{ltens1}{lpar1}
    \ncline{ltens2}{lpar1}
    \ncline{ltens1}{B1}
    \ncline{ltens2}{C1}
    \ncline{ltens2}{D1}
    \pntreemodedotted
    \ncline{B1}{a1}
    \ncline{C1}{a2}
  }{
    \udline{a1}{a2}
  }{
    \ncline{ltens1}{A1}
    \ncline{ltens1}{B1}
    \ncline{ltens2}{C1}
    \ncline{ltens2}{D1}
    \ncline{B1}{a1}
    \ncline{C1}{a2}
    \def\loopvecheight{1.1}
    \udline{a1}{a2}
    \lineanglesheight{A1}{D1}{90}{90}{2}
  }
  $$
  Hence, the cycle in $s'$ goes through $C$, giving us a path from $A$ to $C$,
  not passing through $B$ (see Exercise~\ref{ex:cyclAD}):
  \begin{equation}\label{eq:cyclAC}
    \vcnpnmark{
      \\
      &&&&\na2\rlap{$\lneg$}\\
      &&\na1&&\nC1&&\nD1\\
      &&\nB1&&&\nltens2\\
      \nA1&&&&\nlpar1\\
      &&\nltens1
    }{
      \ncline{ltens1}{A1}
      \ncline{ltens1}{lpar1}
      \ncline{ltens2}{lpar1}
      \ncline{ltens2}{C1}
      \ncline{ltens2}{D1}
      \pntreemodedotted
      \ncline{B1}{a1}
      \ncline{C1}{a2}
    }{
      \lineanglesheight{a1}{a2}{100}{120}{1}
    }{
      \ncline{ltens1}{A1}
      \ncline{ltens1}{lpar1}
      \ncline{ltens2}{lpar1}
      \ncline{ltens2}{C1}
      \lineanglesheight{A1}{C1}{110}{60}{3.5}
    }
  \end{equation}
  By the same argumentation we get a switching $s''$ in $\pi''$ with a path
  from $B$ to $D$, not going through $C$. From $s'$ and $s''$, we can
  now construct a switching $s$ for $\pi$ with a cycle as follows:
  $$
  \vcnpnmark{
    \\
    &&\na1&&\na2\rlap{$\lneg$}\\
    \nA1&&\nB1&&\nC1&&\nD1\\
    &\nltens1&&&&\nltens2\\
    &&&\nlpar1
  }{
    \ncline{ltens1}{A1}
    \ncline{ltens1}{lpar1}
    \ncline{ltens2}{lpar1}
    \ncline{ltens1}{B1}
    \ncline{ltens2}{C1}
    \ncline{ltens2}{D1}
    \pntreemodedotted
    \ncline{B1}{a1}
    \ncline{C1}{a2}
  }{
    \udline{a1}{a2}
  }{
    \ncline{ltens1}{A1}
    \ncline{ltens1}{B1}
    \ncline{ltens2}{C1}
    \ncline{ltens2}{D1}
    \lineanglesheight{A1}{C1}{120}{40}{3.5}
    \lineanglesheight{B1}{D1}{140}{60}{3.5}
  }
  $$
  which contradicts the correctness of $\pi$.
  \qed
\end{proof}

\begin{exercise}\label{ex:cyclAD}
  Explain why we can in \eqref{eq:cyclAD} assume that the cycle does not go
  through $B$ or $C$, and in~\eqref{eq:cyclAC} not through $B$.
\end{exercise}

In our second proof of Theorem~\ref{thm:S-correct} we will also need
the following concept: 

\begin{definition}\label{def:web}
  Let $A$ be a formula. The \emph{relation web} of $A$ is the complete graph,
  whose vertices are the atom occurrences in $A$. An edge between two atom
  occurrences $a$ and $b$ is colored red, if the first common ancestor of them
  in the formula tree is a $\ltens$, and green if it is a $\lpar$.
\end{definition}

\begin{example}
  Consider the formula $(a\lneg\lpar(a\ltens
  a))\lpar(a\lpar(a\lneg\ltens a\lneg))$. Its formula tree is the
  following:
  $$
  \vcnpn{
    \na1\rlap{$\lneg$}&\strut\quad\strut&\na2&&\na3&
    &\na4&\strut\quad\strut&\na5\rlap{$\lneg$}&&\na6\rlap{$\lneg$}\\
    &&&\red\nltens1&&&&&&\red\nltens2\\
    &&\green\nlpar1&&&&&&\green\nlpar2\\
    &&&&&\green\nlpar3
  }{
    \ncline{a1}{lpar1}
    \ncline{a2}{ltens1}
    \ncline{a3}{ltens1}
    \ncline{ltens1}{lpar1}
    \ncline{a4}{lpar2}
    \ncline{a5}{ltens2}
    \ncline{a6}{ltens2}
    \ncline{ltens2}{lpar2}
    \ncline{lpar1}{lpar3}
    \ncline{lpar2}{lpar3}
  }{}
  $$
  The relation web is therefore
  $$
  \def\pathderimode{b}
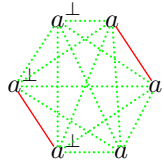
    \begin{psmatrix}[rowsep=3ex,colsep=1ex]
               &\na1\lneg& & \na2\phantom{\lneg}   \\
      \na6\lneg&         & &      &\na3 \\
               &\na5\lneg& & \na4 
      \webrededgemode
      \ncline{a2}{a3}
      \ncline{a5}{a6}
      \webgreenedgemode
      \ncline{a1}{a2}
      \ncline{a1}{a3}
      \ncline{a1}{a4}
      \ncline{a1}{a5}
      \ncline{a1}{a6}
      \ncline{a2}{a4}
      \ncline{a2}{a5}
      \ncline{a2}{a6}
      \ncline{a3}{a4}
      \ncline{a3}{a5}
      \ncline{a3}{a6}
      \ncline{a4}{a5}
      \ncline{a4}{a6}
    \end{psmatrix}
    $$
    where we use regular edges for red and dotted edges for green.
\end{example}

\begin{definition}
  The \emph{degree of freedom} of a formula $A$, is the number of green edges
  in its relation web.
\end{definition}

\begin{proofname}[Second Proof of Theorem~\ref{thm:S-correct}]
  Again, we start by showing that all rules preserve correctness. Here, the
  only interesting case is the switch rule (all others being trivial), which
  does the following transformation somewhere inside the net:
  \begin{equation}\label{eq:switch-correct}
    \vcenter{\hbox{$
	\def\pathderimode{b}
	\begin{psmatrix}[rowsep=3ex,colsep=1ex]
	  \nn1& &\nn2 & &\nn3  &    & \nn4& & \nn5 & & \nn6 \\
	  & &        &\nlpar1& & \qqquad\to\qqquad\strut &     & \nltens2\\
	  & &\nltens1&       & &                    &     &&\nlpar2
	  \pntreemode
	  \ncline{a1}{ltens1}
	  \ncline{a2}{lpar1}
	  \ncline{a3}{lpar1}
	  \ncline{lpar1}{ltens1}
	  \ncline{a4}{ltens2}
	  \ncline{a5}{ltens2}
	  \ncline{a6}{lpar2}
	  \ncline{lpar2}{ltens2}
	\end{psmatrix}$}}
  \end{equation}
  \Bwoc, assume the net on the left is correct, and the one on the right is
  not. First, suppose there is a switching for the second net that is
  cyclic. If that cycle does not contain the $\ltens$-node shown on the right
  in \eqref{eq:switch-correct}, then this cycle is also present in the net on
  the left in \eqref{eq:switch-correct}. If our cycle contains the
  $\ltens$-node, then we can make the same cycle be present in the net on the
  left by switching the $\lpar$-node to the left (i.e., removing the edge to
  the right). Now assume we have a disconnected switching for the net on the
  right. Then the same switching also disconnects the net on the
  left. Contradiction.

  Conversely, assume we have a correct net $\pi$ with conclusion $F$.  For the
  time being, assume that $\pi$ is cut-free.  We proceed by induction on the
  degree of freedom of~$F$. Pick inside~$F$ any pair of atoms that are linked
  together, say $a$ and $a\lneg$. Then $F=S\cons{S_1\cons{a}\lpar
  S_2\cons{a\lneg}}$. \Wolg, we can assume that $S_1\conhole$ and
  $S_2\conhole$ are not $\lpar$-contexts. We have the following cases:
  \begin{itemize}
  \item If $S_1\conhole=S_2\conhole=\conhole$, we can apply the rule
    $\ird$, and proceed by induction hypothesis.
  \item If $S_1\conhole\neq\conhole$ and $S_2\conhole=\conhole$, then
    $F=S\cons{(A\ltens B\cons{a})\lpar a\lneg}$ for some $A$ and $B\conhole$.
    We can apply the switch rule to get $S\cons{A\ltens (B\cons{a}\lpar
    a\lneg)}$, which is still correct (with the same linking as for $F$), but
    has smaller degree of freedom than $F$.  The case where
    $S_1\conhole=\conhole$ and $S_2\conhole\neq\conhole$ is similar.
  \item If $S_1\conhole\neq\conhole$ and $S_2\conhole\neq\conhole$, then,
    \wolg, $F=S\cons{(A\ltens B\cons{a})\lpar(C\cons{a\lneg}\ltens D)}$, for
    some $A$, $B\conhole$, $C\conhole$, $D$.  By Lemma~\ref{lem:pair}, we can
    apply the switch rule, since one of
    $$
    S\cons{A\ltens(B\cons{a}\lpar(C\cons{a\lneg}\ltens D))}
    \quand
    S\cons{((A\ltens B\cons{a})\lpar C\cons{a\lneg})\ltens D}
    $$
    is still correct. Since both of them have smaller degree of freedom than
    $F$, we can proceed by induction hypothesis.
  \end{itemize}
  If $\pi$ contains cuts, we can replace inside $\pi$ all cuts with $\ltens$,
  to get a formula $F'$ such that there is a derivation
  $$
  \vcstrder{\iru}{}{F}{\leaf{F'}}
  $$
  Then $\pi$ becomes a cut-free net with conclusion $F'$, and we can
  proceed as above.
  \qed
\end{proofname}

Note that the two different proofs of Theorem~\ref{thm:S-correct} yield a
stronger version of Theorem~\ref{thm:sq-cos}.

\begin{theorem}\label{thm:sc-cos-2}
  For every sequent calculus proof\proofadjust
  $$
  \sqnsmallderi{A_1,A_2,\ldots,A_n}{}
  $$
  there is a proof in the calculus of structures
  $$
  \strpr{}{}{A_1\lpar A_2\lpar\ldots\lpar A_n}
  $$
  yielding the same proof net, and vice versa.
\end{theorem}

A geometric or graph-theoretic criterion like the one in
Definition~\ref{def:S-correct} and Theorem~\ref{thm:S-correct} is called a
\emph{correctness criterion}. The desired properties are soundness and
completeness, as stated in Theorem~\ref{thm:S-correct}. For $\ufMLL$, the
literature contains quite a lot of such criteria, and it would go far beyond
the scope of this lecture notes to attempt to give a complete survey. But
nonetheless, we will show here two other correctness criteria.

For the next one, we write the pre-proof
nets in a different way:
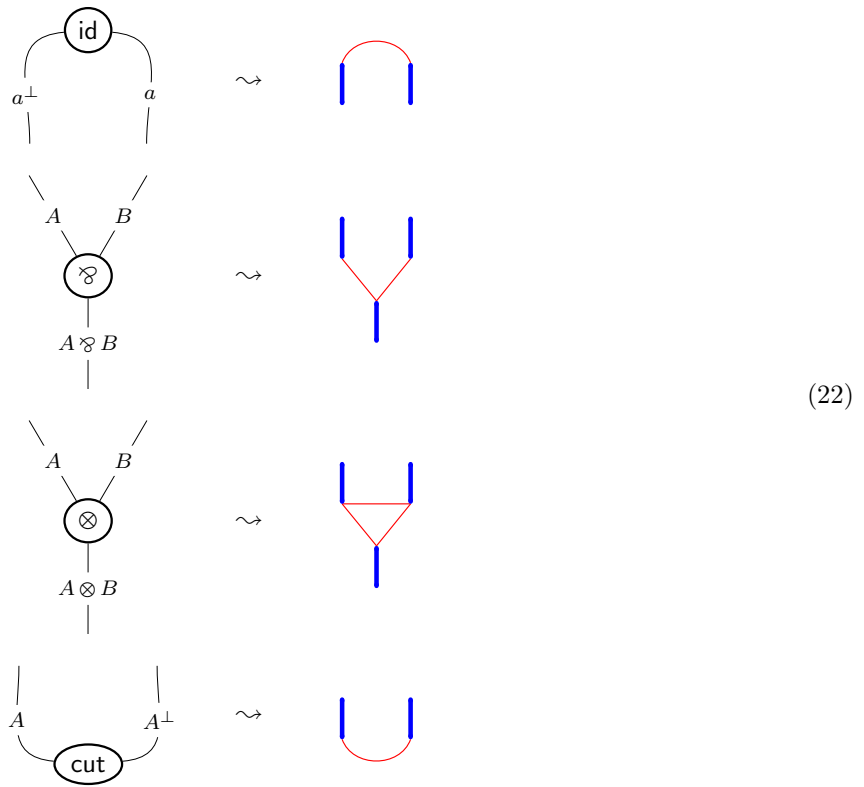
\begin{equation}
  \def\arrayskip{4ex}
  \def\rbrowsep{3ex}
  \begin{array}{c@{\qqquad\leadsto\qqquad}c}
    \idrulenet{a\lneg}{a}
    &
    \vcrbpn{
      \rbv1&&\rbv2\\
      \rbv3&&\rbv4}{
      \rbrededgemode
      \rbid12
      \rbblueedgemode
      \rbedge13
      \rbedge24}
    \\[\arrayskip]
    \vcgpn{
      \gpnout1&\strut&\gpnout2\\
      &\gpnpar3\\
      &\gpnout4 \strut}{
      \treeline{out1}{lpar3}{A}
      \treeline{out2}{lpar3}{B}
      \treeline{out4}{lpar3}{A\lpar B}}
    &
    \vcrbpn{
      \rbv1&&\rbv2\\
      \rbv3&&\rbv4\\
      &\rbv5\\
      &\rbv6}{
      \rbpar534
      \rbblueedgemode
      \rbedge13 
      \rbedge24
      \rbedge56}
    \\[\arrayskip]
    \vcgpn{
      \gpnout1&\strut&\gpnout2\\
      &\gpntens3\\
      &\gpnout4\strut}{
      \treeline{out1}{ltens3}{A}
      \treeline{out2}{ltens3}{B}
      \treeline{out4}{ltens3}{A\ltens B}}
    &
    \vcrbpn{
      \rbv1&&\rbv2\\
      \rbv3&&\rbv4\\
      &\rbv5\\
      &\rbv6}{
      \rbtens534
      \rbblueedgemode
      \rbedge13
      \rbedge24
      \rbedge56}
    \\[\arrayskip]
    \vcgpn{
      \gpnout{1}&\strut&\gpnout{2}\\&\gpncut{1} }{
      \cutleftline{out1}{cut1}{A}
      \cutrightline{out2}{cut1}{A\lneg}
    }
    &
    \vcrbpn{
      \rbv1&&\rbv2\\
      \rbv3&&\rbv4}{
      \rbcut34
      \rbblueedgemode
      \rbedge13
      \rbedge24}
  \end{array}
\end{equation}
We call the resulting graphs \emph{RB-graphs}. The R and B stand for
Regular/Red and Bold/Blue. The main property of these graphs is that the
blue/bold edges (in the following called \emph{B-edges}) provide a bipartition
of the set of vertices, i.e., every vertex in the RB-graph is connected to
exactly one other vertex via a B-edge. The red/regular edges are in the
following called \emph{R-edges}.

Here are the examples from \eqref{eq:exa-pre-pn} and \eqref{eq:exa-pre-pn-2}
written as RB-graphs:
\begin{equation}
  \vcrbpn{
    \\
    \rbv1&&\rbv2&&\rbv3&&\rbv4&&\rbv5&&\rbv6\\
    \rbv{11}&&\rbv{12}&&\rbv{13}&&\rbv{14}&&\rbv{15}&&\rbv{16}\\
    & && &&\rbv{7}  && && \rbv{8}\\
    & && &&\rbv{17} && && \rbv{18}\\
    & && &&         && \rbv{9}\\
    & && &&         && \rbv{19}
  }{
    \rbid14
    \rbid23
    \rbid56
    \rbcut{11}{12}
    \rbtens9{17}{18}
    \rbpar7{13}{14}
    \rbpar8{15}{16}
    \rbblueedgemode
    \rbedge1{11}
    \rbedge2{12}
    \rbedge3{13}
    \rbedge4{14}
    \rbedge5{15}
    \rbedge6{16}
    \rbedge7{17}
    \rbedge8{18}
    \rbedge9{19}
  }
\end{equation}
\begin{equation}
  \vcrbpn{
    \\
    \rbv1&&\rbv2&&\rbv3&&\rbv4&&\rbv5&&\rbv6\\
    \rbv{11}&&\rbv{12}&&\rbv{13}&&\rbv{14}&&\rbv{15}&&\rbv{16}\\
    &\rbv{7} && &&  && && \rbv{8}\\
    &\rbv{17} && && && && \rbv{18}
  }{
    \rbid16
    \rbid23
    \rbid45
    \rbcut{13}{14}
    \rbpar7{11}{12}
    \rbpar8{15}{16}
    \rbblueedgemode
    \rbedge1{11}
    \rbedge2{12}
    \rbedge3{13}
    \rbedge4{14}
    \rbedge5{15}
    \rbedge6{16}
    \rbedge7{17}
    \rbedge8{18}
  }
  \qqqqqquad
  \vcrbpn{
    \\
    \rbv1&&\rbv2&&\rbv3&&\rbv4\\
    \rbv{11}&&\rbv{12}&&\rbv{13}&&\rbv{14}\\
    &\rbv{7} && &&   \rbv{8}\\
    &\rbv{17} && &&  \rbv{18}
  }{
    \rbid14
    \rbid23
    \rbtens7{11}{12}
    \rbtens8{13}{14}
    \rbblueedgemode
    \rbedge1{11}
    \rbedge2{12}
    \rbedge3{13}
    \rbedge4{14}
    \rbedge7{17}
    \rbedge8{18}
  }  
\end{equation}

\begin{definition}\label{def:ae-path}
  Let $G$ be an RB-graph. An \emph{\AE-path} in $G$ is a path whose
  edges are alternating R- and B-edges, and that does not touch any
  vertex more than once. An \emph{\AE-cycle} in $G$ is a \AE-path from
  a vertex to itself, starting with a B-edge and ending with an
  R-edge.
\end{definition}

The A and E stand for ``alternating'' and ``elementary''. The meaning
of ``alternating'' should be clear, and the meaning of ``elementary''
is that the path or cycle must not cross itself.

\begin{definition}\label{def:RB-correct}
  A pre-proof net $\pi$ \emph{obeys the RB-criterion} (or shortly, is
  \emph{RB-correct}) iff its RB-graph $G_\pi$ contains no \AE-cycle and every
  pair of vertices in $G_\pi$ is connected via an \AE-path.
\end{definition}

\begin{theorem}\label{thm:RB-correct}
  A pre-proof net is RB-correct if and only if it is a proof net.
\end{theorem}

\begin{proof}
  We show that a pre-proof net is RB-correct iff it obeys the switching
  criterion, which is easy: If there are two vertices in the RB-graph not
  connected by an \AE-path, then there is a switching yielding a disconnected
  graph, and vice versa. Similarly, the RB-graph contains an \AE-cycle if and
  only if we can provide a switching with a cycle. \qed
\end{proof}

\begin{exercise}
  Work out the details of the previous proof.
\end{exercise}

For the third correctness criterion, we write our nets in yet another way:
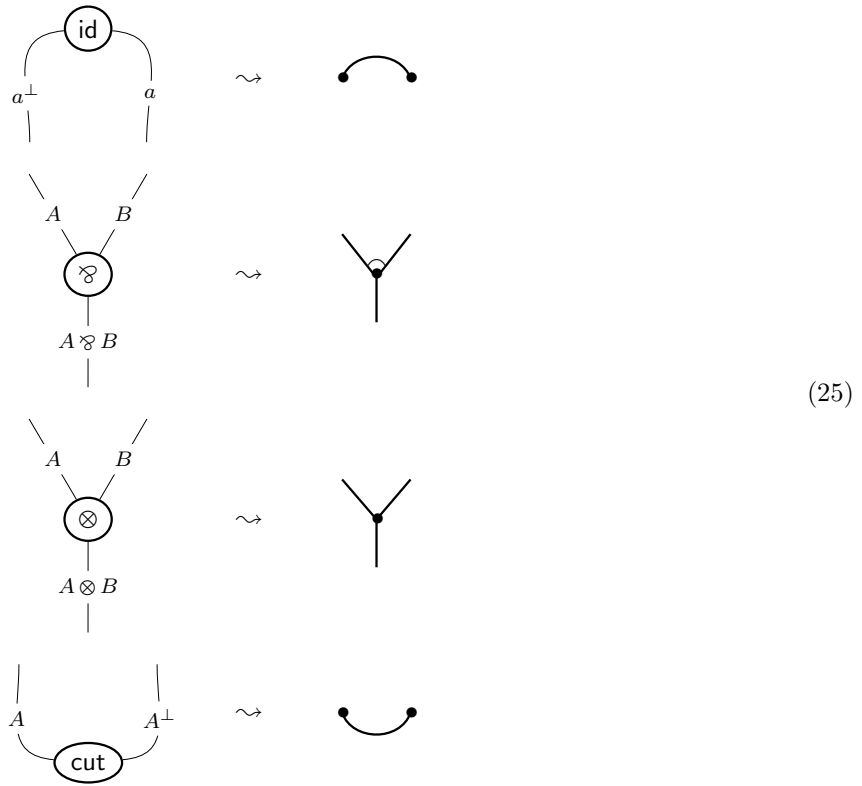
\begin{equation}
  \def\arrayskip{4ex}
  \begin{array}{c@{\qqquad\leadsto\qqquad}c}
    \idrulenet{a\lneg}{a}
    &
    \vccpn{
      \cpnv1&&\cpnv2}{
      \cpnid12}
    \\[\arrayskip]
    \vcgpn{
      \gpnout1&\strut&\gpnout2\\
      &\gpnpar3\\
      &\gpnout4 \strut}{
      \treeline{out1}{lpar3}{A}
      \treeline{out2}{lpar3}{B}
      \treeline{out4}{lpar3}{A\lpar B}}
    &
    \vccpn{
      \cpnf3&&\cpnf4\\
      &\cpnparv5\\
      &\cpnf6}{
      \cpnpar534
      \cpnedge56
    }
    \\[\arrayskip]
    \vcgpn{
      \gpnout1&\strut&\gpnout2\\
      &\gpntens3\\
      &\gpnout4\strut}{
      \treeline{out1}{ltens3}{A}
      \treeline{out2}{ltens3}{B}
      \treeline{out4}{ltens3}{A\ltens B}}
    &
    \vccpn{
      \cpnf3&&\cpnf4\\
      &\cpnv5\\
      &\cpnf6}{
      \cpntens534
      \cpnedge56
    }
    \\[\arrayskip]
    \vcgpn{
      \gpnout{1}&\strut&\gpnout{2}\\&\gpncut{1} }{
      \cutleftline{out1}{cut1}{A}
      \cutrightline{out2}{cut1}{A\lneg}
    }
    &
    \vccpn{
      \cpnv3&&\cpnv4}{
      \cpncut34}
  \end{array}
\end{equation}

Now consider the following two rewriting rules on these graphs:
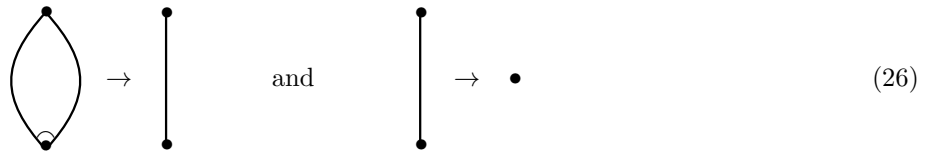
\begin{equation}\label{eq:contraction}
  \vccpn{
    \cpnv1\\
    \\
    \cpnparv2
    }{
    \lineanglesheight{cv1}{cv2}{-132}{132}{1}
    \lineanglesheight{cv1}{cv2}{-48}{48}{1}
  }
  \quad\quadto
  \vccpn{
    \cpnv1\\
    \\
    \cpnv2
    }{
    \cpnedge12 
  }
  \qqqquand
  \vccpn{
    \cpnv1\\
    \\
    \cpnv2
    }{
    \cpnedge12 
  }
  \quadto
  \vccpn{
    \cpnv1\\
    }{
  }
\end{equation}
It is important to note that in the first rule the two edges are between
the same pair of vertices and are connected by an arc at exactly one of the
two vertices. The second rule only applies if the two vertices on the lefthand
side are distinct, and the edge is not connected to another edge by an arc.

\begin{theorem}
  The reduction relation induced by the rules in \eqref{eq:contraction} is
  terminating and confluent.
\end{theorem}

\begin{proof}
  Termination is obvious because at each step the size of the graph is
  reduced. Hence, it suffices to show local confluence to get
  confluence. But this is easy since there are no (proper) critical pairs.
  \qed
\end{proof}

This means that for each pre-proof net we get a uniquely defined reduced
graph, and the question is now how this graph looks like.

\begin{exercise}
  Apply the reduction relation defined in \eqref{eq:contraction} to
  the nets in \eqref{eq:exa-pre-pn} and \eqref{eq:exa-pre-pn-2}.
\end{exercise}

\begin{definition}\label{def:C-correct}
  A pre-proof net \emph{obeys the contraction criterion} if its normal form
  according to the reduction relation defined in
  \eqref{eq:contraction} is
  $$
  \vccpn{\cpnv1}{}
  $$
  i.e., a single vertex without edges.
\end{definition}

At this point rather unsurprisingly, we get:

\begin{theorem}\label{thm:C-correct}
  A pre-proof net obeys the contraction criterion if and only if it is a proof
  net.
\end{theorem}

\begin{proof}
  As before, we show this by showing the equivalence of the switching
  criterion and the contraction criterion. This is easy to see since both
  reductions in \eqref{eq:contraction} preserve and reflect correctness
  according to the switching criterion.  \qed
\end{proof}

Before we leave the topic of correctness criteria, let us make some
important observations on their complexity. The naive implementation
of checking the switching criterion needs exponential time: if there are $n$
par-links in the net, then there are $2^n$ switchings to
check. However, checking the RB-criterion needs only quadratic
runtime. To verify this is an easy graph-theoretic exercise. 
It is also easy to see that checking the contraction criterion can be done in
quadratic time. But it is rather surprising that it can be done in linear time
in the size of the net.\footnote{For references, see Section~\ref{sec:notes}.}
This means that (in the case of $\ufMLL$) when we
go from a formal proof in a deductive system like the sequent calculus
or the calculus of structures (whose correctness can trivially be
checked in linear lime in the size of the proof) to the proof net, we
do not lose any information. The proof net contains the \emph{essence}
of the proof, including the ``deductive information''. Unfortunately,
$\ufMLL$ is (so far) the only logic (except some variants of it), for
which this ideal of proof nets is reached. We come back to this in
Sections \ref{sec:other} and~\ref{sec:cl}.

\subsection{Two-sided proof nets}\label{sec:twosided}

The proof nets we have seen are also called \emph{one-sided proof
nets}. This implies that there is another kind: the so called \emph{two-sided
proof nets}. Their existence is justified by the fact, that systems in the
sequent calculus can also come in a two-sided version. Here is a two-sided
sequent calculus system for $\ufMLL$:
\begin{equation}\label{eq:sc-MLL-twosided}
  \begin{array}{c@{\qquad}c}
    \vcinf{\idr}{\ssqn{A}{A}}{}&
    \vciinf{\cutr}{\ssqn{\Gamma,\Gamma'}{\Delta,\Delta'}}{
      \ssqn{\Gamma}{\Delta,A}}{
      \ssqn{A,\Gamma'}{\Delta'}}\\[\arrayskip]
    \vcinf{\exrl}{
      \ssqn{\Gamma,A,B,\Gamma'}{\Delta}}{
      \ssqn{\Gamma,B,A,\Gamma'}{\Delta}}&
    \vcinf{\exrr}{
      \ssqn{\Gamma}{\Delta,A,B,\Delta'}}{
      \ssqn{\Gamma}{\Delta,B,A,\Delta'}}
    \\[\arrayskip]
    \vciinf{\lparl}{\ssqn{\Gamma,\Gamma',A\lpar B}{\Delta,\Delta'}}{
      \ssqn{\Gamma,A}{\Delta}}{
      \ssqn{\Gamma',B}{\Delta'}}&
    \vcinf{\lparr}{\ssqn{\Gamma}{A\lpar B,\Delta}}{
      \ssqn{\Gamma}{A,B,\Delta}}\\[\arrayskip]
    \vcinf{\ltensl}{\ssqn{\Gamma,A\ltens B}{\Delta}}{
      \ssqn{\Gamma,A,B}{\Delta}}&
    \vciinf{\ltensr}{\ssqn{\Gamma,\Gamma'}{A\ltens B,\Delta,\Delta'}}{
      \ssqn{\Gamma}{A,\Delta}}{
      \ssqn{\Gamma'}{B,\Delta'}}\\[\arrayskip]
    \vciinf{\limpl}{\ssqn{\Gamma,\Gamma',A\limp B}{\Delta,\Delta'}}{
      \ssqn{\Gamma}{A,\Delta}}{
      \ssqn{\Gamma',B}{\Delta'}}&
    \vcinf{\limpr}{\ssqn{\Gamma}{A\limp B,\Delta}}{
      \ssqn{\Gamma,A}{B,\Delta}}\\[\arrayskip]
    \vcinf{\lnegl}{\ssqn{\Gamma,A\lneg}{\Delta}}{\ssqn{\Gamma}{A,\Delta}}&
    \vcinf{\lnegr}{\ssqn{\Gamma}{A\lneg,\Delta}}{\ssqn{\Gamma,A}{\Delta}}    
  \end{array}
\end{equation}
Now sequents are not just lists of formulas, but pairs of
lists of formulas, and these pairs of lists are separated by a
$\sqn{}$.  One reason for using two-sided sequent systems is that one
can treat negation as a proper connective (i.e., it is not pushed down
to the atoms as in the one-sided version) and that one can have
implication as primitive (i.e., there are rules for implication,
instead of just defining $A\limp B=A\lneg\lpar B$).  Of course, the
disadvantage is that we heavily increase the number of rules. Here is
an example of a two-sided derivation using the rules 
in~\eqref{eq:sc-MLL-twosided}:
\begin{equation}\label{eq:exa-sc-twosided}
  \vcenter{\scderitwosided}
\end{equation}
We omitted the instances of the exchange rule in this example.

\begin{figure}[!t]
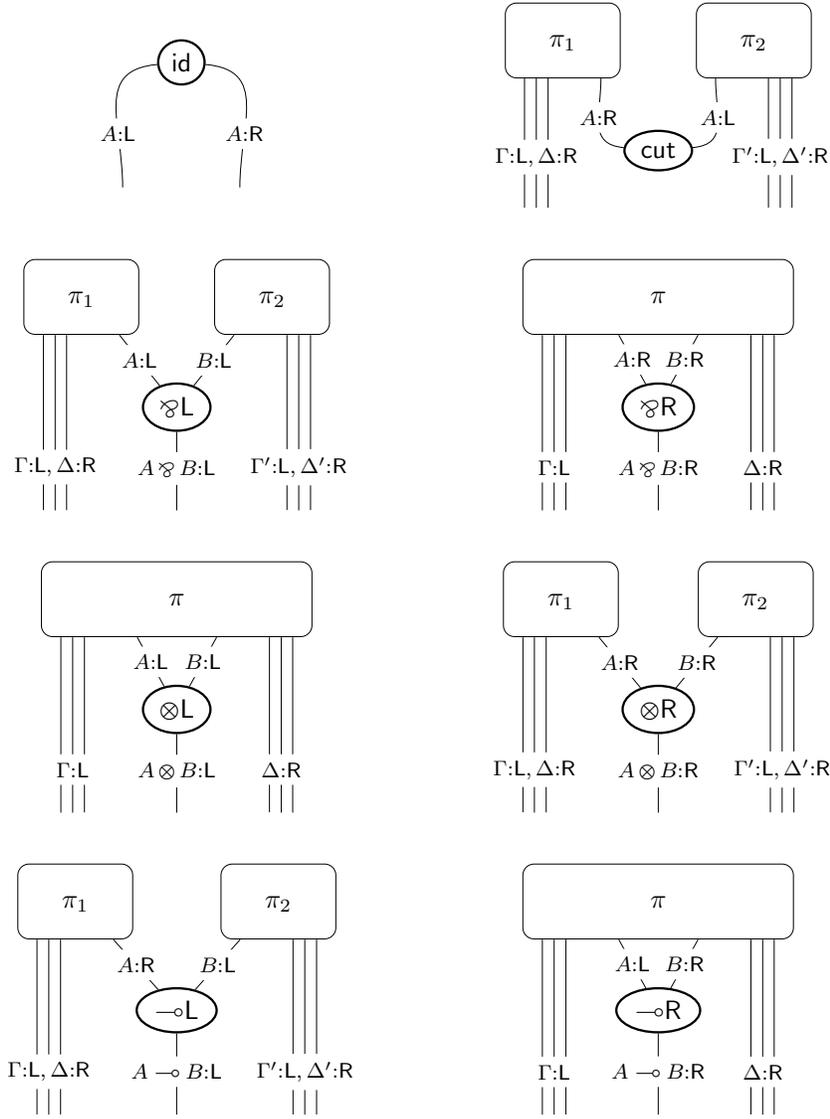

  \def\arrayskip{13ex}
  \def\pnrowsep{7ex}
  \begin{center}
    $$
    \begin{array}{c@{\qqqqquad}c}
      \idrulenet{A\markl}{A\markr}
      &
      \cutrulenet{\pi_1}{\Gamma\markl,\Delta\markr}{A\markr}
		 {\pi_2}{A\markl}{\Gamma'\markl,\Delta'\markr}
      \\[11ex]
      \tensorruleshape{
	\pi_1}{\Gamma\markl,\Delta\markr}{A\markl}{
	\pi_2}{B\markl}{\Gamma'\markl,\Delta'\markr}{
	\gpnparl}{lparl}{A\lpar B\markl}
      &
      \parruleshape{\pi}{
	\Gamma\markl}{A\markr}{B\markr}{\Delta\markr}{
	\gpnparr}{lparr}{A\lpar B\markr}
      \\[\arrayskip]
      \parruleshape{\pi}{
	\Gamma\markl}{A\markl}{B\markl}{\Delta\markr}{
	\gpntensl}{ltensl}{A\ltens B\markl}
      &
      \tensorruleshape{
	\pi_1}{\Gamma\markl,\Delta\markr}{A\markr}{
	\pi_2}{B\markr}{\Gamma'\markl,\Delta'\markr}{
	\gpntensr}{ltensr}{A\ltens B\markr}
      \\[\arrayskip]
      \tensorruleshape{
	\pi_1}{\Gamma\markl,\Delta\markr}{A\markr}{
	\pi_2}{B\markl}{\Gamma'\markl,\Delta'\markr}{
	\gpnimpl}{limpl}{A\limp B\markl}
      &
      \parruleshape{\pi}{
	\Gamma\markl}{A\markl}{B\markr}{\Delta\markr}{
	\gpnimpr}{limpr}{A\limp B\markr}
    \end{array}  
    $$
    \caption{From two-sided sequent calculus to proof nets (sequent calculus
    rule driven)}
    \label{fig:sc-to-pn-1-twosided}
  \end{center}
\end{figure}

We can follow ideology~\ref{ideo:girard} to translate two-sided sequent proofs
into proof nets. But we have to assign to each rule and each formula appearing
in the net a label which indicated whether the rule/formula comes from the
left-hand side or the right-hand side of the sequent. We use here the letters
$\mathsf{L}$ and $\mathsf{R}$. Figure~\ref{fig:sc-to-pn-1-twosided} shows how
the sequent rules in \eqref{eq:sc-MLL-twosided} are translated into proof
nets. We omitted the rules for $\exr$ (compare with
Figure~\ref{fig:sc-to-pn-1}) and $\cdot\lneg$.
Here is the result of
translating~\eqref{eq:exa-sc-twosided}:
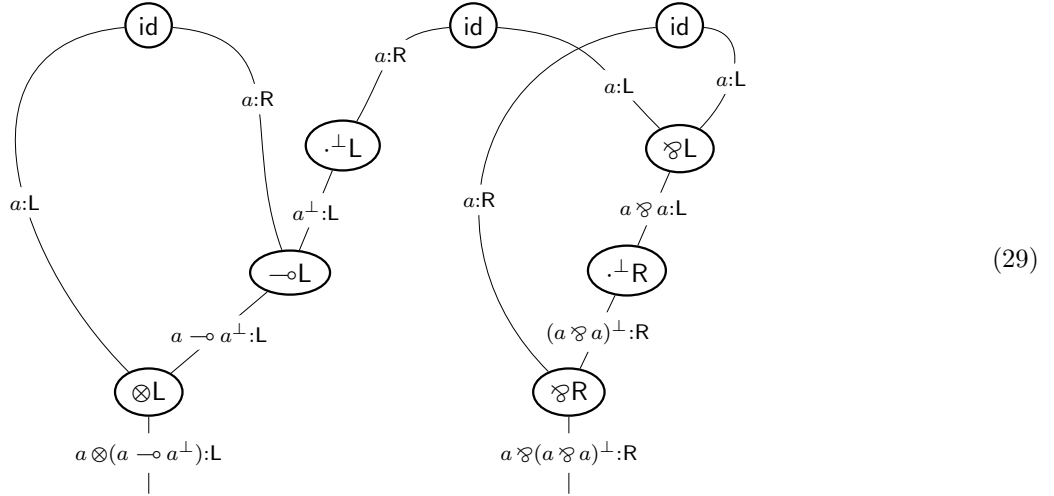
\begin{equation}\label{eq:exa-pn-twosided}
  \vcgpn{
    & \gpnid{1}&\quad&         &\quad&\gpnid{2}&      &\gpnid3\\
    &  & &\hskip2em\gpnnode1{\lnegl}\hskip-2em&&&&\gpnnode3{\lparl}\\
    &  & &\hskip-2em\gpnnode2{\limpl}\hskip2em&&&&\hskip-2em\gpnnode4{\lnegr}\hskip2em\\
    &\gpnnode5{\ltensl}&     &         &     & &\gpnnode6{\lparr}\\
    &\gpnout1  &     &         &     &         &\gpnout2
  }{
    \idleftline{gpn5}{id1}{a\markl}
    \treeline{gpn5}{gpn2}{a\limp a\lneg\markl}
    \treeoutline{gpn5}{out1}{a\ltens(a\limp a\lneg)\markl}
    \gpncurve{gpn2}{id1}{110}{-5}{a\markr}
    \treeline{gpn2}{gpn1}{a\lneg\markl}
    \gpncurve{gpn1}{id2}{60}{185}{a\markr}
    \gpnmycurve{gpn3}{id2}{135}{-5}{.25}{a\markl}
    \idrightline{gpn3}{id3}{a\markl}
    \idleftline{gpn6}{id3}{a\markr}
    \treeline{gpn6}{gpn4}{(a\lpar a)\lneg\markr}
    \treeline{gpn4}{gpn3}{a\lpar a\markl}
    \treeoutline{gpn6}{out2}{a\lpar(a\lpar a)\lneg\markr}
  }  
\end{equation}
Note that the nets in \eqref{eq:3sq-pn1} and
\eqref{eq:exa-pn-twosided} are almost identical from the
graph-theoretical viewpoint. The only differences are that the labels
are different and that there are additional $\lnegl$ and $\lnegr$
nodes in~\eqref{eq:exa-pn-twosided}.  

Of course, we can also use the flow-graph method to obtain a proof net
from~\eqref{eq:exa-sc-twosided}:\setboolean{doflows}{true}
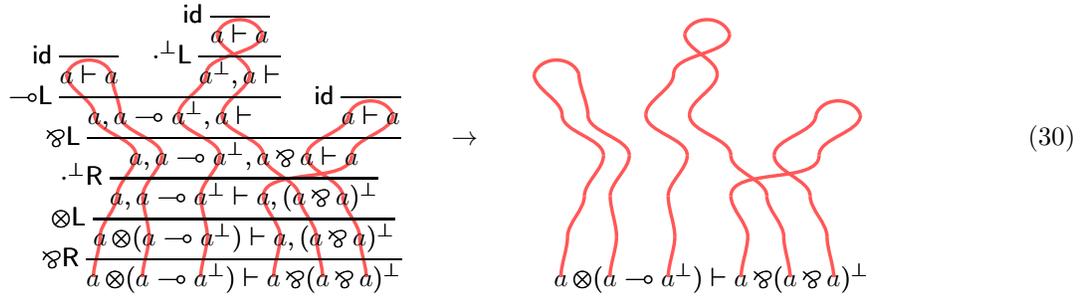
  \begin{equation}\label{eq:exa-pn-twosided-2}
    \hskip-1.5em
    \vctpathderivation{\scderitwosided}{\pathscderitwosided}
    \quad\to\hskip-3em
    \vctpathonlyderivation{\scderitwosided}{\pathscderitwosided}
  \end{equation}
However, observe that from the flow-graph we do not get \emph{a
priori} the information what is left and what is right. This means,
for a proper definition of pre-proof net, we need to find another way
to determine when we may allow an identity link between a pair of
atoms. To do so, we use \emph{polarities}: we assign to each subformula
appearing in the sequent a unique polarity, that is, an element of the
set $\set{\poll,\polr}$, where $\poll$ can be read as left/negative/up
and $\polr$ as right/positive/down. Let $\ssqn{\Gamma}{\Delta}$ be given. Then
\begin{enumerate}[(i)]
\item all formulas in $\Gamma$ get polarity $\poll$, and all formulas
  in $\Delta$ get polarity $\polr$;
\item whenever $A\ltens B$ has polarity~$\poll$ (resp.~$\polr$), then
  $A$ and $B$ also get polarity~$\poll$ (resp.~$\polr$);
\item whenever $A\lpar B$ has polarity~$\poll$ (resp.~$\polr$), then
  $A$ and $B$ also get polarity~$\poll$ (resp.~$\polr$);
\item whenever $A\limp B$ has polarity~$\poll$ (resp.~$\polr$), then
  $A$ gets polarity~$\polr$ (resp.~$\poll$) and $B$ gets
  polarity~$\poll$ (resp.~$\polr$);
\item whenever $A\lneg$ has polarity~$\poll$ (resp.~$\polr$), then
  $A$ gets polarity~$\polr$ (resp.~$\poll$).
\end{enumerate}
Note that only for negation and implication there is a change of
polarity. We can now give the following definition:

\begin{definition}\label{def:twosided-prenet}
  A \emph{two-sided pre-proof net} consists of a list of formula trees,
  which is correctly polarized according to (ii)-(v) above, together
  with a perfect matching of the set of leaves, such that two leaves
  are matched only if they are labeled by the same atom and have
  different polarity. If there are cuts in the net, then the two cut
  formulas have to be identical and must have different polarity.
\end{definition}

As an example we give here the net corresponding
to~\eqref{eq:exa-pn-twosided-2}:
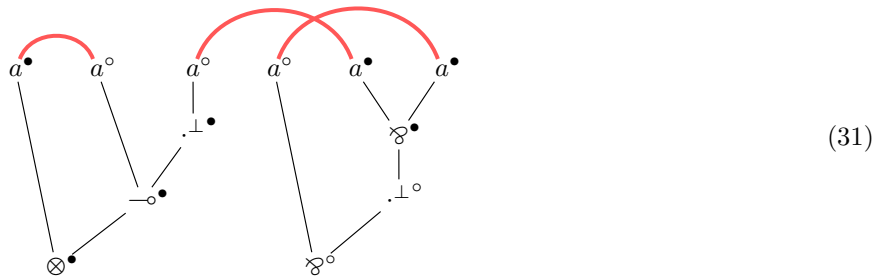
\begin{equation}\label{eq:exa-pn-twosided-3}
  \vcenter{\hbox{$\pathpntwosided$}}
\end{equation}%
where we put the polarity as superscript to nodes in the trees. It
should not come at a surprise that we get (again, up to some trivial
change in the notation) the same proof nets as with the first method.

The correctness criteria are exactly the same as we discussed them for
one-sided proof nets in Section~\ref{sec:correctness}. The only thing to note
is that the $\ltens\markpl$, $\lpar\markpr$, and $\limp\markpr$ behave as the
$\lpar$ in Section~\ref{sec:correctness}, i.e., remove one edge to a child to
get a switching, and the $\ltens\markpr$, $\lpar\markpl$, and $\limp\markpl$
behave as the $\ltens$ in Section~\ref{sec:correctness}. Clearly, the example
in~\eqref{eq:exa-pn-twosided-3} is correct.

\bigskip

Let us now for the time being forget the $\limp$ and $\cdot\lneg$ nodes that
flip the polarity. Then we can draw the two-sided proof net in a two-sided
way. Things with polarity $\poll$ are drawn on the top and things with
polarity $\polr$ are drawn at the bottom. In between we put the identity
links. Since we now have again negated atoms, we have to change the condition
on the identity links: \emph{Two atoms occurrences may be linked, if they are
the same and have different polarity, or if they are dual to each other and
have the same polarity.}  Here are two examples:
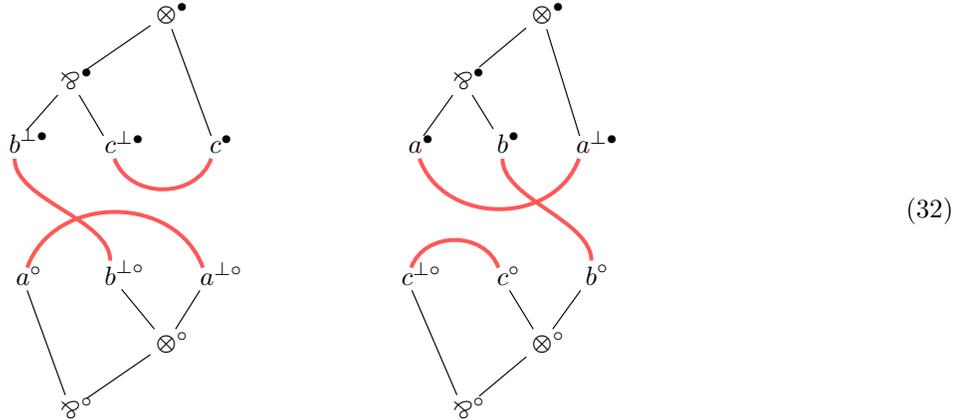
\begin{equation}\label{eq:exa-pn-really-twosided}
  \exapntwosideda
  \qqqqqquad
  \exapntwosidedb
\end{equation}
Note that these are exactly the nets that are obtained via the
flow-graph method from the calculus of structures derivations
in~\eqref{eq:exa-cos}:
\begin{equation}\label{eq:exa-cos-twosided}
  \vctpathderivation{\exacostwosideda}{\pathexacostwosideda}
  \qqqqquad
  \vctpathderivation{\exacostwosidedb}{\pathexacostwosidedb}
\end{equation}
This is the second justification for the existence of two-sided proof
nets. They are directly obtained via flow-graphs from derivations in
the calculus of structures. As before, the correctness criterion does
not change.

Note that from the graph-theoretical point of view, there is now not
much difference between $\ltens\markpr$ and $\lpar\markpl$, between
$\ltens\markpl$ and $\lpar\markpr$, between $a\markpr$ and
$a\lnegpl$, and between $a\markpl$ and $a\lnegpr$ (at
least for $\ufMLL$). For this reason, the nets
in~\eqref{eq:exa-pn-really-twosided} could also be written as
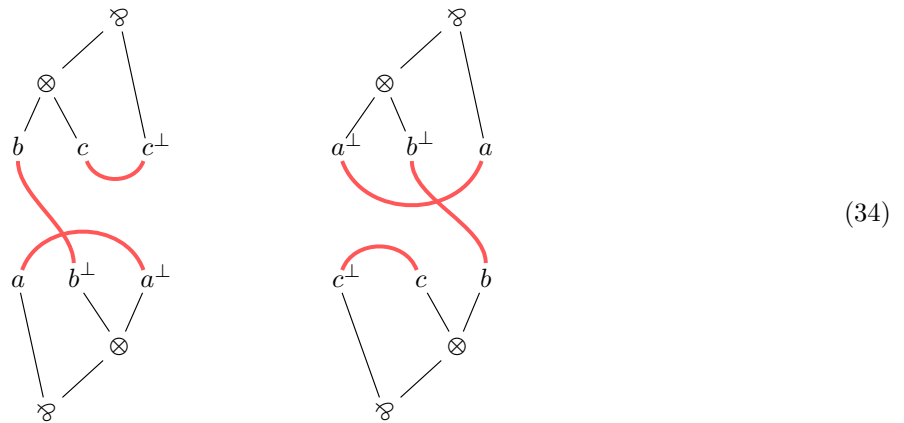
\begin{equation}\label{eq:exa-pn-no-longer-twosided}
  \exapntwosidedavar
  \qqqqqquad
  \exapntwosidedbvar
\end{equation}
Now they are in fact the same net just turned upside down. This is no
surprise since the two derivations in~\eqref{eq:exa-cos-twosided} are
dual to each other.

Let us emphasize that it does not matter in which way the graph is drawn on
the paper. The net on the left
in~(\ref{eq:exa-pn-really-twosided},\ref{eq:exa-pn-no-longer-twosided})
could also be drawn as 
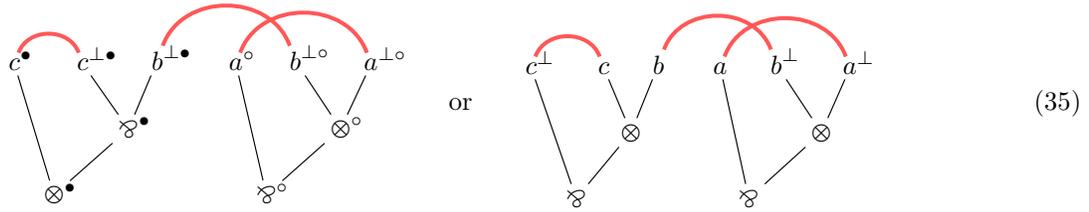
\begin{equation}\label{eq:two-one}
  \exapnonesideda  \qquor  \exapnonesidedavar
\end{equation}
In the case of $\ufMLL$ it makes no difference whether one prefers to put a
proof net in a ``one-sided'' \eqref{eq:two-one} or ``two-sided''
(\ref{eq:exa-pn-really-twosided},\ref{eq:exa-pn-no-longer-twosided}) way on
the paper, and whether one prefers to use the labels $\set{\poll,\polr}$ for
left/right (or up/down) as in (\ref{eq:exa-pn-really-twosided}) or to
live without them as in \eqref{eq:exa-pn-no-longer-twosided}.

This rather trivial observation can be nailed down in the following
theorem which strengthens Theorems \ref{thm:sq-cos}
and~\ref{thm:sc-cos-2}.

\begin{theorem}\label{thm:sc-cos-pn}
  Let $n,m\le 1$ and $A_1,\ldots,A_n,B_1,\ldots,B_m$ be any $\ufMLL$
  formulas. Then the following are equivalent:
  \begin{enumerate}[\rm (i)]
  \item There is a sequent calculus proof\proofadjust
    \begin{equation}\label{eq:sc1}
      \sqnsmallderi{A_1\lneg,\ldots,A_n\lneg,B_1,\ldots,B_m}{}
    \end{equation}
  \item There is a calculus of structures proof
    \begin{equation}\label{eq:cos1}
      \strpr{}{}{A_1\lneg\lpar\cdots\lpar A_n\lneg\lpar 
	B_1\lpar \ldots\lpar B_m}
    \end{equation}
  \item There is a (two-sided) sequent calculus proof\proofadjust
    \begin{equation}\label{eq:sc2}
      \ssqnsmallderi{A_1,\ldots,A_n}{B_1,\ldots,B_m}{}
    \end{equation}
  \item There is a calculus of structures proof
    \begin{equation}\label{eq:cos2}
      \strder{}{}{B_1\lpar \ldots\lpar B_m}{
	\leaf{A_1\ltens\cdots\ltens A_n}} 
    \end{equation}
  \end{enumerate}
  Furthermore all of \eqref{eq:sc1}--\eqref{eq:cos2} can be
  constructed such that they all yield the same proof net.
\end{theorem}

This section can be summarized a follows: Two-sided proof nets are a
special kind of one-sided proof nets, and one-sided proof nets are
a special kind of two-sided proof nets. Whether one prefers the
one-sided or the two-sided version is a matter of personal taste and
also a matter of economics in notation.

This observation is \emph{not} restricted to the special case of
$\ufMLL$. Every theory of two-sided proof nets can trivially transformed into
a theory of one-sided proof nets, and vice versa. By using polarities as
described above, the one-sided version can usually be presented in a more
compact way than the two-sided version. For this reason, we will in the
following stay in the one-sided world. 

\subsection{Cut elimination}\label{sec:cutelim}

Let us now briefly discuss an issue that usually excites proof theorists: In a
well-designed deduction system, every formula/sequent that is derivable, can
also be derived without using the cut-rule. Proving this fact is usually a
highly nontrivial task and involves (depending on the logic in question) very
sophisticated proof theory. 

But here we do not have time and space to go into the details of cut
elimination and its important consequences. We discuss
here only a very simple logic, namely $\ufMLL$, for which cut
elimination is easy. For the sequent calculus it is stated as follows:

\begin{theorem}\label{thm:sc-cutelim}
  For every proof \proofadjust
  $$
  \sqnsmallderi{\Gamma}{\Pi}
  $$
  using the rules in \eqref{eq:sc-mll} there is a proof\proofadjust 
  $$
  \sqnsmallderi{\Gamma}{\Pi'}
  $$
  of the same conclusion, that does not use the $\cutr$-rule.
\end{theorem}

\begin{proof}[ (Sketch)]
  We do only sketch the direct proof within the sequent calculus,
  because, even in the simple case of $\ufMLL$, carrying out all the
  details is quite tiresome and boring. But the basic idea is simple:
  The cut rule is permuted up in the proof until it disappears. Most
  cases are trivial rule permutations similar to the one in
  \eqref{eq:rule-permut}. But there are two so called \emph{key cases}
  which are not trivial. Here is the first one: 
  \proofadjust
  $$
    \qlapm{
  \ddernote{\cutr}{}{
    \sqn{\Gamma,\Delta,\Gamma',\Delta'}}{
    \root{\lpar}{\sqn{\Gamma,A\lpar B,\Delta}}{
      \leaf{\sqnsmallderi{\Gamma,A, B,\Delta}{\Pi_1}}}}{
    \rroot{\ltens}{\sqn{\Gamma',B\lneg\ltens A\lneg,\Delta'}}{
      \leaf{\sqnsmallderi{\Gamma',B\lneg}{\Pi_2}}}{
      \leaf{\sqnsmallderi{A\lneg,\Delta'}{\Pi_3}}}}
  \quad
  \leadsto
  \quad
  \ddernote{\cutr}{}{
    \sqn{\Gamma,\Delta,\Gamma',\Delta'}}{
    \rroot{\cutr}{
      \sqn{\Gamma,A,\Delta,\Gamma'}}{
      \leaf{\sqnsmallderi{\Gamma,A, B,\Delta}{\Pi_1}}}{
      \leaf{\sqnsmallderi{\Gamma',B\lneg}{\Pi_2}}}}{
    \leaf{\sqnsmallderi{A\lneg,\Delta'}{\Pi_3}}}
  }
  $$
  where some instances of the $\exr$-rule have been omitted. The second
  key case lets the cut disappear when it meets an identity rule at the top of
  the proof:\proofadjust
  $$
  \ddernote{\cutr}{}{\sqn{A\lneg,\Gamma}}{
    \root{\idr}{\sqn{A\lneg,A}}{
      \leaf{}}}{
    \leaf{\sqnsmallderi{A\lneg,\Gamma}{\Pi_1}}}
  \qquad
  \leadsto
  \qquad
  \sqnsmallderi{A\lneg,\Gamma}{\Pi_1}
  $$
  The difficult part of a cut elimination proof is usually to show termination
  of this ``permuting the cut up'' business, which is done by coming up with a
  clever induction measure. We leave that to the reader.
  \qed
\end{proof}

\begin{exercise}
  Complete the proof of Theorem~\ref{thm:sc-cutelim}. The following things
  still have to be done: 
  \begin{enumerate}
  \item Find a correct way of dealing with the $\exr$-rule. (Hint: Design a
    ``super-cut-rule'', that has exchange built in.)
  \item Show the termination of the process of permuting up the cut, i.e., find
    the right induction measure.
  \item Show that after the termination, the resulting proof is indeed
    cut-free.
  \end{enumerate}
  If you have seen a cut elimination argument before, then this exercise
  should not be too hard for you. If you have never before seen some cut
  elimination, you will learn a lot about it by doing this exercise. In any
  case, this (rather painful) exercise will help you to admire the beauty of
  proof nets.
\end{exercise}

Let us now turn to cut elimination in the calculus of structures, where it
means that not only the cut (i.e., the rule $\iru$), but the whole up-fragment
(i.e., all rules with the $\uparrow$ in the name) are not needed for
provability.

\begin{theorem}\label{thm:cos-cutelim}
  If there is a proof 
  $$
  \strpr{\set{\ird,\iru,\comrd,\comru,\assrd,\assru,\swir}}{\Pi}{A}
  $$
  using all rules in Figure~\ref{fig:cos-mll}, then there is a proof 
  $$
  \strpr{\set{\ird,\comrd,\assrd,\swir}}{\Pi}{A}
  $$
  that has the same conclusion $A$ and that does not use any up-rule. 
\end{theorem}

\begin{proof}[ (Sketch)]
  The easy trick of permuting the cut up does not work in the presence of deep
  inference. For this reason completely different techniques have to be used
  to eliminate the up-fragment. But as in the sequent calculus, carrying out
  all the details is technical and boring. For this reason we will give here
  only a sketch.  The first observation to make is that every up-rule
  $$
  \vcinf{\ruleu}{S\cons{A\lneg}}{S\cons{B\lneg}}
  $$ 
  can be replaced by a derivation using only
  $\ird,\iru,\swir,\ruled$:
  \begin{equation}\label{eq:iru}
  \vcdernote{\iru}{}{S\cons{A\lneg}}{
    \root{\swir}{S\cons{A\lneg\lpar(B\ltens B\lneg)}}{
      \root{\ruled}{S\cons{(A\lneg\lpar B)\ltens B\lneg}}{
	\root{\ird}{S\cons{(A\lneg\lpar A)\ltens B\lneg}}{
	  \leaf{S\cons{B\lneg}}}}}}
  \end{equation}
  This means that only the rule $\iru$ needs to be eliminated. The second
  observation is, that this rule can be reduced to its atomic
  version:
  $$
  \begin{array}{c@{\qquad\leadsto\qquad}c}
    \vcinf{\iru}{}{A\ltens A\lneg} 
    &
    \vcinf{\atiru}{}{a\ltens a\lneg} 
    \\[\arrayskip]
    \vcinf{\iru}{S\cons{B}}{S\cons{B\lpar(A\ltens A\lneg)}} 
    &
    \vcinf{\atiru}{S\cons{B}}{S\cons{B\lpar(a\ltens a\lneg)}} 
  \end{array}
  $$ This\footnote{By duality, we can do the same trick to the rule
  $\ird$. Compare with what we did in \eqref{eq:sc-id-to-atom} to the
  $\idr$-rule in the sequent calculus.}  is done by systematically
  replacing
  \begin{equation}\label{eq:atiru}
    \qlapm{
    \vcinf{\iru}{S\cons{B}}{
      S\cons{B\lpar((A\lpar C)\ltens(C\lneg\ltens A\lneg))}} 
    \quad\mbox{by}\quad
    \vcdernote{\iru}{}{S\cons{B}}{
      \root{\iru}{S\cons{B\lpar(A\ltens A\lneg)}}{
	\root{\swir}{S\cons{B\lpar((A\lpar(C\ltens C\lneg))\ltens A\lneg)}}{
	  \root{\assru}{
	    S\cons{B\lpar(((A\lpar C)\ltens C\lneg)\ltens A\lneg)}}{
	    \leaf{S\cons{B\lpar((A\lpar C)\ltens(C\lneg\ltens A\lneg))}}}}}}
    \qqquad}
  \end{equation}
  Now the rules to eliminate are $\atiru$, $\assru$, and $\comru$ (see
  Exercise~\ref{ex:comru}).\footnote{It might seem silly to first explain that
  only the rule $\iru$ needs to be eliminated, and then say that also the
  other up-rules need to be taken care of. The reason is that with the trick
  in \eqref{eq:iru} \emph{all} up-rules can be removed, but by reducing the
  rule $\iru$ to it atomic version, only \emph{some} of them are
  reintroduced. This is crucial for more sophisticated logics than $\ufMLL$.}
  For this, we use Guglielmi's powerful \emph{splitting lemma}, which says
  that whenever there is a proof
  \begin{equation}\label{eq:splittingin}
    \vcstrpr{\set{\ird,\comrd,\assrd,\swir}}{\Pi}{S\cons{C\lpar(A\ltens B)}}
  \end{equation}
  then there are formulas $K_A$ and $K_B$, such that for every formula $C$, we
  have derivations
  \begin{equation}\label{eq:splittingout}
    \!\!\!
    \vcstrder{\set{\ird,\comrd,\assrd,\swir}}{\Pi_1}{S\cons{C}}{
      \leaf{K_A\lpar K_B}}
    \qquad,\qquad
    \vcstrpr{\set{\ird,\comrd,\assrd,\swir}}{\Pi_2}{K_A\lpar A}
    \qquad,\qquad
    \vcstrpr{\set{\ird,\comrd,\assrd,\swir}}{\Pi_3}{K_B\lpar B}
    \quad
  \end{equation}
  Additionally, we get a derivation
  \begin{equation}\label{eq:splittingoutb}
    \vcstrder{\set{\ird,\comrd,\assrd,\swir}}{\Pit_1}{S\cons{C\lpar X}}{
      \leaf{K_A\lpar K_B\lpar X}}
  \end{equation}
  for every formula $X$.
  The crucial point of this lemma is that it speaks only about the
  down-fragment.
  Furthermore we need a variant of the splitting lemma, which says that
  whenever we have
  $$
  \vcstrpr{\set{\ird,\comrd,\assrd,\swir}}{\Pi}{C\lpar a}
  \quad,\quad
  \mbox{ then we also have }
  \quad
  \vcstrder{\set{\ird,\comrd,\assrd,\swir}}{\Pi'}{C}{\leaf{a\lneg}}
  \quadfs
  $$
  Now we can remove the rule $\atiru$ starting with the topmost instance as
  follows. What we have is 
  $$
  \dernote{\atiru}{}{S\cons{B}}{
    \stempr{\set{\ird,\comrd,\assrd,\swir}}{\Pi}{
      S\cons{B\lpar(a\ltens a\lneg)}}}
  $$
  Via the splitting lemma, we can get
  $$
  \vcstrder{\set{\ird,\comrd,\assrd,\swir}}{\Pi_1}{S\cons{B}}{
    \leaf{K_a\lpar K_{a\lneg}}}
  \qquad,\qquad
  \vcstrpr{\set{\ird,\comrd,\assrd,\swir}}{\Pi_2}{K_a\lpar a}
  \qquad,\qquad
  \vcstrpr{\set{\ird,\comrd,\assrd,\swir}}{\Pi_3}{K_{a\lneg}\lpar a\lneg}
  $$
  Now we can get from $\Pi_2$ and $\Pi_3$ the following:
  $$
  \vcstrder{\set{\ird,\comrd,\assrd,\swir}}{\Pi_2'}{K_a}{\leaf{a\lneg}}
  \qqquand
  \vcstrder{\set{\ird,\comrd,\assrd,\swir}}{\Pi_3'}{K_{a\lneg}}{\leaf{a}}
  $$
  now we can build:
  $$
  \vcstrder{\set{\ird,\comrd,\assrd,\swir}}{\Pi_1}{S\cons{B}}{
    \stem{\set{\ird,\comrd,\assrd,\swir}}{\Pi_2'\lpar\Pi_3'}{
      K_a\lpar K_{a\lneg}}{
      \root{\atird}{a\lneg\lpar a}{
	\leaf{}}}}
  $$
  which gives us a proof for $S\cons{B}$ that does not use any up-rules.
  \qed
\end{proof}

\begin{exercise}\label{ex:comru}
  Explain, why not only the rules $\atiru$ and $\assru$, but also the rule
  $\comru$ needs to be eliminated after doing \eqref{eq:atiru}.
\end{exercise}

\begin{exercise}
  Complete the proof of Theorem~\ref{thm:cos-cutelim}. Things that remain to
  be done:
  \begin{enumerate}
  \item Show how the rules $\assru$ and $\comru$ can be eliminated in
    a similar way as $\atiru$. Hint: Use $\Pit_1$ instead of $\Pi_1$. For
    $\comru$ plug in $B\ltens A$ for $X$. For $\assru$, you need to apply
    splitting twice.
  \item Prove the splitting lemma. This is the really hard part. Hint:
    First show the lemma for $S\conhole=\conhole$ (shallow
    splitting). To do so, proceed by induction on the length of $\Pi$
    and the size of $C\lpar(A\ltens B)$. Then show the lemma for
    arbitrary context $S\conhole$. For this, proceed by induction on
    $S\conhole$.
  \item Show that the rule
    $$
    \vcinf{\iru}{}{A\ltens A\lneg}
    $$ is not needed for provability, i.e., it is impossible to prove
    ``nothing''. Hint: You need a variant of splitting, saying, if
    there is a proof
    $$
    \vcstrpr{\set{\ird,\comrd,\assrd,\swir}}{\Pi}{A\ltens B}
    $$
    then there are proofs
    $$
    \vcstrpr{\set{\ird,\comrd,\assrd,\swir}}{\Pi_2}{A}
    \qquand
    \vcstrpr{\set{\ird,\comrd,\assrd,\swir}}{\Pi_3}{B}
    $$
    i.e., $S\conhole$, $C$, $K_A$, and $K_B$ are all ``empty''.
  \end{enumerate}
\end{exercise}

Proving cut elimination via splitting in the calculus of structures is in the
unit-free case a little more messy than in the case with units because many
cases need to be considered separately that could be treated together in the
presence of units. 

Let us now see how proof nets deal with the problem
of cut elimination. Of course, the main point to make here is that cut
elimination will become considerably simpler:

Consider the following reduction rules on pre-proof nets with cuts:
\begin{equation}\label{eq:pn-cut-red-simple}
  \vcgpn{
    &\gpnid1&\quad&\quad&\gpnout2\\
    \gpnout3&\strut&\strut&\gpncut4}{
    \idleftoutline{out3}{id1}{A}
    \gpncurve{id1}{cut4}{-5}{175}{A\lneg}
    \cutrightoutline{out2}{cut4}{A}}
  \qqquad\leadsto\qqquad
  \vcgpn{
    \gpnout1\strut\\
    \gpnout2\strut}{
    \treeline{out1}{out2}{A}}
\end{equation}
and
\begin{equation}\label{eq:pn-cut-red-compound}
  \vcgpn{
    \gpnout1&\strut&\gpnout2&\strut&\gpnout3&\strut&\gpnout4\\
    &\gpnpar5&&&&\gpntens6\\
    &&&\gpncut7}{
    \cutleftline{lpar5}{cut7}{A\lpar B}
    \cutrightline{ltens6}{cut7}{B\lneg\ltens A\lneg}
    \treeline{lpar5}{out1}{A}
    \treeline{lpar5}{out2}{B}
    \treeline{ltens6}{out3}{B\lneg}
    \treeline{ltens6}{out4}{A\lneg}}
  \qqquad\leadsto\qqquad
  \vcgpn{
    \gpnout1&\strut&\gpnout2&\strut&\gpnout3&\strut&\gpnout4\\
    &&&\gpncut7\\
    &&&\gpncut8}{
    \cutleftoutline{out2}{cut7}{B}
    \cutrightoutline{out3}{cut7}{B\lneg}
    \cutleftoutline{out1}{cut8}{A}
    \cutrightoutline{out4}{cut8}{A\lneg}}
\end{equation}

\begin{theorem}\label{thm:cut-red-norm}
  The cut reduction relation defined by \eqref{eq:pn-cut-red-simple} and
  \eqref{eq:pn-cut-red-compound} terminates and is confluent.
\end{theorem}

\begin{proof}
  Showing termination is trivial because in every reduction step the size of
  the net decreases. For showing confluence, note that the only possibility
  for making a critical pair is when two cuts want to reduce with the same
  identity link. Then the situation must be of the shape:
  $$
  \vcgpn{
    &\gpnid1&\quad&\quad&\quad&\gpnid2&\quad&\quad&\quad&\gpnid3\\
    \gpnout1&\quad&\quad&\gpncut1&\quad&\quad&\quad&\gpncut2&
    \quad&\quad&\gpnout2}{
    \idleftoutline{out1}{id1}{A}
    \gpncurve{id1}{cut1}{-5}{175}{A\lneg}
    \gpncurve{id2}{cut1}{185}{5}{A}
    \gpncurve{id2}{cut2}{-5}{175}{A\lneg}
    \gpncurve{id3}{cut2}{185}{5}{A}
    \idrightoutline{out2}{id3}{A\lneg}}
  $$
  But no matter in which order and with which identity we reduce the cuts, the
  final result will always be
  $$
  \idrulenet{A\lneg}{A}
  $$
  Hence we also have confluence.
  \qed
\end{proof}

However, it could happen, that we end up in a situation like
$$
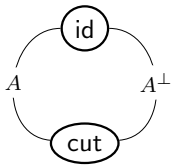
\vcgpn{
  \gpnid1\\
  \gpncut1}{
  \gpncurve{id1}{cut1}{0}{5}{A\lneg}
  \gpncurve{id1}{cut1}{180}{175}{A}
  }
$$ where we cannot reduce any further. That something like this cannot
happen if we start out with a correct net is ensured by the following
theorem, which says that the cut reduction preserves
correctness.

\begin{theorem}\label{thm:cut-correct}
  Let $\pi$ and $\pi'$ be pre-proof nets such that $\pi$ reduces to $\pi'$ via
  the reductions \eqref{eq:pn-cut-red-simple} and
  \eqref{eq:pn-cut-red-compound}. If $\pi$ is correct, then so is $\pi'$.
\end{theorem}

\begin{proof}
  For proving this, let us use the RB-correctness criterion. Written
  in terms of RB-graphs, the two reduction rules look as follows:
  \begin{equation}\label{eq:rb-cut-red-1}
    \vcrbpn{\rbv1&\rbv2&\rbv3&\rbv4&\rbv5&\rbv6}{
      \rbedge23 \rbedge45
      \rbblueedgemode
      \rbedge12 \rbedge34 \rbedge56
    }
    \qqquad\leadsto\qqquad
    \vcrbpn{\rbv1&\rbv2}{
      \rbblueedgemode
      \rbedge12 
    }
  \end{equation}
  and
  \begin{equation}\label{eq:rb-cut-red-2}
    \vcrbpn{\rbv1&\rbv2&&&&&\rbv7&\rbv8\\
      &&\rbv3&\rbv4&\rbv5&\rbv6&&\\
      \rbv{11}&\rbv{12}&&&&&\rbv{17}&\rbv{18}}{
      \rbtens32{12} \rbedge45 \rbpar67{17}
      \rbblueedgemode
      \rbedge12 \rbedge34 \rbedge56 \rbedge78 
      \rbedge{11}{12} \rbedge{17}{18}
    }
    \qqquad\leadsto\qqquad
    \vcrbpn{\rbv1&\rbv2&\rbv3&\rbv4\\ \\
      \rbv5&\rbv6&\rbv7&\rbv8}{
      \rbedge23 \rbedge67
      \rbblueedgemode
      \rbedge12 \rbedge34 \rbedge56 \rbedge78 
    }
  \end{equation}
  That the first rule preserves RB-correctness is obvious because it just
  shortens an existing path. For the second rule, we proceed \bwoc.
  First, assume that the graph on the right contains an \AE-cycle, while the
  one on the left does not. There are three possibilities:
  \begin{enumerate}
  \item The \AE-cycle does not contain one of the new
    B-R-B-paths. Then the same cycle is also present on the
    left. Contradiction.
  \item The \AE-cycle contains exactly one of the new
    B-R-B-paths. Then, as before, the same cycle is also present on
    the left. Contradiction.
  \item The \AE-cycle contains both of the new B-R-B-paths. Then we
    can construct an \AE-cycle on the left that comes in at the upper
    left corner, goes down through the $\ltens$-link, and goes out at
    the lower left corner. Again, we get a contradiction.
  \end{enumerate}
  That \AE-path connectedness is preserved is shown in a similar way.
  \qed
\end{proof}

\begin{exercise}
  Complete the proof of Theorem~\ref{thm:cut-correct}, i.e., show that
  if we apply \eqref{eq:rb-cut-red-2} to an RB-correct net, then in
  the result every pair of vertices is connected by an
  \AE-path. Hint~1: Note that the two rightmost vertices in
  \eqref{eq:rb-cut-red-2} must be connected by an \AE-path that does not touch
  the new B-R-B-paths
  (why?). Hint~2: You will need the fact that the first net is also
  \AE-cycle free.
\end{exercise}

The important point of Theorem~\ref{thm:cut-correct} is that it allows us to
give short proofs of Theorems \ref{thm:sc-cutelim} and~\ref{thm:cos-cutelim}:
Let $\Pi$ be a proof with cuts in $\ufMLL$ given in the sequent calculus or
the calculus of structures. We can translate $\Pi$ into a proof net $\pi$, as
described in Sections~\ref{sec:sc-to-pn-1}--\ref{sec:cos-to-pn} and remove the
cuts from the proof net as described above. This gives us a proof net $\pi'$,
which we can translate back to the sequent calculus or the calculus of
structures. This works because removing the cuts from the proof net preserves
the property of being correct (i.e., being a proof net), and translating back
does not introduce any new cuts.

This raises an important question: Suppose we start out with a proof $\Pi$
with cuts in $\ufMLL$ (given in sequent calculus or the calculus of
structures). Now we could first remove the cuts as sketched out in the proofs
of Theorems \ref{thm:sc-cutelim} and~\ref{thm:cos-cutelim}, and then translate
the resulting cut-free proof $\Pi'$ into a proof net $\pi_1'$. Alternatively,
we could first translate $\Pi$ into a proof net $\pi$, and then remove the
cuts from $\pi$, to obtain the cut-free proof net $\pi_2'$. Do we get the same
result? Is $\pi_1'=\pi_2'$? 

The answer is of course {\bf yes}.
To see this, note that the cut reduction steps in the sequent calculus either
preserve the proof net (if the cut is just permuted up via a trivial rule
permutation) or do exactly the same as the cut reduction steps for proof nets.

The same is true for the calculus of structures. The proof of the splitting
lemma is designed such that it preserves the net. To make this formally
precise would go beyond the scope of these lecture notes, but by comparing
Figures \ref{fig:cos-to-pn-cut} and~\ref{fig:cos-to-pn-cutcut} the reader
should get an idea.

\bigskip

We can summarize this by the following commuting diagram:
\begin{equation}\label{eq:cutcomm}
  \psset{linecolor=black,linewidth=.1ex,nodesep=1ex,arrows=->}
  \begin{psmatrix}[rowsep=15ex,colsep=10em]
    \rnode{1}{\parbox{7em}{proof with cuts\\ (in SC or CoS)}} & 
    \rnode{2}{\parbox{9em}{proof net with cuts}}\\
    \rnode{3}{\parbox{7em}{cut-free proof\\ (in SC or CoS)}} & 
    \rnode{4}{\parbox{9em}{cut-free proof net}}
  \end{psmatrix}
  \ncline12 \ncline34
  \ncline13 \Bput{\parbox{7em}{cut elimination\\ (in SC or CoS)}}
  \ncline24 \Aput{\parbox{7em}{cut elimination\\ (in proof nets)}}
\end{equation}

Our basic introduction into the theory of proof nets for $\ufMLL$ is
now finished. However, a very important and fundamental question has
not yet been mentioned: 

\begin{para}{Big Question}\label{big:cut}
  Let $\pi$ and $\pi'$ be two proof nets such
  that $\pi'$ is obtained from $\pi$ by applying some cut reduction
  steps. Do $\pi$ and $\pi'$ represent the \emph{same} proof?
\end{para}

By comparing \eqref{eq:3sq-pn1} and \eqref{eq:3sq-pn4}, one might be
tempted to say no. But by looking at Figures~\ref{fig:sc-to-pn-2-cut}
and~\ref{fig:cos-to-pn-cut}, one is tempted to say yes.  Furthermore,
from the viewpoint of proof nets it makes no difference, whether we
eliminate the cut from
\proofadjust
$$
\vcddernote{\cutr}{}{\sqn{A\lneg,C}}{
  \leaf{\sqnsmallderi{A\lneg,B}{\Pi_1}}}{
  \leaf{\sqnsmallderi{B\lneg,C}{\Pi_2}}}
$$
in the sequent calculus or whether we perform the the composition
$$
\vcstrder{\Pi_2}{}{C}{\stem{\Pi_1}{}{B}{\leaf{A}}}
\qquadto
\vcstrder{}{}{C}{\leaf{A}}
$$ in the calculus of structures (no matter whether we use one-sided
or two-sided proof nets, cf.~Section~\ref{sec:twosided}).

In the following section we are going to give another justification
for the ``yes'', which is independent from proof nets, sequent
calculus, calculus of structures, or any other way of presenting
proofs.

\subsection{*-Autonomous categories (without units)}\label{sec:star}

In this section we will introduce the concept of *-autonomous categories. We
do not presuppose any knowledge of category theory. We introduce what we need
on the way along. It is not much anyway.  The basic idea is to give an
abstract algebraic theory of proofs, which is based on the following
postulates about proofs:
\begin{enumerate}[(i)]
\item\label{it:first} for every proof $f$ of conclusion $B$ from hypothesis
  $A$ (denoted by $f\colon A\to B$) and every proof $g$ of conclusion $C$ from
  hypothesis $B$ (denoted by $g\colon B\to C$) there is a uniquely defined
  composite proof $g\fcomp f$ of conclusion $C$ from hypothesis $A$ (denoted
  by $g\fcomp f\colon A\to C$),
\item this composition of proofs is associative,
\item for each formula $A$ there is an identity proof $\idf[A]\colon A\to A$
  such that for $f\colon A\to B$ we have $f\fcomp\idf[A]=f=\idf[B]\fcomp f$,
  i.e, it behaves as identity w.r.t.\ composition.
\end{enumerate}
These axioms say no more and no less than that the proofs are the
arrows in a category whose objects are the formulas. 
Let us now add more axioms that are specific to logic and do not hold
in general in categories:
\begin{enumerate}[(i)]
\setcounter{enumi}{3}
\item\label{it:bifun} Whenever we have a formula $A$ and formula $B$, then
  $A\ltens B$ is another formula. For two proofs $f\colon A\to C$ and $g\colon
  B\to D$ we have a uniquely defined proof $f\ltens g\colon A\ltens B\to
  C\ltens D$, such that for all $h\colon C\to E$ and $k\colon D\to F$, we have
  \begin{equation}\label{eq:bifun}
    (h\ltens k)\fcomp(f\ltens g)=(h\fcomp f)\ltens(h\fcomp g)\colon 
    A\ltens B\to E\ltens F
    \quad.   
  \end{equation}
\end{enumerate}
Using category theoretical language, Axiom \eqref{it:bifun} just says that
$\ltens$ is a bifunctor. What does this mean? Consider the
following two derivations
(using the notation from the calculus of structures):
\begin{equation}\label{eq:formA}
  \vcstrder{C\ltens g}{}{C\ltens D}{
    \stem{f\ltens B}{}{C\ltens B}{
      \leaf{A\ltens B}}}
  \qqquand
  \vcstrder{f\ltens D}{}{C\ltens D}{
    \stem{A\ltens g}{}{A\ltens D}{
      \leaf{A\ltens B}}}
\end{equation}
In the left one we use first $f$ to go from $A$ to $C$, and do nothing
to $B$,\footnote{More precisely, it is the identity $\idf[B]$ taking
us from $B$ to $B$.} and then use $g$ to go from $B$ to $D$ (and do
nothing to $C$). In the right derivation, we first use $g$ to go from
$B$ to $D$, and then $f$ to go from $A$ to $C$. Equation
\eqref{eq:bifun} says that the two derivations with premise $A\ltens
B$ and conclusion $C\ltens D$ in \eqref{eq:formA} represent the same
proof, denoted by $f\ltens g$.  Mathematicians came up with a very
clever way of writing an equation between objects as in
\eqref{eq:formA}, namely, via \emph{commuting diagrams}. Instead of
writing the two derivations in \eqref{eq:formA} and saying they are
equal, we write:
$$
\mbox{The diagram }
\quad
\vcdiagsquaredown{{A\ltens B}{f\ltens B}}{{C\ltens B}{C\ltens g}}
{{A\ltens g}{A\ltens D}}{{f\ltens D}{C\ltens D}}
\quad
\mbox{ commutes .}
$$ From the proof theoretical viewpoint, this equation is indeed
wanted. The difference between the two derivations in \eqref{eq:formA}
is an artefact of syntactic bureaucracy. The kind of bureaucracy in
exhibited in \eqref{eq:formA} is called \emph{bureaucracy of type
A}. This implies that there must also be a \emph{bureaucracy of type
B}. Consider the following two derivations:
\begin{equation}\label{eq:formB}
  \vcdernote{\swir}{}{(A'\ltens B)\lpar C}{
    \stem{f\ltens (B\lpar C)}{}{A'\ltens(B\lpar C)}{
      \leaf{A\ltens(B\lpar C)}}}
  \qqquand
  \vcstrder{(f\ltens B)\lpar C}{}{(A'\ltens B)\lpar C}{
    \root{\swir}{(A\ltens B)\lpar C}{
      \leaf{A\ltens(B\lpar C)}}}
\end{equation}
In the left one we first use the proof $f$, taking us from $A$ to $A'$
(and doing nothing to $B$ and $C$), and then we apply the switch
rule. In the derivation on the right we first apply the switch rule,
and then do $f$. Clearly the two are essentially the same and should
be identified eventually. Let us write this as commuting diagram:
\begin{equation}\label{eq:natur}
  \vcdiagsquaredown{{A\ltens(B\lpar C)}{f\ltens (B\lpar C)}}
  {{A'\ltens(B\lpar C)}{\swir[A',B,C]}}
  {{\swir[A,B,C]}{(A\ltens B)\lpar C}}
  {{(f\ltens B)\lpar C}{(A'\ltens B)\lpar C}}
\end{equation}
Using category theoretical language, equation \eqref{eq:natur} says
precisely that the morphism $\swir[A,B,C]\colon A\ltens(B\lpar
C)\to(A\ltens B)\lpar C$ is \emph{natural in $A$}. Of course, in the
end, we should have that switch is natural in all three arguments.

Before we can continue with our list of axioms, we need another
category theoretical concept. Suppose we have two formulas $A$ and $B$
and proofs $f\colon A\to B$ and $g\colon B\to A$. If we have for some
reason that $f\fcomp g=\idf[B]$ and $g\fcomp f=\idf[A]$, then we say
that $A$ and $B$ are \emph{isomorphic}. In this case $f$ and $g$ are
\emph{isomorphisms}. The following axiom shows two examples:

\begin{enumerate}[(i)]
\setcounter{enumi}{4}
\item\label{it:monoidal} For all formulas $A$, $B$, and $C$, we
  postulate the existence of proofs
  \begin{equation}
    \begin{array}{r@{\colon}l}
      \assoc[A,B,C]&A\ltens(B\ltens C)\to (A\ltens B)\ltens C\\[2ex]
      \twist[A,B]&A\ltens B\to B\ltens A
    \end{array}
  \end{equation}
  which are isomorphisms, and which are natural in all
  arguments,\footnote{At this point you should start to see why it
  makes sense to use the category theoretical language. Without it, we
  would have, for example, to postulate for all formulas $A$, $B$, and
  $C$ another proof $\assoc[A,B,C]^{-1}$ such that the two derivations
  $$
  \vcstrder{\assoc[A,B,C]^{-1}}{}{A\ltens(B\ltens C)}{
    \stem{\assoc[A,B,C]}{}{(A\ltens B)\ltens C}{
      \leaf{A\ltens(B\ltens C)}}}
  \qquand
  \vcstrder{\assoc[A,B,C]}{}{(A\ltens B)\ltens C}{
    \stem{\assoc[A,B,C]^{-1}}{}{A\ltens(B\ltens C)}{
      \leaf{(A\ltens B)\ltens C}}}
  $$ are both doing nothing (i.e., are equal to the identity
  proof). Furthermore, we would need a lot of equations in the form
  of~\eqref{eq:formB}, in order to express the naturality.}  and which
  obey the following equations:
  \begin{equation}\label{eq:pentagon}
    \vcdiagpentagondown
	{{A\ltens(B\ltens (C\ltens D))}{\id[A]\ltens \assoc[B,C,D]}}
	{{A\ltens ((B\ltens C)\ltens D)}{\assoc[A,B\ltens C,D]}}
	{{(A\ltens (B\ltens C))\ltens D}{\;\assoc[A,B,C]\ltens \id[D]}}
	{{\assoc[A,B,C\ltens D]}{(A\ltens B)\ltens (C\ltens D)}}
	{{\assoc[A\ltens B,C,D]}{\qqqlapm{((A\ltens B)\ltens C)\ltens D}}}
  \end{equation}
  \begin{equation}\label{eq:hexagon}
    \vcdiaghexagondown
	{{A\ltens(B\ltens C)}{\assoc[A,B,C]}}
	{{(A\ltens B)\ltens C}{\twist[A\ltens B,C]}}
	{{C\ltens (A\ltens B)}{\assoc[C,A,B]}}
	{{\id[A]\ltens \twist[B,C]}{A\ltens (C\ltens B)}}
	{{\assoc[A,C,B]}{(A\ltens C)\ltens B}}
	{{\twist[A,C]\ltens\id[B]}{(C\ltens A)\ltens B}}
  \end{equation}
  \begin{equation}\label{eq:triangle}
    \vcdiagtriangledown
	{{A\ltens B}{\twist[A,B]}}
	{{B\ltens A}{\twist[B,A]}}
	{{\idf[A\ltens B]}{A\ltens B}}
  \end{equation}
\end{enumerate}
What we have defined so far, could be called a \emph{symmetric monoidal
category without unit}. This terminology is not standard, because the notion
has not much been used in mathematics. What is standard is the notion of
\emph{monoidal category} and \emph{symmetric monoidal category} (the first one
being without the $\twist$), which additionally have a distinguished
\emph{unit object} $\lone$ and natural isomorphisms
$\lunit[A]\colon\lone\ltens A\to A$ and $\runit[A]\colon A\ltens\lone\to A$
obeying the equations
\proofadjust
\begin{equation}\label{eq:unit}
  \vcdiagtriangledown
      {{A\ltens(\lone\ltens B)}{\assoc[A,\lone,B]}}
      {{(A\ltens\lone)\ltens B}{\runit[A]\ltens \id[B]}}
      {{\id[A]\ltens \lunit[B]}{A\ltens B}}
  \quand
  \vcdiagtriangledown
  {{\lone\ltens A}{\twist[\lone,A]}}
  {{A\ltens\lone}{\runit[A]}}
  {{\lunit[A]}{A}}
\end{equation}

An important property of monoidal categories is MacLane's
\emph{coherence theorem}. Stated in terms of proofs, it says the
following:

\begin{theorem}\label{thm:coherence}
  Let $n\ge 1$ and $A_1,\ldots,A_n$ be formulas. Now let $B$ and $C$
  be formulas built from $A_1,\ldots,A_n$ by using $\ltens$ such that
  every $A_i$ appears exactly once in $B$ and $C$. If Axioms
  \eqref{it:first}--\eqref{it:monoidal} hold, then all proofs from $B$
  to $C$ formed with the available data are equal. This proof always
  exists, is an isomorphism, and is natural in all $n$ arguments.
\end{theorem}

We will not give a proof here. 

\bigskip

For being able to really speak about logic and proofs, we need negation, which
is introduced by the following axioms:

\begin{enumerate}[(i)]
\setcounter{enumi}{5}
\item\label{it:contra} For every formula $A$ there is another formula
  $A\lneg$, and for every proof $f\colon A\to B$, there is another proof
  $f\lneg\colon B\lneg\to A\lneg$ such that $1_A\lneg=\idf[A\lneg]\colon
  A\lneg\to A\lneg$ and such that $(g\fcomp f)\lneg=f\lneg\fcomp g\lneg\colon
  C\lneg\to A\lneg$ for every $f\colon A\to B$ and $g\colon B\to C$.
\item\label{it:strict} For every formula $A$ and proof $f\colon A\to B$ we
  have that $A\lnegneg=A$ and $f\lnegneg=f$. (More precisely, the mapping
  $A\lnegneg\to A$ is the identity on $A$).
\end{enumerate}
Spoken in category theoretical terms, Axiom \eqref{it:contra} says that
$(-)\lneg$ is a \emph{contravariant endofunctor}. With this, we can define the
$\lpar$ via $A\lpar B=(A\lneg\ltens B\lneg)\lneg$.  Axiom \eqref{it:strict}
says that if we flip around a derivation twice, we get back where we started
from.\footnote{What we impose here is also called \emph{strictness}, and does
usually not hold. For example, the double dual of a vector space is usually
not the space itself. Even in the finite dimensional case we only have a
natural isomorphism between $A$ and $A\lnegneg$.} It also allows us to
conclude that the $\lpar$ that we just defined has the same properties as
postulated for the $\ltens$ in \eqref{it:bifun} and~\eqref{it:monoidal}, i.e.,
it is a bifunctor and carries a monoidal structure (without unit).

\begin{exercise}\label{ex:par}
  Formulate the statements of Axioms \eqref{it:bifun}
  and~\eqref{it:monoidal} for the $\lpar$ defined via $A\lpar
  B=(A\lneg\ltens B\lneg)\lneg$, and show that they follow from
  \eqref{it:first}--\eqref{it:strict}.
\end{exercise}

Before stating our final postulates about proofs, let us introduce the
following notation. For two formulas $A$ and $B$, we write $\Hom(A,B)$ for the
set of proofs from $A$ to $B$, and we write $\repr{\fakeone}(B)$ for the set
of proofs of $B$ that have no premise.\footnote{The reason for this notation
is the following: $\Hom(A,B)$ is in fact the value of the functor $\Hom(-,-)$
in two arguments. The functor $\Hom(A,-)$ in one argument is also written as
$\repr{A}$. If there is a proper unit $\lone$ then the proofs of $A$ are the
elements of the set $\Hom(\lone,A)$, i.e., $\repr{\lone}$ is a functor mapping
every formula to its set of proofs. In $\repr{\fakeone}$, the unit is
\emph{virtual}.}
\begin{enumerate}[(i)]
\setcounter{enumi}{7}
\item\label{it:star} For all formulas $A$, $B$, and $C$, there is a bijection
  \begin{equation}\label{eq:star}
    \varphi\colon\Hom(A\ltens B,C)\to\Hom(A,B\lneg\lpar C)
  \end{equation}
  which is natural in all three arguments.
\item\label{it:fake} For all formulas $A$ and $B$, we have a bijection 
  \begin{equation}\label{eq:fake}
    \varphi\colon\repr{\fakeone}(A\lneg\lpar B)\to\Hom(A,B)
  \end{equation}
  which is natural in both arguments and respects the monoidal structure.
\end{enumerate}
In the case with units, Axiom \eqref{it:star} would complete the
definition of a \emph{*-autonomous category}. It essentially says that
we are allowed to do \emph{currying} and \emph{uncurrying}. To see
this, note that linear logic knows the connective $\limp$, standing
for \emph{linear implication}, defined via $A\limp B=A\lneg\lpar
B$.\footnote{As in classical logic, ``$A$ implies $B$'' is the same as
``not $A$ or $B$''.} Equation \eqref{eq:star} now says that we can
jump freely back and forth between proofs $A\ltens B\to C$ and $A\to
B\limp C$.\footnote{If you have never seen currying, think of a
function $f$ in two arguments, denoted by $f\colon A\times B\to
C$. This is essentially the same as a function $f'\colon A\to B\to C$,
taking and argument from the set $A$ and returning a function $B\to C$
which asks for an element of $B$ to finally return the result in $C$.}

Since we do not have units, we also need \eqref{it:fake}, which says
that the proofs of $A\limp B$ are the same as the proofs $A\to B$. To
be precise, we need to give additional equation saying that
$\repr{\fakeone}$ is a functor, i.e., every proof $f\colon A\to B$ is
mapped to a function
$\repr{\fakeone}(f)\colon\repr{\fakeone}(A)\to\repr{\fakeone}(B)$ such
that composition and identity are preserved. Furthermore, the
$\repr{\fakeone}$ needs to go well along with the monoidal structure,
to say what that means exactly would take us too far astray.  But to
give you an idea of the problem, let us figure out how we could
construct a proof $B\to (A\lpar A\lneg)\ltens B$, corresponding to the
rule $\ird$ in Figure~\ref{fig:cos-mll}, by using the axioms
\eqref{it:first}--\eqref{it:fake}. If we had a unit $\lone$ together
with the equations~\eqref{eq:unit}, then it would be easy: we could
start out with $\lunit[A]\colon\lone\ltens A\to A$,
apply~\eqref{eq:star} to get
$$
\hlunit[A]=\varphi(\lunit[A])\colon\lone\to A\lpar A\lneg
$$ By \eqref{it:bifun}, we can form a proof
$\hlunit[A]\ltens\idf[B]\colon\lone\ltens B\to(A\lpar A\lneg)\ltens
B$. We can precompose this with $\lunit[B]^{-1}\colon B\to\lone\ltens B$,
to get
$$
(\hlunit[A]\ltens\idf[B])\fcomp\lunit[B]^{-1}\colon 
B\to (A\lpar A\lneg)\ltens B
$$ Constructing this map without using the unit requires heavy
category theoretical machinery that we are not going to show here.
See Section~\ref{sec:notes} for references.

\begin{exercise}\label{ex:switch}
  We mentioned switch in \eqref{eq:formB} and \eqref{eq:natur} but we
  did not postulate it in \eqref{it:first}--\eqref{it:fake}. 
  In this exercise you are asked to
  construct $\swir[A,B,C]\colon A\ltens(B\lpar C)\to (A\lpar B)\ltens
  C$, corresponding to the switch rule in Figure~\ref{fig:cos-mll}, by
  using \eqref{it:first}--\eqref{it:star}.  Hint: Start
  with the identity $B\ltens C\to B\ltens C$ and
  apply~\eqref{eq:star}. You will also need the associativity of
  $\lpar$ that you have constructed in Exercise~\ref{ex:par}.
\end{exercise}

The wonderful point of Axioms \eqref{it:first}--\eqref{it:fake} is
that they precisely describe the mathematical structure spanned by
cut-free proof nets for $\ufMLL$. This means two things:

First, the proof nets for $\ufMLL$ that we discussed in the previous
sections form a category: The objects are the formulas and the maps
$A\to B$ are the cut-free proof nets with conclusion
$\sqn{A\lneg,B}$,\footnote{Or, equivalently, the two-sided proof nets
where $A$ has polarity $\poll$ and $B$ has polarity $\polr$.}  and the
composition $g\fcomp f$ of two maps $f\colon A\to B$ and $g\colon B\to
C$ is defined by eliminating the cut from
\proofadjust
$$
\vcddernote{\cutr}{}{\sqn{A\lneg,C}}{
  \leaf{\sqnsmallderi{A\lneg,B}{f}}}{
  \leaf{\sqnsmallderi{B\lneg,C}{g}}}
$$
In the calculus of structures, this corresponds to performing the composition
$$
\vcstrder{}{}{C}{\stem{}{}{B}{\leaf{A}}}
\qquadto
\vcstrder{}{}{C}{\leaf{A}}
$$
It is easy to verify that this category, denoted by $\PN$, obeys
\eqref{it:first}--\eqref{it:fake} (where $\repr{\fakeone}(A)$ is just the set
of cut-free proof nets with conclusion $A$).

Second, the the category $\PN$ is the free category with this
property. This means that whenever there is a category $\cC$, obeying
\eqref{it:first}--\eqref{it:fake}, then there is a uniquely defined functor
(i.e., map that preserves all the structure defined in
\eqref{it:first}--\eqref{it:fake}) from $\PN$ to $\cC$.

\begin{theorem}\label{thm:freestar}
  The category of cut-free proof nets for $\ufMLL$, with arrow composition
  defined by cut elimination, forms the free *-autonomous category without
  units \rp{generated from the set $\cA$ of propositional variables}.
\end{theorem}

Another way of seeing this is that we can trivially translate proofs in the
sequent calculus or the calculus of structures into the free *-autonomous
category (without units), by simply following the
syntax. Theorem~\ref{thm:freestar} says, that if we do this to two proofs
$\Pi_1$ and $\Pi_2$, we get the same map in the category, if and only if
$\Pi_1$ and $\Pi_2$ yield the same proof net after cut elimination.

In other words, if you have no objections against any of the Axioms
\eqref{it:first}--\eqref{it:fake}, you must answer the Big
Question~\ref{big:cut} with {\bf yes}.

But there is also a {\bf but}: Let us emphasize that this yes is valid
only for $\ufMLL$. What we have said in this section does {\bf not}
allow us to draw any conclusions about any other logic.

\subsection{Notes}\label{sec:notes}

As already mentioned in the introduction, the terminology of ``proof nets''
and ``bureaucracy'' is due to Girard. He introduced proof nets along with
sequent calculus presentation for linear logic in~\cite{girard:87}. He
essentially followed Ideology~\ref{ideo:girard} for obtaining his proof
nets. The terminology ``coherence'' is due to MacLane. In~\cite{maclane:63} he
proves the ``coherence theorem'' for symmetric monoidal categories. See
also~\cite{maclane:71}.  The concept of coherence graph is based in the work
of Eilenberg, Kelly, and
MacLane~\cite{eilenberg:kelly:fun-cal,kelly:maclane:71}, who also provided the
acyclicity condition and observed that it is preserved by composition, i.e.,
cut elimination. The observation that cut elimination is composition in a
category is due to Lambek \cite{lambek:68,lambek:69}.  The terminology
``flow-graph'' is due to Buss~\cite{buss:91}.\footnote{Strictly speaking,
  coherence graphs and flow graphs are not the same thing. But in the simple
  case of $\ufMLL$, the two notions coincide.} 

The calculus of structures has been discovered by
Guglielmi~\cite{guglielmi:02}, who initiated the systematic proof theoretic
investigation of the concept of deep inference. For more details on this see
\cite{guglielmi:strassburger:01,brunnler:tiu:01,dissvonlutz,brunnler:phd}.

The notion of ``correctness criterion'' is also due to
Girard. In~\cite{girard:87} he gave the ``long-trip-criterion'' that we did
not present here. The splitting tensor theorem (our
Lemma~\ref{lem:splittingtensor}) also first appeared in~\cite{girard:87}. The
proof given in Section~\ref{sec:correctness} follows the presentation of
Bellin and van de Wiele in~\cite{bellin:wiele:subnets}, who also give a proof
of Theorem~\ref{thm:trivial} and discuss in more detail the relation between
proof nets and trivial rule permutations. Another well-written short
discussion on this issue can be found in \cite{lafont:95}.  Our second proof
of Theorem~\ref{thm:S-correct} (i.e., the one using the calculus of
structures) follows the presentation in \cite{dissvonlutz}. However, the
result is already implicit present in the work of \cite{devarajan:etal:99} and
\cite{retore:97}. A different way of proving Theorem~\ref{thm:S-correct} via
the calculus of structures is presented in~\cite{joinet:PNswitch}.

The switching criterion (Definition~\ref{def:S-correct} and
Theorem~\ref{thm:S-correct}) is due to Danos and Regnier
\cite{danos:regnier:89}. For this reason the switching criterion is in the
literature also called Danos-Regnier-criterion or DR-criterion. However, the
contraction criterion is also due to Danos and Regnier\footnote{It first
appears in Danos' thesis \cite{danos:phd}, but he insists that it is joint
work with Regnier.}  and should therefore also be called DR-criterion. See
\cite{moot:phd,puite:phd} for a more recent investigation of the contraction
criterion. That (a version of) the contraction criterion can be checked in
linear time in the size of the net has been discovered by
Guerrini~\cite{guerrini:99}.  The RB-correctness criterion has been found by
Retor{\'e} \cite{retore:phd,retore:99:b,retore:03}, who provided a detailed
analysis of proof nets using RB-graphs in various forms.

The concept of two-sided proof nets must be considered as folklore. In
several early papers on proof nets the possibility of a two-sided version
is mentioned, but the details are never carried out because the
one-sided version is more economic. In the literature, two-sided proof
nets are used when for some reason the authors want to avoid the use of
negation (e.g., \cite{BCST,fuhrmann:pym:oecm}). The concept of
polarities has to be attributed to Lamarche (e.g., \cite{lamarche:95}). In
\cite{lamarche:contexts} he develops the algebraic theory
of polarities and structural contexts in full detail.

The notion of cut elimination has been developed by Gentzen
\cite{gentzen:34,gentzen:35}. For a variant of linear logic it has first been
proved by Lambek~\cite{lambek:58}. For full
linear logic (sequent calculus and proof
nets) it has been proved by Girard~\cite{girard:87}. For linear logic
presented in the calculus of structures, the first direct proof of cut
elimination was also based on rule permutation (similar to the sequent
calculus) \cite{dissvonlutz,str:MELL}. The idea of using splitting as sketched
in the proof of Theorem~\ref{thm:cos-cutelim} is due to
Guglielmi~\cite{guglielmi:02}.

*-Autonomous categories have been discovered by Barr~\cite{barr:79}. That
there is a relation to linear logic was discovered immediately after the
introduction of linear logic (see, e.g.,
\cite{lafont:88,seely:89}). Blute~\cite{blute:93} was the first to note that
the category of proof nets is actually the free *-autonomous category without
units. However, no complete proof was given; there was no proper definition of
a *-autonomous category without units. That there is in fact a non-trivial
mathematical problem to give such a definition was observed only 12 years
later, but then by three research groups independently at the same time
\cite{lam:str:05:freebool,dosen:petric:05,houston:etal:unitless} (see also
Sections \ref{sec:units} and~\ref{sec:cl-rob}). The most in-depth treatment
is~\cite{houston:etal:unitless}. We used here the notation
of~\cite{lam:str:05:freebool}.

The terminology of ``Formalism A'' and ``Formalism B'' is due to
Guglielmi~\cite{guglielmi:A,guglielmi:B}. See also
\cite{hughes:DI-cat,mckinley:SKS-cat,str:deepnet,str:medial} for the relation
between deep inference and category theory.

\section{Other fragments of linear logic}\label{sec:other}

In this section we will very briefly inspect how proof nets for larger
fragments of linear logic look like. We will first look at the
so-called exponentials (which are modalities) and the additives (which
are a second pair of conjunction/disjunction). Then we will go back to
the purely multiplicative fragment and play with variations of it.

\subsection{Multiplicative exponential linear logic (without units)}
\label{sec:MELL}

\begin{figure}[!t]
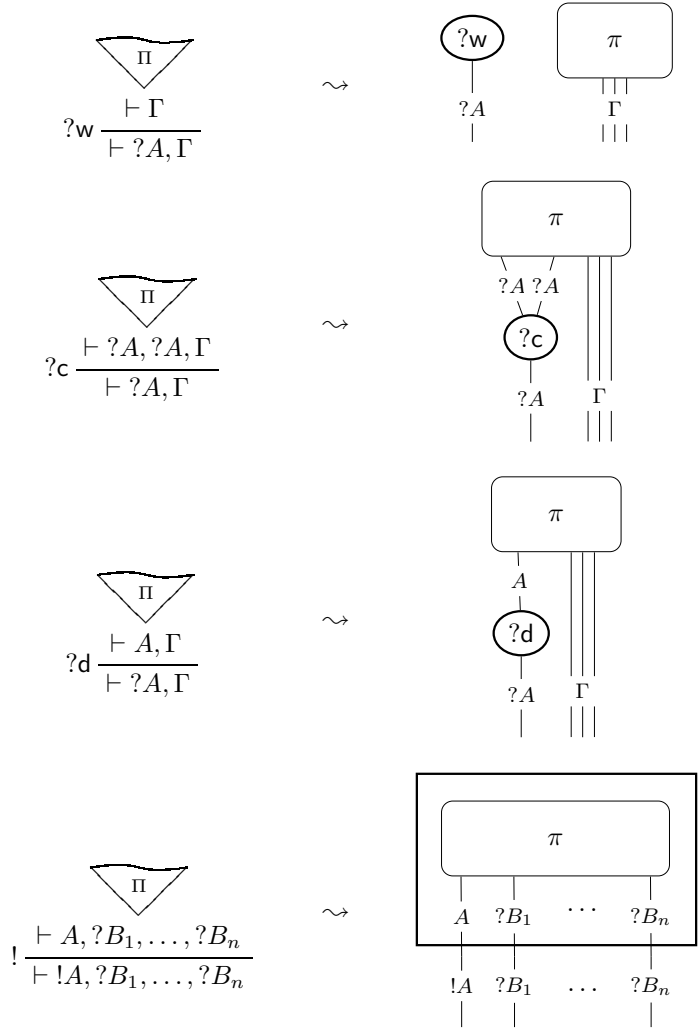

  \def\arrayskip{12ex}
  \def\pnrowsep{7ex}
  \begin{center}
    $$
    \begin{array}{c@{\qquad\leadsto\qquad}c}
      \vcinf{\lweakr} {\sqn{\lwn A,\Gamma}} {\sqnsmallderi{\Gamma}{\Pi}}
      &
      \vcgpn{
	\begin{array}{c@{\qquad}c}
	  \gpnnode{wn1}{\lweakr}&
	  \gpnbox{\pi}{\quad\gpbbunchboxout{2}\quad}\\[\pnrowsep]
	  \gpnout{1}&{\gpbbunchout{2}}
 	 \end{array}
       }{
 	\bunchboxoutline{2}{\Gamma}
	\myoutline{.5}{gpnwn1}{out1}{\lwn A}
       }
      \\[7ex]
      \vcinf{\lconr} {
	\sqn{\lwn A,\Gamma}} {
	\sqnsmallderi{\lwn A, \lwn A,\Gamma}{\Pi}}
      &
      \vcgpn{
	\begin{array}{c}
	  \gpnbox{\pi}{\gpnboxout3\qquad\gpnboxout4\quad\;
	    \gpbbunchboxout{2}}\\[\pnrowsep]
	  \gpnnode{1}{\lconr}\qquad\;\strut\\[\pnrowsep]
	  \qquad\gpnout{1}\qquad{\gpbbunchout{2}}
 	 \end{array}
       }{
	\treeline{box3}{gpn1}{\lwn A}
	\treeline{box4}{gpn1}{\lwn A}
 	\mybunchboxoutline{.7}{2}{\Gamma}
	\treeoutline{gpn1}{out1}{\lwn A}
       }
      \\[\arrayskip]
      \vcinf{\lderr} {\sqn{\lwn A,\Gamma}} {\sqnsmallderi{A,\Gamma}{\Pi}}
      &
      \vcgpn{
	\begin{array}{c}
	  \gpnbox{\pi}{\;\gpnboxout3\qquad\gpbbunchboxout{2}\;}
	  \\[\pnrowsep]
	  \gpnnode{1}{\lderr}\qqquad\strut\\[\pnrowsep]
	  \qquad\gpnout{1}\qquad{\gpbbunchout{2}}
 	 \end{array}
       }{
	\treeline{box3}{gpn1}{A}
 	\mybunchboxoutline{.7}{2}{\Gamma}
	\treeoutline{gpn1}{out1}{\lwn A}
       }
      \\[\arrayskip]
      \vcinf{\loc}{
	\sqn{\loc A,\lwn B_1,\ldots,\lwn B_n}}{
	\sqnsmallderi{A,\lwn B_1,\ldots,\lwn B_n}{\Pi}}
      &
      \vcgpn{
	\begin{array}{c}
	  \gpnbangbox{\;
	    \gpnbox{\pi}{
	      \gpnboxout1\qquad
	      \gpnboxout2\qquad\smash{\lower3ex\hbox{$\ldots$}}\qquad
	      \gpnboxout3}
	    \\[\pnrowsep]
	    \gpnboxout4\qquad
	    \gpnboxout5\qquad\smash{\lower3ex\hbox{$\ldots$}}\qquad\gpnboxout6
	    \\[-\pnrowsep]
	  }{}
	  \\[\pnrowsep]
	  \\[\halfpnrowsep]
	  \quad\gpnout{4}\qquad\gpnout{5}\qqqqquad\gpnout{6}
 	\end{array}
      }{
	\treeline{box1}{box4}{A}
	\treeline{box2}{box5}{\lwn B_1}
	\treeline{box3}{box6}{\lwn B_n}
	\boxoutline{4}{\loc A}
	\boxoutline{5}{\lwn B_1}
	\boxoutline{6}{\lwn B_n}
      }
    \end{array}
    $$
    \caption{From sequent calculus to proof nets: exponentials}
    \label{fig:sc-to-pn-exp}
  \end{center}
\end{figure}

The formulas of unit-free multiplicative exponential linear logic
($\ufMELL$) are generated by the syntax
$$
\cF\grammareq\cA\mid\cA\lneg\mid\cF\lpar\cF\mid\cF\ltens\cF\mid
\loc\cF\mid\lwn\cF
$$ where everything is as in Section~\ref{sec:sc-mll}. The modalities
are dual to each other:
$$
(\loc A)\lneg=\lwn A\lneg
\qqquad
(\lwn A)\lneg=\loc A\lneg
$$ The inference rules in the sequent calculus are the same as in
\eqref{eq:sc-mll} plus the ones for the new modalities:
\begin{equation}\label{eq:sc-exp}
  \vcinf{\lweakr} {\sqn{\lwn A,\Gamma}} {\sqn{\Gamma}}
  \quad\;\;
  \vcinf{\lconr} {\sqn{\lwn A,\Gamma}} {\sqn{\lwn A, \lwn A,\Gamma}}
  \quad\;\;
  \vcinf{\lderr} {\sqn{\lwn A,\Gamma}} {\sqn{A,\Gamma}}
  \quad\;\;
  \vcinf{\loc}{
    \sqn{\loc A,\lwn B_1,\ldots,\lwn B_n}}{
    \sqn{A,\lwn B_1,\ldots,\lwn B_n}}
\end{equation}
where in the $\loc$-rule $n\ge0$.  In Figure~\ref{fig:sc-to-pn-exp} we show
how these rules are translated into proof nets according to
Ideology~\ref{ideo:girard}. This is exactly the way how Girard introduced them
in \cite{girard:87}. Note that for dealing with the $\loc$-modality, we do not
exactly follow Ideology~\ref{ideo:girard}. Instead, the concept of \emph{box}
around a proof net is introduced. Such a box always has a main door, the
formula $\loc A$, and auxiliary doors, which are all occupied by
$\lwn$-formulas.

The correctness of such a proof net with boxes is defined as in
Section~\ref{sec:correctness} ($\lconr$ behaves as $\lpar$) with the
difference, that each box has to be treated separately. However, the
correctness criteria do only work if no instance of $\lweakr$ is
present.\footnote{There is a general problem with multiplicative proof nets
when some form of weakening is around. We come back to this in
Sections \ref{sec:units} and~\ref{sec:cl-rob}.}

Cut elimination is not as simple as for $\ufMLL$. There are now not only more
reduction rules, we also have the problem that
showing termination is no longer trivial because the size of the net
can increase during the reduction process. This is due to the presence of
contraction. In the sequent calculus, the problematic case appears when
\proofadjust
$$
\vcddernote{\cutr}{}{\sqn{\Gamma,\lwn B_1,\ldots,\lwn B_n}}{
  \root{\lconr}{\sqn{\Gamma,\lwn A}}{
    \leaf{\sqnsmallderi{\Gamma,\lwn A,\lwn A}{\Pi_1}}}}{
  \root{\loc}{\sqn{\loc A,\lwn B_1,\ldots,\lwn B_n}}{
    \leaf{\sqnsmallderi{A,\lwn B_1,\ldots,\lwn B_n}{\Pi_2}}}}
$$
is reduced to
\proofadjust
$$
\vcdernote{\lconr}{}{\sqn{\Gamma,\lwn B_1,\ldots,\lwn B_n}}{
  \root{\lconr}{\vdots}{
    \rroot{\cutr}{\sqn{\Gamma,\lwn B_1,\ldots,\lwn B_n,
	\lwn B_1,\ldots,\lwn B_n}}{
      \rroot{\cutr}{\sqn{\Gamma,\lwn A,\lwn B_1,\ldots,\lwn B_n}}{
	  \leaf{\sqnsmallderi{\Gamma,\lwn A,\lwn A}{\Pi_1}}}{
	\root{\loc}{\sqn{\loc A,\lwn B_1,\ldots,\lwn B_n}}{
	  \leaf{\sqnsmallderi{A,\lwn B_1,\ldots,\lwn B_n}{\Pi_2}}}}}{
      \root{\loc}{\sqn{\loc A,\lwn B_1,\ldots,\lwn B_n}}{
	\leaf{\sqnsmallderi{A,\lwn B_1,\ldots,\lwn B_n}{\Pi_2}}}}}}
$$    
where the proof $\Pi_2$ has been duplicated. In terms of proof nets, this
means that the whole box with $\pi_2$ inside is duplicated.

\begin{exercise}
  Visualize this reduction in terms of proof nets by using the translation of
  Figure~\ref{fig:sc-to-pn-exp}.
\end{exercise}

In spite of this nasty behavior, we have the
following theorem, which we are not going to prove her. The interested reader
is referred to~\cite{girard:87,danos:phd,joinet:SN-MELL}.

\begin{theorem}
  Cut elimination for $\ufMELL$ proof nets is terminating, confluent,
  and preserves correctness.
\end{theorem}

So far, there are no proof nets for $\ufMELL$ that follow
Ideology~\ref{ideo:coherence}. One of the reasons might be that the
sequent calculus rules in \eqref{eq:sc-exp} do not allow to properly
trace the modalities in the derivation. However, in the calculus of
structures, this becomes very natural (see
\cite{dissvonlutz,str:MELL,guglielmi:strassburger:02:big}). 

\begin{para}{Open Research Problem}
  Find for $\ufMELL$ a notion of proof nets without boxes, that is
  based on Ideology~\ref{ideo:coherence}.
\end{para}

\subsection{Multiplicative additive linear logic (without units)}
\label{sec:MALL}

Let us now turn to unit-free multiplicative additive linear logic
($\ufMALL$). The formulas of $\ufMALL$ are generated by the syntax
$$
\cF\grammareq\cA\mid\cA\lneg\mid\cF\lpar\cF\mid\cF\ltens\cF\mid
\cF\lplus\cF\mid\cF\lwith\cF
$$ where again everything is as in Section~\ref{sec:sc-mll}. The
two new connectives are dual to each other:
$$
(A\lwith B)\lneg=B\lneg\lplus A\lneg
\qqquad
(A\lplus B)\lneg=B\lneg\lwith A\lneg
$$ The inference rules in the sequent calculus are the same as in
\eqref{eq:sc-mll} plus the ones for the new connectives:
\begin{equation}\label{eq:sc-add}
  \vciinf{\lwith} {\sqn{A\lwith B,\Gamma}} {
    \sqn{A,\Gamma }} {\sqn{B,\Gamma}}
  \qquad
  \vcinf{\lplus_1} {\sqn{A\lplus B,\Gamma}} {\sqn{A,\Gamma}}
  \qquad
  \vcinf{\lplus_2} {\sqn{A\lplus B,\Gamma}} {\sqn{B,\Gamma}}
\end{equation}
In his first proposal for proof nets for $\ufMALL$, Girard
\cite{girard:87} used boxes (as for the exponentials). The problem
with this approach is that it distinguishes between the two proofs
\proofadjust
\begin{equation}\label{eq:with-with}
  \vcddernote{\lwith}{}{
    \sqn{A\lwith B,C\lwith D,\Gamma}}{
    \rroot{\lwith}{\sqn{A,C\lwith D,\Gamma}}{
      \leaf{\sqnsmallderi{A,C,\Gamma}{\Pi_1}}}{
      \leaf{\sqnsmallderi{A,D,\Gamma}{\Pi_2}}}}{
    \rroot{\lwith}{\sqn{B,C\lwith D,\Gamma}}{
      \leaf{\sqnsmallderi{B,C,\Gamma}{\Pi_3}}}{
      \leaf{\sqnsmallderi{B,D,\Gamma}{\Pi_4}}}}
\end{equation}
and
\proofadjust
\begin{equation}\label{eq:with-with-1}
  \vcddernote{\lwith}{}{
    \sqn{A\lwith B,C\lwith D,\Gamma}}{
    \rroot{\lwith}{\sqn{A\lwith B,C,\Gamma}}{
      \leaf{\sqnsmallderi{A,C,\Gamma}{\Pi_1}}}{
      \leaf{\sqnsmallderi{B,C,\Gamma}{\Pi_3}}}}{
    \rroot{\lwith}{\sqn{A\lwith B,D,\Gamma}}{
      \leaf{\sqnsmallderi{A,D,\Gamma}{\Pi_2}}}{
      \leaf{\sqnsmallderi{B,D,\Gamma}{\Pi_4}}}}
\end{equation}

In his second proposal \cite{girard:96:PN}, Girard proposed the notion of
\emph{monomial proof nets} (still following Ideology~\ref{ideo:girard}), where
he introduces the notion of slice and attaches a monomial (boolean) weight
to the identity links.

Here we will sketch a very recent proposal by Hughes and van
Glabbeek \cite{hughes:glabbeek:03}, which follows
Ideology~\ref{ideo:coherence}. Remember that in
Section~\ref{sec:sc-to-pn-2}, we observed that the additional graph
structure that captures the essence of the proof for $\ufMLL$ consist
of a linking, which is just a pairing of dual atom occurrences.

The discovery of Hughes and van Glabbeek was that for $\ufMALL$, this
additional graph structure is a \emph{set of linkings}, where a
linking is again, simply a pairing of dual atom occurrences. But this
time, a single linking need not to be exhaustive (i.e., it does not
necessarily pair up everyone).

\begin{figure}[!t]
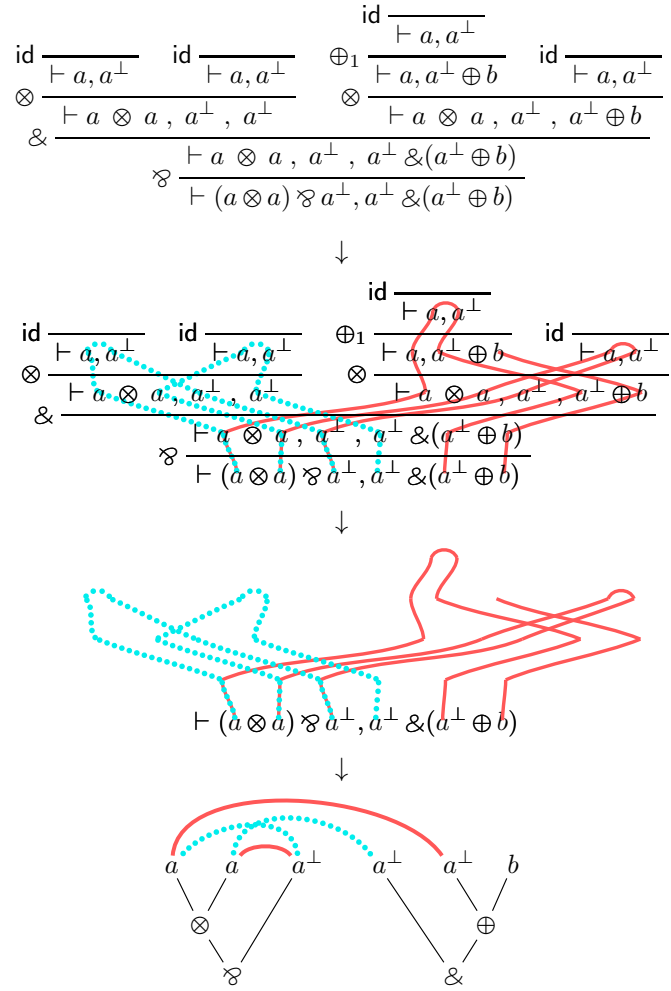

  \def\arrayskip{2ex}
  \def\thefirstcomma{\qqqqquad\qquad,\qqqqquad\qquad}
  \def\thesecondcomma{\hskip1.5em,\hskip1.5em}
  \begin{center}
    $$
    \begin{array}{c}
      \exascMALL
      \\[\arrayskip]
      \downarrow
      \\[-1ex]
      \tpathderivation{\exascMALL}{\pathexascMALL}
      \\[\arrayskip]
      \downarrow
      \\[-3ex]
      \tpathonlyderivation{\exascMALL}{\pathexascMALL}
      \\[\arrayskip]
      \downarrow
      \\[\arrayskip]
      \pathpnexaMALL
    \end{array}
    $$
    \caption{From sequent calculus to MALL proof nets via coherence graphs}
    \label{fig:exaMALL}
  \end{center}
\end{figure}

We can extract the proof net from the sequent calculus proof in the same way
as in Section~\ref{sec:sc-to-pn-2}. But when we encounter a $\lwith$-rule, we
have to separate the two linkings of the two branches, and we have to keep
track of which pairing belongs to which linking. Figure~\ref{fig:exaMALL}
shows an example (taken from \cite{hughes:glabbeek:03}) where we do this by
choosing different line style/colors.

In \cite{hughes:glabbeek:03}, Hughes and van Glabbeek provide a correctness
criterion and a cut elimination that preserves the correctness. Part of the
correctness criterion is that each linking itself has to be multiplicatively
correct (in the sense of Section~\ref{sec:correctness}).
Unfortunately, we do here not have time and space to go
into the details. However, from the viewpoint of the identity of
proofs, there is an important observation to make about these proof
nets: The following two sequent proofs are mapped to the same proof
net for all possible $\Pi_1$, $\Pi_2$, and $\Pi_3$:  \proofadjust
\begin{equation}\label{eq:with-tens-1}
  \vcddernote{\lwith}{}{\sqn{\Gamma,A\lwith B,C\ltens D,\Delta}}{
    \rroot{\ltens}{\sqn{\Gamma,A,C\ltens D,\Delta}}{
      \leaf{\sqnsmallderi{\Gamma,A,C}{\Pi_1}}}{
      \leaf{\sqnsmallderi{D,\Delta}{\Pi_3}}}}{
    \rroot{\ltens}{\sqn{\Gamma,B,C\ltens D,\Delta}}{
      \leaf{\sqnsmallderi{\Gamma,B,C}{\Pi_2}}}{
      \leaf{\sqnsmallderi{D,\Delta}{\Pi_3}}}}
\end{equation}
\begin{equation}\label{eq:with-tens-2}
  \vcddernote{\ltens}{}{\sqn{\Gamma,A\lwith B,C\ltens D,\Delta}}{
    \rroot{\lwith}{\sqn{\Gamma,A\lwith B,C}}{
      \leaf{\sqnsmallderi{\Gamma,A,C}{\Pi_1}}}{
      \leaf{\sqnsmallderi{\Gamma,B,C}{\Pi_2}}}}{
    \leaf{\sqnsmallderi{D,\Delta}{\Pi_3}}}
\end{equation}
In other words, the proof nets identify the two sequent proofs
\eqref{eq:with-tens-1} and~\eqref{eq:with-tens-2}. Is this wanted from the
proof theoretical perspective?  At the current state of the art there is no
definite answer.

An important point against the identification is that
in~\eqref{eq:with-tens-1} the subproof $\Pi_3$ appears twice, while
in~\eqref{eq:with-tens-2} it appears only once, which means that with the
identification of \eqref{eq:with-tens-1} and~\eqref{eq:with-tens-2} it becomes
difficult to speak about the size of proofs.

An important point in favor of the identification comes from algebra, where
adding the connectives $\lwith$ and $\lplus$ means adding cartesian products
and coproducts to the axioms of *-autonomous categories. If we want that
$\lwith$ behaves as cartesian product in our category of proofs, we have to
identify \eqref{eq:with-tens-1} and~\eqref{eq:with-tens-2}.

Maybe in the end both worlds have their right to exist. In any case, the
observation above leads to the following:

\begin{para}{Open Research Problem}
  Find for $\ufMALL$ a notion of proof nets that does not identify
  \eqref{eq:with-tens-1} and~\eqref{eq:with-tens-2}. A good starting point
  could be to consider flow-graphs in the calculus of structures for 
  $\MALL$ (see \cite{strassburger:02,dissvonlutz}).
  Of course, the problem is
  to find the right correctness criterion and the right notion of cut
  elimination.
\end{para}

\subsection{Intuitionistic multiplicative linear logic (without unit)}
\label{sec:IMLL}

Formulas of intuitionistic multiplicative linear logic\footnote{This is the
  logic that Kelly and MacLane studied in their seminal paper
  \cite{kelly:maclane:71}. Of course, they did not use that name. But they did
  have the correctness criterion of proof nets, called coherence graphs. They
  only had the acyclicity condition and not the connectedness condition
  because the unit was present. They proved their results via cut elimination
  (they used that terminology). They even envisaged the notion of polarity.}
  without unit ($\ufIMLL$) are generated by the syntax:
\begin{equation}\label{eq:form-IMLL}
  \cF\grammareq\cA\mid\cF\limp\cF\mid\cF\ltens\cF
\end{equation}
Note that there is no $\lpar$ and no $\cdot\lneg$.
The sequent calculus for $\ufIMLL$ is usually given in two-sided form:
\begin{equation}\label{eq:sc-IMLL}
  \begin{array}{c@{\qquad}c}
    \vcinf{\idr}{\ssqn{A}{A}}{}&
    \vciinf{\cutr}{\ssqn{\Gamma,\Gamma'}{C}}{
      \ssqn{\Gamma}{A}}{
      \ssqn{A,\Gamma'}{C}}\\[\arrayskip]
    \multicolumn{2}{c}{
      \vcinf{\exrl}{
	\ssqn{\Gamma,A,B,\Gamma'}{C}}{
	\ssqn{\Gamma,B,A,\Gamma'}{C}}}
    \\[\arrayskip]
    \vcinf{\ltensl}{\ssqn{\Gamma,A\ltens B}{C}}{
      \ssqn{\Gamma,A,B}{C}}&
    \vciinf{\ltensr}{\ssqn{\Gamma,\Gamma'}{A\ltens B}}{
      \ssqn{\Gamma}{A}}{
      \ssqn{\Gamma'}{B}}\\[\arrayskip]
    \vciinf{\limpl}{\ssqn{\Gamma,\Gamma',A\limp B}{C}}{
      \ssqn{\Gamma}{A}}{
      \ssqn{\Gamma',B}{C}}&
    \vcinf{\limpr}{\ssqn{\Gamma}{A\limp B}}{
      \ssqn{\Gamma,A}{B}}
  \end{array}
\end{equation}
Note that the right-hand side of each sequent contains at most one
formula. Apart from that, the rules are exactly the same as
in~\eqref{eq:sc-MLL-twosided}. We can therefore simply take our notion
of two-sided proof nets from Section~\ref{sec:twosided}.  The
correctness criterion remains unchanged. This can be used to prove
that $\ufIMLL$ is a conservative extension of $\ufMLL$, i.e., an
$\ufIMLL$ formula is provable in $\ufIMLL$ if and only if it is
provable in $\ufMLL$.\footnote{This is not true for full
intuitionistic linear logic with respect to linear logic. Also
classical logic is certainly not a conservative extension of
intuitionistic logic, since there are formulas, like Peirce's law
$((A\to B)\to A)\to A)$ that are provable in classical logic, but not
in intuitionistic logic.}

We can use polarities to provide a one-sided version of proof nets for
$\ufIMLL$. For this we define the set of negative and positive formulas as
follows:
\begin{equation}\label{eq:form-IMLL-posneg}
  \begin{array}{r@{\;\grammareq\;}l}
    \cF\markpl&\cA\lneg\mid\cF\markpl\lpar\cF\markpl\mid
    \cF\markpl\ltens\cF\markpr\mid\cF\markpr\ltens\cF\markpl\\[2ex]
    \cF\markpr&\cA\phantom{\lneg}\mid\cF\markpr\ltens\cF\markpr\mid
    \cF\markpr\lpar\cF\markpl\mid\cF\markpl\lpar\cF\markpr    
  \end{array}
\end{equation}
We can now define a proof net for $\ufIMLL$ to be a proof net for $\ufMLL$ in
which at most one of the conclusions is in $\cF\markpr$ and all other
conclusions are in $\cF\markpl$. See \cite{lamarche:retore:96}
or~\cite{lamarche:contexts} for further details.

\subsection{Cyclic linear logic (without units)}
\label{sec:cLL}

In Section~\ref{sec:sc-mll} we defined sequents to be lists of
formulas, but we also had the exchange rule which made the order of
the formulas irrelevant for provability, and indeed, the two
connectives $\lpar$ and $\ltens$ were commutative.  In this section we
discuss what happens if we drop the exchange rule. As expected, the
two connectives $\lpar$ and $\ltens$ will lose their property of being
commutative, and we enter the realm of non-commutative logics. But not
everything works as one might expect...

Recall that in Observation~\ref{obs:exchange} we noted that in the
proof nets we always get a crossing of edges when we apply the
exchange rule.  Let us hence change the correctness criterion. We keep
everything as in Section~\ref{sec:correctness}, but we add as
condition that there must be no crossing of edges in the proof net
(in graph theoretical terminology: the graph has to be
\emph{planar}). Hence, our running examples from Sections
\ref{sec:sc-to-pn-1}--\ref{sec:cos-to-pn} are not correct in the sense
of this new criterion. And indeed, we used the exchange rule in all
sequent proofs.

Luckily, there is another proof net for our favorite sequent in which
there are no crossings:
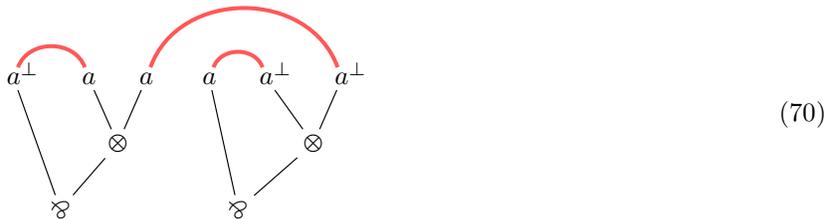
\begin{equation}\label{eq:3sq-planar}
  \vcenter{\hbox{$\pathpnplanar$}}
\end{equation}
It clearly obeys the correctness criterion.

\begin{exercise}
  Give a sequent calculus proof in the system in \eqref{eq:sc-mll}
  that translates into the net \eqref{eq:3sq-planar}. Then try to find
  such a proof that does not use the exchange rule.
\end{exercise}

Here we are in for a surprise. It is very easy to find a sequent
calculus proof for \eqref{eq:3sq-planar}, but it is impossible to find
one that does not use the $\exr$-rule. What went wrong?

There are two explanations. The first is based on the following
observation. If we have a planar proof net, i.e., a list of formula trees
together with a perfect matching of their atom occurrences such that there is
no crossing among the links that connect the atoms, then we can from that data
not tell \emph{a priori} which formula is the first in the list. We have to
think of the formulas to be grouped in a \emph{cycle} around the graph with
the axiom links. We now have to think of a proof net not as something like
$$
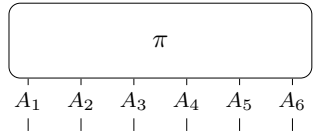
\vcgpnonlybox{\pi}{
  \gpnboxout1\qquad\gpnboxout2\qquad\gpnboxout3\qquad
  \gpnboxout4\qquad\gpnboxout5\qquad\gpnboxout6}{
  \gpnout1\qquad\gpnout2\qquad\gpnout3\qquad
  \gpnout4\qquad\gpnout5\qquad\gpnout6}{	
  \boxoutline{1}{A_1}
  \boxoutline{2}{A_2}
  \boxoutline{3}{A_3}
  \boxoutline{4}{A_4}
  \boxoutline{5}{A_5}
  \boxoutline{6}{A_6}
}
$$
but as something like
$$
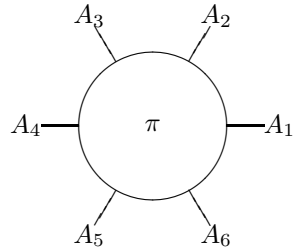
\xy\xygraph{!{/r4pc/:}
*{\pi}*\cir<7ex>{}
!P6"B"{~<{-}~>{}{\phantom{A_0}}}
"B1"{A_1} "B2"{A_2} "B3"{A_3} "B4"{A_4} "B5"{A_5} "B6"{A_6}
} 
\endxy
$$
To go along with this behavior, we cannot just drop the exchange rule, but
have replace it by the \emph{cyclic exchange rule}:
$$
\vcinf{\cyclr}{\sqn{A,\Gamma}}{\sqn{\Gamma,A}}
$$ where $\Gamma$ is an arbitrary list of formulas.  This is also the reason
why the name \emph{cyclic linear logic} has been chosen for this logic. It has
been investigated in detail by Yetter \cite{yetter:90}.

We can now give a sequent proof that translates into \eqref{eq:3sq-planar}:
{\footnotesize
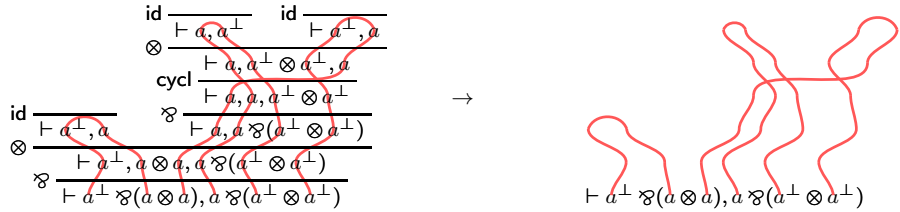
$$
\clapm{
\vctpathderivation{\scderiplanar}{\pathscderiplanar}
\qquad\to
\vctpathonlyderivation{\scderiplanar}{\pathscderiplanar}
}
$$}%
But there is still this disturbing fact that there are crossings in the
flow-graph but no crossings in the net.

\begin{figure}[!t]
  \begin{center}
    $$
    \begin{array}{c@{\qqquad}c}
      \vcinf{\ird}{A\lneg\lpar A}{} 
      &
      \vcinf{\iru}{}{A\ltens A\lneg} 
      \\[\arrayskip]
      \vcinf{\ird}{S\cons{(A\lneg\lpar A)\ltens B}}{S\cons{B}} 
      &
      \vcinf{\iru}{S\cons{B}}{S\cons{(A\ltens A\lneg)\lpar B}} 
      \\[\arrayskip]
      \vcinf{\ird}{S\cons{B\ltens(A\lneg\lpar A)}}{S\cons{B}} 
      &
      \vcinf{\iru}{S\cons{B}}{S\cons{B\lpar(A\ltens A\lneg)}} 
      \\[\arrayskip]
      \vcinf{\assrd}{S\cons{(A\lpar B)\lpar C}}{S\cons{A\lpar(B\lpar C)}}
      &
      \vcinf{\assru}{S\cons{(A\ltens B)\ltens C}}{S\cons{A\ltens (B\ltens C)}}
      \\[\arrayskip]
      \multicolumn{2}{c}{
	\vcinf{\swir}{S\cons{(A\ltens B)\lpar C}}{S\cons{A\ltens(B\lpar C)}}}
      \\[\arrayskip]
      \multicolumn{2}{c}{
	\vcinf{\swir}{S\cons{A\lpar(B\ltens C)}}{S\cons{(A\lpar B)\ltens C}}}
    \end{array}
    $$
    \caption{A system for cyclic $\ufMLL$ in the calculus of structures}
    \label{fig:cos-cll}
  \end{center}
\end{figure}

This leads us to our second explanation: The sequent calculus is
simply too rigid to fully capture the phenomenon. Recall that in
Observation~\ref{obs:subtle} we noted that in the calculus of
structures we can provide a derivation such that crossings appear in
the flow-graph only if they also appear in the net. This means we
should now be able to come up with a derivation that has no crossings
in the flow-graph, and hence does not need the exchange rule nor a
cyclic replacement. Of course, we need to adapt the system a little
bit: we need another version of $\ird$ and $\iru$, namely the one in
\eqref{eq:ir} and we need two versions of switch. Note that since
we do not have $\comrd$ nor $\comru$, we cannot generate the different
four versions of switch as in \eqref{eq:switch}. The new system is
shown in Figure~\ref{fig:cos-cll} (both versions of switch are
self-dual).

Here is now our favorite example without crossings:
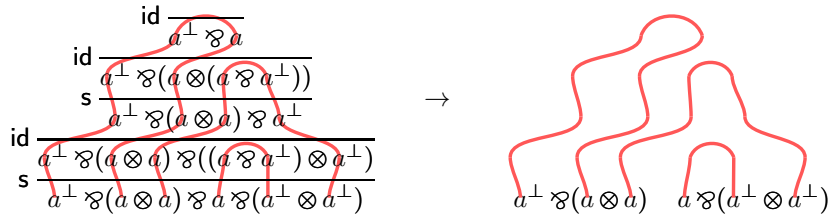
\proofadjust
$$
\qlapm{
\vctpathderivation{\cosderiplanar}{\pathcosderiplanar}
\quad\to
\vctpathonlyderivation{\cosderiplanar}{\pathcosderiplanar}
}
$$
That deep inference is powerful enough to get rid of the cyclic exchange rule
has first been observed by Di~Gianantonio \cite{digianantonio:04}. But he did
not use proof nets to show this fact.

By combining the last two sections, i.e., proof nets for cyclic linear logic
and intuitionistic linear logic, we can obtain proof nets for the Lambek
calculus~\cite{lambek:58}. We leave the details as an exercise. The interested
reader is referred to \cite{lamarche:retore:96} or~\cite{lamarche:contexts},
which contain a systematic treatment.

It is also possible to combine usual commutative $\ufMLL$ and cyclic $\ufMLL$
together in a single system. Then there is a par/tensor pair which is
commutative and one which is non-commutative. This is done
in~\cite{abrusci:ruet:00}.

\subsection{Multiplicative linear logic with units}
\label{sec:units}

Let us now finally discuss the problem of the units. We stick to the
multiplicative fragment of linear logic. The formulas of $\MLL$ are
generated by
$$
\cF\grammareq\cA\mid\cA\lneg\mid\cF\lpar\cF\mid\cF\ltens\cF\mid
\lbot\mid\lone
$$ where again everything is as in Section~\ref{sec:sc-mll}. The
two units are dual to each other:
$$
\lbot\lneg=\lone
\qqqqquad
\lone\lneg=\lbot
$$ The inference rules in the sequent calculus are the same as in
\eqref{eq:sc-mll} plus the ones for the units:
\begin{equation}\label{eq:sc-units}
  \vcinf{\lbot} {\sqn{\Gamma,\lbot,\Delta}} {\sqn{\Gamma,\Delta}}
  \qqqqquad
  \vcinf{\lone} {\sqn{\lone}} {}
\end{equation}

If we now follow the argumentation of avoiding trivial rule
permutations as in \eqref{eq:rule-permut}, we have to identify the
following three derivations in the sequent calculus:
$$
\vcenter{\ddernote{\ltens}{}{\lbot,\Gamma,A\ltens B,\Delta}
{\root{\lbot}{\lbot,\Gamma,A}{\leaf{\Gamma,A}}}{\leaf{\quad B,\Delta}}}
\quad\longleftrightarrow\quad
\vcenter{\dernote{\lbot}{}{\lbot,\Gamma,A\ltens B,\Delta}
{\rroot{\ltens}{\Gamma,A\ltens
    B,\Delta}{\leaf{\Gamma,A\quad}}{\leaf{\quad B,\Delta}}}}
\quad\longleftrightarrow\quad
\vcenter{\ddernote{\ltens}{}{\lbot,\Gamma,A\ltens B,\Delta}
{\leaf{\Gamma,A\quad}}{\root{\lbot}{\lbot,B,\Delta}{\leaf{B,\Delta}}}}
$$ because they are obtained from each other by permuting the
$\lbot$-rule under and over the $\ltens$-rule. Furthermore, following
the equations that are imposed on proofs by the axioms of *-autonomous
categories (with units) these three proofs have to be identified.

However, if we follow the idea of having a graph as proof net, we have
to attach the $\lbot$ somewhere. But there is no canonical place to do
so. This is explained in further detail in
\cite{lam:str:freestar}. That there is this problem with the
$\lbot$-rule, and in fact with every kind of weakening rule has
already been observed in \cite{girard:87,lafont:95,girard:96:PN} and
others.\footnote{To be precise, one should note that the problem has already
  been noted in \cite{kelly:maclane:71}.}

\begin{figure}[!t]
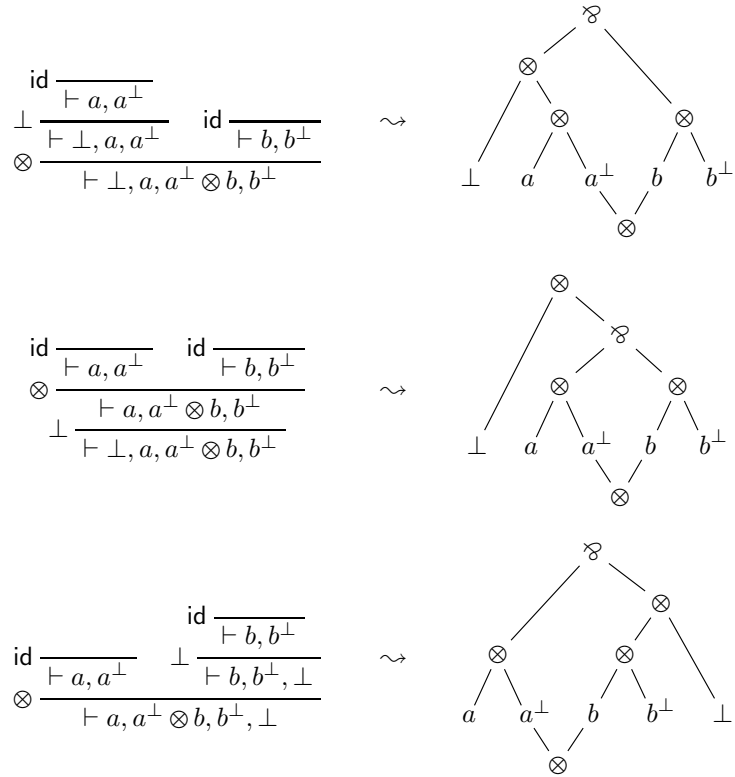

  \def\arrayskip{12ex}
  \begin{center}
    $$
    \begin{array}{c@{\qquad\leadsto\qquad}c}
      \vcddernote{\ltens}{}{\sqn{\lbot,a,a\lneg\ltens b,b\lneg}}{
	\root{\lbot}{\sqn{\lbot,a,a\lneg}}{
	  \root{\idr}{\sqn{a,a\lneg}}{
	    \leaf{}}}}{
	\root{\idr}{\sqn{b,b\lneg}}{
	  \leaf{}}}
      &
      \vcnpn{
	&&&&\nlpar1\\
	&&\nltens4\\
	&&&\nltens2&&&&\nltens3\\
	\nlbot1&\pa&\na1 &\pa&\na2\rlapm{\lneg}&\qlap{}&\mb1&\pa&\mb2\lneg\\
	&&&&&\nltens1
      }{
	\ncline{lbot1}{ltens4}
	\ncline{lpar1}{ltens4}
	\ncline{ltens1}{a2}
	\ncline{ltens1}{b1}
	\ncline{ltens4}{ltens2}
	\ncline{lpar1}{ltens3}
	\ncline{ltens2}{a1}
	\ncline{ltens2}{a2}
	\ncline{ltens3}{b1}
	\ncline{ltens3}{b2}
      }{}
      \\[\arrayskip]
      \vcdernote{\lbot}{}{\sqn{\lbot,a,a\lneg\ltens b,b\lneg}}{
	\rroot{\ltens}{\sqn{a,a\lneg\ltens b,b\lneg}}{
	  \root{\idr}{\sqn{a,a\lneg}}{
	    \leaf{}}}{
	  \root{\idr}{\sqn{b,b\lneg}}{
	    \leaf{}}}}
      &
      \vcnpn{
	&&&\nltens4\\
	&&&&&\nlpar1\\
	&&&\nltens2&&&&\nltens3\\
	\nlbot1&\pa&\na1 &\pa&\na2\rlapm{\lneg}&\qlap{}&\mb1&\pa&\mb2\lneg\\
	&&&&&\nltens1
      }{
	\ncline{lbot1}{ltens4}
	\ncline{lpar1}{ltens4}
	\ncline{ltens1}{a2}
	\ncline{ltens1}{b1}
	\ncline{lpar1}{ltens2}
	\ncline{lpar1}{ltens3}
	\ncline{ltens2}{a1}
	\ncline{ltens2}{a2}
	\ncline{ltens3}{b1}
	\ncline{ltens3}{b2}
      }{}
      \\[\arrayskip]
      \vcddernote{\ltens}{}{\sqn{a,a\lneg\ltens b,b\lneg,\lbot}}{
	\root{\idr}{\sqn{a,a\lneg}}{
	  \leaf{}}}{
	\root{\lbot}{\sqn{b,b\lneg,\lbot}}{
	  \root{\idr}{\sqn{b,b\lneg}}{
	    \leaf{}}}}
      &
      \vcnpn{
	&&&&\nlpar1\\
	&&&&&&\nltens4\\
	&\nltens2&&&&\nltens3\\
	\na1 &\pa&\na2\rlapm{\lneg}&\qlap{}&\mb1&\pa&\mb2\lneg&\pa&\nlbot1\\
	&&&\nltens1
      }{
	\ncline{lbot1}{ltens4}
	\ncline{lpar1}{ltens4}
	\ncline{ltens1}{a2}
	\ncline{ltens1}{b1}
	\ncline{ltens4}{ltens3}
	\ncline{lpar1}{ltens2}
	\ncline{ltens2}{a1}
	\ncline{ltens2}{a2}
	\ncline{ltens3}{b1}
	\ncline{ltens3}{b2}
      }{}      
    \end{array}
    $$
    \caption{From sequent calculus to $\MLL$ proof nets with units}
    \label{fig:MLL-unit}
  \end{center}
\end{figure}

The solution to the problem is instead of considering graphs as proof nets, to
consider equivalence classes of such graphs. There are now three different
proposals in the literature to do so. In \cite{BCST} the authors use a
two-sided version of proof nets and attach the $\lbot$ to the edges in the
graph (see also Section~\ref{sec:cl-rob}). We will show here the approach
taken by \cite{str:lam:04:CSL,lam:str:freestar} which clearly follows
Ideology~\ref{ideo:coherence}. The additional graph structure attached to the
sequent forest does no longer consist of only the identity links, but of
another formula tree. Then the resulting graph behaves as ordinary unit-free
proof net and obeys the usual correctness criteria. 
The basic idea is the following. The ordinary proof net
$$
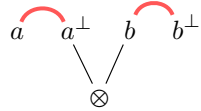
\vcnpn{
    \\[-2ex]
  \na1 &\pa&\na2\rlapm{\lneg}&\qlap{}&\mb1&\pa&\mb2\lneg\\
  &&&\nltens1
    \\[-2ex]
}{
  \ncline{ltens1}{a2}
  \ncline{ltens1}{b1}
}{
  \udline{a1}{a2}
  \udline{b1}{b2}
}
$$
is now written as 
$$
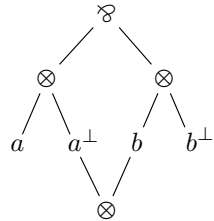
\vcnpn{
  &&&\nlpar1\\
  &\nltens2&&&&\nltens3\\
  \na1 &\pa&\na2\rlapm{\lneg}&\qlap{}&\mb1&\pa&\mb2\lneg\\
  &&&\nltens1
}{
  \ncline{ltens1}{a2}
  \ncline{ltens1}{b1}
  \ncline{lpar1}{ltens2}
  \ncline{lpar1}{ltens3}
  \ncline{ltens2}{a1}
  \ncline{ltens2}{a2}
  \ncline{ltens3}{b1}
  \ncline{ltens3}{b2}
}{}
$$ i.e., the identity links are replaced by $\ltens$-nodes which are connected
by a $\lpar$.\footnote{If there are more than two axiom links we either use
$n$-ary $\lpar$ or take the equivalence class modulo associativity and
commutativity of $\lpar$ in the linking tree. This is not problematic since we
have to take equivalence classes anyway.}
If we now have to attach a $\lbot$, we attach it via a $\ltens$ to the root of
linking tree of the proof to which we apply the $\lbot$-rule.
Figure~\ref{fig:MLL-unit} shows three examples. Since all three of them are
the same proof according to what has been said above, we have to put them in
the same equivalence class. To achieve this, we consider linking trees
equivalent modulo the following equation:
\begin{equation}\label{eq:bot}
  \vcnpn{
    &&\nltens1\\
    &&&\nlpar1\\
    \nlbot1&\pa&\nP1&\pa&\nQ1
  }{
    \ncline{lbot1}{ltens1}
    \ncline{lpar1}{ltens1}
    \ncline{P1}{lpar1}
    \ncline{Q1}{lpar1}
  }{}
  \qquad=\qquad
  \vcnpn{
    &&\nlpar1\\
    &\nltens1\\
    \nlbot1&\pa&\nP1&\pa&\nQ1
  }{
    \ncline{lbot1}{ltens1}
    \ncline{lpar1}{ltens1}
    \ncline{P1}{ltens1}
    \ncline{Q1}{lpar1}
  }{}
\end{equation}
provided going from right to left in~\eqref{eq:bot} does not destroy the
correctness.\footnote{Observe that going to left to right cannot destroy
correctness because this corresponds to an application of the switch rule in
the calculus of structures, which always preserves correctness (see second
proof of Theorem~\ref{thm:S-correct}).}

The details of this kind of proof nets are carried out in
\cite{str:lam:04:CSL,lam:str:freestar}, where it is also shown that they form
the free *-autonomous category (in the usual sense, with units). 

Let us note here only that the linking tree of an $\MLL$ proof net can be seen
as the up-side in a two-sided unit-free proof net, as we discussed it in
Section~\ref{sec:twosided}, provided we read the units $\lbot$ and $\lone$ as
ordinary atoms. Here is an example:
$$
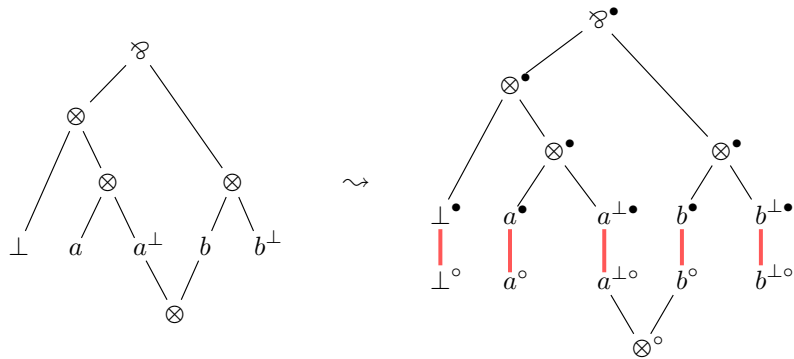
\vcnpn{
  &&&&\nlpar1\\
  &&\nltens4\\
  &&&\nltens2&&&&\nltens3\\
  \nlbot1&\pa&\na1 &\pa&\na2\rlapm{\lneg}&\qlap{}&\mb1&\pa&\mb2\lneg\\
  &&&&&\nltens1
}{
  \ncline{lbot1}{ltens4}
  \ncline{lpar1}{ltens4}
  \ncline{ltens1}{a2}
  \ncline{ltens1}{b1}
  \ncline{ltens4}{ltens2}
  \ncline{lpar1}{ltens3}
  \ncline{ltens2}{a1}
  \ncline{ltens2}{a2}
  \ncline{ltens3}{b1}
  \ncline{ltens3}{b2}
}{}
\qquad\leadsto\qquad
\vcnpn{
  &&&&\nlpar1\markpl\\
  &&\nltens4\markpl\\
  &&&\nltens2\markpl&&&&\nltens3\markpl\\
  \nlbot1\markpl&\pa&\na1\markpl&\pa&\na2\rlapm{\lnegpl}&\qlap{}&
  \mb1\markpl&\pa&\mb2\lnegpl\\
  \nlbot2\markpr&\pa&\na3\markpr&\pa&\na4\rlapm{\lnegpr}&\qlap{}&
  \mb3\markpr&\pa&\mb4\lnegpr\\
  &&&&&\nltens1\markpr
}{
  \ncline{lbot1}{ltens4}
  \ncline{lpar1}{ltens4}
  \ncline{ltens1}{a4}
  \ncline{ltens1}{b3}
  \ncline{ltens4}{ltens2}
  \ncline{lpar1}{ltens3}
  \ncline{ltens2}{a1}
  \ncline{ltens2}{a2}
  \ncline{ltens3}{b1}
  \ncline{ltens3}{b2}
}{
  \uline{lbot2}{lbot1}
  \uline{a3}{a1}
  \uline{a4}{a2}
  \uline{b3}{b1}
  \uline{b4}{b2}
}
$$

An alternative approach to the problem of the units has recently proposed by
Hughes. In~\cite{hughes:simple-mult} he defines a notion for proof nets for
$\MLL$ (also following Ideology~\ref{ideo:coherence}) where he attaches the
$\lbot$ to some atom in the proof. This has the advantage that there is no
need to consider equivalence classes of proof nets for defining cut
elimination. However, for constructing the free *-autonomous category he also
needs equivalence classes~\cite{hughes:freestar}.

\begin{para}{Open Research Problem}
  Find for a notion of proof nets that is also able to accommodate the additive
  units $\ltop$ and $\lzero$ of linear logic.
\end{para}

\subsection{Mix}\label{sec:mix}

In order to avoid the problems with the units, one can simply ignore
them, as we did in the whole first part of these notes. An alternative
is to add the rules
\begin{equation}\label{eq:sc-mix}
  \vciinf{\mixr}{\sqn{\Gamma,\Delta}}{\sqn{\Gamma}}{\sqn{\Delta}}
  \qquand
  \vcinf{\mixzr}{\sqn{}}{}
\end{equation}
called \emph{mix} and \emph{nullary mix}. Adding these two rule to
\eqref{eq:sc-mll} or \eqref{eq:sc-mll}+\eqref{eq:sc-units} is
equivalent to saying that $\lbot=\lone$. By doing this, the
correctness criteria shown in Section~\ref{sec:correctness} change
only slightly: The acyclicity condition remains, but the connectedness
condition has to be dropped. A detailed analysis for this can be found
in \cite{fleury:retore:94}.

\subsection{Pomset logic and BV}\label{sec:pomset}

Before we leave the realm of linear logic, let us show another
variation, which has a particularly nice theory of proof nets. It
has been introduced by Retor{\'e}~\cite{retore:97} under the name
\emph{pomset logic}.  The set of formulas is generated via
$$
\cF\grammareq\cA\mid\cA\lneg\mid\cF\lpar\cF\mid\cF\ltens\cF\mid
\cF\lseq\cF
$$ Negation is defined as in Section~\ref{sec:sc-mll}. The new
connective $\lseq$ is non-commutative and self-dual, i.e.,
$$
(A\lseq B)\lneg=A\lneg\lseq B\lneg
$$ 
We will also allow to write $A\lcoseq B$ for $B\lseq A$.

For defining proof nets we use Retor{\'e}'s RB-graph
presentation. The $\lpar$, the $\ltens$, the cut and the identity are
translated into RB-graphs as in Section~\ref{sec:correctness}. The new
connective is translated as follows:
$$
    \vcgpn{
      \gpnout1&\strut&\gpnout2\\
      &\gpnseq3\\
      &\gpnout4 \strut}{
      \treeline{out1}{lseq3}{A}
      \treeline{out2}{lseq3}{B}
      \treeline{out4}{lseq3}{A\lseq B}}
    \qquad\leadsto\qquad
    \vcrbpn{
      \rbv1&&\rbv2\\
      \rbv3&&\rbv4\\
      &\rbv5\\
      &\rbv6}{
      \rbseq534
      \rbblueedgemode
      \rbedge13 
      \rbedge24
      \rbedge56}
$$ The correctness criterion hast to be changed such that in an
\AE-cycle we can walk through a red edge with an arrow only in the
direction of the arrow. Through undirected edges we can still walk
either way. Furthermore, the connectedness condition has to be
dropped.\footnote{The reason for this is that because of the
self-duality of the new connective $\lseq$ the two units $\lbot$ and $\lone$
are identified.}
Here are 4 examples:
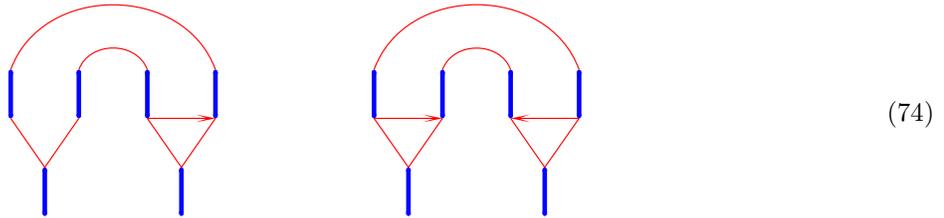
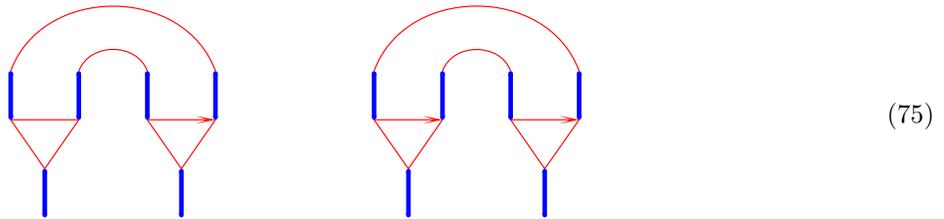
\begin{equation}\label{eq:exa-pomset-ok}
  \vcrbpn{
    \\
    \rbv1&&\rbv2&&\rbv3&&\rbv4\\
    \rbv{11}&&\rbv{12}&&\rbv{13}&&\rbv{14}\\
    &\rbv{7} && &&   \rbv{8}\\
    &\rbv{17} && &&  \rbv{18}
  }{
    \rbid14
    \rbid23
    \rbpar7{11}{12}
    \rbseq8{13}{14}
    \rbblueedgemode
    \rbedge1{11}
    \rbedge2{12}
    \rbedge3{13}
    \rbedge4{14}
    \rbedge7{17}
    \rbedge8{18}
  }    
  \qqqqqquad
  \vcrbpn{
    \\
    \rbv1&&\rbv2&&\rbv3&&\rbv4\\
    \rbv{11}&&\rbv{12}&&\rbv{13}&&\rbv{14}\\
    &\rbv{7} && &&   \rbv{8}\\
    &\rbv{17} && &&  \rbv{18}
  }{
    \rbid14
    \rbid23
    \rbseq7{11}{12}
    \rbcoseq8{13}{14}
    \rbblueedgemode
    \rbedge1{11}
    \rbedge2{12}
    \rbedge3{13}
    \rbedge4{14}
    \rbedge7{17}
    \rbedge8{18}
  }    
\end{equation}
\begin{equation}\label{eq:exa-pomset-not-ok}
  \vcrbpn{
    \\
    \rbv1&&\rbv2&&\rbv3&&\rbv4\\
    \rbv{11}&&\rbv{12}&&\rbv{13}&&\rbv{14}\\
    &\rbv{7} && &&   \rbv{8}\\
    &\rbv{17} && &&  \rbv{18}
  }{
    \rbid14
    \rbid23
    \rbtens7{11}{12}
    \rbseq8{13}{14}
    \rbblueedgemode
    \rbedge1{11}
    \rbedge2{12}
    \rbedge3{13}
    \rbedge4{14}
    \rbedge7{17}
    \rbedge8{18}
  }    
  \qqqqqquad
  \vcrbpn{
    \\
    \rbv1&&\rbv2&&\rbv3&&\rbv4\\
    \rbv{11}&&\rbv{12}&&\rbv{13}&&\rbv{14}\\
    &\rbv{7} && &&   \rbv{8}\\
    &\rbv{17} && &&  \rbv{18}
  }{
    \rbid14
    \rbid23
    \rbseq7{11}{12}
    \rbseq8{13}{14}
    \rbblueedgemode
    \rbedge1{11}
    \rbedge2{12}
    \rbedge3{13}
    \rbedge4{14}
    \rbedge7{17}
    \rbedge8{18}
  }    
\end{equation}
The two examples in~\eqref{eq:exa-pomset-ok} fulfill the correctness
criterion, while the two examples in~\eqref{eq:exa-pomset-not-ok} do not.

Note that we can define cut elimination as before, but we have to make
sure that the arrows go in the right direction. Here is the new
reduction rule that has to be added to the ones in
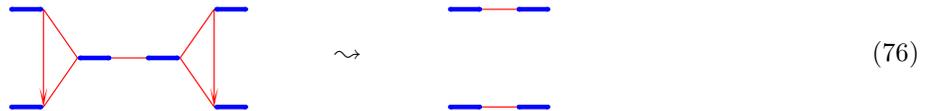
\eqref{eq:rb-cut-red-1} and~\eqref{eq:rb-cut-red-2}:
  \begin{equation}\label{eq:rb-cut-red-3}
    \vcrbpn{\rbv1&\rbv2&&&&&\rbv7&\rbv8\\
      &&\rbv3&\rbv4&\rbv5&\rbv6&&\\
      \rbv{11}&\rbv{12}&&&&&\rbv{17}&\rbv{18}}{
      \rbseq32{12} \rbedge45 \rbseq67{17}
      \rbblueedgemode
      \rbedge12 \rbedge34 \rbedge56 \rbedge78 
      \rbedge{11}{12} \rbedge{17}{18}
    }
    \qqquad\leadsto\qqquad
    \vcrbpn{\rbv1&\rbv2&\rbv3&\rbv4\\ \\
      \rbv5&\rbv6&\rbv7&\rbv8}{
      \rbedge23 \rbedge67
      \rbblueedgemode
      \rbedge12 \rbedge34 \rbedge56 \rbedge78 
    }
  \end{equation}

With the same methods as in Section~\ref{sec:cutelim}, we can show the
following theorem:

\begin{theorem}
  Cut elimination for pomset logic proof nets is terminating,
  confluent, and preserves correctness.
\end{theorem}

You might already have wondered why for pomset logic we did not
introduce the sequent calculus system before introducing the proof
nets, as we did it with all other logics so far. The reason is
simple. There is no sequent calculus system for pomset logic.

This of course questions the label ``logic'' for the object we have
here, if there is not even an ordinary deductive system for it. And
what is the meaning of cut elimination in that respect?

\begin{figure}[!t]
  \begin{center}
    $$
    \begin{array}{c@{\qqquad}c}
      \vcinf{\ird}{S\cons{A\lneg\lpar A}}{S\cons{\un}} 
      &
      \vcinf{\iru}{S\cons{\un}}{S\cons{A\ltens A\lneg}} 
      \\[\arrayskip]
      \vcinf{\comrd}{S\cons{B\lpar A}}{S\cons{A\lpar B}}
      &
      \vcinf{\comru}{S\cons{B\ltens A}}{S\cons{A\ltens B}}
      \\[\arrayskip]
      \vcinf{\assrd}{S\cons{(A\lpar B)\lpar C}}{S\cons{A\lpar(B\lpar C)}}
      &
      \vcinf{\assru}{S\cons{(A\ltens B)\ltens C}}{S\cons{A\ltens (B\ltens C)}}
      \\[\arrayskip]
      \vcinf{\sassrd}{S\cons{(A\lseq B)\lseq C}}{S\cons{A\lseq(B\lseq C)}}
      &
      \vcinf{\sassru}{S\cons{A\lseq(B\lseq C)}}{S\cons{(A\lseq B)\lseq C}}
      \\[\arrayskip]
      \multicolumn{2}{c}{
	\vcinf{\swir}{S\cons{(A\ltens B)\lpar C}}{S\cons{A\ltens(B\lpar C)}}
      }
      \\[\arrayskip]
      \vcinf{\seqrd}{
	S\cons{(A\lseq B)\lpar(C\lseq D)}}{
	S\cons{(A\lpar C)\lseq(B\lpar D)}}
      &
      \vcinf{\seqru}{
	S\cons{(A\lseq B)\ltens(C\lseq D)}}{
	S\cons{(A\ltens C)\lseq(B\ltens D)}}
    \end{array}
    $$
     \caption{System $\SBV$ in the calculus of structures}
    \label{fig:SBV}
   \end{center}
\end{figure}

Fortunately, there is deep inference and the calculus of structures.
The system shown in Figure~\ref{fig:SBV} has been introduced by
Guglielmi~\cite{guglielmi:02} who called it~$\SBV$. For that system, we extend
our language for formulas with another generator, the unit $\un$. Furthermore,
we consider formulas to be equivalent modulo the smallest congruence relation
generated by the equations
\begin{equation}\label{eq:equiv}
  A\lpar\un=A=\un\lpar A\qqquad A\ltens\un=A=\un\ltens A\qqquad 
  A\lseq\un=A=\un\lseq A
\end{equation}
The reason for this is to avoid to have five different versions of $\ird$ and
$\iru$.\footnote{Remember that for cyclic linear logic we had already three of
them.}  Furthermore, we want the following derivation to be valid:
\begin{equation}\label{eq:mixder}
  \vcdernote{=}{}{A\lpar B}{
    \root{\seqrd}{(A\lseq\un)\lpar(\un\lseq B)}{
      \root{=}{(A\lpar\un)\lseq(\un\lpar B)}{
	\root{=}{A\lseq B}{
	  \root{\seqru}{(A\ltens\un)\lseq(\un\ltens B)}{
	    \root{=}{(A\lseq\un)\ltens(\un\lseq B)}{
	      \leaf{A\ltens B}}}}}}}
  \qqquad\leftrightsquigarrow\qqquad
  \vcdernote{\seqrd}{}{A\lpar B}{
    \root{\seqru}{A\lseq B}{
      \leaf{A\ltens B}}}
\end{equation}
Here we implicitly use the ``fake inference rule'' 
$$
\vcinf{=}{S\cons{A}}{S\cons{B}}
$$
where $A=B$ according to the equations in~\eqref{eq:equiv}. Without it, we
would need several more variants of $\seqrd$ and $\seqru$, and the system
would become rather big. The fragment of~$\SBV$
without the up-rules (i.e., without the rules with the $\uparrow$ in
the name) is called~$\BV$. As usual in the calculus of structures, the
up-fragment corresponds to the cut, and we have for $\BV$ the cut elimination
theorem (proved by Guglielmi~\cite{guglielmi:02}):

\begin{theorem}\label{thm:bv-cutelim}
  Let $A$ be any pomset formula. Then,
  $$
  \mbox{for every derivation}\quad
  \vcstrder{\SBV}{}{A}{\leaf{\un}}
  \quad
  \mbox{there is a derivation}\quad
  \vcstrder{\BV}{}{A}{\leaf{\un}}
  \quad.
  $$
\end{theorem}

The proof uses the technique of \emph{splitting} and works essentially the
same way as we have sketched it for Theorem~\ref{thm:cos-cutelim}. The logic
that is defined by $\BV$ is the first logic that definitely needs deep
inference. Alwen Tiu~\cite{tiu:msc,tiu:deep} has shown that there cannot be a
shallow inference system (in particular, no sequent calculus system) defining
the logic of~$\BV$.

\begin{figure}[!t]
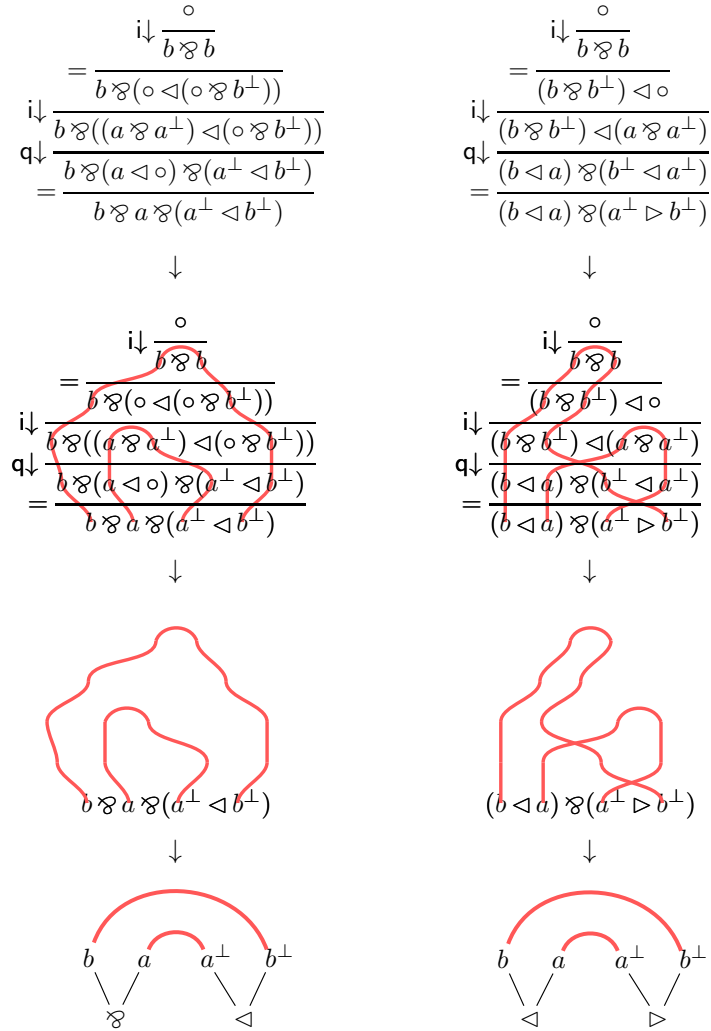

  \def\arrayskip{2ex}
  \begin{center}
    $$
    \qlapm{\begin{array}{c@{\qqqqquad}c}
      \bvderione
      &
      \bvderitwo
      \\[\arrayskip]
      \downarrow&\downarrow
      \\[\arrayskip]
      \qlapm{\pathderivation{\bvderione}{\pathbvderione}\qquad\;}
      &
      \qlapm{\pathderivation{\bvderitwo}{\pathbvderitwo}\qquad\;}
      \\[\arrayskip]
      \downarrow&\downarrow
      \\
      \qlapm{\pathonlyderivation{\bvderione}{\pathbvderione}\qquad\;}
      &
      \qlapm{\pathonlyderivation{\bvderitwo}{\pathbvderitwo}\qquad\;}
      \\[\arrayskip]
      \downarrow&\downarrow
      \\[3ex]
      \quad\pathbvpnone
      &
      \quad\pathbvpntwo
    \end{array}}
    $$
    \caption{From $\BV$ to proof nets}
    \label{fig:BV-to-pn}
  \end{center}
\end{figure}

We can apply the flow-graph method (see Section~\ref{sec:cos-to-pn}
to obtain proof nets from proofs in $\BV$. Or, if you prefer, you can obtain
two-sided proof nets from derivations in $\SBV$. Everything that has been said
in Section~\ref{sec:twosided} does also apply here. Figure~\ref{fig:BV-to-pn}
shows two small examples of proof nets obtained from $\BV$ derivations. They,
in fact, correspond to the two RB-graphs in \eqref{eq:exa-pomset-ok}.

The reason for mentioning all this is the following theorem:

\begin{theorem}\label{thm:BV-pomset}
  Let $A$ and $B$ be any pomset logic formulas. Then
  \begin{equation}\label{eq:BV-SBV}
    \vcstrder{\SBV}{}{B}{\leaf{A}}
    \qqquad\mbox{ if and only if }\qqquad
    \vcstrder{\BV}{}{A\lneg\lpar B}{\leaf{\un}}
  \end{equation}
  and if these derivations exist, then they have the same proof net, and this
  proof net obeys the pomset logic correctness criterion, when written as
  RB-graph.
\end{theorem}

\begin{proof}[ (Sketch)]
  Let us start with the first part of the theorem which is easy, provided we
  have cut elimination: From
  $$
  \vcstrder{\SBV}{}{B}{\leaf{A}}
  $$
  we can build
  $$
  \vcstrder{\SBV}{}{A\lneg\lpar B}{\root{\ird}{A\lneg\lpar A}{\leaf{\un}}}
  $$
  Now we apply Theorem~\ref{thm:bv-cutelim} to get 
  \begin{equation}\label{eq:bvproof}
    \vcstrder{\BV}{}{A\lneg\lpar B}{\leaf{\un}}    
  \end{equation}
  The other direction is even easier. From \eqref{eq:bvproof}, we can directly
  build
  $$
  \dernote{=}{}{B}{
    \root{\iru}{\un\lpar B}{
      \root{\swir}{(A\ltens A\lneg)\lpar B}{
	\stem{\BV}{}{A\ltens(A\lneg\lpar B)}{
	  \root{=}{A\ltens\un}{
	    \leaf{A}}}}}}
  $$ Note that this part of the proof did not use the fact that we are
  speaking about $\BV$. It in fact holds for any logical system in the
  calculus of structures, provided we have the switch rule and a cut
  elimination that preserves the flow-graph.\footnote{Roughly speaking, this
  is the same as saying that the proofs in the logic form some sort of
  *-autonomous category.}

  Let us now come to the second more interesting and more difficult part or
  the theorem. There are two ways to show that every $\BV$ proof net obeys the
  pomset logic criterion. The first is to show that every rule in $\BV$
  preserves the criterion (and that the unique net with $\un$ as conclusion is
  correct). This has been done already by Retor{\'e}~\cite{retore:99}. The
  second way, which has been used
  in~\cite{dissvonlutz,guglielmi:strassburger:02:big} is based on the
  following claims:
  \begin{itemize}[\bf Claim 1:]
  \item Let $n\ge 1$ and atoms $a_1,\ldots,a_n$ and formulas $A_1,\ldots,A_n$
    be given, such that for all $i\in\set{1,\ldots,n}$, we have
    $A_i=a_i\lneg\ltens a_{i+1}$ or $A_i=a_i\lneg\lseq a_{i+1}$ (where we
    count indices modulo $n$).  Then there is no derivation
    \begin{equation}\label{eq:BVnoderi}
      \vcstrder{\BV}{}{A_1\lpar\cdots\lpar A_n}{\leaf{\un}}        
    \end{equation}
  \end{itemize}
  For proving the claim we proceed by way of contradiction and by induction
  on~$n$. We have to perform a case analysis how the rules of $\BV$ could be
  applied inside $A_1\lpar\cdots\lpar A_n$. We also need the concept of a
  quasi-subformula. We say that $A$ is a \emph{quasi-subformula} of $B$, if
  $A$ can be obtained from $B$ by replacing some atom occurrences by
  $\un$.\footnote{When $A$ is a quasi-subformula of $B$, then the relation
  web (see Definition~\ref{def:web}) of $A$ is a subweb of the relation web of
  $B$.}
  \begin{itemize}[\bf Claim 2:]
  \item Let $\pi$ be a pomset logic pre-proof net with conclusion $B$, and let
  $A$ be a quasi-subformula of $B$ such that whenever an atom occurrence is
  replaced by $\un$, then also its mate (according to the linking in $\pi$) is
  replaced by $\un$. Then we have the following:
  $$
  \mbox{If}\qquad
  \vcstrder{\BV}{}{B}{\leaf{\un}}        
  \qquad\mbox{then}\qquad
  \vcstrder{\BV}{}{A}{\leaf{\un}}
  \quad.
  $$
  \end{itemize}
  This is easy to see because the proof of $B$ remains valid if we replace
  pairs of atoms by $\un$.
  \begin{itemize}[\bf Claim 3:]
  \item Let $\pi$ be a pomset logic pre-proof net with conclusion $B$. Then
    $\pi$ does not fulfill the pomset logic correctness criterion if and only
    if $B$ has a quasi-subformula $A_1\lpar\cdots\lpar A_n$, where all $A_i$
    are as in Claim~1, and all pairs $a_i$, $a_i\lneg$ are paired up in the
    linking of $\pi$.
  \end{itemize}
  Claims 1--3 together imply the result.
  \qed
\end{proof}

\begin{remark}
  The Claim 3 in the proof above states a second correctness criterion for
  $\BV$/pomset logic, which does not need RB-graphs, and which first appears
  in~\cite{dissvonlutz}. Claims 1 and~2 in the proof above show that $\BV$
  derivations obey this criterion. Proving Claim~3 then means showing that the
  two criteria are indeed equivalent.
\end{remark}

\begin{exercise}
  Complete the proof of the second part of Theorem~\eqref{thm:BV-pomset},
  i.e., show that every proof net coming from a $\BV$ derivation obeys
  the pomset logic correctness criterion.
\end{exercise}

\begin{para}{Open Research Problem}
  Prove the converse. I.e., show that for every correct pomset logic proof net
  there is a $\BV$ derivation having it as flow graph.
\end{para}

Solving this problem would finally (because of Tiu's result~\cite{tiu:deep})
show that it is indeed impossible to give a sequent calculus system for pomset
logic.

\section{Intuitionistic logic}

We will not speak about intuitionistic logic in this course. But I
will mention here some important facts related to the title of the
lecture notes, because these facts have had a strong impact on the
development of proof nets for classical logic that we will discuss in
the next section.

The inference rules in natural deduction for implication in
intuitionistic logic
$$
\vcinf{\impI}{\ssqn{\Gamma}{A\to B}}{\ssqn{\Gamma,A}{B}}
\qqquand
\vciinf{\impE}{\ssqn{\Gamma}{B}}{\ssqn{\Gamma}{A\to B}}{\ssqn{\Gamma}{A}}
$$
are the same as the typing rules 
$$
\vcinf{\absr}{
  \ssqn{\Gamma}{\lambda x.u\colon A\to B}}{
  \ssqn{\Gamma,x\colon A}{u\colon B}}
\qqquand
\vciinf{\appr}{
  \ssqn{\Gamma}{uv\colon B}}{
  \ssqn{\Gamma}{u\colon A\to B}}{
  \ssqn{\Gamma}{v\colon A}}
$$ for the simply typed $\lambda$-calculus.  This is the basis for the
so-called \emph{Curry-Howard-correspondence} (also known as
\emph{formulas-as-types-correspondence} and
\emph{proofs-as-programs-corres\-pon\-dence}). It is also called
``isomorphism'' because the normalization in natural deduction
\cite{prawitz:65} does the same as $\beta$-reduction in the
$\lambda$-calculus.\footnote{One could also argue that we have here
just two different syntactic presentations of the same mathematical
objects.}  If we add conjunction to the logic (or equivalently product
types to the $\lambda$-calculus) we can use the proofs in natural
deduction (or equivalently $\lambda$-terms) for specifying morphism in
\emph{cartesian closed categories} (short: \emph{CCCs}). What makes
things interesting is the fact that the identity forced on proofs by
the notion of normalization in natural deduction (or equivalently the
identity forced on $\lambda$-terms by
normalization\footnote{$\beta$-reduction, $\eta$-expansion, and
$\alpha$-conversion}) is exactly the same as the identity of morphism
that is determined by the axioms of CCCs. For further details on
this see \cite{lambek:scott:86}.  Of course, this simple observation
has been extended to more expressive logics and larger type systems
(e.g., System F \cite{girard:72}, calculus of constructions
\cite{coquand:huet:88}, \ldots).

Since there is already a well-understood canonical and
bureaucracy-free presentation of proofs in intuitionistic logic,
namely terms in the (simply) typed $\lambda$-calculus, the concept of
``proof nets for intuitionistic logic'' is not well investigated. To
my knowledge there are only two proposals in that direction. The
first, by Lamarche~\cite{lamarche:essential} uses the encoding of
intuitionistic logic into (intuitionistic) linear logic (the
multiplicative exponential fragment). The second, by
Horbach~\cite{horbach:msc}, restricts a class of proof nets for
classical logic (that we will discuss in the next section) to
intuitionistic logic. Of course, this cannot be as simple as for
$\ufIMLL$ (see Section~\ref{sec:IMLL}) because classical logic is
\emph{not} a conservative extension of intuitionistic logic.


\section{Classical Logic}\label{sec:cl}

It is surely very tempting to extend the beautiful connection between
deductive system, $\lambda$-calculus, and category theory from
intuitionistic logic to classical logic.

We get from intuitionistic logic to classical logic by adding the law
of excluded middle (i.e., $A\cor\cneg A$), or equivalently, an
involutive negation (i.e., $\cneg{\cneg A}=A$). Adding this to a
Cartesian closed category $\cC$, means adding a contravariant functor
$\fneg\colon\cC\to\cC$ such that $\cneg{\cneg A}\isom A$ and
$\widecneg{(A\cand B)}\isom\cneg A\cor\cneg B$ where $A\cor B=\cneg
A\cimp B$. However, if we do this we get a collapse: all proofs of the
same formula are identified, which leads to a rather boring proof
theory. This observation is due to Andr{\'e} Joyal, and a detailed
proof and discussion can be found in \cite{lambek:scott:86} and in the
appendix of \cite{girard:LC}.

We will not show the category theoretic proof of the collapse. But we
will explain the phenomenon in terms of the sequent calculus (the
argumentation is due to Yves Lafont \cite[Appendix
B]{girard:etal:89}). Suppose we have two proofs of the formula $B$ in
some sequent calculus system:
\proofadjust
$$
\sqnsmallderi{B}{\Pi_1}
\qqquand
\sqnsmallderi{B}{\Pi_2}
$$
Then we can with the help of the rules weakening, contraction, and cut
$$
\vcinf{\weakr}{\sqn{\Gamma,A}}{\sqn{\Gamma}}
\qqquad
\vcinf{\conr}{\sqn{\Gamma,A}}{\sqn{\Gamma,A,A}}
\qqquad
\vciinf{\cutr}{\sqn{\Gamma,\Delta}}{\sqn{\Gamma,A}}{\sqn{\cneg A,\Delta}}
$$
form the following proof of $B$
\proofadjust
\begin{equation}\label{eq:weak-weak}
  \vcdernote{\conr}{}{\sqn{B}}{
    \rroot{\cutr}{\sqn{B,B}}{
      \root{\weakr}{\sqn{B,A}}{
	\leaf{\sqnsmallderi{B}{\Pi_1}}}}{
      \root{\weakr}{\sqn{A\lneg,B}}{
	\leaf{\sqnsmallderi{B}{\Pi_2}}}}}
\end{equation}
If we eliminate the cut from this proof, we get either
\proofadjust
\begin{equation}\label{eq:lafont}
  \vcdernote{\conr}{}{\sqn{B}}{
    \root{\weakr}{\sqn{B,B}}{
      \leaf{\sqnsmallderi{B}{\Pi_1}}}}
  \qqqquor
  \vcdernote{\conr}{}{\sqn{B}}{
    \root{\weakr}{\sqn{B,B}}{
      \leaf{\sqnsmallderi{B}{\Pi_2}}}}
\end{equation}
depending on a nondeterministic choice.  Now note that one can hardly
find a reason why for any proof $\Pi$, the two proofs
\proofadjust
\begin{equation}\label{eq:conweak}
  \vcdernote{\conr}{}{\sqn{B}}{
    \root{\weakr}{\sqn{B,B}}{
      \leaf{\sqnsmallderi{B}{\Pi}}}}
  \qqqquand
  \vcenter{\sqnsmallderi{B}{\Pi}}
\end{equation}
should be distinguished. After all, duplicating a formula and
immediately afterwards deleting one copy is not doing
much. Also the laws of category theory tell us to identify the
two. 

On the other hand, if we want the nice relationship between deductive
system and category theory, we need a confluent cut elimination, which
means we have to equate the two proofs
in~\eqref{eq:lafont}. Consequently, by~\eqref{eq:conweak}, we have to
equate $\Pi_1$ and $\Pi_2$. Since there was no initial condition on
$\Pi_1$ and $\Pi_2$, we conclude that any two proofs of $B$ must be
equal.

But the problem with weakening, which could in fact be solved by using
mix (see Sections \ref{sec:mix} and~\ref{sec:cl-simple}), is not the only
one. We run into similar problems with the contraction rule. If we try
to eliminate the cut from
\proofadjust
\begin{equation}\label{eq:cont-cont}
  \vcddernote{\cutr}{\quadfs}{\sqn{\Gamma,\Delta}}
           {\root{\conr}{\sqn{\Gamma,A}}
             {\leaf{\sqnsmallderi{\Gamma,A,A}{\Pi_1}}}}
           {\root{\conr}{\sqn{\cneg A,\Delta}}
             {\leaf{\sqnsmallderi{\cneg A,\cneg A,\Delta}{\Pi_2}}}}
\end{equation}
we again have to make a nondeterministic choice. In the sections
below, we will see a concrete example for this.

\bigskip

There are several possibilities to cope with these problems. The
easiest is to say that there cannot be any good proof theory for
classical logic, and stop thinking about the problem\footnote{and
making sure that also other people do not think about the
problem...}. A more serious and more difficult approach is of course
to say that we have to drop some of the axioms, i.e., some of the
equations that we would like to hold between proofs in classical
logic. But which ones should go?

There are now essentially two different approaches, and both have
their advantages and disadvantages.

\begin{enumerate}
\item The first says that the axioms of cartesian closed categories
  are essential and cannot be dispensed with. Instead, one sacrifices
  the duality between $\cand$ and $\cor$. The motivation for this
  approach is that a proof system for classical logic can now be seen
  as an extension of the $\lambda$-calculus and the notion of
  normalization does not change.  With this approach one has a term
  calculus for proofs, namely Parigot's $\lambda\mu$-calculus
  \cite{parigot:92} and a denotational semantics \cite{girard:LC}.  An
  important aspect is the computational meaning in terms of
  continuations \cite{thielecke:phd,streicher:reus:98}. There is a
  well explored category theoretical axiomatization
  \cite{selinger:01}, and, of course, a theory of proof nets
  \cite{laurent:99,laurent:03}, which is based on the proof nets for
  $\MELL$ (see Section~\ref{sec:MELL}). Although these proof nets are
  very important from the ``normalization-as-computation'' point of
  view, we will not discuss them here because at the current state of
  the art it is not clear how they can be used to identify proofs in
  various deductive systems for classical logic (sequent calculus,
  resolution, tableaux, \ldots).
\item The second approach considers the perfect symmetry between $\cand$ and
  $\cor$ to be an essential facet of Boolean logic, that cannot be dispensed
  with. Consequently, the axioms of cartesian closed categories and the close
  relation to the $\lambda$-calculus have to be sacrificed. It is much less
  clear than in the first approach, what the category theoretical
  axiomatization~\cite{dosen:petric:coherence-book,fuhrmann:pym:oecm,lam:str:05:freebool,mckinley:SKS-cat,str:medial,lamarche:gap}
  should be, and how a theory of proof nets should look like.  There are now
  two different versions in the literature, one following
  Ideology~\ref{ideo:girard} (Section~\ref{sec:cl-rob}), and the other
  following Ideology~\ref{ideo:coherence} (Sections \ref{sec:cl-simple}
  and~\ref{sec:cl-ext}).
\end{enumerate}


\subsection{Sequent calculus rule based proof nets}\label{sec:cl-rob}

Classical logic can be obtained from multiplicative linear logic by
adding the rules of contraction and weakening. But for obvious reasons
we will here change the notation and use the symbols $\cand$ and
$\cor$ instead of $\ltens$ and $\lpar$. Negation will be denoted by
$\cneg\cdot$, instead of $\cdot\lneg$. In other words, our formulas
are generated by the syntax
\begin{equation}\label{eq:form-CL}
  \cF\grammareq\cA\mid\cneg\cA\mid\cF\cor\cF\mid\cF\cand\cF
\end{equation}
Note that as before, we ignore the units\footnote{Note, that in
classical logic, this does not change the logic because we can pick a
distinguished atom, say $p_0$, which may not appear in the formulas,
and define $\ctrue=p_0\cor\cneg p_0$ and $\cfalse=p_0\cand\cneg p_0$.}
and push negation to the atoms. Here are the inference rules in the
sequent calculus:
\begin{equation}\label{eq:sc-CL}
  \begin{array}{c}
    \vcinf{\idr}{\sqn{\cneg A,A}}{}
    \qquad
    \vciinf{\cutr}{\sqn{\Gamma,\Delta}}{
      \sqn{\Gamma,A}}{\sqn{\cneg A,\Delta}}
    \\[\arrayskip]
    \vcinf{\exr}{\sqn{\Gamma,B,A,\Delta}}{\sqn{\Gamma,A,B,\Delta}}
    \qquad
    \vcinf{\weakr}{\sqn{A,\Gamma}}{\sqn{\Gamma}}
    \qquad
    \vcinf{\conr}{\sqn{A,\Gamma}}{\sqn{A,A,\Gamma}}
    \\[\arrayskip]
    \vcinf{\cor}{\sqn{A\cor B,\Gamma}}{\sqn{A,B,\Gamma}}
    \qquad
    \vciinf{\cand}{\sqn{\Gamma,A\cand B,\Delta}}{
      \sqn{\Gamma,A}} {\sqn{B,\Delta}}
  \end{array}
\end{equation}
Note that there are various different sequent calculus systems for
classical propositional logic, starting with the one by
Gentzen~\cite{gentzen:34}. For a systematic treatment, the reader is
referred to~\cite{troelstra:schwichtenberg:00}. The motivation for
choosing the one in \eqref{eq:sc-CL} is that it allows for the
simplest notion for proof nets according to
Ideology~\ref{ideo:girard}. When we look at
Ideology~\ref{ideo:coherence} and the notion of flow-graphs in the
next sections, it does not matter which sequent system is taken as
starting point.

The rules in \eqref{eq:sc-CL} are essentially the same as the ones in
\eqref{eq:sc-mll} plus the ones for contraction and weakening. In
order to obtain proof nets according to Ideology~\ref{ideo:girard}, we
can therefore proceed as shown in Figures \ref{fig:sc-to-pn-1}
and~\ref{fig:sc-to-pn-exp}.

\begin{exercise}
  Give the translation of the rules in \eqref{eq:sc-CL} into proof
  nets as it is done in Figures \ref{fig:sc-to-pn-1}
  and~\ref{fig:sc-to-pn-exp}.
\end{exercise}

Here is an example of a sequent calculus proof
\proofadjust
\begin{equation}\label{eq:sc-cont-cont}
  \vcddernote{\cutr}{}{\sqn{\cneg b\cand a,\cneg a\cand\cneg b, 
      b\cand a,\cneg a\cand b}}{
    \root{\conr}{\sqn{\cneg b\cand a,\cneg a\cand\cneg b,b}}{
      \rroot{\cand}{\sqn{\cneg b\cand a,\cneg a\cand\cneg b,b,b}}{
	\rroot{\cand}{\sqn{\cneg b\cand a,\cneg a,b}}{
	  \root{\idr}{\sqn{\cneg b,b}}{
	    \leaf{}}}{
	  \root{\idr}{\sqn{a,\cneg a}}{
	    \leaf{}}}}{
	\root{\idr}{\sqn{\cneg b,b}}{
	  \leaf{}}}}}{
    \root{\conr}{\sqn{\cneg b, b\cand a,\cneg a\cand b}}{
      \rroot{\cand}{\sqn{\cneg b,\cneg b,\cneg b\cand a,\cneg a\cand\cneg b}}{
	\root{\idr}{\sqn{\cneg b,b}}{
	  \leaf{}}}{
	\rroot{\cand}{\sqn{\cneg b,a,\cneg a\cand b}}{
	  \root{\idr}{\sqn{a,\cneg a}}{
	    \leaf{}}}{
	  \root{\idr}{\sqn{\cneg b,b}}{
	    \leaf{}}}}}}
\end{equation}
and its translation into a proof net:
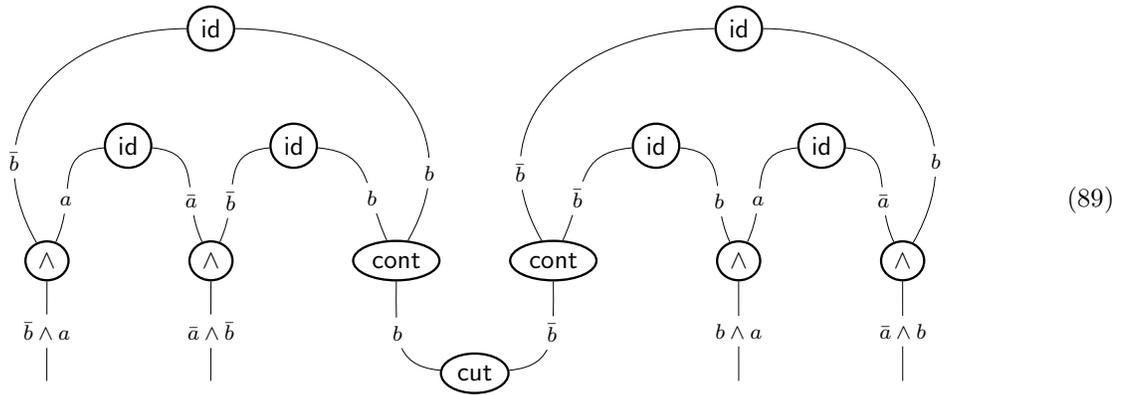
\begin{equation}\label{eq:pn-cont-cont}
  \vcgpn{
    &&\gpnid{1}&&&&&&\gpnid{4}\\
    & \gpnid2&&\gpnid3&&&&\gpnid5&&\gpnid6&\\
    \gpnand{1}&&\gpnand{2}&&\gpnnode1{\conr} &&\gpnnode2{\conr}
    &&\gpnand{3}&&\gpnand{4}\\
    \gpnout1 &&\gpnout2 &&&\clapm{\gpncut1}&&&\gpnout3 &&\gpnout4
  }{
    \gpnmycurve{cand1}{id2}{65}{185}{.3}{a}
    \gpnmycurve{cand2}{id2}{115}{-5}{.3}{\cneg a}
    \gpnmycurve{cand2}{id3}{65}{185}{.3}{\cneg b}
    \gpnmycurve{gpn1}{id3}{115}{-5}{.3}{b}
    \gpnmycurve{cand1}{id1}{120}{180}{.25}{\cneg b}
    \gpnmycurve{gpn1}{id1}{60}{0}{.2}{b}
    \gpnmycurve{cand3}{id6}{65}{185}{.3}{a}
    \gpnmycurve{cand4}{id6}{115}{-5}{.3}{\cneg a}
    \gpnmycurve{cand3}{id5}{115}{-5}{.3}{b}
    \gpnmycurve{gpn2}{id5}{65}{185}{.3}{\cneg b}
    \gpnmycurve{cand4}{id4}{60}{0}{.25}{b}
    \gpnmycurve{gpn2}{id4}{120}{180}{.2}{\cneg b}
    \cutleftline{gpn1}{cut1}{b}
    \cutrightline{gpn2}{cut1}{\cneg b}
    \treeline{cand1}{out1}{\cneg b\cand a}
    \treeline{cand2}{out2}{\cneg a\cand \cneg b}
    \treeline{cand3}{out3}{b\cand a}
    \treeline{cand4}{out4}{\cneg a\cand b}
  }  
\end{equation}

That this can easily be done has already been noted by Girard in the
appendix of \cite{girard:LC}, where he also explained the problems
with this approach that we will discuss below. In~\cite{robinson:03},
Robinson carries out the details. He uses a two-sided version of the
set of rules in~\eqref{eq:sc-CL}, but with what we learned in
Section~\ref{sec:twosided}, it is an easy exercise to translate back
and forth between the one- and two-sided version.

Robinson also proves the soundness and completeness of the switching
correctness criterion (see Definition~\ref{def:S-correct} and
Theorem~\ref{thm:S-correct}). For the acyclicity condition this is not
surprising since the rules in~\eqref{eq:sc-CL} are the same as
in~\eqref{eq:sc-mll} (recall that contraction behaves as $\lpar$). The
connectedness condition is obtained simply by attaching a weakening
somewhere, similarly as we did it with $\lbot$ in
Section~\ref{sec:units}. The price to pay is that now certain proofs
are distinguished that should be identified according the
rule-permutability-argument. To see a very simple example, consider
the following three sequent calculus proofs (we systematically omit
the exchange rule).
\proofadjust
\begin{equation}\label{eq:scweak}
  \vcddernote{\cand}{}{\sqn{c,\cneg a,a\cand b,\cneg b}}{
    \root{\weakr}{\sqn{c,\cneg a,a}}{
      \root{\idr}{\sqn{\cneg a,a}}{
	\leaf{}}}}{
    \root{\idr}{\sqn{b,\cneg b}}{
      \leaf{}}}
  \qquad
  \vcdernote{\weakr}{}{\sqn{c,\cneg a,a\cand b,\cneg b}}{
    \rroot{\cand}{\sqn{\cneg a,a\cand b,\cneg b}}{
      \root{\idr}{\sqn{\cneg a,a}}{
	\leaf{}}}{
      \root{\idr}{\sqn{b,\cneg b}}{
	\leaf{}}}}
  \qquad
  \vcddernote{\cand}{}{\sqn{c,\cneg a,a\cand b,\cneg b}}{
    \root{\idr}{\sqn{\cneg a,a}}{
      \leaf{}}}{
    \root{\weakr}{\sqn{c,b,\cneg b}}{
      \root{\idr}{\sqn{b,\cneg b}}{
	\leaf{}}}}
\end{equation}
They differ from each other only via some trivial rule permutation,
and should therefore be identified. But according to \cite{robinson:03}
they can be translated into five different proof nets. Two of them are
shown below:
\begin{equation}
  \vcgpn{
    \quad&&\gpnid1&&\gpnid2\\
    &\qlapm{\gpnnode1{\weakr}}&&\qqqlapm{\gpnand1}\\
    \gpnout1&&\gpnout2&\gpnout3&&&\gpnout4
  }{
    \gpnmycurve{gpn1}{id1}{90}{185}{.3}{\cneg a}
    \gpnmycurve{cand1}{id1}{115}{-5}{.3}{a}
    \gpnmycurve{cand1}{id2}{65}{185}{.3}{b}
    \def\gpncurveheight{.5}
    \gpnmycurve{out4}{id2}{90}{-5}{.3}{\cneg b}
    \treeline{gpn1}{out1}{c}
    \treeline{gpn1}{out2}{\cneg a}
    \treeline{cand1}{out3}{a\cand b}
  }
\end{equation}
and
\begin{equation}
  \vcgpn{
    &&&&\gpnid2\\
    &\gpnid1&&\clapm{\gpnnode1{\weakr}}\\
    &&\qqlapm{\gpnand1}\\
    \gpnout1&&\gpnout2&&\gpnout3&&\gpnout4
  }{
    \gpnmycurve{cand1}{id1}{120}{-5}{.3}{a}
    \gpnmycurve{gpn1}{id2}{90}{185}{.3}{b}
    \def\gpncurveheight{.5}
    \gpnmycurve{out4}{id2}{90}{-5}{.3}{\cneg b}
    \def\gpncurveheight{.4}
    \gpnmycurve{out1}{id1}{90}{185}{.3}{\cneg a}
    \treeline{gpn1}{out3}{c}
    \treeline{gpn1}{cand1}{b}
    \treeline{cand1}{out2}{a\cand b}
  }
\end{equation}
\begin{exercise}
  What are the other three proof nets corresponding to~\eqref{eq:scweak}?
  Compare the nets with the ones in Figure~\ref{fig:MLL-unit}.
\end{exercise}

Let us now come to cut elimination. We have already seen in the introductory
part of Section~\ref{sec:cl}, that there is a problem with weakening.  The
short discussion in Section~\ref{sec:MELL} suggests that there might also be a
problem with contraction. Observe that in Section~\ref{sec:MELL}, the
contraction rule could appear only on one side of the cut, never at both sides
at the same time (because only $\lwn$-formulas could be contracted, never
$\loc$-formulas). This is the reason why in the end we get confluence of cut
elimination for $\ufMELL$ proof nets. However, for classical logic the
situation is different. Contraction can appear on both sides of the cut, as it
is shown in the example in \eqref{eq:sc-cont-cont}
and~\eqref{eq:pn-cont-cont}. For typesetting reasons, let us use the more
compact notation (as we also did in Section~\ref{sec:correctness}):
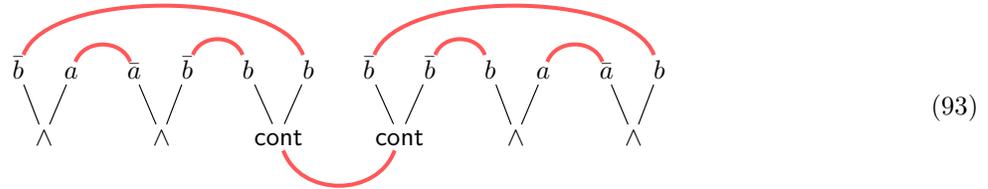
\begin{equation}\label{eq:pn-cont-cont-short}
   \vcnpn{%
     \\
     \cneg{\mb1}&&\na1&&\cneg{\na2}&&\cneg{\mb2}&&
     \mb3&&\mb4&&\cneg{\mb5}&&\cneg{\mb6}&&
     \mb7&&\na3&&\cneg{\na4}&&\mb8\\
     &\ncand1&&\qlapm{}&&\ncand2&&\qlapm{}&&\nconr1&&\qlapm{}&&
     \nconr2&&\qlapm{}&&\ncand3&&\qlapm{}&&\ncand4
     \\
   }{%
     \ncline{b1}{cand1}
     \ncline{a1}{cand1}
     \ncline{b2}{cand2}
     \ncline{a2}{cand2}
     \ncline{b3}{conr1}
     \ncline{b4}{conr1}
     \ncline{b5}{conr2}
     \ncline{b6}{conr2}
     \ncline{b7}{cand3}
     \ncline{a3}{cand3}
     \ncline{b8}{cand4}
     \ncline{a4}{cand4}
   }{%
     \duline{conr1}{conr2}
     \udline{a1}{a2}
     \udline{b2}{b3}
     \udline{a3}{a4}
     \udline{b6}{b7}
     \def\loopvecheight{.5}    
     \udline{b1}{b4}
     \udline{b5}{b8}
   }
\end{equation}
Let us stress that \eqref{eq:pn-cont-cont} and~\eqref{eq:pn-cont-cont-short}
are only different notations for the \emph{same} proof net. Note that we have
here an example for the general case in~\eqref{eq:cont-cont}.
If we want to eliminate the cut from~\eqref{eq:pn-cont-cont-short}, we
have to make a nondeterministic choice, which subproof we duplicate.
As outcome we get either
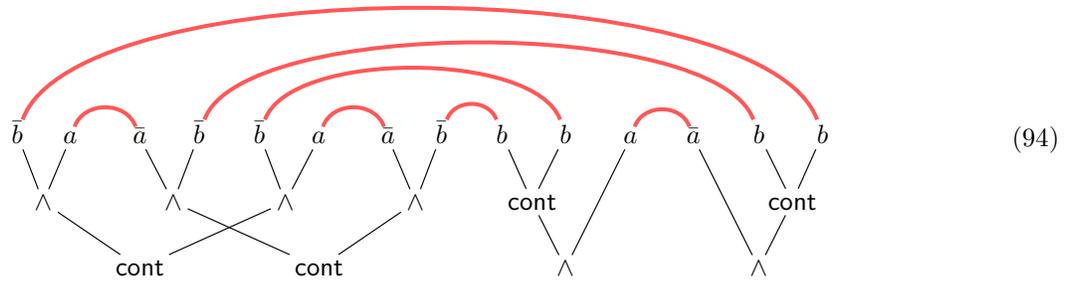
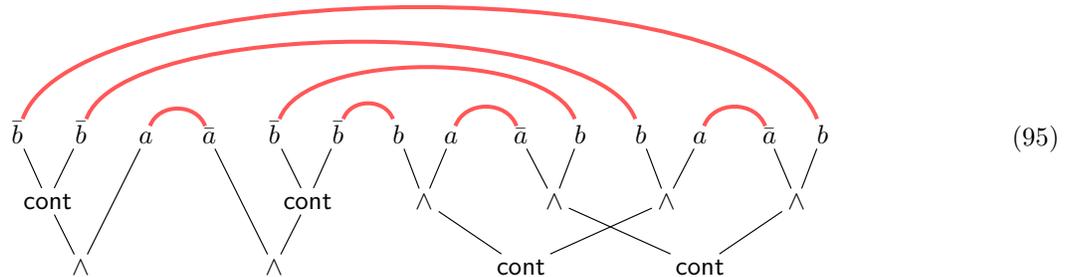
\begin{equation}\label{eq:pn-cont-out1}
   \vcnpn{%
     \\ \\
     \cneg{\mb1}&&\na1&&\cneg{\na2}&&\cneg{\mb2}&&
     \cneg{\mb3}&&\na3&&\cneg{\na4}&&\cneg{\mb4}&&  
     \mb5&&\mb6&&\na5&&\cneg{\na6}&&\mb7&&\mb8\\
     &\ncand1&&\qlapm{}&&\ncand2&&\qlapm{}&&\ncand6&&\qlapm{}&&\ncand7
     &&\qlapm{}&&\nconr1&&\qlapm{}&&\qlapm{}&&\qlapm{}&&\nconr2\\
     &&&&\nconr3&&&&&&\nconr4&&&&&&&&\ncand3&&&&&&\ncand4
   }{%
     \ncline{b1}{cand1}
     \ncline{a1}{cand1}
     \ncline{b2}{cand2}
     \ncline{a2}{cand2}
     \ncline{b3}{cand6}
     \ncline{a3}{cand6}
     \ncline{b4}{cand7}
     \ncline{a4}{cand7}
     \ncline{cand1}{conr3}
     \ncline{cand6}{conr3}
     \ncline{cand2}{conr4}
     \ncline{cand7}{conr4}
     \ncline{b5}{conr1}
     \ncline{b6}{conr1}
     \ncline{b7}{conr2}
     \ncline{b8}{conr2}
     \ncline{conr1}{cand3}
     \ncline{a5}{cand3}
     \ncline{conr2}{cand4}
     \ncline{a6}{cand4}
   }{%
     \udline{a1}{a2}
     \udline{a3}{a4}
     \udline{a5}{a6}
     \udline{b4}{b5}
     \def\loopvecheight{.5}    
     \udline{b3}{b6}
     \def\loopvecheight{.4}    
     \udline{b2}{b7}
     \udline{b1}{b8}
   }
\end{equation}
or
\begin{equation}\label{eq:pn-cont-out2}
   \vcnpn{%
     \\ \\
     \cneg{\mb5}&&\cneg{\mb6}&&\na5&&\cneg{\na6}&&\cneg{\mb7}&&\cneg{\mb8}&&
     \mb1&&\na1&&\cneg{\na2}&&\mb2&&\mb3&&\na3&&\cneg{\na4}&&\mb4\\
     &\nconr1&&\qlapm{}&&\qlapm{}&&\qlapm{}&&\nconr2&
     &\qlapm{}&
     &\ncand1&&\qlapm{}&&\ncand2&&\qlapm{}&&\ncand6&&\qlapm{}&&\ncand7\\
     &&\ncand3&&&&&&\ncand4&&&&&&&&\nconr3&&&&&&\nconr4
   }{%
     \ncline{b1}{cand1}
     \ncline{a1}{cand1}
     \ncline{b2}{cand2}
     \ncline{a2}{cand2}
     \ncline{b3}{cand6}
     \ncline{a3}{cand6}
     \ncline{b4}{cand7}
     \ncline{a4}{cand7}
     \ncline{cand1}{conr3}
     \ncline{cand6}{conr3}
     \ncline{cand2}{conr4}
     \ncline{cand7}{conr4}
     \ncline{b5}{conr1}
     \ncline{b6}{conr1}
     \ncline{b7}{conr2}
     \ncline{b8}{conr2}
     \ncline{conr1}{cand3}
     \ncline{a5}{cand3}
     \ncline{conr2}{cand4}
     \ncline{a6}{cand4}
   }{%
     \udline{a1}{a2}
     \udline{a3}{a4}
     \udline{a5}{a6}
     \udline{b8}{b1}
     \def\loopvecheight{.5}    
     \udline{b7}{b2}
     \def\loopvecheight{.4}    
     \udline{b6}{b3}
     \udline{b5}{b4}
   }
\end{equation}

In the appendix of \cite{girard:LC}, Girard argues that for this
reason it is impossible to have a confluent notion of cut elimination
for proof nets for classical logic. Of course, his argumentation is
valid only for proof nets following Ideology~\ref{ideo:girard}.

\begin{exercise}\label{ex:pn-sc-rob}
  Show that \eqref{eq:pn-cont-out1} and~\eqref{eq:pn-cont-out2} obey
  the switching criterion, and give the sequent proofs corresponding
  to \eqref{eq:pn-cont-out1} and~\eqref{eq:pn-cont-out2}.
\end{exercise}

Although there is no confluent cut elimination, F\"uhrmann and
Pym~\cite{fuhrmann:pym:oecm} managed to give a category theoretical
axiomatization for the proof identifications made by these proof nets. This
could be done because the authors dropped the ``equation''
$$
\mbox{cut elimination}
\quad=\quad
\mbox{arrow composition in the category of proofs}
$$
and added an order enrichment instead. They defined for two proofs $f,g\colon
A\to B$ that $f\preccurlyeq g$ iff $f$ reduces to $g$ via cut
elimination. See~\cite{mckinley:SKS-cat} for relating this to the
calculus of structures.

\subsection{Flow graph based proof nets (simple version)}
\label{sec:cl-simple}

\begin{figure}[!t]
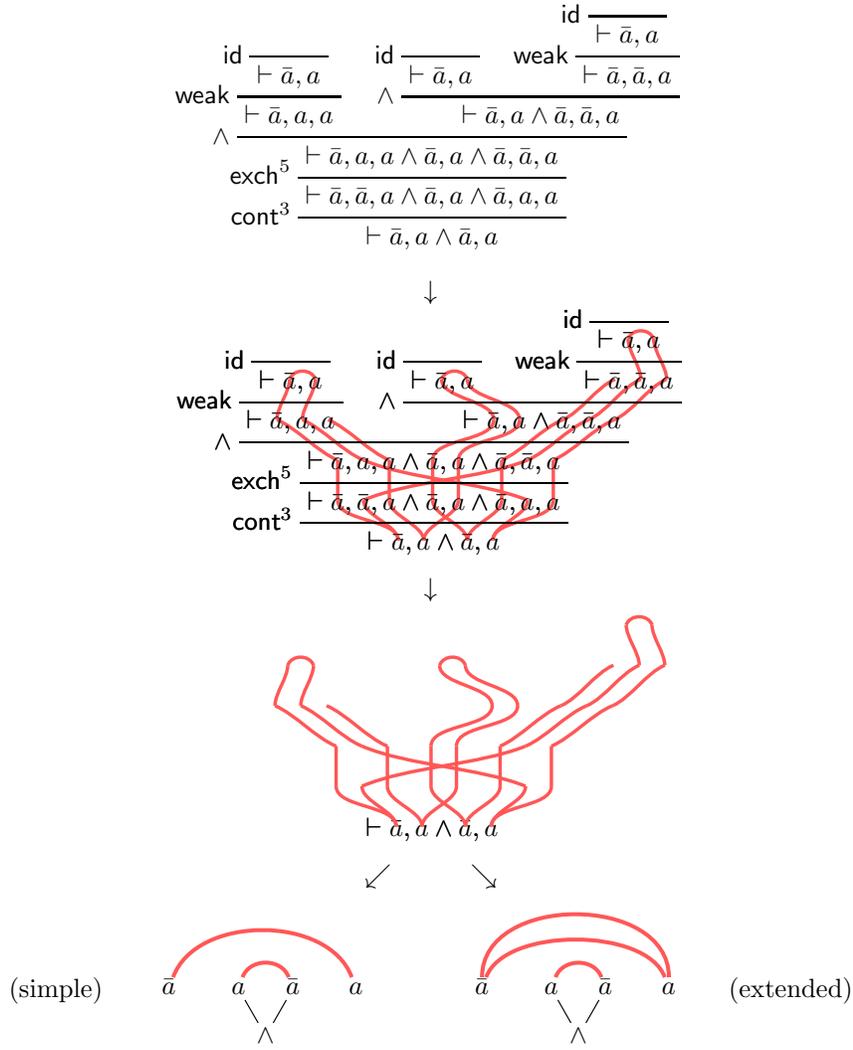

  \def\arrayskip{2ex}
  \begin{center}
    \vskip-2ex
    $$
    \begin{array}{c}
      \exaclflow
      \\[\arrayskip]
      \downarrow
      \\[-2ex]
      \tpathderivation{\exaclflow}{\pathexaclflow}
      \\[\arrayskip]
      \downarrow
      \\[-3ex]
      \tpathonlyderivation{\exaclflow}{\pathexaclflow}
      \\[\arrayskip]
      \swarrow \qqquad \searrow
      \\[4ex]
      \mbox{(simple)}\qquad
      \pathpnclflowsimple \qqqquad  \pathpnclflow
      \qquad\mbox{(extended)}
    \end{array}
    $$
    \caption{From sequent calculus to classical logic proof nets via
    flow graphs}
    \label{fig:exascflow}
  \end{center}
\end{figure}

In the previous section we have seen what happens if we naively carry
the approach of Section~\ref{sec:sc-to-pn-1} to classical logic. Now
we are going to see what happens if we naively carry the approach of
Section~\ref{sec:sc-to-pn-2} to classical logic. We can define a
pre-proof net as in Definition~\ref{def:pre-pn}, but now the identity
links do not provide a perfect matching for the set of leaves of the
sequent forest. It can happen that some atoms have no mate, i.e., live
celibate, and that some atoms have more than one mate, i.e., live
polygamous. It could even happen that there are two or more identity
links between a pair of atoms, as the example (on the right) in
Figure~\ref{fig:exascflow} shows.  This example also shows that we can
now completely forget about the correctness criteria that we have seen
in Section~\ref{sec:correctness}.  But most importantly, this example
shows that for classical logic it makes a huge difference whether we
follow Ideology~\ref{ideo:girard} or Ideology~\ref{ideo:coherence} to
obtain proof nets. This also means, that the term ``proof net'' should
be used with care, because it is not necessarily clear what a proof
net for classical logic actually is.

Let us, for the time being, restrict ourselves to those
nets that have \emph{at most one} identity link between any
pair of atoms, and call them \emph{simple prenets}:

\begin{definition}
  A \emph{simple prenet} is a sequent forest $\Gamma$, possibly with
  cuts, together with a symmetric, irreflexive binary relation $P$ on
  its set of leaves, such that whenever two atom occurrences are
  related by $P$ then they must be dual to each other. We will use the
  notation $\prfnet{P}{\Gamma}$.
\end{definition}

As in Section~\ref{sec:correctness}, we have to treat cuts as special
formulas (see page~\pageref{def:pre-pn}). 

In the left of Figure~\ref{fig:exascflow} we have shown an example how
to translate a sequent calculus proof into a simple prenet. We draw
the flow-graph as we did in Section~\ref{sec:sc-to-pn-2} and forget
how often a pair of atoms is connected to each other.
Here is an example of a simple prenet with cut. It is the one obtained
from~\eqref{eq:sc-cont-cont}:
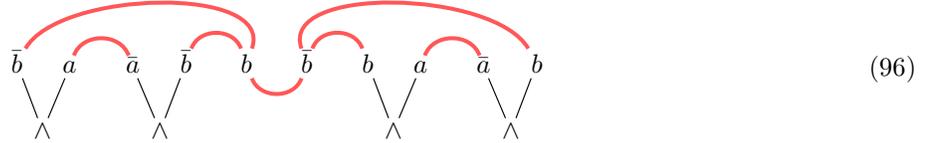
\begin{equation}\label{eq:simple-pn-cont-cont}
   \vcnpn{%
     \\
     \cneg{\mb1}&&\na1&&\cneg{\na2}&&\cneg{\mb2}&&
     \mb3&&\cneg{\mb5}&&
     \mb7&&\na3&&\cneg{\na4}&&\mb8\\
     &\ncand1&&\qlapm{}&&\ncand2&&\qlapm{}&&\qlapm{}&&
     \qlapm{}&&\ncand3&&\qlapm{}&&\ncand4
   }{%
     \ncline{b1}{cand1}
     \ncline{a1}{cand1}
     \ncline{b2}{cand2}
     \ncline{a2}{cand2}
     \ncline{b7}{cand3}
     \ncline{a3}{cand3}
     \ncline{b8}{cand4}
     \ncline{a4}{cand4}
   }{%
     \duline{b3}{b5}
     \udline{a1}{a2}
     \udline{b2}{b3}
     \udline{a3}{a4}
     \udline{b5}{b7}
     \lineanglesheight{b1}{b3}{60}{70}{.6}
     \lineanglesheight{b5}{b8}{110}{120}{.6}
   }
\end{equation}

We will call a simple prenet \emph{sequentializable} if it can be
obtained via the
flow-graph method from the sequent calculus system in~\eqref{eq:sc-CL}
extended by the $\mixr$-rule
$$
\vciinf{\mixr}{\sqn{\Gamma,\Delta}}{\sqn{\Gamma}}{\sqn{\Delta}}
$$

\begin{exercise}\label{ex:sc-simple-pn}
  Draw the simple prenets that are obtained from the sequent calculus proofs
  corresponding to the proof nets in \eqref{eq:pn-cont-out1}
  and~\eqref{eq:pn-cont-out2}. Hint:  You can do this exercise without doing
  Exercise~\ref{ex:pn-sc-rob}.
\end{exercise}

We are now going to define a correctness criterion
for simple prenets.

\begin{definition}\label{def:pruning}
  Let $\Gamma$ be a sequent. A \emph{conjunctive pruning} of $\Gamma$
  is the subforest obtained $\Gamma$ by deleting one of the two
  immediate subformulas for every $\cand$ and for every cut.  Let
  $\pi=\prfnet{P}{\Gamma}$ be a simple prenet. we call the prenet
  $\pi'=\prfnet{P'}{\Gamma'}$ a \emph{conjunctive pruning} of $\pi$ if
  $\Gamma'$ is a conjunctive pruning of $\Gamma$ and $P'$ is the
  restriction of $P$ to (the set of leaves of)
  $\Gamma'$.\footnote{Strictly speaking, $\pi'$ is not a simple prenet
  because $\Gamma'$ is not a sequent because every $\cand$ in
  $\Gamma'$ is unary. For this reason $\pi'$ is not a
  subprenet in the sense of Definition~\ref{def:subnet}.}
\end{definition}

Here are two examples for conjunctive prunings of
\eqref{eq:simple-pn-cont-cont}:
$$
   \vcnpn{%
     \\
     \prune{\cneg{\mb1}}&&\na1&&\cneg{\na2}&&\prune{\cneg{\mb2}}&&
     \prune{\mb3}&&\cneg{\mb5}&&
     \mb7&&\prune{\na3}&&\cneg{\na4}&&\prune{\mb8}\\
     &\ncand1&&\qlapm{}&&\ncand2&&\qlapm{}&&\qlapm{}&&
     \qlapm{}&&\ncand3&&\qlapm{}&&\ncand4
   }{%
     \prune{
       \ncline{b1}{cand1}
       \ncline{b2}{cand2}
       \ncline{a3}{cand3}
       \ncline{b8}{cand4}
       }
     \ncline{a1}{cand1}
     \ncline{a2}{cand2}
     \ncline{b7}{cand3}
     \ncline{a4}{cand4}
   }{%
     \prune{
       \duline{b3}{b5}
       \lineanglesheight{b1}{b3}{60}{70}{.6}
       \lineanglesheight{b5}{b8}{110}{120}{.6}
       \udline{b2}{b3}
       \udline{a3}{a4}
       }
     \udline{a1}{a2}
     \udline{b5}{b7}
   }
$$
and
$$
   \vcnpn{%
     \\
     \cneg{\mb1}&&\prune{\na1}&&\prune{\cneg{\na2}}&&\cneg{\mb2}&&
     \mb3&&\prune{\cneg{\mb5}}&&
     \mb7&&\prune{\na3}&&\cneg{\na4}&&\prune{\mb8}\\
     &\ncand1&&\qlapm{}&&\ncand2&&\qlapm{}&&\qlapm{}&&
     \qlapm{}&&\ncand3&&\qlapm{}&&\ncand4
   }{%
     \prune{
       \ncline{a1}{cand1}
       \ncline{a2}{cand2}
       \ncline{a3}{cand3}
       \ncline{b8}{cand4}
       }
     \ncline{b1}{cand1}
     \ncline{b2}{cand2}
     \ncline{b7}{cand3}
     \ncline{a4}{cand4}
   }{%
     \prune{
       \duline{b3}{b5}
       \udline{a1}{a2}
       \udline{a3}{a4}
       \udline{b5}{b7}
       \lineanglesheight{b5}{b8}{110}{120}{.6}
       }
     \udline{b2}{b3}
     \lineanglesheight{b1}{b3}{60}{70}{.6}
   }
$$

\begin{exercise}
  How many others are there?
\end{exercise}

\begin{definition}
  A simple prenet $\prfnet{P}{\Gamma}$ is called \emph{correct} (or,
  \emph{obeys the pruning condition}, if every conjunctive pruning
  $\prfnet{P'}{\Gamma'}$ of $\pi$ contains at least one identity link,
  i.e., $P'$ is not empty.
\end{definition}

Clearly, the example in \eqref{eq:simple-pn-cont-cont} is
correct. Here is an example, which is not correct, because there is a
pruning, in which all links disappear:
$$
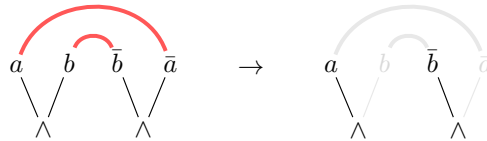
\vcnpn{
  \\
  \na1&\pa&\mb1&\pa&\cneg{\mb2}&\pa&\nna2\\
  &\ncand1&&&&\ncand2
}{
  \ncline{cand1}{a1}
  \ncline{cand1}{b1}
  \ncline{cand2}{a2}
  \ncline{cand2}{b2}
}{
  \udline{a1}{a2}
  \udline{b1}{b2}
}
\qquadto
\vcnpn{
  \\
  \na1&\pa&\prune{\mb1}&\pa&\cneg{\mb2}&\pa&\prune{\nna2}\\
  &\ncand1&&&&\ncand2
}{
  \ncline{cand1}{a1}
  \prune{
    \ncline{cand1}{b1}
    \ncline{cand2}{a2}}
  \ncline{cand2}{b2}
}{
  \prune{
    \udline{a1}{a2}
    \udline{b1}{b2}
  }
}
$$

Interestingly, also in the correctness criterion given by Hughes and
van Glabbeek in~\cite{hughes:glabbeek:03} for $\ufMALL$ (see
Section~\ref{sec:MALL}) the pruning condition plays a role.

\begin{theorem}
  A simple prenet $\pi$ is correct, if and only if it is sequentializable.
\end{theorem}

\begin{proof}
  First, observe that the $\idr$-rule produces correct prenets and that all
  rules in~\eqref{eq:sc-CL}, including $\mixr$, preserve correctness. For the
  rules $\cor$, $\exr$, $\weakr$, $\conr$, and $\mixr$, this is obvious. Let
  us show it here for the $\cand$-rule. Let $\pi_1$ and $\pi_2$ be the simple
  prenets obtained from\proofadjust
  $$
  \sqnsmallderi{\Gamma,A}{\Pi_1}
  \qquand
  \sqnsmallderi{B,\Delta}{\Pi_2}
  $$
  respectively. Let us assume by induction hypothesis that they are correct,
  and let $\pi$ be the simple prenet obtained from\proofadjust
  $$
  \vcddernote{\cand}{}{\sqn{\Gamma,A\cand B,\Delta}}{
    \leaf{\sqnsmallderi{\Gamma,A}{\Pi_1}}}{
    \leaf{\sqnsmallderi{B,\Delta}{\Pi_2}}}
  $$
  Let $\pi'$ be a conjunctive pruning of $\pi$. If it removes $B$ from $A\cand
  B$, then $\pi'$ contains a link because $\pi_1$ is correct, and if $\pi'$
  removes $A$, then it must contain a link because $\pi_2$ is correct. For the
  $\cutr$-rule it is similar.

  Conversely, let $\pi=\prfnet{P}{\Delta}$ be a correct prenet. we will proceed
  by induction on the size of $\Delta$ (that is, the number of $\cand$,
  $\cor$, atoms, and cuts appearing in it) to construct a a corresponding
  sequent proof using the rules in~\eqref{eq:sc-CL} plus $\mixr$.
  \begin{itemize}
  \item If $\Delta$ contains a formula $A\cor B$, then we can apply the
    $\cor$-rule and proceed by induction hypothesis.
  \item If $\Delta$ contains a formula $A\cand B$, i.e., $\Delta=\Gamma,A\cand
    B$, then we can form three correct simple prenets
    $\pi_1=\prfnet{P'}{\Gamma,A}$, and $\pi_2=\prfnet{P''}{\Gamma,B}$, and
    $\pi_3=\prfnet{P}{\Gamma,A,B}$, where $P'$ and $P''$ are the restrictions
    of $P$ to $\Gamma,A$ and $\Gamma,B$, respectively. Since $\pi_1$, and
    $\pi_2$, and $\pi_3$ are all correct, we get three sequent calculus
    proofs\proofadjust
    $$
    \sqnsmallderi{\Gamma,A}{\Pi_1}
    \qqqquad
    \sqnsmallderi{\Gamma,B}{\Pi_2}
    \qqqquad
    \sqnsmallderi{\Gamma,A,B}{\Pi_3}
    $$ from which we can form the following proof (we omit the instances of
    $\exr$):\proofadjust
    \begin{equation}\label{eq:seq-cand}
      \vcdernote{\conr}{}{\sqn{\Gamma,A\cand B}}{
	\root{\conr}{\vdots}{
	  \rroot{\cand}{\sqn{\Gamma,\Gamma,\Gamma,A\cand B,A\cand B}}{
	    \rroot{\cand}{\sqn{\Gamma,\Gamma,A\cand B,A}}{
	      \leaf{\sqnsmallderi{\Gamma,A}{\Pi_1}}}{
	      \leaf{\sqnsmallderi{\Gamma,A,B}{\Pi_3}}}}{
	    \leaf{\sqnsmallderi{\Gamma,B}{\Pi_2}}}}}
    \end{equation}
    which translates into $\pi$. Let us make three remarks about that case:
    \begin{itemize}
    \item Note that we made crucial use of the fact that we forget how often
      an identity link is used inside the proof.
    \item The proof $\Pi_3$ is needed for keeping the links that cross $A\cand
      B$. If there is in $\pi$ no link between an atom in $A$ and an atom in
      $B$, then we do not need $\Pi_3$ and could replace \eqref{eq:seq-cand}
      by\proofadjust
      $$
      \vcdernote{\conr}{}{\sqn{\Gamma,A\cand B}}{
	\root{\conr}{\vdots}{
	  \rroot{\cand}{\sqn{\Gamma,\Gamma,A\cand B}}{
	    \leaf{\sqnsmallderi{\Gamma,A}{\Pi_1}}}{
	    \leaf{\sqnsmallderi{\Gamma,B}{\Pi_2}}}}}
      $$
    \item For dealing with cuts we proceed similarly. But to be formally
      precise, we need to allow to apply the contraction rule not only to
      formulas in the sequent, but also to cuts.
    \end{itemize}
  \item If $\Delta=\Gamma,A$ such that no link is coming out of $A$, then we
    apply the weakening rule and proceed by induction hypothesis.
  \item The only remaining case is where all formulas in $\Delta$ are
    atoms. Then our sequent calculus proof is obtained by a sufficient number
    of instances of $\idr$, $\conr$, $\exr$, and $\mixr$.
    \qed
  \end{itemize}
\end{proof}

We will use the term \emph{simple proof net} for the correct (i.e.,
sequentializable) simple prenets. They have been introduced
in~\cite{lam:str:05:naming}. However, the idea of looking at paired
atom occurrences in classical logic is much older. In
\cite{andrews:76} Andrews applied the ``flow-graph'' method to
resolution proofs and called the result \emph{matings}. He had, in
fact, exactly the same correctness criterion as we have seen
above. Independently, Bibel~\cite{bibel:81} investigated the same
objects and called them \emph{connection proofs}. Also the term
{matrix proofs} is used because formulas have been written in form of
matrices. 

However, both authors, Andrews and Bibel, did not allow the linking of two
atoms that are connected in the tree by a conjunction\footnote{That means they
are connected by a red edge in the relation web, see
Definition~\ref{def:web}.}. From the viewpoint of provability/correctness, this
perfectly makes sense. Note that such links (we have an example in
Figure~\ref{fig:exascflow}) disappear in \emph{every} conjunctive pruning. So,
why having them in the first place?

The reason is cut elimination, i.e., the composition of proofs. We
define it in the same way as for $\ufMLL$ in
Section~\ref{sec:cutelim}. In the case of a cut on a compound formula,
we do the obvious thing:
\begin{equation}\label{eq:cutred-comp}
  \vcnpn{
    \nA1&\pa&\nB1&\pa&\nnB2&\pa&\nnA2\\
    &\ncand1&&&&\ncor1\\
  }{
    \ncline{A1}{cand1}
    \ncline{B1}{cand1}
    \ncline{A2}{cor1}
    \ncline{B2}{cor1}
  }{
    \duline{cand1}{cor1}
  }
  \qqquadto
  \vcnpn{
    \nA1&\pa&\nB1&\pa&\nnB2&\pa&\nnA2\\
  }{ }{
    \duline{A1}{A2}
    \duline{B1}{B2}
  }
\end{equation}
To see that this preserves correctness, note that a conjunctive pruning of the
cut on the left yields either $A$ or $B$ or $\cneg B\cor\cneg A$, and a
conjunctive pruning of the cut on the right yields either $A,B$ or $A,\cneg B$
or $\cneg A,B$ or $\cneg A,\cneg B$. Hence, every conjunctive pruning of the
reduced simple prenet contains a conjunctive pruning of the non-reduced simple
prenet. Therefore it most contain at least one link.  

For the cut reduction on atomic cuts, we have to be careful, since
the atoms can be connected to many other atoms (or no other
atoms). Instead of simply having:
$$
\vcnpn{
  \\[-2ex]
  \nna1&\pa&\na2&\pa&\nna3&\pa&\na4
  \\[-2ex]
}{}{
  \udline{a1}{a2}
  \udline{a3}{a4}
  \duline{a2}{a3}
}
\qqquadto
\vcnpn{
  \\[-2ex]
  \nna1&\pa&\na2
  \\[-2ex]
}{}{
  \udline{a1}{a2}
}
$$
the reduction looks as follows:
$$
\vcnpn{
  \\
  \nna1&\pa&\nna2&\pa&\cdots&\pa&\nna3&\pa&
  \na4&\pa&\nna5&\pa&
  \na6&\pa&\cdots&\pa&\na7
  \\[-2ex]
}{}{
  \lineanglesheight{a1}{a4}{60}{80}{.7}
  \lineanglesheight{a2}{a4}{60}{100}{.6}
  \lineanglesheight{a3}{a4}{70}{120}{.7}
  \lineanglesheight{a5}{a6}{70}{110}{.7}
  \lineanglesheight{a5}{a7}{90}{120}{.65}
  \duline{a4}{a5}
}
\qquadto
\vcnpn{
  \\
  \nna1&\pa&\nna2&\pa&\cdots&\pa&\nna3&\pa&
  \na6&\pa&\cdots&\pa&\na7
  \\[-2ex]
}{}{
  \lineanglesheight{a1}{a6}{70}{80}{.7}
  \lineanglesheight{a2}{a6}{70}{100}{.6}
  \lineanglesheight{a3}{a6}{70}{120}{.7}
  \lineanglesheight{a1}{a7}{90}{80}{.7}
  \lineanglesheight{a2}{a7}{90}{100}{.6}
  \lineanglesheight{a3}{a7}{90}{120}{.7}
}
$$
If one of the two cut atoms is celibate, no link remains:
$$
\vcnpn{
  \\
  \nna1&\pa&\nna2&\pa&\cdots&\pa&\nna3&\pa&
  \na4&\pa&\nna5
  \\[-2ex]
}{}{
  \lineanglesheight{a1}{a4}{60}{80}{.7}
  \lineanglesheight{a2}{a4}{60}{100}{.6}
  \lineanglesheight{a3}{a4}{70}{120}{.7}
  \duline{a4}{a5}
}
\qquadto
\vcnpn{
  \\
  \nna1&\pa&\nna2&\pa&\cdots&\pa&\nna3
  \\[-2ex]
}{}{}
$$
If the two cut atoms are linked together, then this link is ignored
in the reduction (and, of course, removed with the cut):
$$
\vcnpn{
  \\
  \nna1&\pa&\nna2&\pa&\cdots&\pa&\nna3&\pa&
  \na4&\pa&\nna5&\pa&
  \na6&\pa&\cdots&\pa&\na7
  \\[-2ex]
}{}{
  \lineanglesheight{a1}{a4}{60}{80}{.7}
  \lineanglesheight{a2}{a4}{60}{100}{.6}
  \lineanglesheight{a3}{a4}{70}{120}{.7}
  \lineanglesheight{a4}{a5}{60}{110}{.7}
  \lineanglesheight{a5}{a6}{70}{110}{.7}
  \lineanglesheight{a5}{a7}{90}{120}{.65}
  \duline{a4}{a5}
}
\qquadto
\vcnpn{
  \\
  \nna1&\pa&\nna2&\pa&\cdots&\pa&\nna3&\pa&
  \na6&\pa&\cdots&\pa&\na7
  \\[-2ex]
}{}{
  \lineanglesheight{a1}{a6}{70}{80}{.7}
  \lineanglesheight{a2}{a6}{70}{100}{.6}
  \lineanglesheight{a3}{a6}{70}{120}{.7}
  \lineanglesheight{a1}{a7}{90}{80}{.7}
  \lineanglesheight{a2}{a7}{90}{100}{.6}
  \lineanglesheight{a3}{a7}{90}{120}{.7}
}
$$
The atomic cut reduction also preserves correctness. If we could in the
reduced simple prenet construct a conjunctive pruning that does not contain an
identity link, then the same pruning would not contain a link in the
non-reduced net, where one cut atom is always removed.

\begin{theorem}
  Cut elimination for simple proof nets preserves correctness, is
  confluent, and terminating.
\end{theorem}

\begin{proof}
  Preservation of correctness has already been shown above. Termination is
  obvious since each step reduces the size of the simple proof net.  For
  confluence we only need to consider atomic cuts since the reduction of
  compound cuts does not create critical pairs. Let $\pi$ be a simple proof
  net with atomic cuts and let $\pi'$ be the result of reducing these atomic
  cuts. Then there is a link between two dual atom occurrences $a$ and $\cneg
  a$ in $\pi'$, if either that link is already present in $\pi$, or there is
  an alternating link-cut-link-cut-\ldots-link path in $\pi$ that connects $a$
  and $\cneg a$. This is independent from the order in which the atomic cuts
  are reduces.
  \qed
\end{proof}

The natural question that arises now is: How does this confluent cut
elimination relate to the non-confluent cut elimination in the sequent
calculus? 

Let us look again at the two problematic cases \eqref{eq:weak-weak}
and~\eqref{eq:cont-cont}. The problem with weakening~\eqref{eq:weak-weak} can
easily be solved by using the $\mixr$-rule:
\proofadjust
$$
\vcddernote{\cutr}{}{\sqn{\Gamma,\Delta}}{
  \root{\weakr}{\sqn{\Gamma,A}}{
    \leaf{\sqnsmallderi{\Gamma}{\Pi_1}}}}{
  \root{\weakr}{\sqn{\cneg A,\Delta}}{
    \leaf{\sqnsmallderi{\Delta}{\Pi_2}}}}
\qqquadto
\vcddernote{\mixr}{}{\sqn{\Gamma,\Delta}}{
  \leaf{\sqnsmallderi{\Gamma}{\Pi_1}}}{
  \leaf{\qquad\sqnsmallderi{\Delta}{\Pi_2}}}
$$
Both subproofs $\Pi_1$ and $\Pi_2$ are kept in the reduced net. With the help
of mix, this can also be done in the sequent calculus, and in simple proof
nets it is done in the same way.

For the contraction case~\eqref{eq:cont-cont} the situation is less
obvious. Consider again the simple proof net
in~\eqref{eq:simple-pn-cont-cont}, which corresponds to the sequent calculus
proof in~\eqref{eq:sc-cont-cont}. If we apply the cut elimination for simple
proof nets, we obtain the following result:
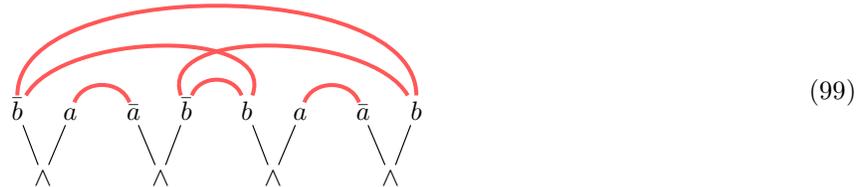
\begin{equation}\label{eq:simple-pn-cont-cont-res}
   \vcnpn{%
     \\[4ex]
     \cneg{\mb1}&&\na1&&\cneg{\na2}&&\cneg{\mb2}&&
     \mb7&&\na3&&\cneg{\na4}&&\mb8\\
     &\ncand1&&\qlapm{}&&\ncand2&&\qlapm{}&
     &\ncand3&&\qlapm{}&&\ncand4
   }{%
     \ncline{b1}{cand1}
     \ncline{a1}{cand1}
     \ncline{b2}{cand2}
     \ncline{a2}{cand2}
     \ncline{b7}{cand3}
     \ncline{a3}{cand3}
     \ncline{b8}{cand4}
     \ncline{a4}{cand4}
   }{%
     \udline{a1}{a2}
     \udline{a3}{a4}
     \udline{b2}{b7}
     \lineanglesheight{b1}{b7}{60}{70}{.65}
     \lineanglesheight{b2}{b8}{110}{120}{.65}
     \lineanglesheight{b1}{b8}{90}{90}{.6}
   }
\end{equation}
If you did Exercise~\ref{ex:sc-simple-pn}, you will notice that this is exactly
the simple proof net obtained from the sequent proofs corresponding to
\eqref{eq:pn-cont-out1} and~\eqref{eq:pn-cont-out2}. However, let us emphasize
that this correspondence also makes crucial use of the fact that we
deliberately forget how often an identity link is used in the proof.
If we kept this information, the proofs in \eqref{eq:pn-cont-out1}
and~\eqref{eq:pn-cont-out2} would be represented by
$$
   \vcnpn{%
     \\[4ex]
     \cneg{\mb1}&&\na1&&\cneg{\na2}&&\cneg{\mb2}&&
     \mb7&&\na3&&\cneg{\na4}&&\mb8\\
     &\ncand1&&\qlapm{}&&\ncand2&&\qlapm{}&
     &\ncand3&&\qlapm{}&&\ncand4
   }{%
     \ncline{b1}{cand1}
     \ncline{a1}{cand1}
     \ncline{b2}{cand2}
     \ncline{a2}{cand2}
     \ncline{b7}{cand3}
     \ncline{a3}{cand3}
     \ncline{b8}{cand4}
     \ncline{a4}{cand4}
   }{%
     \udline{a3}{a4}
     \lineanglesheight{a1}{a2}{100}{80}{1.2}
     \lineanglesheight{a1}{a2}{70}{110}{.8}
     \udline{b2}{b7}
     \lineanglesheight{b1}{b7}{60}{70}{.7}
     \lineanglesheight{b2}{b8}{110}{120}{.7}
     \lineanglesheight{b1}{b8}{90}{90}{.6}
   }
\qquand
   \vcnpn{%
     \\[4ex]
     \cneg{\mb1}&&\na1&&\cneg{\na2}&&\cneg{\mb2}&&
     \mb7&&\na3&&\cneg{\na4}&&\mb8\\
     &\ncand1&&\qlapm{}&&\ncand2&&\qlapm{}&
     &\ncand3&&\qlapm{}&&\ncand4
   }{%
     \ncline{b1}{cand1}
     \ncline{a1}{cand1}
     \ncline{b2}{cand2}
     \ncline{a2}{cand2}
     \ncline{b7}{cand3}
     \ncline{a3}{cand3}
     \ncline{b8}{cand4}
     \ncline{a4}{cand4}
   }{%
     \lineanglesheight{a3}{a4}{100}{80}{1.2}
     \lineanglesheight{a3}{a4}{70}{110}{.8}
     \udline{a1}{a2}
     \udline{b2}{b7}
     \lineanglesheight{b1}{b7}{60}{70}{.7}
     \lineanglesheight{b2}{b8}{110}{120}{.7}
     \lineanglesheight{b1}{b8}{90}{90}{.6}
   }
$$
respectively. See~\cite{lam:str:05:naming} for further details.

\begin{figure}[!t]
  \begin{center}
    $$
    \begin{array}{c@{\qqquad}c}
      \vcinf{\ird}{\cneg A\cor A}{} 
      &
      \vcinf{\iru}{}{A\cand\cneg A} 
      \\[\arrayskip]
      \vcinf{\ird}{S\cons{(\cneg A\cor A)\cand B}}{S\cons{B}} 
      &
      \vcinf{\iru}{S\cons{B}}{S\cons{B\cor(A\cand\cneg A)}} 
      \\[\arrayskip]
      \vcinf{\comrd}{S\cons{B\cor A}}{S\cons{A\cor B}}
      &
      \vcinf{\comru}{S\cons{B\cand A}}{S\cons{A\cand B}}
      \\[\arrayskip]
      \vcinf{\assrd}{S\cons{(A\cor B)\cor C}}{S\cons{A\cor(B\cor C)}}
      &
      \vcinf{\assru}{S\cons{(A\cand B)\cand C}}{S\cons{A\cand (B\cand C)}}
      \\[\arrayskip]
      \multicolumn{2}{c}{
	\vcinf{\swir}{S\cons{(A\cand B)\cor C}}{S\cons{A\cand(B\cor C)}}
      }
      \\[\arrayskip]
      \multicolumn{2}{c}{
	\vcinf{\mixr}{S\cons{A\cor B}}{S\cons{A\cand B}}
      }
      \\[\arrayskip]
      \multicolumn{2}{c}{
	\vcinf{\medr}{
	  S\cons{(A\cor C)\cand(B\cor D)}}{
	  S\cons{(A\cand B)\cor(C\cand D)}}
      }
      \\[\arrayskip]
      \vcinf{\conrd}{S\cons{A}}{S\cons{A\cor A}}
      &
      \vcinf{\conru}{S\cons{A\cand A}}{S\cons{A}}
      \\[\arrayskip]
      \vcinf{\weakrd}{S\cons{B\cor A}}{S\cons{B}}
      &
      \vcinf{\weakru}{S\cons{B}}{S\cons{A\cand B}}
    \end{array}
    $$
    \caption{A system for classical logic in the calculus of structures}
    \label{fig:cos-cl}
  \end{center}
\end{figure}

It should be clear that simple proof nets are not particularly connected to
the sequent calculus. We can obtain them in the same way from proofs presented
in the calculus of structures. Figure~\ref{fig:cos-cl} shows a deductive
system for classical logic in the calculus of structures. It is a (unit-free)
variation of system~$\SKS$, presented in~\cite{brunnler:tiu:01}. The rules are
the same as the ones for $\ufMLL$ in Figure~\ref{fig:cos-mll}. As in the
sequent calculus, we add the rules for contraction, weakening, and mix. We
also add the $\medr$-rule, called \emph{medial}. As it is the case with
$\mixr$, it is is not necessary from the viewpoint of provability, but its
presence gives the system a much nicer proof-theoretic
behavior. See~\cite{brunnler:tiu:01,brunnler:phd,str:medial,lamarche:gap} for
further details.

Most of the theory on the relation between proof nets, sequent calculus and
calculus of structures, that has been developed in Section~\ref{sec:mll}, can
be ported easily to classical logic. In particular, we can obtain
Theorem~\ref{thm:sc-cos-pn} and the commutativity of
\eqref{eq:cutcomm}, provided we restrict ourselves to simple proof nets.

\begin{para}{Open Research Problem}
  Find a category theoretical axiomatization that generates the same
  identification for proofs as it is done by simple proof nets for classical
  logic. (This is meant in the same sense as *-autonomous categories provide
  the axiomatization for proof nets for $\MLL$, and cartesian closed
  categories for typed $\lambda$-terms.)
\end{para}

There is already preliminary research in this direction
in~\cite{lam:str:05:naming,lam:str:05:freebool,str:deepnet,lamarche:gap,str:medial},
but there is still no satisfactory solution.

\subsection{Flow graph based proof nets (extended version)}
\label{sec:cl-ext}

The first problem with simple proof nets is that they are too simple. When
speaking about the identity of proofs one should also take into account the
size of proofs. But simple proof nets have a size at most quadratic in the
size of the conclusion sequent. This means they are not able to observe any
kind of complexity, not even the exponential blow-up related to cut
elimination.

The second problem with simple proof nets is that they are too simple. They
omit too much important information about the proof. Checking their
correctness takes exponential time, which is not faster than trying to prove
the conclusion from scratch. But checking a proof should be a feasible task,
taking only linear time in the size of the proof.

A naive solution could be to simply keep track of how often an identity link
is used in the proof, i.e., allow more than one link between a pair of dual
atoms, as shown on the right in Figure~\ref{fig:exascflow}. Let us call these
new objects \emph{extended prenets}. 

The first problem we encounter now is not
only to define what \emph{correctness} means, but also, what
\emph{sequentializable} means. To see the difficulty, consider again the
simple proof net in~\eqref{eq:simple-pn-cont-cont-res}. Now, all of the
sudden, there is no sequent calculus proof that has it a flow-graph. However,
we can find a proof in the calculus of structures, in the system shown in 
Figure~\ref{fig:cos-cl}:
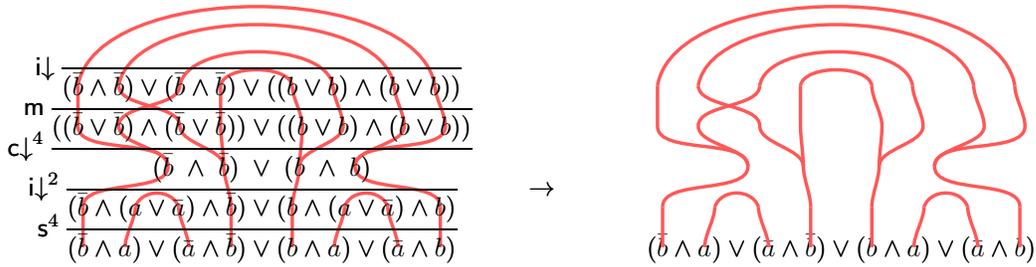
$$
  \tpathderivation{\medialderi}{\pathmedialderi}
  \qquad
  \def\derpfeilimmedial{\llapm{\black\to\qquad}}
  \tpathonlyderivation{\medialderi}{\pathmedialderi}
$$

This means that for extended prenets, it depends on the chosen deductive
system whether a given prenet is sequentializable or not.

\begin{para}{Open Research Problem}
  Find a good notion of ``sequentializability'' and a corresponding
  correctness criterion for extended prenets.
\end{para}

The second, more serious problem comes with cut
elimination. In~\cite{lam:str:05:naming} it is explained, how cut elimination
for extended prenets has to look like. For a cut on compound formulas it is
the same as for simple proof nets. But for an atomic cut, we now have to
multiply the number of edges. For example
$$
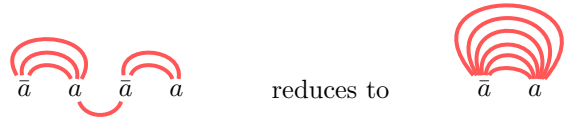
\vcnpn{
  \\
  \nna1&\pa&\na2&\pa&\nna3&\pa&\na4\\
}{}{
  \lineanglesheight{a1}{a2}{120}{60}{1.8}
  \lineanglesheight{a1}{a2}{100}{80}{1.2}
  \lineanglesheight{a1}{a2}{70}{110}{.8}
  \lineanglesheight{a3}{a4}{100}{80}{1.2}
  \lineanglesheight{a3}{a4}{70}{110}{.8}
  \duline{a2}{a3}
}
\qqquad\mbox{ reduces to }\qqquad
\vcnpn{
  \\
  \nna1&\pa&\na2\\
}{}{
  \lineanglesheight{a1}{a2}{130}{50}{3.5}
  \lineanglesheight{a1}{a2}{120}{60}{2.8}
  \lineanglesheight{a1}{a2}{110}{70}{2.2}
  \lineanglesheight{a1}{a2}{100}{80}{1.6}
  \lineanglesheight{a1}{a2}{85}{95}{1.2}
  \lineanglesheight{a1}{a2}{70}{110}{.6}
}
$$ 
If there are already some links between the remaining pair of atoms, then
these links have to be added. For example
$$
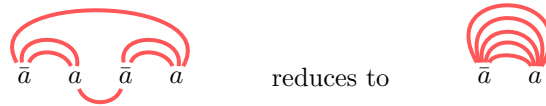
\vcnpn{
  \\
  \nna1&\pa&\na2&\pa&\nna3&\pa&\na4\\
}{}{
  \lineanglesheight{a1}{a4}{120}{60}{1}
  \lineanglesheight{a1}{a2}{100}{80}{1.2}
  \lineanglesheight{a1}{a2}{70}{110}{.8}
  \lineanglesheight{a3}{a4}{100}{80}{1.2}
  \lineanglesheight{a3}{a4}{70}{110}{.8}
  \duline{a2}{a3}
}
\qqquad\mbox{ reduces to }\qqquad
\vcnpn{
  \\
  \nna1&\pa&\na2\\
}{}{
  \lineanglesheight{a1}{a2}{120}{60}{2.8}
  \lineanglesheight{a1}{a2}{110}{70}{2.2}
  \lineanglesheight{a1}{a2}{100}{80}{1.6}
  \lineanglesheight{a1}{a2}{85}{95}{1.2}
  \lineanglesheight{a1}{a2}{70}{110}{.6}
}
$$ 
Of course, we cannot say whether this preserves correctness, since we do
not know what correctness means. But the good news is that cut reduction is
still terminating, and that we can get an exponential blow-up in the size of
the proof when doing cut elimination.

The bad news is that cut elimination is no longer confluent. This has already
been observed in~\cite{lam:str:05:naming}. The following example is taken
from~\cite{horbach:msc}. Depending on which cut in
$$
  \vcnpn{
    \\[-2ex]
    \nna1&\pa&\na2&\pa&\nna3&\pa&\na4&\pa&\nna5&\pa&\na6
    \\[-2ex]
  }{}{
    \udline{a1}{a2}
    \udline{a3}{a4}
    \lineanglesheight{a2}{a5}{80}{100}{.7}
    \lineanglesheight{a3}{a6}{90}{100}{.7}
    \duline{a2}{a3}
    \duline{a4}{a5}
  }
$$
we reduce first, we get either 
$$
  \vcnpn{
    \\[-2ex]
    \nna1&\pa&\na2&\pa&\nna3&\pa&\na4
    \\[-2ex]
  }{}{
    \udline{a1}{a2}
    \udline{a2}{a3}
    \udline{a3}{a4}
    \lineanglesheight{a1}{a4}{90}{90}{.7}
    \duline{a2}{a3}
  }
\qqquor
  \vcnpn{
    \\[-2ex]
    \nna1&\pa&\na2&\pa&\nna3&\pa&\na4
    \\[-2ex]
  }{}{
    \udline{a1}{a2}
    \udline{a2}{a3}
    \udline{a3}{a4}
    \duline{a2}{a3}
  }
$$
If we reduce the remaining cut, we get 
$$
  \vcnpn{
    \\[-2ex]
    \nna3&\pa&\na6
  }{}{
    \lineanglesheight{a3}{a6}{100}{80}{1.2}
    \lineanglesheight{a3}{a6}{70}{110}{.8}
  }
\qqquor
  \vcnpn{
    \\[-2ex]
    \nna3&\pa&\na6
  }{}{
    \udline{a3}{a6}
  }
$$ respectively. The basic reason of this non-confluence is that we do
not have the acyclicity condition anymore, and it depends on the order
of the cut reductions how often ``the flow-graph runs through a
cycle''. In~\cite{horbach:msc}, Horbach proposes a way of keeping
track of the cycles in order to get confluence for cut elimination on
extended prenets representing intuitionistic proofs.

An alternative possible solution could be to redefine cut
elimination such that no longer all atomic cuts are
reduced. In~\cite{str:deepnet,str:medial} it is shown how this can be done
such that we get a confluent cut elimination which corresponds to composition
of derivations in the calculus of structures:
$$
\vcstrder{}{}{C}{\stem{}{}{B}{\leaf{A}}}
\qquadto
\vcstrder{}{}{C}{\leaf{A}}
$$
Which still allows to keep the equation
$$
\mbox{cut elimination}
\quad=\quad
\mbox{arrow composition in the category of proofs}
$$ But then cut elimination for extended prenets does no longer correspond to
cut elimination in the sequent calculus. That it is indeed impossible to make
the diagram~\eqref{eq:cutcomm} commute for the sequent calculus and extended
prenets has already been shown with the example
in~\eqref{eq:simple-pn-cont-cont-res} in the previous section. But it is not
known whether we can make it commute for the calculus of structures.

\begin{para}{Open Research Problem}
  Find for the calculus of structures (possibly for the system in
  Figure~\ref{fig:cos-cl}) a cut elimination procedure, possibly based on
  splitting (as done in the proof of Theorem~\ref{thm:cos-cutelim}) such that
  it behaves in the same way as the cut elimination for extended prenets,
  i.e., such that diagram~\eqref{eq:cutcomm} commutes.
\end{para}

In any case, what has been said in this section shows that for classical logic
the answer to the Big Question~\ref{big:cut} on page~\pageref{big:cut} is no
longer an obvious yes. In fact, the answer might be {\bf No!}

\begin{para}{Open Research Problem}
  Make a nice theory out of this mess.
\end{para}

\bibliographystyle{alpha} \bibliography{my_lit}

\end{document}